\documentclass[twocolumn]{aastex62}
\pdfoutput=1

\usepackage[figuresright]{rotating}
\usepackage{makecell}
\usepackage{gensymb}
\usepackage{soul}
\usepackage{xcolor}

\usepackage{algorithm}
\usepackage[noend]{algpseudocode}
\usepackage{tabto}
\makeatletter

\makeatletter
\renewcommand{\fnum@algorithm}{\fname@algorithm}
\makeatother

\makeatletter
\algnewcommand{\LineComment}[1]{\Statex \hskip\ALG@thistlm \(\triangleright\) #1}
\makeatother

\makeatletter
\def\algbackskip{\hskip-\ALG@thistlm}
\makeatother

\usepackage{dirtree, array, booktabs}

\definecolor{deepblue}{rgb}{0,0,0.5}
\definecolor{deepred}{rgb}{0.6,0,0}
\definecolor{deepgreen}{rgb}{0,0.5,0}
\definecolor{mygreen}{rgb}{0,0.6,0}
\definecolor{mygray}{rgb}{0.5,0.5,0.5}
\definecolor{mymauve}{rgb}{0.58,0,0.82}

\usepackage{listings}
\newcommand\pythonstyle{\lstset{
    language=Python,
    basicstyle=\ttfamily\small,
    commentstyle=\color{mygray},
    morekeywords={self},              
    keywordstyle=\ttb\color{deepblue},
    emph={MyClass,__init__},          
    emphstyle=\ttb\color{deepred},    
    stringstyle=\color{deepblue},
    frame=tb,                         
    showstringspaces=false
}}

\DeclareFixedFont{\ttb}{T1}{txtt}{bx}{n}{12} 
\DeclareFixedFont{\ttm}{T1}{txtt}{m}{n}{12}

\newcommand{\TNM}[1]{\tablenotemark{#1}}
\newcommand{\TNT}[2]{\tablenotetext{#1}{#2}}

\newcommand\sqdeg{\;\mathrm{deg}^2}

\newcommand\NStatTypeAll{36333 }

\newcommand\NStatHostAll{19268 }

\newcommand\NStatTypeCC{8437 }

\newcommand\NStatHostCC{5629 }

\newcommand\NStatTypeCarich{21 }

\newcommand\NStatHostCarich{15 }

\newcommand\NStatTypeFRB{14 }

\newcommand\NStatHostFRB{14 }

\newcommand\NStatTypeGRB{1621 }

\newcommand\NStatTypeGRBUnspec{20 }

\newcommand\NStatHostGRB{22 }
\newcommand\NStatHostGRBUnspec{0 }

\newcommand\NStatTypeGW{1 }

\newcommand\NStatHostGW{1 }

\newcommand\NStatTypeII{6685 }

\newcommand\NStatTypeIIUnspec{5023 }

\newcommand\NStatHostII{4605 }
\newcommand\NStatHostIIUnspec{3514 }

\newcommand\NStatTypeIIL{55 }

\newcommand\NStatHostIIL{34 }

\newcommand\NStatTypeIIP{918 }

\newcommand\NStatHostIIP{688 }

\newcommand\NStatTypeIIb{323 }

\newcommand\NStatHostIIb{235 }

\newcommand\NStatTypeIIn{681 }

\newcommand\NStatHostIIn{373 }

\newcommand\NStatTypeIap{15435 }

\newcommand\NStatTypeIapUnspec{14534 }

\newcommand\NStatHostIap{8093 }
\newcommand\NStatHostIapUnspec{7558 }

\newcommand\NStatTypeIaCSM{25 }

\newcommand\NStatHostIaCSM{9 }

\newcommand\NStatTypeIacx{4 }

\newcommand\NStatHostIacx{3 }

\newcommand\NStatTypeIacxp{57 }

\newcommand\NStatHostIacxp{38 }

\newcommand\NStatTypeIaes{6 }

\newcommand\NStatHostIaes{5 }

\newcommand\NStatTypeIadc{8 }

\newcommand\NStatHostIadc{8 }

\newcommand\NStatTypeIaT{398 }

\newcommand\NStatHostIaT{177 }

\newcommand\NStatTypeIabg{135 }

\newcommand\NStatHostIabg{91 }

\newcommand\NStatTypeIaaa{6 }

\newcommand\NStatHostIaaa{3 }

\newcommand\NStatTypeIb{757 }

\newcommand\NStatTypeIbUnspec{363 }

\newcommand\NStatHostIb{546 }
\newcommand\NStatHostIbUnspec{270 }

\newcommand\NStatTypeIbc{315 }

\newcommand\NStatHostIbc{250 }

\newcommand\NStatTypeIbn{51 }

\newcommand\NStatHostIbn{27 }

\newcommand\NStatTypeIc{710 }

\newcommand\NStatTypeIcUnspec{672 }

\newcommand\NStatHostIc{481 }
\newcommand\NStatHostIcUnspec{452 }

\newcommand\NStatTypeIcBL{137 }

\newcommand\NStatHostIcBL{73 }

\newcommand\NStatTypeKilonova{1 }

\newcommand\NStatHostKilonova{1 }

\newcommand\NStatTypeLGRB{1486 }

\newcommand\NStatHostLGRB{21 }

\newcommand\NStatTypeSE{1619 }

\newcommand\NStatHostSE{1123 }

\newcommand\NStatTypeSGRB{115 }

\newcommand\NStatHostSGRB{1 }

\newcommand\NStatTypeSLSN{230 }

\newcommand\NStatTypeSLSNUnspec{13 }

\newcommand\NStatHostSLSN{29 }
\newcommand\NStatHostSLSNUnspec{1 }

\newcommand\NStatTypeSLSNI{154 }

\newcommand\NStatHostSLSNI{20 }

\newcommand\NStatTypeSLSNII{61 }

\newcommand\NStatHostSLSNII{8 }

\newcommand\NStatTypeSLSNR{6 }

\newcommand\NStatHostSLSNR{2 }

\newcommand\NStatTypeTDE{99 }

\newcommand\NStatTypeTDEUnspec{56 }

\newcommand\NStatHostTDE{93 }
\newcommand\NStatHostTDEUnspec{54 }

\newcommand\NStatTypeUVOptTDE{22 }

\newcommand\NStatHostUVOptTDE{19 }

\newcommand\NStatTypeXrayTDE{21 }

\newcommand\NStatHostXrayTDE{20 }

\newcommand\NStatTypeSN{23703 }

\newcommand\NStatEvents{36333 }

\newcommand\NStatConfirmed{18140 }

\newcommand\NStatHostByNames{13753 }

\newcommand\NStatHostByCoord{4480 }

\newcommand\NStatPrimaryCand{18100 }

\newcommand\NStatGALEXanybandConf{9125 }
\newcommand\NStatGALEXanybandCand{8161 }

\newcommand\NStatGALEXallbandConf{6791 }
\newcommand\NStatGALEXallbandCand{5993 }

\newcommand\NStatSDSSConf{12018 }
\newcommand\NStatSDSSCand{10998 }

\newcommand\NStatSDSSHighSNRgriConf{9845 }
\newcommand\NStatSDSSHighSNRgriCand{9711 }

\newcommand\NStatSDSSHighSNRallConf{6575 }
\newcommand\NStatSDSSHighSNRallCand{6783 }

\newcommand\NStatMPAJHUConf{3811 }
\newcommand\NStatMPAJHUCand{3709 }

\newcommand\NStatLSConf{14576 }
\newcommand\NStatLSCand{13958 }

\newcommand\NStatPSConf{14446 }
\newcommand\NStatPSCand{13797 }

\newcommand\NStatPSCConf{10707 }
\newcommand\NStatPSCCand{8117 }

\newcommand\NStatXSCConf{9419 }
\newcommand\NStatXSCCand{5529 }

\newcommand\NStatPSConlyConf{1368 }
\newcommand\NStatPSConlyCand{2613 }

\newcommand\NStatAllWISEConf{15061 }
\newcommand\NStatAllWISECand{13060 }

\newcommand\NStatunWISEConf{16444 }
\newcommand\NStatunWISECand{15208 }

\newcommand\NStatDESConf{5549 }
\newcommand\NStatDESCand{3031 }

\newcommand\NStatLASConf{5036 }
\newcommand\NStatLASCand{2886 }

\newcommand\NStatLASfullConf{3620 }
\newcommand\NStatLASfullCand{2258 }

\newcommand\NStatATLASConf{1616 }
\newcommand\NStatATLASCand{1436 }

\newcommand\NStatVHSConf{5664 }
\newcommand\NStatVHSCand{2318 }

\newcommand\NStatSkyMapperConf{6896 }
\newcommand\NStatSkyMapperCand{4218 }

\newcommand\NStatSCOSConf{12307 }
\newcommand\NStatSCOSCand{12038 }

\newcommand\NStatHyperLEDAConf{12728 }
\newcommand\NStatHyperLEDACand{8465 }

\newcommand\NStatNSAConf{5638 }
\newcommand\NStatNSACand{4044 }

\definecolor{darkblue}{rgb}{0,0,0.5}
\newcommand{\rev}[1]{\textcolor{darkblue}{#1}}

\definecolor{eclipseStrings}{RGB}{42,0.0,255}
\definecolor{eclipseKeywords}{RGB}{127,0,85}
\colorlet{numb}{magenta!60!black}

\graphicspath{{./}{figs/}}

\received{\today}
\revised{}
\accepted{}

\submitjournal{ApJS}

\shorttitle{Cross-identification of Transient Host Galaxies}

\shortauthors{Qin et al.}

\begin{document}

\renewcommand{\DTstyle}{\textrm}

\title{Linking Extragalactic Transients and their Host Galaxy Properties: \\
Transient Sample, Multi-Wavelength Host Identification, and Database Construction}

\correspondingauthor{Yu-Jing Qin}
\email{qinyj@email.arizona.edu}

\author[0000-0003-3658-6026]{Yu-Jing Qin}
\affiliation{Department of Astronomy and Steward Observatory, University of Arizona, 933 N Cherry Ave, Tucson, AZ 85721}

\author{Ann Zabludoff}
\affiliation{Department of Astronomy and Steward Observatory, University of Arizona, 933 N Cherry Ave, Tucson, AZ 85721}

\author{Marina Kisley}
\affiliation{Department of Computer Science, University of Arizona, 1040 4th St, Tucson, AZ 85721}

\author{Yuantian Liu}
\affiliation{Department of Physics, University of Arizona, 1118 E. 4th Street, PO Box 210081, Tucson, AZ 85721}

\author[0000-0001-7090-4898]{Iair Arcavi}
\affiliation{The School of Physics and Astronomy, Tel Aviv University, Tel Aviv 69978, Israel}
\affiliation{CIFAR Azrieli Global Scholars program, CIFAR, Toronto, Canada}

\author[0000-0002-8568-9518]{Kobus Barnard}
\affiliation{Department of Computer Science, University of Arizona, 1040 4th St, Tucson, AZ 85721}

\author[0000-0002-2517-6446]{Peter Behroozi}
\affiliation{Department of Astronomy, University of Arizona, 933 N Cherry Ave, Tucson, AZ 85721}

\author[0000-0002-4235-7337]{K. Decker French}
\affiliation{Observatories of the Carnegie Institute for Science, 813 Santa Barbara St. Pasadena, CA 91001}
\affiliation{Hubble Fellow}

\author{Curtis McCully}
\affiliation{Las Cumbres Observatory}

\author{Nirav Merchant}
\affiliation{Data Science Institute, University of Arizona, BSRL Building (\#242), 1230 N. Cherry Ave, Tucson, AZ 85721}

\begin{abstract}

Understanding the preferences of transient types for host galaxies with certain characteristics is key to studies of transient physics and galaxy evolution, as well as to transient identification and classification in the LSST era.
Here we describe a value-added database of extragalactic transients---supernovae, tidal disruption events, gamma-ray bursts, and other rare events---and their host galaxy properties.
Based on reported coordinates, redshifts, and host galaxies (if known) of events, we cross-identify their host galaxies or most likely host candidates in various value-added or survey catalogs, and compile the existing photometric, spectroscopic, and derived physical properties of host galaxies in these catalogs.
This new database covers photometric measurements from the far-ultraviolet to mid-infrared. 
Spectroscopic measurements and derived physical properties are also available for a smaller subset of hosts.
For our \NStatEvents unique events, we have cross-identified \NStatHostByNames host galaxies using host names, plus \NStatHostByCoord using host coordinates.
Besides those with known hosts, there are \NStatPrimaryCand transients with newly identified host candidates.
This large database will allow explorations of the connections of transients to their hosts, including a path toward transient alert filtering and probabilistic classification based on host properties.
\end{abstract}


\section{Introduction} \label{sec:introduction}

Extragalactic transients cover a wide range of energetic, fast, and usually cataclysmic events observed across the electromagnetic spectrum, as well as the recently revealed gravitational wave and neutrino windows.
These events include, but are not limited to, supernovae (SNe), gamma-ray bursts (GRBs), tidal distribution events (TDEs), fast radio bursts (FRBs), and gravitational wave events (GWs).
With better observational facilities and computational technologies, our understanding of these transient phenomena has advanced considerably in the new century.
However, their connection to the evolution history and observational properties of their host galaxies remains to be further explored. %

Transients are stellar or supermassive black hole phenomena in nature. They occur on much smaller spatial scales than the typical size of galaxies. They are also ephemeral compared to timescales of galaxy evolution.
Notwithstanding the dramatic contrast in spatial and temporal scales, there are examples of well-established transient-host connections:
\begin{enumerate}
    \item Core-collapse supernovae (CC SNe) show stronger preference for late type, star-forming galaxies than thermonuclear Type-Ia supernovae (SNe Ia) \citep[e.g.,][]{vandenBergh59, Tammann78, Oemler79, Sullivan06, Li11, Foley13, Graur17};
    \item Compared to ``normal'' SNe Ia, SNe Ia-91bg tend to occur in early-type hosts, while SNe Ia-91T and Ia-02cx favor late-types \citep[][and references therein]{Taubenberger17};
    \item The hosts of short-duration gamma-ray bursts (SGRBs) are brighter, larger in size, more metal-rich, and, on average, more quiescent than typical hosts of long-duration gamma-ray bursts (LGRBs) \citep{Berger09, Fong10, Lyman17}.
    \item The rate of TDEs is dramatically boosted in post-starburst galaxies \citep{French16}, although the underlying driving mechanism is subject to debate \citep{Or18, Stone18};
    \item The recently discovered electromagnetic counterpart of gravitational wave event GW170817 \citep{Abbott17}, also detected as a SGRB, lies in an early-type galaxy with shell-like merger features;
    \item The known hosts of FRBs show rather diverse properties, with a possible excess of galaxies in the ``green valley,'' or galaxies with harder ionizing radiation than provided by star formation alone \citep[e.g.,][]{Heintz20}.
\end{enumerate}
This synopsis is certainly not exhaustive but nevertheless demonstrates that some transients prefer certain host environments.
There may be still other connections that, while subtle, could be discovered with a large, well-curated database gleaned from archived surveys.

To a certain extent, such transient-host connections can be interpreted within our existing picture of transient physics and galaxy evolution.
From a statistical and empirical perspective, the event rates of optical supernovae and their counterparts at other wavelengths depend on the recent star formation history and the delay time distribution of supernova explosion since star formation;
theoretically, the preferred channel and mechanism of the explosion could also depend on metallicity and dynamical environment, which directly shape the evolutionary path of progenitors and indirectly affect stellar binarity and initial mass function.
For TDEs, the event rates not only are related to the integrated stellar populations of the host but also regulated by the concentration of star formation, availability of gas near the galactic nucleus, and/or the disturbed central stellar and gas dynamics in the host.
Due to the uncertainties in our theoretical models, it is unclear to what extent the rates of transients depend on the global properties of their hosts.
Potential biases in transient surveys further complicate the interpretation of the observed transient-host connections.
Nevertheless, the study of transient-host connections points to a better-refined view of transient phenomena and galaxy evolution, including the physics of massive stars, interacting binaries, and compact objects and the role of supermassive black holes and active nuclei in galaxy evolution. %

Transient-host connections are also of practical interest.
Modern transient surveys, particularly the Legacy Survey of Space and Time (LSST) at Vera C. Rubin Observatory \citep{Ivezic19}, are expected to deliver nearly 10 million transient alerts every night in real-time.
For extragalactic events, over a year of operation, LSST is estimated to detect about 0.2 million SNe Ia, a smaller but comparable sample of CC SNe, and orders of magnitude fewer other uncommon transients inundated in the torrent of alerts\footnote{LSST Science Book v2, Chapter 8 \& Chapter 11.}.
Selecting a subset of potentially interesting events would be essential for any study that requires follow-up observations using other facilities. 
Often this must be done \textit{quickly} in the early phase of transient evolution, when evanescent phenomena may give us valuable insights into their physical models.
Host properties thus open up a new avenue towards automatized alert filtering and classification, even ``pre-classification,'' i.e., assigning event rates or probabilities to galaxies in the field based on their properties, before any alerts arrive.
Indeed, the viability and reliability of host-based transient classification have been clearly demonstrated by the early conceptual work of \citet{Foley13} and recent attempts of \citet{Gagliano20}.
Even photometric data alone may effectively distinguish potential hosts of certain rare transients \citep[e.g.,][]{French18}.
Meanwhile, new-generation transient brokers for Zwicky Transient Facility \citep[ZTF;][]{Bellm19} and the future LSST, such as ANTARES \citep{Gautham18}, ALeRCE \citep{Forster20}, Lasair \citep{Smith19}, MARS\footnote{\url{mars.lco.global}} and Fink \citep{Moller20}, are all capable of cross-matching transient alerts with external source catalogs, providing the necessary infrastructure for host-based event filtering and classification.
Given this potential to provide a baseline estimate of transient probabilities (i.e., a categorical prior function of different types) or rates independent of any transient characteristics, the power of host properties must be fully utilized now for imminent transient surveys. %

Exploring the connections of extragalactic transient and their host properties entails a complete census of known transient hosts, including their properties measured by various sky surveys across the electromagnetic spectrum.
Existing samples and catalogs of transient host galaxies, such as the time-controlled sample of Lick Observatory Supernova Search \citep[LOSS;][]{Leaman11, Li11}, also the host samples of SDSS-II Supernova Survey \citep{Sako18} and DES Supernova Programme \citep{Wiseman20}, usually focus on a dedicated transient survey or a specific class of transients. The recently released \texttt{GHOST} database \citep{Gagliano20}, which has a considerably larger transient sample, is mostly limited to the photometric properties of supernova hosts obtained by the Pan-STARRS survey.
There is no single catalog to date of consistently measured host properties, across a wide wavelength range, for a large and up-to-date sample of transient events.
Reasons include:
\begin{enumerate}
    \item Host galaxies are not always reported by transient surveys or summarized in catalogs or platforms. Consequently, a considerable fraction of transients in the literature, even with accurate sky coordinates, redshifts, and classifications, have no easily accessible host data of any kind. 
    \item Beyond that, new transients are continuously being discovered at ever-improving efficiencies, posing an immediate challenge to human-based host identifying and reporting.
    Automated host identifying algorithms, using either catalogs or images, are favored in this situation. However, only a few such algorithms have been developed up to date. \citep[e.g.,][]{Sullivan06, Gupta16, Sako18, Gagliano20}.
    \item Even for transients with identifiable hosts, obtaining their properties involves substantial efforts. Some surveys conduct dedicated observations of hosts. Accessing archival data of public surveys, a more viable approach for many other cases, also requires host matching, quality control, and data compilation.
\end{enumerate}

Constructing a transient host database with better coverage of previous transient events and available host properties remains a crucial step for the in-depth exploration and analysis of transient-host connections.
Here we present such a new transient-host database.
We find counterparts of host galaxies in external catalogs using the reported host name designations or coordinates, as presented in the data sources.
For events without reported host galaxies, we identify their best host candidates within nearby extragalactic objects, cross-matched in various external catalogs.
Our goal here is to locate the correct counterparts of transient host galaxies in various catalogs, which entails cross-matching sources in multiple catalogs, and, when necessary, ranking potential host galaxies to identify the most likely candidates.
Therefore, we refer to the process as the \textit{cross-identification} of transient host galaxies.
We then collate properties of our identified host galaxies and host candidates, including photometry across UV to IR wavelengths, morphology, spectral lines and indices, and derived physical parameters such as stellar mass, metallicity, and star formation rate provided in those catalogs. %

This first paper describes our database of transient host galaxies and the workflow to construct it.
In Section \ref{sec:transientsample}, we review our data sources for transient events (published catalogs and online services) and the procedure to create a sample of unique transient events.
In Section \ref{sec:hostcrossid}, we describe our method to search host properties in external catalogs and collect their information. We also discuss our strategy to identify new host galaxies for those without hosts reported in our data sources.
We then describe the detailed object selection criteria and data coverage in each external catalog in Section \ref{sec:extcatalogs} and present the basic statistics of our transient-host pairs in Section \ref{sec:hostpropstat}.
Finally, we discuss our software implementation and data format in Section \ref{sec:datastructure} and summarize in Section \ref{sec:summary}.

\section{The transient sample} \label{sec:transientsample}

This section describes our selected upstream data sources for transient events and the procedure to assemble a transient sample.
We also present the basic statistics in these upstream data sources and the compiled transient sample as of June 1, 2021, i.e., the cut-off date of transient events for the initial release of our database presented in this paper.

\begin{deluxetable}{lrrl}
\tablecaption{Upstream Data Sources for Transients \label{tab:transientdatasource}}
\tablecolumns{4}
\tablenum{1}
\tablewidth{0pt}
\tablehead{
\colhead{Source} &
\multicolumn{2}{c}{Number} &
\colhead{Type} \\
\cline{2-3} 
\colhead{} &
\colhead{Total} &
\colhead{Used} &
\colhead{}
}
\startdata
The Open Supernova Catalog                      & 82864     & 34547     & SN  \\
\citet{French20} Review                         & 43        & 43        & TDE \\
The Open TDE Catalog                            & 97        & 96        & TDE \\
\textit{Swift} GRB Catalog                      & 1651      & 1493      & GRB \\
\textit{Fermi}/LAT GRB Catalog                  & 145       & 78        & GRB \\
\textit{INTEGRAL}/IBAS GRB Catalog              & 140       & 140       & GRB \\
The Fast Radio Burst Catalogue                  & 118       & 13        & FRB \\
LIGO Detections                                 & 12        & 1         & GW  \\
The Open Kilonova Catalog                       & 6         & 6         & KN  \\
\hline
GRB Host Studies (GHostS)\tablenotemark{a}      & 235       & --        & GRB \\
Jochen Greiner's GRB list\tablenotemark{a}      & 2226      & --        & GRB \\ 
Gamma-Ray Burst Online Index\tablenotemark{a}   & 2082      & --        & GRB \\
GRBweb Database\tablenotemark{a}                & 7559      & --        & GRB \\
FRB Host Galaxy Database\tablenotemark{b}       & 19        & --        & FRB \\
Taubenberger (2017) Review\tablenotemark{c}     & 60        & --        & SN  \\
\enddata
\tablenotetext{a}{Ancillary data source for GRB localization, $T_{90}$, and redshifts.}
\tablenotetext{b}{Ancillary data source for FRB host association.}
\tablenotetext{c}{Ancillary data source for SNe Ia classification.}
\tablecomments{Number counts include all records in the original data source (\textit{Total}) and the number of records that contributed any data in our sample (\textit{Used}). Besides \citet{French20}, other reference sources are actively updated. Numbers are as of June 1, 2021.}
\end{deluxetable}

\subsection{Upstream data sources for transients}\label{sec:transientdatasource}

Existing transient events are detected by various surveys or missions, reported at multiple circulars or platforms, and cataloged by several sites or numerous individual publications.
There is no single data source that covers all previously known transient events. 
Therefore, we choose a few representative data sources with relatively complete records for each major transient type, organized in a machine-readable format for easier access.
Below we discuss our transient data sources for each major type.

\subsubsection{Supernovae}

Our supernova records are imported from the Open Supernova Catalog \citep[OSC;][]{Guillochon17}\footnote{\url{sne.space}, \url{github.com/astrocatalogs/supernovae}}.
The OSC compiles supernovae in numerous published catalogs, individual works, and online services, notably the Transient Name Server\footnote{\url{www.wis-tns.org}}, Padova-Asiago Supernova Catalog \citep{Barbon99}\footnote{\url{graspa.oapd.inaf.it/cgi-bin/sncat.php}}, the Latest Supernova website\footnote{\url{www.rochesterastronomy.org/supernova.html}} \citep{GalYam13}, the Weizmann Interactive Supernova data REPository (WISeREP)\footnote{\url{wiserep.weizmann.ac.il/}}, and the CfA Supernova Data Archive\footnote{\url{www.cfa.harvard.edu/supernova/SNarchive.html}}, etc.
Apart from the basic information (or ``metadata'') of supernovae and supernova candidates, the catalog also compiles their light curves, spectra, X-ray, and radio measurements.
For every single event, all related data are self-contained in a single file, where the historical versions of each file are traceable via their \texttt{git}-based version control solution.

Although great efforts are made in \citet{Guillochon17} to collect and homogenize supernova data in various sources, for our transient sample, there are a few issues to be resolved when importing their supernova records.
First, the supernova basic data (or ``metadata'') in OSC -- such as classification, sky coordinates, and redshifts -- trace back to multiple automatically-updated or manually-maintained reference sources.
As a result, slightly different, inconsistent, or even erroneous values are often provided for the same data field (or column), requiring further selection and validation.
Secondly, not all the records in OSC are directly detected and reliably classified supernovae.
{There are a large fraction of transient candidates in OSC without classification.}
There are also supernova remnants and historical supernovae (i.e., discovered before photometric techniques became available) in the Milky Way or nearby galaxies.
{Long-duration GRBs in the \textit{CGRO}/BATSE Catalog and \url{grbcatalog.com}, even without associated supernovae detected, are also included for the association of LGRBs with some core-collapse supernovae.}
Thirdly, the type labels (including major types and detailed subtypes) in OSC can sometimes be ambiguous, incomplete, and even inconsistent.
Reliable classification requires high-quality light curves and spectra covering critical phases of evolution, which are not always available.
The criteria and technique for transient classification could also differ among survey programs or research groups.
Even the classification scheme is evolving as our knowledge of supernova physics advances. 
As a result, events are not classified consistently among different sources.

For our work, we only select a subset of records in OSC, with further data curation for clarity and consistency.
{Records with classification, redshift {\it or} host information of any kind (names, or coordinates) are selected.}
{Supernova candidates and unclassified events are omitted unless their redshifts or hosts are also reported.}
{Those temporarily ignored candidates and unclassified events will be added back to our database during future updates once they are confirmed or classified in OSC.}
We also skip ancient supernovae and supernova remnants, even if their types are indirectly inferred in the literature.
{Furthermore, we only select GRBs in OSC with associated supernovae detected (i.e., classified as supernovae at the same time).}
Similarly, we omit records labeled as TDEs at this stage.
TDEs and GRBs are added via other data sources (discussed later).

The snapshot of OSC we use here has 82864 records, where 26868 records are classified (i.e., with type labels assigned), and 21592 records have host galaxy information, either name designations or coordinates.
As new records and data are added into OSC by programs and users regularly, we set up an automatic script to pull the latest version of OSC hosted at \url{github.com} and update our transient records accordingly.
Our database does not include detailed supplementary data of each event in OSC, including transient light curves, spectra, X-ray, or radio fluxes.

\subsubsection{Tidal disruption events}

Currently, there are only a few dozen tidal disruption events reported in the literature.
For the initial release of our database, we use the summarized list in the review of \citet{French20} as our primary data source.
Additionally, we use the Open TDE Catalog\footnote{\url{tde.space}}, a close sibling of the Open Supernova Catalog, as our secondary data source\footnote{{For future updates of the database, we may switch our primary data source to the Transient Name Server.}}.
Like OSC, the Open TDE Catalog is designed as a complete sample of known TDEs, including likely ones, in the literature or transient circulars, where basic data are compiled from various reference sources.
This catalog also includes unconfirmed or possible events where alternative explanations or classifications may exist, which requires manual inspection and selection.

To incorporate these records into the database, we combined events in both data sources into a single, unified list of TDEs.
We manually match the events in these two data sources by name designations (including variants) to ensure that the combined list is free of duplicates.
To verify our name-based matching and to further identify unique events in each data source, we also spatially cross-match events using an angular distance threshold of 5''.
Finally, we use the metadata (name, coordinate, redshift, and host galaxy) in \citet{French20} whenever available, but also keeping other names and aliases in the Open TDE catalog.

We also unified the type labels in our combined list of TDEs.
The type labels within the Open TDE Catalog are still to be standardized. Some events are classified by their inferred progenitors or physical scenarios (e.g., ``MS+SMBH''), while others are only labeled as ``TDE'' or ``TDE?''.
We first assign ``\texttt{TDE}'' labels to all the events in the combined list.
For events in the list of \citet{French20}, we further assign labels to indicate their detected wavelengths, like ``\texttt{UVOptTDE}'' for events detected in UV or optical bands and ``\texttt{XrayTDE}'' for X-ray detected events.
Possible and likely X-ray detected events in \citet{French20} are labeled as ``\texttt{PossibleXrayTDE}'' and ``\texttt{LikelyXrayTDE},'' based on the samples of \citet{Auchettl17}.
There are other proposed classification schemes in the literature \citep[e.g.,][]{vanVelzen20}, which could be included in the future release of our database.

There are 43 TDEs in the list of \citet{French20} and 97 events in the Open TDE catalog (as of June 1, 2021). Excluding 41 common ones, the combined list has 98 unique events.

\subsubsection{Gamma-ray bursts} \label{sec:grbdatasources}

We use the following catalogs of individual GRB missions for their relatively better localization accuracy than other catalogs.

\begin{itemize}
  \item[-] \textit{Swift} Catalog\footnote{\url{swift.gsfc.nasa.gov/archive/grb_table/fullview/}} \citep{Lien16}. This is a catalog of events detected by the Burst Alert Telescope (BAT) on \textit{Swift}.
  BAT localizes GRBs to arcminute-level accuracy.
  Some events also have afterglow detection by the X-ray Telescope (XRT) and UV/Optical Telescope (UVOT) on board, which further improves the positioning accuracy to arcsecond-level.
  \item[-] \textit{Fermi}/LAT Catalog\footnote{\url{fermi.gsfc.nasa.gov/ssc/observations/types/grbs/lat_grbs/table.php}} \citep{Ackermann13}.
  This catalog includes events detected by the Large Area Telescope (LAT) on \textit{Fermi}, with arcminute-level positioning accuracy.
  This does not include events only detected by the Gamma-ray Burst Monitor (GBM), which has a larger error circle (degree-level in the best case).
  \item[-] \textit{INTEGRAL}/IBIS Catalog \citep{Vianello09, Bosnjak14} and later events localized with the \textit{INTEGRAL} Burst Alert System\footnote{\url{ibas.iasf-milano.inaf.it}} \citep{Mereghetti03}.
  These events are detected by the Imager on-Board the \textit{INTEGRAL} Satellite (IBIS), with a typical positional error of about $2'$. 
  For high signal-to-noise events, the accuracy may reach sub-arcminute level.
\end{itemize}

The catalogs listed above certainly do not cover all well-localized GRBs.
Other missions, such as \textit{HETE}-2, \textit{BeppoSAX}, and \textit{AGILE}, also report events that are well-localized via afterglows or by associated supernovae, usually with other facilities. We do not use their original catalogs due to the lack of precise coordinates there.

We also accessed several ancillary data sources for GRBs, including 
the catalog of the GRB Host Studies (GHostS) Project\footnote{\url{www.grbhosts.org}} \citep{Savaglio06a, Savaglio06b}, 
the GRB list maintained by Jochen Greiner\footnote{\url{https://www.mpe.mpg.de/~jcg/grbgen.html}}, 
{the Gamma-Ray Burst Online Index (GRBOX)}\footnote{\url{https://sites.astro.caltech.edu/grbox/grbox.php}},
{and the GRBweb catalog}\footnote{\url{https://user-web.icecube.wisc.edu/~grbweb_public/}}.
{These ancillary data sources summarize GRB properties from multiple missions and follow-up programs, including data reported to GCN circulars}\footnote{\url{gcn.gsfc.nasa.gov/gcn3_archive.html}} {that are not machine-readable at the moment.}
{These ancillary data sources provide us 1) potentially better coordinates than reported in the three mission catalogs above; 2) more well-localized GRBs that are not reported in those mission catalogs; 3) most importantly, GRB redshifts measured from afterglows, host galaxies, or associated supernovae.}

We use the latest online version of these mission catalogs and ancillary data sources to ensure that our GRB records are up-to-date.
We combine records from these reference sources into a single list, during which we cross-match GRBs by their standardized names to eliminate duplicates.
{For each event, we choose the best sky coordinate (i.e., with the smallest 90\% error circle) within these reference sources. Redshift and $T_{90}$ (i.e., the time when $90\%$ photons arrive), when available, are usually consistent across these reference sources. If not, we take the median value of all reported measurements.}

Events in our list are classified as GRB by default.
{According to their $T_{90}$ values, we further classify them into long-duration GRBs (LGRBs; $T_{90}\geq2\,\text{s}$) and short-duration GRBs (SGRBs; $T_{90}<2\,\text{s}$).} Events without $T_{90}$ in any of our data sources are only classified as GRB.
The traditional long-short dichotomy might be further refined using the spectral hardness \citep[e.g.,][]{Zhang12}, which remains a part of our future work.

{It should be emphasized that GRBs are, on average, detected at higher redshifts than other transients.
The drastically decreased completeness of survey catalogs, combined with the large error in coordinates, makes the host hard or even impossible to be identified in some cases.}
{Therefore, we use a rather conservative sample selection and catalog accessing strategy to reduce the fraction of misidentified hosts.
We only include GRBs with a $90\%$ error radius under $5''$ in our sample. Meanwhile, the default search radius for GRBs without redshifts is reduced to minimize the confusion due to other nearby galaxies (Section \ref{sec:hostcrossid}).}
{Even with more accurate coordinates and smaller search radii, the chance remains that some GRB hosts are undetected or incorrectly identified. Checking quality control metrics (Section \ref{sec:qualitycontrolmetrics}) is always encouraged when using their compiled host properties.}

{Besides nine GRBs with associated optical supernovae in OSC, we have 7939 unique GRBs from our selected data sources, where 1621 of them have a $90\%$ error radius better than $5''$ and are thus included in the transient sample.}

\subsubsection{Other rare transients}

Our database also includes rare transients such as fast radio bursts (FRBs) and gravitational wave events (GWs).
These rare events are generally poorly localized, so we only include 14 FRBs and one GW event with host galaxies reported in the literature out of over 100 FRBs and nine GWs.
These events include
GW170817 \citep{Abbott17}, 
FRB121102 \citep{Chatterjee17}, 
FRB180916 \citep{Marcote20}, 
FRB180924 \citep{Bannister19}, 
FRB181112 \citep{Prochaska19}, 
FRB190102 \citep{Macquart20}, 
FRB190523 \citep{Ravi19}, 
FRB190608 \citep{Marcote20}, and
FRB190614 \citep{Law20}.
\footnote{\citet{Heintz20} and \citet{Bhandari20} have recently reported the host galaxies of FRB190611, FRB190711, FRB190714, FRB191001, FRB200434, which will be included in the next release of the database.}
\footnote{FRB150418 \citep{Keane16} is not included as its ``afterglow,'' with which the host was located, is likely a coincident variable source \citep{Williams16}.
Also, the case of FRB171020 \citep{Mahony18} is not included due to its indefinitive host association.}
Their metadata are referred from the Fast Radio Burst Catalogue \citep[FRBCAT;][]{Petroff16} and the list of LIGO detections\footnote{\url{www.ligo.org/detections.php}}.
We also refer to the FRB Host database\footnote{\url{frbhosts.org}} \citep{Heintz20} for host association of known FRBs.
For the rest of the events, we do not attempt to identify their host candidates.

Besides the one associated with GW170817, there are several more kilonova or ``macronova'' events (including candidates) reported in the literature.
We label the event associated with GW170817 as ``\texttt{Kilonova}.'' Events listed in the Open Kilonova Catalog\footnote{\url{kilonova.space}} (another sibling of OSC), mostly short GRBs, are labeled as kilonova candidates (``\texttt{PossibleKilonova}'').
These kilonova candidates include GRB050709 \citep{Jin16}, GRB060614 \citep{Yang15}, GRB080503 \citep{Perley09}, GRB130603B \citep{Fong14b} and GRB150101B \citep{Troja18}.

\subsubsection{Remarks on data source selection}

There are a few more actively maintained catalogs and online services for extragalactic transients,
including the Transient Name Server\footnote{\url{wis-tns.weizmann.ac.il}}, which cross-identifies transient events;
the Astronomer's Telegram (ATel)\footnote{\url{www.astronomerstelegram.org}} where initial discovery, confirmation, and classification of transients are posted;
and the GCN circular\footnote{\url{gcn.gsfc.nasa.gov}} where high energy transients are reported and archived.
Many sky surveys or space missions also produce their own catalogs of detected events and candidates.
There are also catalogs that are no longer being actively updated (e.g., the supernova catalog of \citealt{Lennarz12}).

We use the data sources above as they are relatively complete (including events in multiple missions, surveys, and catalogs), and data are summarized and tabulated in a machine-readable form.
The basic statistics of our data sources are summarized in Table \ref{tab:transientdatasource}.
We may include other transient data sources for transients with records in an easily accessible format in future releases.

\begin{deluxetable}{lll}
\tablecaption{Type Hierarchy and Number Counts of Transients\label{tab:eventstatistics}}
\tablecolumns{3}
\tablenum{2}
\tablewidth{0pt}
\tablehead{
\colhead{Type} &
\colhead{Total} &
\colhead{Host}
}
\startdata
\DTsetlength{0.5em}{0.8em}{0.2em}{0.6pt}{1.6pt}
\begin{minipage}[t]{0.675\linewidth}\raggedright
\dirtree{%
.1 Full Sample.
.2 Thermonuclear SNe (\texttt{Ia}).
.3 \texttt{Ia} (unspecified).
.3 \texttt{Ia-91bg}.
.3 \texttt{Ia-91T}.
.3 \texttt{Ia-02cx}.
.3 \texttt{Ia-CSM}.
.3 \texttt{Ia-09dc}.
.3 \texttt{Ia-02es}.
.3 \texttt{Ia-99aa}.
.3 \texttt{Ia-00cx}.
.3 \texttt{Ca-rich}.
.2 Core-collapse SNe (\texttt{CC}).
.3 Stripped-envelope SNe (\texttt{SE}).
.4 Type Ib-like (\texttt{Ib}).
.5 \texttt{Ib} (unspecified).
.5 \texttt{Ibn}.
.5 \texttt{IIb}.
.4 Type Ic (\texttt{Ic}).
.5 \texttt{Ic} (unspecified).
.5 \texttt{Ic-BL}.
.4 \texttt{Ib/c}.
.3 H-rich Core-collapse SNe (\texttt{II}).
.4 \texttt{II} (unspecified).
.4 \texttt{II P}.
.4 \texttt{II L}.
.4 \texttt{IIn}.
.3 Superluminous SNe (\texttt{SLSN}).
.4 \texttt{SLSN} (unspecified).
.4 \texttt{SLSN-R}.
.4 \texttt{SLSN-I}.
.4 \texttt{SLSN-II}.
.2 Tidal Disruption Events (\texttt{TDE}).
.3 \texttt{TDE} (unspecified).
.3 \texttt{UVOptTDE}.
.3 \texttt{XrayTDE} (+ likely events).
.2 Gamma-ray Bursts (\texttt{GRB}).
.3 \texttt{GRB} (unspecified).
.3 \texttt{SGRB}.
.3 \texttt{LGRB}.
.2 Fast Radio Bursts (\texttt{FRB}).
.2 \texttt{Kilonova} (optical only).
.2 Gravitational Wave (\texttt{GW}).
}
\end{minipage} &
\DTsetlength{0pt}{0pt}{0pt}{0pt}{0pt}
\begin{minipage}[t]{0.125\linewidth}\raggedleft
\dirtree{%
.1 \NStatTypeAll .          
.1 \NStatTypeIap .          
.1 \NStatTypeIapUnspec .    
.1 \NStatTypeIabg .         
.1 \NStatTypeIaT .          
.1 \NStatTypeIacxp .        
.1 \NStatTypeIaCSM .        
.1 \NStatTypeIadc .         
.1 \NStatTypeIaes .         
.1 \NStatTypeIaaa .         
.1 \NStatTypeIacx .         
.1 \NStatTypeCarich .         
.1 \NStatTypeCC .           
.1 \NStatTypeSE .           
.1 \NStatTypeIb .           
.1 \NStatTypeIbUnspec .     
.1 \NStatTypeIbn .          
.1 \NStatTypeIIb .          
.1 \NStatTypeIc .           
.1 \NStatTypeIcUnspec .     
.1 \NStatTypeIcBL .         
.1 \NStatTypeIbc .          
.1 \NStatTypeII .           
.1 \NStatTypeIIUnspec .     
.1 \NStatTypeIIP .          
.1 \NStatTypeIIL .          
.1 \NStatTypeIIn .          
.1 \NStatTypeSLSN .         
.1 \NStatTypeSLSNUnspec .   
.1 \NStatTypeSLSNR .        
.1 \NStatTypeSLSNI .        
.1 \NStatTypeSLSNII .       
.1 \NStatTypeTDE .          
.1 \NStatTypeTDEUnspec .    
.1 \NStatTypeUVOptTDE .     
.1 \NStatTypeXrayTDE .      
.1 \NStatTypeGRB .          
.1 \NStatTypeGRBUnspec .    
.1 \NStatTypeSGRB .         
.1 \NStatTypeLGRB .         
.1 \NStatTypeFRB .          
.1 \NStatTypeKilonova .     
.1 \NStatTypeGW .           
}
\end{minipage} &
\DTsetlength{0pt}{0pt}{0pt}{0pt}{0pt}
\begin{minipage}[t]{0.125\linewidth}\raggedleft
\dirtree{%
.1 \NStatHostAll .          
.1 \NStatHostIap .          
.1 \NStatHostIapUnspec .    
.1 \NStatHostIabg .         
.1 \NStatHostIaT .          
.1 \NStatHostIacxp .        
.1 \NStatHostIaCSM .        
.1 \NStatHostIadc .         
.1 \NStatHostIaes .         
.1 \NStatHostIaaa .         
.1 \NStatHostIacx .         
.1 \NStatHostCarich .       
.1 \NStatHostCC .           
.1 \NStatHostSE .           
.1 \NStatHostIb .           
.1 \NStatHostIbUnspec .     
.1 \NStatHostIbn .          
.1 \NStatHostIIb .          
.1 \NStatHostIc .           
.1 \NStatHostIcUnspec .     
.1 \NStatHostIcBL .         
.1 \NStatHostIbc .          
.1 \NStatHostII .           
.1 \NStatHostIIUnspec .     
.1 \NStatHostIIP .          
.1 \NStatHostIIL .          
.1 \NStatHostIIn .          
.1 \NStatHostSLSN .         
.1 \NStatHostSLSNUnspec .   
.1 \NStatHostSLSNR .        
.1 \NStatHostSLSNI .        
.1 \NStatHostSLSNII .       
.1 \NStatHostTDE .          
.1 \NStatHostTDEUnspec .    
.1 \NStatHostUVOptTDE .     
.1 \NStatHostXrayTDE .      
.1 \NStatHostGRB .          
.1 \NStatHostGRBUnspec .    
.1 \NStatHostSGRB .         
.1 \NStatHostLGRB .         
.1 \NStatHostFRB .          
.1 \NStatHostKilonova .     
.1 \NStatHostGW .           
}
\end{minipage} \\
\enddata
\tablecomments{
\textit{Total} indicates the number of events with a certain type label, including events classified as its sub-types.
\textit{Host} indicates the number events with known host galaxies (names or coordinates).
Rows with ``unspecified'' in \textit{Type} exclude events with sub-type labels under the same parent type.
\rev{Transient types here are not mutually exclusive. Some types may have common physical origin, and the same event may receive multiple types.}}
\end{deluxetable}

\begin{figure}
\centering
\includegraphics[height=7in]{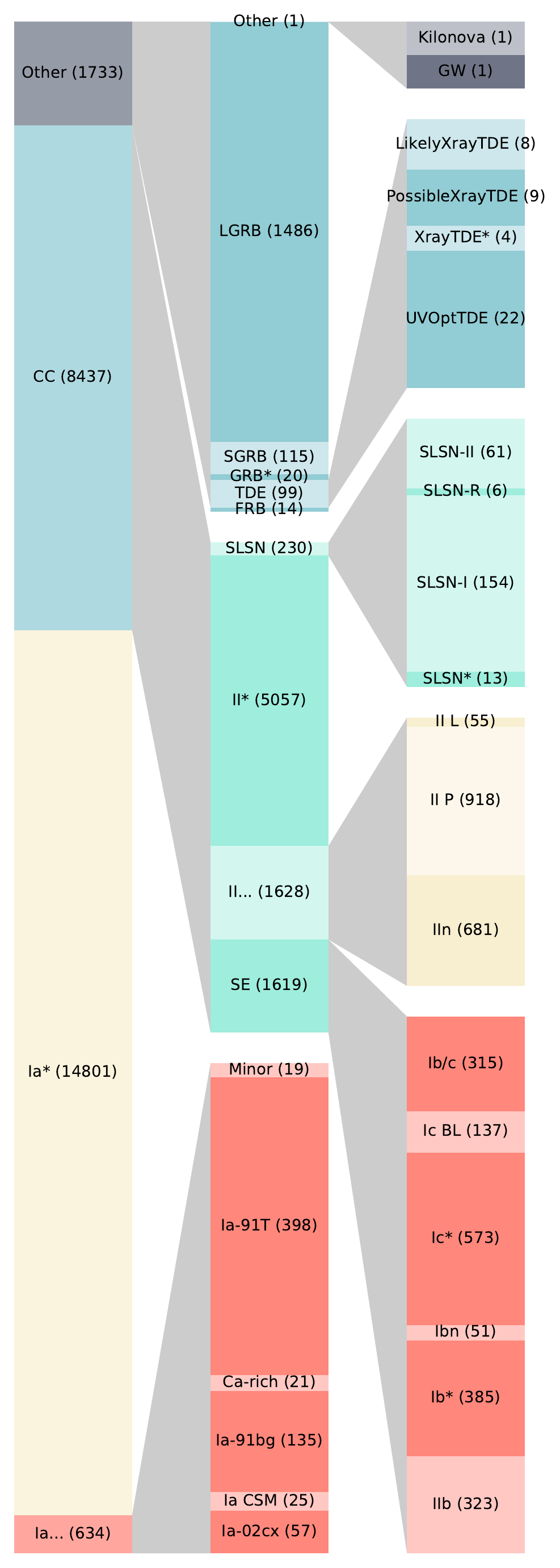}
\caption{
The number of transients under each type label, illustrated with stacked bars to show the relative sample sizes.
Here, labels with ellipses (\ldots) only count events \textit{with} any subtype classification under this parent type, while labels with asterisks (*) count events \textit{without} any further refined subtype label in the hierarchy (i.e., ``unspecified'' as defined in Table \ref{tab:eventstatistics}).
Events with multiple incompatible labels (i.e., labels that are not subtypes of each other) are accounted for multiple times.
Type Ia supernova is the most representative type in our sample, followed by core-collapse supernovae (Type Ib, Ic, II). Gamma-ray bursts dominate the remaining $7\%$ of classified events.
Relatively rare events like superluminous supernovae (SLSNe) and tidal disruption events (TDEs) only account for about $1\%$ of all records.
\label{fig:typeplot}}
\end{figure}

\subsection{Construction of transient sample}

Having transient records selected from those upstream data sources, we combine these records into a single, unified transient sample.
We only keep their metadata for each event, including sky coordinates, redshift, type labels, host name, and/or coordinate (if available).
We take the following approaches to ensure that events in the combined transient sample are unique, consistently classified, and having their best-available metadata compiled.

\subsubsection{Ensuring the uniqueness of events}

We first make our transient sample unique to each event.
A transient can be independently detected, with unique identifiers assigned by different surveys, missions, or research groups, in their own naming conventions.
When events are reported to circulars like TNS and ATel, there could be another assigned designation (e.g., ``AT'' prefix + year + alphabetic identifier) for the event.
Finally, when confirmed with follow-up observations, there might be an ``official'' name assigned using the naming convention of the community.
These aliases and designations should point to the same event in our transient sample.
Meanwhile, future events could be detected by multiple messengers, and they should point to the same entry in the database, provided that clear associations are established.
This is already partly done within OSC but not across all our upstream data sources.

To make the table unique to each event, besides the preferred or most commonly used designations, we also keep a list of aliases for each event. 
When a new event was reported at our transient data sources, we first search the alias lists of existing events to check if this is already assigned to a record.
We create a new record for this event only when all its names are new in our transient sample.
When maintaining alias lists, we also considered variants of transient designations in the literature.
For example, starting on Jan 1, 2010, the first GRB detected on a day is always labeled as ``A,'' even there is no other GRB on that day \citep{Barthelmy09}. However, such a change of naming convention is not strictly obeyed in the literature.
There are also minor variations like with or without whitespaces or dash (e.g., ASASSN vs. ASAS-SN).
We created possible variants of event names, even though they do not strictly follow the naming conventions of certain transient communities or survey programs.
This effectively prevents duplicates when the database is updated in the future.

Finally, we create unique identifiers for each transient event in our database (``\texttt{\_id}''), which are used consistently everywhere for host cross-matching and host candidate identification.

\subsubsection{Finding the best-available metadata}

When assembling the transient sample using several different data sources, we may also have multiple coordinates or redshifts reported for the same event besides multiple type labels.
Finding the best available sky coordinates and redshifts would thus be necessary for our later host cross-matching or identification.
We only use the ``most referred'' coordinates for supernovae, i.e., the sky coordinates that are most widely used in various data sources.
We do this by pairing RA/Dec by their data sources and counting the number of primary references linked to each coordinate pair reported in OSC.
When available, we always prefer the spectroscopic redshift of the host galaxy over the spectroscopic redshift of the supernovae, as we are assembling a catalog of transient host galaxies.
Photometric redshifts of any kind have the lowest priority.
For redshift values without types indicated, we assume that they are the spectroscopic redshifts of the transients.
Finally, when more than one redshift of the same priority level is reported, we take the median value.

Most GRBs have no associated supernovae detected, except for a handful of cases.
We always use the coordinates with the smallest $90\%$ error circle for these events.
We also include positional accuracy ($90\%$ error radius in arcseconds) for GRBs.
Many GRBs have X-ray or UV/optical afterglows detected, which provides more accurate localization (arcsecond-level) than coarse coordinates determined with wide-angle, coded mask detectors (usually arcminute level in the best cases).
When available, we use the most accurate position determined using X-ray or optical afterglows.
Finally, when associated optical supernovae are detected, we prefer supernova coordinates and redshifts over those reported for GRB for the same reason.

Commonly assumed as nuclear events, the reported transient coordinates of TDEs are usually their host coordinates in optical wavelengths.
Instead, a small subset of TDEs may only have X-ray determined transient coordinates reported, with relatively poor localization accuracy.
We take the as-reported positions as their \textit{nominal} host coordinates when matching objects in other catalogs (see Section \ref{sec:hostcrossid}).
Using the host coordinates of their optical counterparts in our final compiled summary table is always recommended over their nominal host coordinates.

Finally, the metadata of other rare events, including FRBs and GWs, are generally unique and consistent across different reference sources, so we do not make a further selection.

\begin{figure*}
\includegraphics[width=0.495\linewidth]{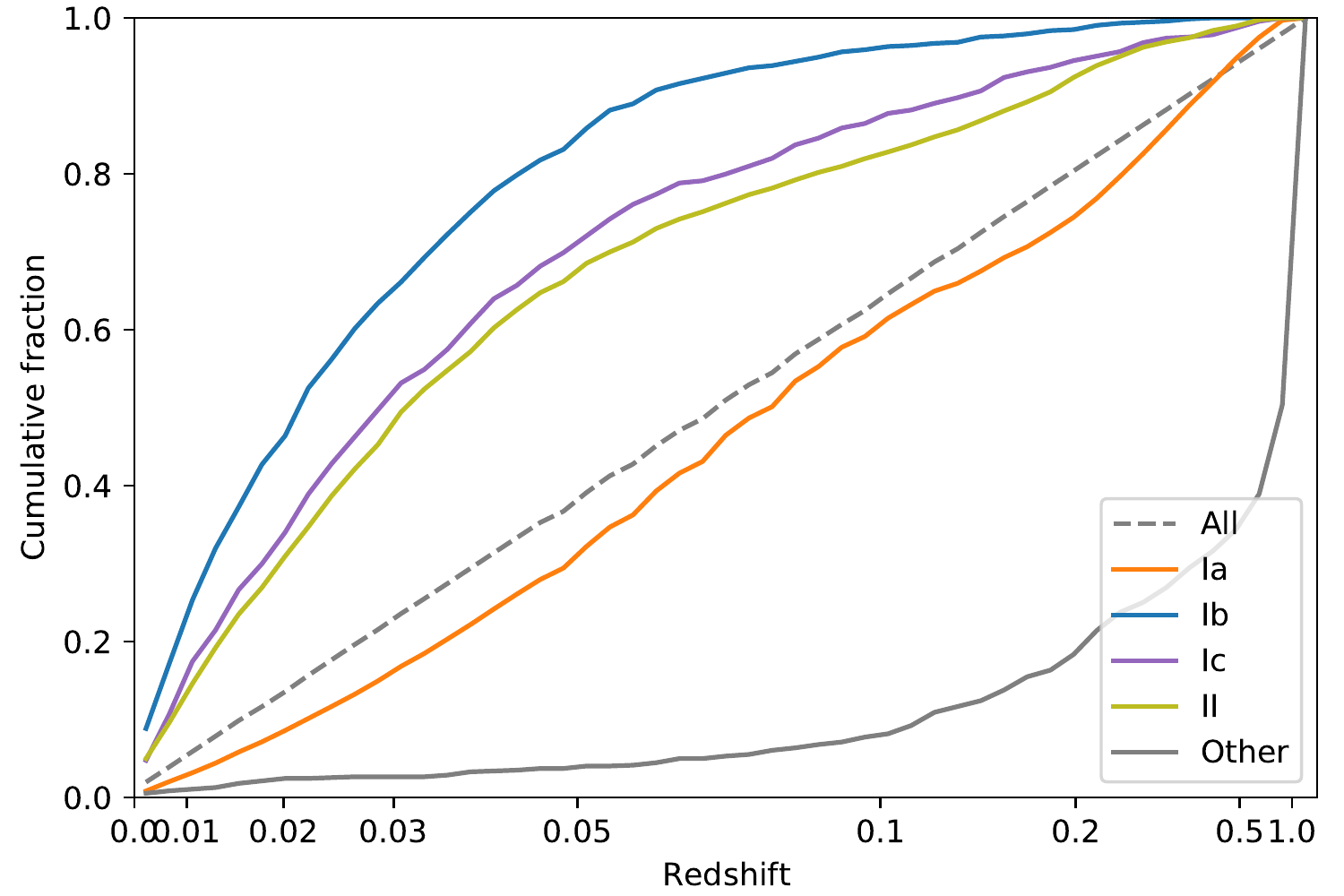}
\includegraphics[width=0.495\linewidth]{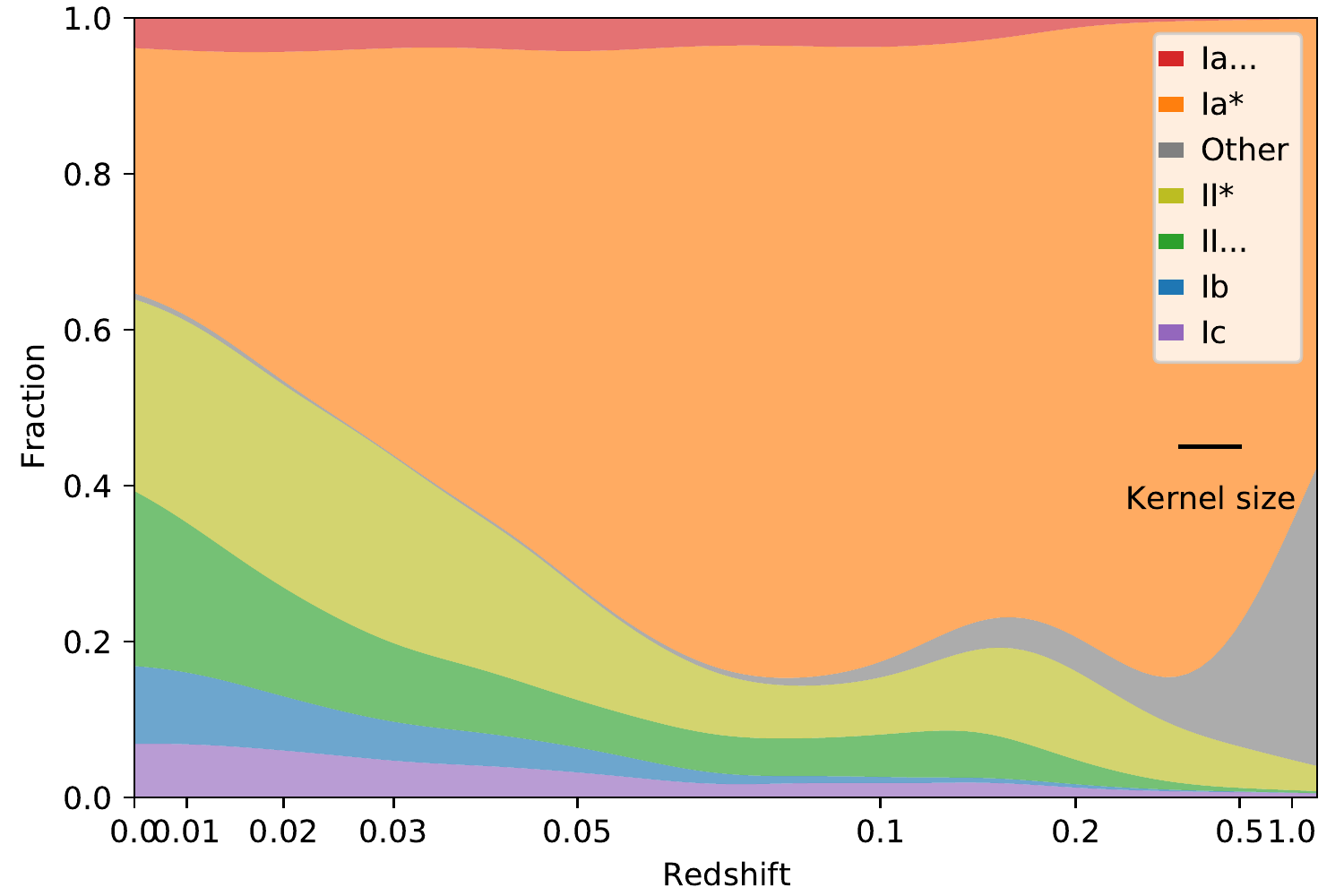}
\caption{
(Left) The cumulative distribution of transient redshifts in our database. The redshift axis is stretched so that events are uniformly distributed along the axis.
Major supernova types, including Type-Ia, Type-II, and stripped-envelope (Ib, Ic), are highlighted here.
Compared to Type-II and SE SNe, SNe Ia generally have higher redshifts in our sample, likely due to their higher luminosities. Meanwhile, SE SNe are distributed at even lower redshifts than Type-II.
{Other transients, mainly GRBs with redshifts reported, are distributed at the higher redshift side of our sample.}
(Right) The fraction of each major type across the redshift range, with similar, stretched redshift so that the shaded area scales with number counts.
Labels with ellipses count events of any subtype under this parent type, while labels with asterisks only include events without any subtype labels.
A large fraction of events at low redshift are core-collapse supernovae, and the relative fraction of stripped-envelope supernovae (SN Ib and Ic) within CC SNe drops with increasing the redshift.
The contribution of SNe Ia increases with redshift and peaks near $z\simeq 0.1$, while subtypes of SNe Ia are mostly at the lower redshifts.
\label{fig:zredtype1}}
\end{figure*}

\subsubsection{Standardization of the classification scheme}

To create a master sample of transients, one needs coherent and consistent classification for events observed in various wavelengths and time scales. This is a non-trivial task when working with diverse reference sources and archival records of historical events due to their minor discrepancies of classification schemes.
The existing classification scheme for transients, particularly for supernovae, is mostly phenomenological. Supernova types are primarily assigned based on spectral features and sometimes the shape of light curves. Meanwhile, the grouping of subtypes (taxonomy or ``hierarchy'' of supernova types) is usually motivated by physical models.
Even most data sources follow this well-established classification scheme, the detailed type notations and \textit{label systems} (i.e., the union of possible types labels) are not identical, especially when new phenomena are discovered or potentially new types are identified.
From the viewpoint of a retrospective meta-analysis, there are a few issues to be resolved when unifying transient types in different reference sources into a single, consistent scheme.

First, the notations (or type labels) for transient classification could vary in different reference sources, especially for unusual transients that do not naturally fit the existing classification scheme or newly identified subtypes that lack a standard notation in the community.
When the complexity of supernova phenomena cannot be described in our existing scheme, hybrid, intermediate, or transitional types are occasionally used in the literature to designate these indecisive cases.
For example, SNe Ia with circumstellar interaction have been once classified as ``Ia/IIn'' or ``Ian'' by some observers for their strong narrow Balmer line emission, a definitive signature for Type IIn supernovae, before they are recognized as a sub-group of SNe Ia \citep[Ia-CSM or Ia-02ic,][]{Silverman13, Silverman16}.
Similarly, Type-IIb supernovae, whose hydrogen lines -- for which Type-II is assigned -- fades at later stages until they eventually become SNe Ib-like. Before the popularization of ``IIb'' designation, these events occasionally receive ``II/Ib'' classification for such transitional behavior.
More commonly, several type notations are used interchangeably to refer to the same newly identified sub-group of events (e.g., Ia-06gz and Ia-09dc), which should be unified during data compilation.

Second, the label system could be different across reference sources, requiring a unified classification scheme.
Some reference sources may use complex or fine-grained classification schemes, including more refined subtypes, while others may only limited to some major types.
When combining type labels in multiple reference sources, as a result, there could be inconsistencies due to the classification schemes themselves.
For example, the same supernova could be classified as ``Ia-91bg'' in some reference sources and as ``Ia'' in other reference sources with a more limited range of type labels.
As a result, in the combined sample, some ``Ia-91bg" events may have coexisting ``Ia'' labels assigned, but other ``Ia-91bg'' events may not, leading to inconsistent labeling of events.
Ideally, we would use the best-refined subtype label of each event whenever possible so that transients are always uniquely and unambiguously classified to the finest details.
However, in a combined sample of transients, the same event could have multiple, incompatible subtype labels, i.e., labels not along the same branch of the classification tree.
This can result from disagreements among reference sources or independent detection in multiple wavelengths or by multiple messengers.
Therefore, we allow the same transient event to have multiple labels. Meanwhile, we choose a hierarchical classification scheme so that variations of classification schemes can naturally fit in.

Our convectional, phenomenologically-based classification scheme for transients is not strictly connected to the driving physical mechanisms, including progenitors, exploding mechanisms, and circumstellar environments. 
When constructing a hierarchical classification scheme, instead of adhering to a purely phenomenological scheme, we group transient types based on their physical origin whenever known.
For example, SNe IIb are commonly considered as close siblings of hydrogen-poor SNe Ib and Ic (or ``stripped-envelope supernovae''), rather than a subtype of Type-II supernova \citep[e.g.,][]{Filippenko93}.
Similarly, Ca-rich events, named after their strong nebular-phase Calcium lines, have been previously classified as a sub-group of SNe Ib by their maximum-light spectra.
Later studies revealed their thermonuclear origin -- they closely resemble SNe Ia, rather than collapsing massive stars like SNe Ib \citep[e.g.,][]{Perets10}.
These subtypes are preferably grouped by their physical origin instead of spectral signatures.
To attain consistency of classification when combining labels from reference sources with different classification schemes, we assign each event its best-refined subtype (there can be multiple) \textit{and} all their \textit{physical} parent labels so that each label in our transient sample points to a complete subset of events, including its physical subtypes.

Third, another source of inconsistency is the group of so-called peculiar events.
Supernovae with extreme properties or unusual signatures are labeled as peculiar events in our conventional classification scheme (e.g., ``Pec'' in many catalogs).
However, this all-inclusive label is not used consistently in different reference sources.
For example, a few well-defined subtypes of SN Ia, such as Ia-91T and Ia-91bg, are sometimes (but not always) labeled as  ``Ia Pec.'' These robustly-defined, relatively uncommon subtypes have characteristic observational signatures and are potentially related to the diverse channels, progenitors, or environments of SNe Ia.
Many other peculiar supernovae are just classified by template-matching algorithms, where the ``Pec'' label is inherited from the best-matching template, without other detailed and definitive type labels.
Occasionally, individual papers classify events as peculiar without a new label for their uncommon properties. These unspecified peculiar supernovae could just be outliers of ordinary events, but they may also contain new subtypes that are yet to be discovered.
Given the context-specific meaning of ``peculiar supernovae,'' we use ``Pec'' as a flag for relatively uncommon subtypes \textit{and} individual events with unusual properties instead of a subtype.

The standardization of the label system and subtype hierarchy must be done on a case-by-case basis. We discuss our detailed strategy in Appendix \ref{appendix:typelabels}.
The hierarchy of transient type labels is listed in Table \ref{tab:eventstatistics}.

\subsubsection{Remarks on transient classification}

We discussed the standardization of the classification scheme in the previous subsection. Such a procedure ensures a unified label system and coherent subtype hierarchy within the database.
However, transients in our combined sample are detected and classified by numerous survey programs and researchers over decades, using a wide variety of data and techniques.
Therefore, the results may suffer from possible ambiguity or uncertainty, potential methodological bias, and even human subjectivity.

First, transients are primarily classified by their photometric or spectroscopic signatures, while the quality, wavelength range, and temporal coverage of data are sometimes insufficient to precisely and definitively subtype those events.
For example, some supernova subtypes are characterized by their early-time or nebular phase spectral features, whereas a high-quality spectrum is not always available when they are far before or after maximum.
As a workaround for time-intensive spectroscopic classification, efforts have been made to classify transients purely based on light curves of limited epochs and filters, either with or without redshifts \cite[e.g.,][]{Frieman08, Kessler10, Lochner16}. The results of such classifications bear even larger uncertainties.

Second, the technique and criteria used could also bias the results.
Several active transient follow-up programs rely on template-matching techniques to classify events, where a spectral library of various transient types and phases are cross-correlated with the observed spectra to determine the most-alike types \citep[e.g.,][]{Smartt15}.
Such classifications and dating of transients are implicitly affected by the coverage of types and phases in the template library and possibly even by the quality and representativeness of individual spectra.
The aforementioned photometric classification is another example of how technique and criteria may affect the results.

Third, a non-negligible fraction of classifications are visually done by experienced observers in an empirical and perhaps subjective manner. The results may therefore vary from one person to another.

Indeed, the ambiguities, biases, and inconsistencies cannot be fully controlled or corrected without tracing back the original reference sources or conducting a comprehensive reanalysis of archival data.
However, proper curation of existing classification does, at least, guarantee the consistency of type labels \textit{within} our database.

\begin{figure}
\centering
\includegraphics[width=\linewidth]{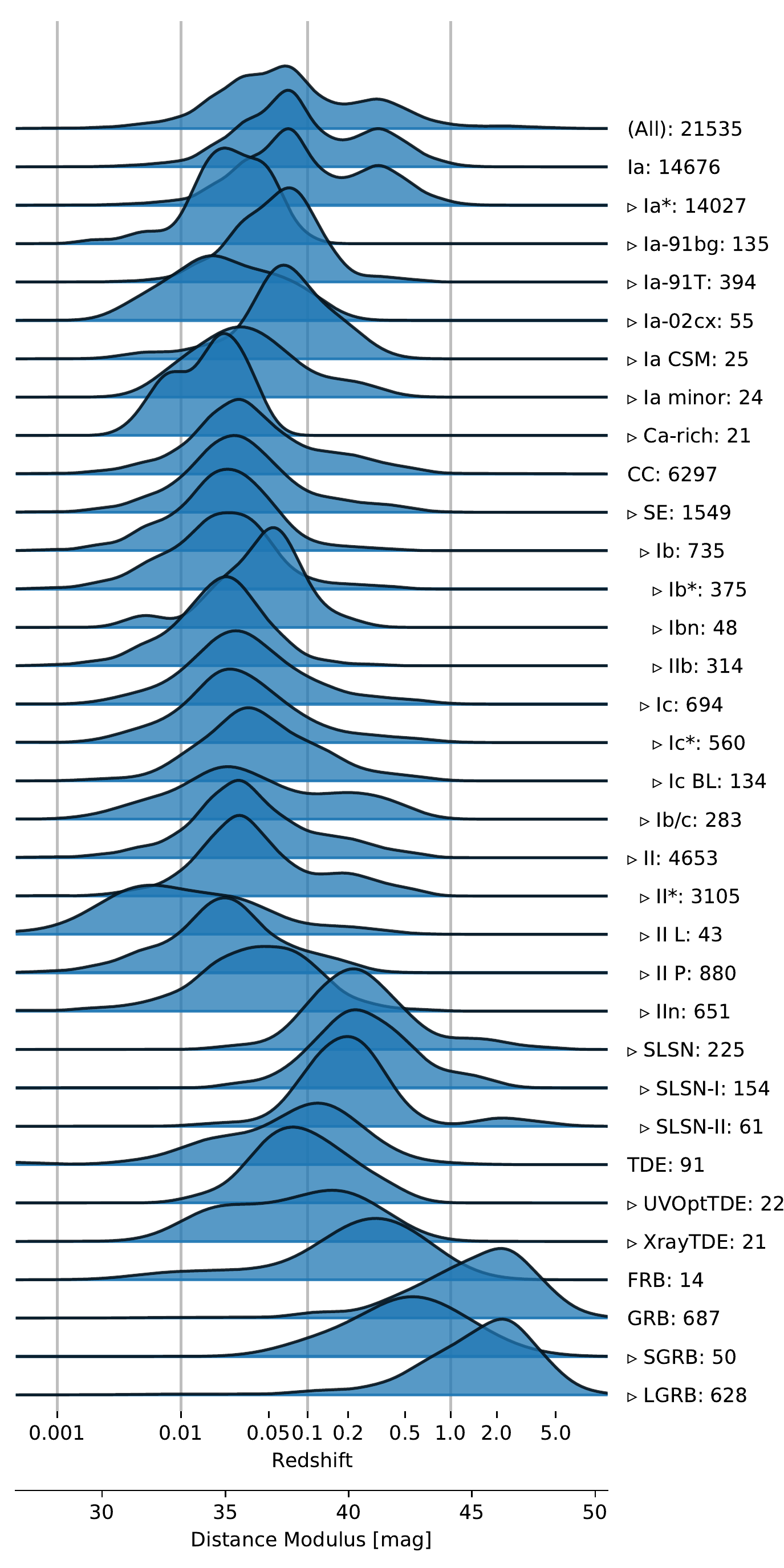}
\caption{
The distribution of redshift and luminosity distance for each type, shown as normalized density with respect to distance modulus. Labels with asterisks are events that are not further classified as any subtype of this parent type, i.e., ``unspecified'' events in Table \ref{tab:eventstatistics}.
Luminous events, like SNe Ia and SLSNe, are generally populated at the higher redshift side. Meanwhile, relative to their parent types, luminous subtypes (e.g., Ia-91T, Ia CSM, Ibn) are usually at higher redshifts side, while faint subtypes (e.g., Ia-91bg, Ia-02cx, Ca-rich) are usually at lower redshifts.
However, the different distributions cannot be fully attributed to the classical Malmquist bias. Some types require detailed observations (e.g., densely sampled light curves, spectra) to be classified, implicitly biasing them to lower redshifts. Also, our transients are compiled from various surveys and spontaneous discoveries that do not share the same detection efficiencies.
\label{fig:zredtype2}}
\end{figure}

\begin{figure}
\centering
\includegraphics[width=\linewidth]{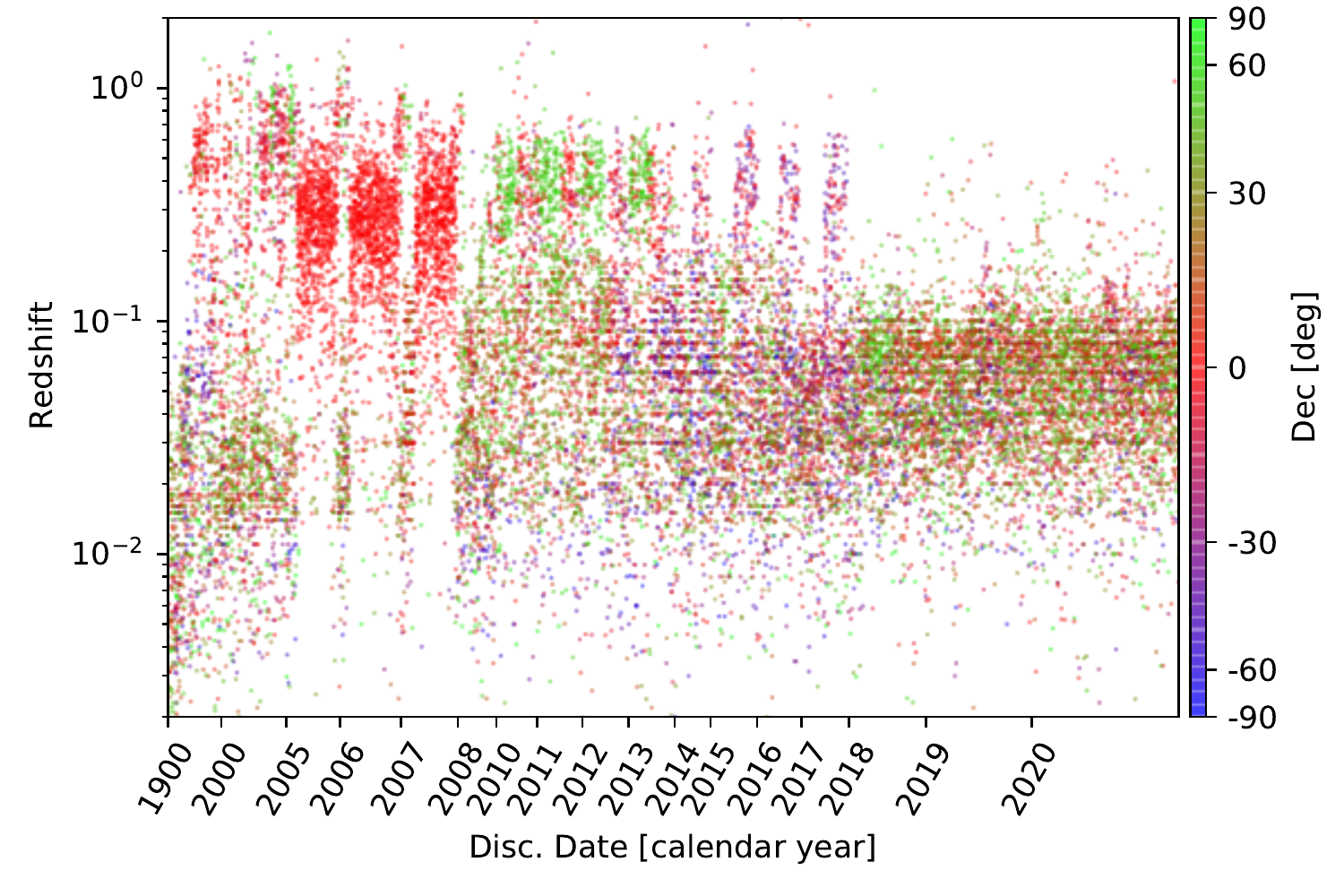}
\caption{Year of discovery vs. redshift of our transient events, color-coded with the declination angle. The time axis is stretched so that transients are uniformly distributed, and the vertical axis is in uniform scale of cosmological distance modulus.
The transient sample is a mixture of shallow, wide-field surveys and deep, high-redshift surveys with limited sky coverage.
There exists a low-redshift, spatially scattered component peaked at $z\simeq0.05$, with another higher redshift, spatially more localized component around $z\simeq0.5$.
Consequently, the redshift or luminosity distance distribution of our events is double-peaked, as illustrated in Figure \ref{fig:zredtype2}.
\label{fig:disctime}}
\end{figure}

\subsection{Transient sample statistics}

In this version of our database, the transient sample has \NStatEvents unique events, including \NStatTypeSN supernovae, \NStatTypeGRB GRBs, \NStatTypeTDE TDEs, \NStatTypeFRB FRBs, and \NStatTypeGW GW event.
The remaining ones are not explicitly classified yet. These are mostly candidate events with either potential host or redshift reported.
The numbers above do not sum up to the total number of classified events in the transient sample, as some events are detected in multiple wavelengths or messengers and thus are double-counted.
Also, as we have applied a series of selection criteria, the number counts do not match their corresponding upstream data sources.
Number statistics by type labels are summarized in Table \ref{tab:eventstatistics} and Figure \ref{fig:typeplot}.

More than half of the events in our database are Type Ia supernovae, where $4\%$ of them are further classified as some subtype of SNe Ia or marked as peculiar SNe Ia.
These $4\%$ SNe Ia include well-defined subtypes like Ia-91T, Ia-91bg, Ia-02cx, and other less commonly detected subtypes like Ia-09dc peculiar events without clearly defined subtypes.
Core-collapse supernovae (``\texttt{CC}''), including Ib, Ic, II, and their subtypes, are the second most abundant major class of events in our sample, accounting for nearly one-third of our transient events.
We further divide CC SNe into hydrogen-poor stripped-envelope SNe (``\texttt{SE}'', including Ib, Ic, IIb) and hydrogen-rich Type-II SNe.
SE SNe accounts about $20\%$ of CC SNe, which is slightly lower than the common estimate of volumetric SE-to-CC ratio (e.g., $36.5_{-5.4}^{+5.5}\%$ in \citealt{Smith11}, $30.4_{-4.9}^{+5.0}\%$ in \citealt{Shivvers17})
Most CC SNe are Type-II, but only $24\%$ of Type-II SNe are further classified into subtypes by their light curve shapes or spectral signatures.
The remaining $5\%$ of events in our transient sample are mostly GRBs, and the majority are LGRBs.
Other rare events, such as TDEs and superluminous supernovae (``\texttt{SLSN}'' under CC SNe), only account for about $1\%$ of all events.

Within the transient sample, 28995 events have redshifts reported in their upstream data sources. These are spectroscopic redshifts of either transient or their host galaxies. Very occasionally, host photometric redshifts are used.
The redshift distributions of major transient types are quite distinct from each other (Figure \ref{fig:zredtype1}, left). Type Ia supernovae are distributed over the entire redshift range, with an extended high-redshift tail, likely due to their higher luminosities.
Type-II and stripped-envelope supernovae are concentrated at lower redshifts, and particularly, SE SNe are at even lower redshifts than Type-II SNe.
Non-SN events, primarily GRBs with redshifts available in our sample, also extend to very high redshifts compared to other transients.
Regarding the fractions of major transient types (Figure \ref{fig:zredtype1}, right), most events at the lower redshift side ($z\lesssim 0.02$) are CC SNe, and the fraction of SE SNe within CC SNe drops toward higher redshifts.
At medium redshifts ($z\gtrsim 0.02$), SNe Ia quickly dominate the local fraction of events, within which SNe Ia with detailed subtypes are mostly distributed at the lower redshift side.

We further inspected the redshift and luminosity distance distributions of transients across sub-types as defined in Table \ref{tab:eventstatistics}, calculated using \textit{WMAP} 9-year cosmological parameters \citep{Hinshaw13}.
Luminosity distances are presented as distance moduli to be more closely related to the detectability of transients and can better illustrate the biases and heterogeneity in our database (Figure \ref{fig:zredtype2}).
{Notably, SNe Ia in our sample have a double-peaked distance modulus distribution, where the first peak appears between $z\sim 0.05$ and $0.1$, with a secondary peak between $z\sim 0.2$ and $0.5$.}
Given the large fraction and extended redshift distribution of SNe Ia in our transient sample, the full-sample redshift distribution (in distance modulus) is also double-peaked.
Regarding subtypes of SNe Ia, we notice that subluminous Ia-91bg and Ia-02cx are skewed towards lower redshifts than overluminous Ia-91T and Ia CSM.
Ca-rich transients are distributed similarly to other low-luminosity SNe Ia.
Within CC SNe, SE SNe (Ib, IIb, Ic) have very similar, if not slightly lower, redshifts compared to Type-II SNe, which dominate the total number of CC SNe.
Superluminous supernovae (``\texttt{SLSN}'') are clearly skewed towards higher redshifts, in which hydrogen-poor SLSNe (\texttt{SLSN-I}) are distributed further than hydrogen-rich SLSNe (\texttt{SLSN-II}).
Meanwhile, Type II L are at the lower redshift end of Type II, although their peak magnitudes are not fainter than Type II P, which dominate the number count of photometrically classified Type II.
For other subtypes of CC SNe, the differences are more nuanced. Type Ibn SNe are at the higher redshift tail of SE SNe, likely due to their higher peak luminosities. Similarly, relatively brighter SNe Ic BL (SNe Ic with broad emission lines) are at slightly higher redshifts than normal SNe Ic.
Finally, rare events like TDEs and FRBs are distributed at higher redshifts than CC SNe, and GRBs have the highest average redshift compared to other major types.

The double-peaked distance modulus distribution for SNe Ia could be a direct result of our sample heterogeneity. These transients are discovered by various surveys with significantly different sky coverage, cadence, and sensitivity. Consequently, the spatial-temporal coverage of our sample is far more complicated than a single survey.
As illustrated in Figure \ref{fig:disctime}, our events can be loosely grouped into low-redshift and high-redshift groups. The low-redshift group is mostly associated with wide-area, shallow surveys, with steadily improving depth over the years. Meanwhile, the high-redshift group is primarily contributed by a few deeper surveys that are more localized in their sky coverage. This is evident in their declination angles on the sky, where low-redshift events have scattered and well-mixed declination angles, while high-redshift events only have a few discrete ranges of values (i.e., their survey fields) that are related to their time of discovery.

\begin{sidewaysfigure*}
\centering
\includegraphics[width=0.875\linewidth]{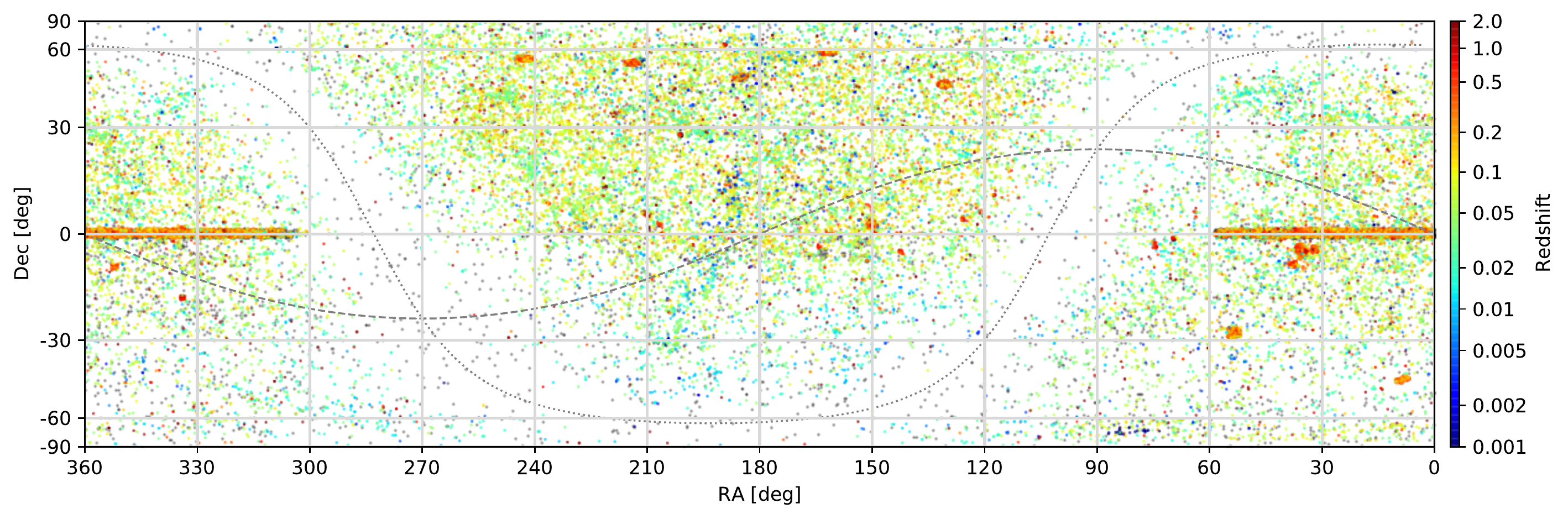} \\
\includegraphics[width=0.875\linewidth]{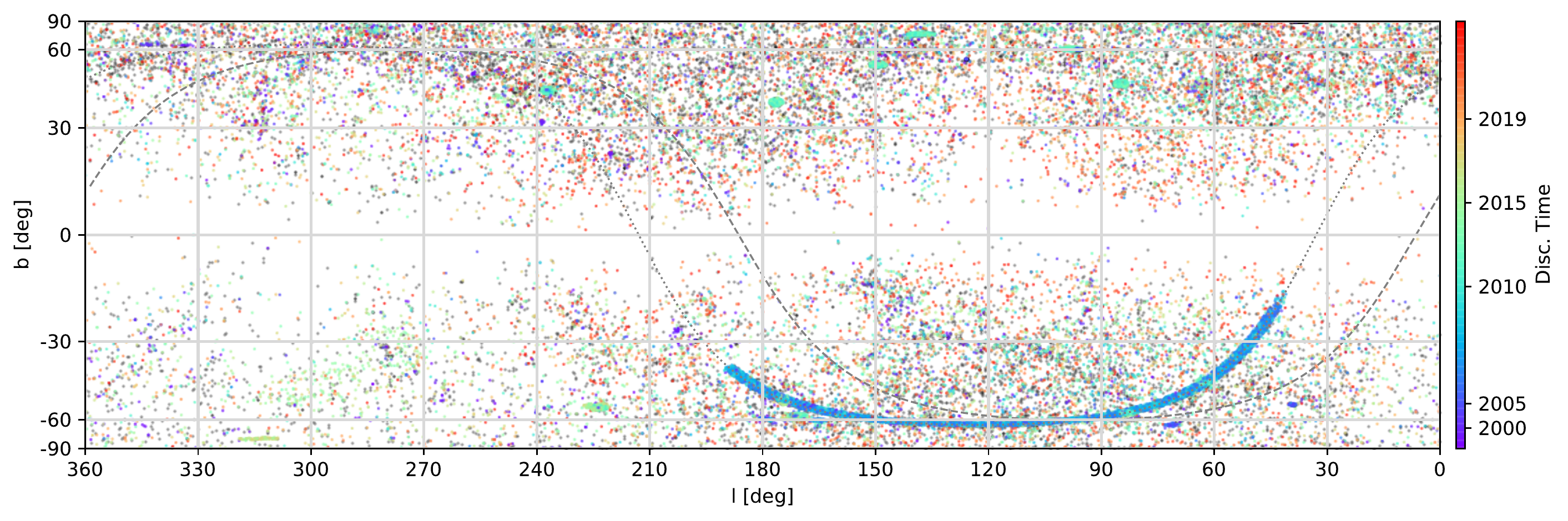}
\caption{(Top) Equatorial coordinates of transients in Lambert cylindrical equal-area projection. Colors indicate their redshifts, and gray points are events without redshifts. The dashed and dotted lines trace ecliptic and Galactic planes, respectively.
The fields of several deep supernova surveys are clearly visible, but they only contribute a limited fraction of events. Large-area surveys discover most events.
(Bottom) Similar to the Top panel, but showing Galactic coordinates instead. The discovery dates of events are color-coded with uniform number density over the color axis. Gray points indicate events without discovery dates. The dashed and dotted lines trace equatorial and ecliptic planes, respectively.
The timespans of major deep supernova surveys, particularly the SDSS-II supernova survey near the equatorial plane, are clearly visible.
\label{fig:skydistribution}}
\end{sidewaysfigure*}

Finally, the all-sky map of our transient events clearly shows the footprints of a few major deep and narrow-field supernova surveys (Figure \ref{fig:skydistribution}), notably the SDSS Supernova Survey \citep{Frieman08}, the Supernova Legacy Survey \citep{Astier06}, the ESSENCE Supernova Survey \citep{Miknaitis07}, the High-Z Supernova Search \citep{Schmidt98}, and the Supernova Cosmology Project.
These deep supernova surveys, however, only contribute a limited fraction of events in our transient sample. Wide-field transient surveys contribute most records.
Labeling transient records by individual survey programs could be interesting for certain studies, but this cannot be achieved easily due to the incompleteness of information in our upstream data sources.

The number statistics, local fraction, and redshift distribution indicate that our transient sample is highly heterogeneous.
Multiple factors could contribute to such heterogeneity.
First, there could be a classical Malmquist bias, where intrinsically brighter events can be observed at greater distances.
A possible second-order effect is that slowly-evolving events have a longer time window for detection and classification and are thus easier to be identified than those that rise and fade quickly.
Second, the criteria or techniques for classification may also introduce some bias, as many subtypes require high-quality spectra or even spectra at a certain phase of evolution to identify, which are often unavailable for distant, faint, or rapidly-fading objects.
Third, and perhaps most importantly, our master transient list is a compilation of various data sources, including spontaneous discoveries and transient surveys with vastly different detection efficiencies defined by their survey design (e.g., sky area, field-of-view, single-epoch sensitivity, and cadence).
Therefore, the number count, relative fraction by redshift, and luminosity distance distribution do not represent their true cosmological event rates or that of a sensitivity-limited survey.

\onecolumngrid \clearpage
\twocolumngrid

\section{Host Cross-identification} \label{sec:hostcrossid}

To compile the properties of host galaxies, one needs to identify their counterparts in other datasets and then access their observed or derived properties.
The procedure includes three steps: searching for potential counterparts of host galaxies in other datasets, matching potential counterparts across datasets to ensure their spatial association and correspondence relationship, and identifying the true counterparts of host galaxies.

Through this Section, we use ``sources'' for records in other datasets, ``objects'' for actual galaxies and stars in the sky, ``groups'' for clumps of spatially-associated sources in other datasets matched by our algorithm, and ``fields'' for patches of sky near transient coordinates containing the host and other objects.

We aim to cross-match sources in other datasets into groups correctly, then identify the counterparts of known hosts or rank cross-matched groups to identify the best host candidates when necessary. Therefore, we refer to the entire procedure as ``cross-identification.''
{We use ``known hosts'' for host galaxies cross-matched using existing host names or coordinates, and ``newly-identified hosts'' (or simply ``new hosts'') for host galaxies identified by our ranking algorithm.
Many ``newly-identified'' hosts are probably apparent to previous authors or observers, and some may have been identified elsewhere. However, these hosts are not reported to our upstream data sources. Our work provides independent host cross-identification for these events.}

In the following subsections, we describe the detailed procedure of host cross-identification. The workflow is outlined in Figure \ref{fig:fullworkflow}.

\subsection{Accessing sources in external catalogs}

We search for the counterparts of host galaxies in other datasets, include value-added catalogs (VACs) and survey catalogs, which we collectively refer to as ``external catalogs.''
VACs are online astronomical databases including NED, SIMBAD, and HyperLEDA. They collect the observed and derived properties of objects from multiple surveys and in the literature. More importantly, they are directly accessible using common name designations of astronomical objects.
Survey catalogs are high-level data products of individual photometric or spectroscopic surveys, either hosted at web services or provided as static files containing directly measured properties of objects.
Generally, without name-resolving services like Sesame\footnote{\url{cds.u-strasbg.fr/cgi-bin/Sesame}}, data of individual objects in survey catalogs are only accessible using their sky coordinates.
External catalogs we use in this work are summarized in Tables \ref{tab:photocatalogs} and \ref{tab:othercatalogs}. The number statistics and source selection criteria of these catalogs are discussed in Section \ref{sec:extcatalogs}.

{When either host name or coordinate is known, finding the counterparts of the host in external catalogs becomes a trivial problem.
Even some survey catalogs are not directly accessible using host names alone, the corresponding sky coordinates of these names are still easily obtainable from VACs.}
However, host names or coordinates are only known for a fraction of events. About half of all events, even having redshift or classification, do not have host name or coordinate reported.
Subject to the availability of host information, we use two different approaches to search for potential counterparts of hosts in external catalogs.
We outline the two approaches here. The details are discussed in Appendix \ref{appendix:externalcatalogs}.

\subsubsection{Events with known hosts}

Nearly half of transients in our sample have host galaxies reported. Their host galaxies are usually identified by experienced human observers, whereas the involvement of automated programs is increasing in recent years \citep[e.g.,][]{Gupta16, Sako18, Gagliano20}.
Often host galaxy names are reported, with or without the corresponding coordinates. Occasionally only host coordinates are provided.
Whenever possible, we use host names to access their properties in VACs to ensure reliable matching.
Meanwhile, many survey catalogs are only accessible with sky coordinates. We resolve the existing host names in VACs to find their corresponding host coordinates.
Even some events already have host coordinates reported along with host names, to avoid possible errors, we would like to update their as-reported host coordinates with the name-resolved coordinates in VACs.

{To ensure that the host names and coordinates we use are reliable, we conducted a systematic quality check of their original reference sources.
Only host names and coordinates from reliable reference sources are used in this work. The procedure of quality check and the selection of reference sources are described in Appendix \ref{appendix:hostinfoquality}.}
For events with known host names, we first resolve their names in NED and SIMBAD to find the best-available host coordinates. The procedure is described in Appendix \ref{appendix:resolving}.
Once succeeded, we use the name-resolved host coordinates to access other catalogs and find host properties.
We use the existing, as-reported host coordinates when no host name can be resolved, but host coordinates are given to search external catalogs.
{When searching external catalogs, we use a fixed radius of 15'' for name-resolved host coordinates and 30'' for as-reported coordinates.
Such large search radii minimize the chance of missing the true host due to ambiguous or inaccurate coordinates, a common situation for irregular, disturbed, or well-resolved large host galaxies.}
Even using known host coordinates, the returned sources do not always correspond to our known host galaxies. {Other non-host objects may also populate the field given our relatively large search radii.}
Therefore, we always perform a local cross-matching to ensure that the host properties we compiled are complete and uncontaminated (Section \ref{sec:xmatch}).
Finally, we use transient coordinates to access nearby sources for events without any host coordinate, either name-resolved or as-reported.

\subsubsection{Events without host information}

For events without host galaxies reported, we aim to identify their best host candidates and then compile their properties in external catalogs.
We search transient coordinates in VACs and survey catalogs with flexible, per-case search radii to enclose catalog sources that match the true hosts.
Transient coordinates may have significant angular offsets to their hosts. Therefore, the search radius for each event must be adjusted to optimize the chance to enclose the true host.
We determine the search radius primarily based on transient redshifts.
{For events with redshifts reported, we use angular distance corresponding to a projected distance of $45$ kpc at the reported redshift to search catalog sources.} When there is no redshift reported, we use a default search radius of $30''$.
{Specifically, for GRBs without known redshifts, we use 3 times the 90\% error radius as the search radius. Due to this conservative search radius, we may miss their true hosts, but this also reduces misidentified hosts.}
{We set lower and upper limits of search radius to $15''$ and $2'$, but for GRBs, the lower and upper limits are $5''$ and $15''$.}
These catalog sources are then cross-matched into spatially associated groups (i.e., host candidates) with compiled properties (Section \ref{sec:xmatch}).

\begin{figure*}[ht!]
\centering
\includegraphics[width=0.875\textwidth]{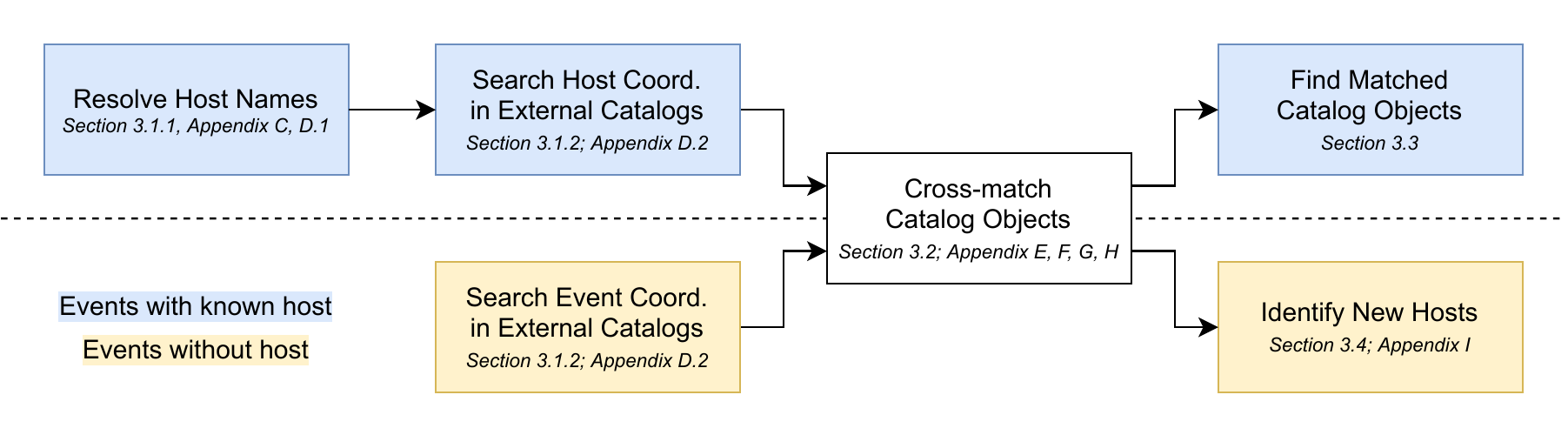}
\caption{The general procedure to compile host properties for transient events.
For events with known hosts, we resolve host names in NED and SIMBAD to obtain the best-available host coordinates and search the coordinates in external catalogs. The catalog sources that correspond to our known hosts are selected with a multi-catalog cross-matching process to avoid mismatching.
For events without hosts reported, we search transient coordinates in external catalogs with larger cone radii, cross-match all nearby catalog sources with a similar approach, and identify the most likely host candidates within cross-matched groups using a trained ranking function.
The subsections are indicated on each block. 
\label{fig:fullworkflow}}
\end{figure*}

\subsection{Cross-matching objects} \label{sec:xmatch}

To compile the properties of host galaxies in external catalogs, we need to ensure that only robustly detected catalog sources that spatially coincide with our best-available host coordinates are used.
Meanwhile, when identifying new host galaxies, we also need to select a group of catalog sources in the field with confirmed spatial association and readily compiled properties, rather than treating catalog sources as individual objects and choosing the best host candidates from them.
We need to spatially cross-match sources we obtained in external catalogs and establish their correspondent relationship for both purposes.

Cross-matching more than 20 catalogs with a wide latitude of angular resolution, sensitivity, and a mixture of photometric techniques is challenging.
Several factors make the problem complicated.
The sensitivities (or depths) of sky surveys are vastly different, where sources detected in one survey could be absent in other catalogs. Meanwhile, there could also be significant field-to-field variations of source densities and survey coverage.
Some sources in certain catalogs could be artifacts or duplicates of existing sources in the same catalog that should be removed. Finally, even true and unique sources could be foreground stars rather than distant galaxies that could host our transients.
Therefore, when cross-matching catalog sources, the criteria should be optimized for every catalog and tuned in each field.
Artifacts and duplicates of existing sources should be excluded to ensure that cross-matched catalog sources are genuine and unique. Foreground stars should be flagged whenever possible to facilitate the identification of new host candidates.

Catalog sources are cross-matched once their coordinates coincide within a certain threshold.
Source coordinates bear random errors depending on the signal-to-noise ratio of detection and the angular resolution of the survey (optical and pixel sampling), along with systematic errors inherited from survey-wide astrometric calibration.
For resolved sources such as our host galaxies, the measured positions also depend on the photometric technique used and the deblending of detected sources. Irregular, disturbed, or well-resolved galaxies are particularly affected by this issue. The criterion for spatial coincidence is therefore essential for any cross-matching algorithm.
The conventional way to cross-match two catalogs relies on a constant angular distance threshold, which is heuristically tuned to maximize the number of reliably matched pairs while not causing mismatches. One may also use the position errors of sources when provided and match sources in a probabilistic way.
Matching multiple catalogs, however, is a substantially more challenging problem.
For example, \cite{Pineau17} use $\chi^2$ test in combinatorial analysis to cross-match multiple catalogs, while \cite{Salvato18} establish the correspondence of sources with Bayesian statistics.
These algorithms either rely on detailed source properties or require extensive calibration with large samples and are thus unfeasible given our large number of external catalogs.
Due to data availability issues, we attempt not to elaborate on the detailed source properties (astrometric, geometric or photometric) when cross-matching. Also, the algorithm must be tuned and optimized using existing fields only, rather than a significant portion of the full catalog.

We cross-match these catalogs simultaneously in a single run using connected components in undirected graphs, where connectivity (i.e., adjacency matrix) is determined by pairwise angular distances, predefined per-catalog astrometric tolerances, and per-field matching thresholds.
Here, the astrometric tolerance of each catalog is a predefined global constant reflecting the average positional error of sources in this catalog, while the matching threshold in each field represents an ``acceptance threshold'' of cross-matching that can be fine-tuned locally to compensate the field-to-field variation of source density and catalog coverage.
The algorithm is a simple generalization of the two-catalog matching case, which is outlined below and described as pseudo-code in Appendix \ref{appendix:xidcode}:
\begin{enumerate}
\item[-] \textit{Selecting catalog sources.} We first select true and unique sources in each catalog in the vicinity of each queried coordinate (or ``field''). The detailed criteria are discussed in Section \ref{sec:extcatalogs}. Real sources with potential quality issues (e.g., saturated in the image, close to a bright star, or field edge) are preserved. We leave such quality control of measured properties for those who use our data.
\item[-] \textit{Calculating pairwise distances.} We project selected catalog sources in the field onto a local 2-d Cartesian frame, in which sky coordinates are converted to directional angular offsets to the queried coordinate. Then we calculate the pairwise angular distances of catalog sources, excluding pairs within the same catalog that are always supposed to be disconnected.
\item[-] \textit{Finding connected source pairs.} The pairwise angular distances are further normalized with their corresponding pairwise astrometric tolerances. Here the pairwise astrometric tolerance for any two catalogs is defined as the square roots of their quadratically summed, per-catalog astrometric tolerances. Source pairs with normalized angular distances below the per-field matching threshold are considered as ``connected.''
\item[-] \textit{Finding groups of connected sources.} We finally split sources into interconnected groups, in which any two sources are either directly connected or can be indirectly connected via a ``chain'' of connected sources. 
Each interconnected group of catalog sources is considered a cross-matched group. Isolated sources that are not connected to other sources are also considered as single-source ``cross-matched'' groups.
\end{enumerate}
In analogous to a conventional probabilistic two-catalog cross-matching problem, assuming that per-catalog astrometric tolerances are isotropic positional errors of sources, the normalized distance is related to the Mahalanobis distance of centroid position distributions for a source pair.
Not surprisingly, the values of per-catalog astrometric tolerances and per-field matching thresholds are critical to the accuracy of cross-matching. We choose per-catalog astrometric tolerances using a global optimization method, and the per-field matching threshold is tuned with a similar approach, which is discussed in Appendix \ref{appendix:astromtol}.
To further assess the quality of cross-matching, we introduced several metrics for cross-matched groups, which are discussed in Appendix \ref{appendix:xmatchqc}.

Before finding the known host galaxies or best host candidates and compiling their properties, some issues and complications remain with these cross-matched groups.
First, a substantial fraction of catalog sources is actually Milky Way stars, rather than distant galaxies that could be transient hosts. These foreground stars may become interlopers during host identification and contaminate the compiled host properties if not handled properly. The situation worsens for transients at lower Galactic latitudes. Therefore, foreground stars must be properly labeled and excluded when necessary.
Another issue is that some catalogs may contribute more than one source in a single cross-matched group. Indeed, we do not directly connect sources contributed by the same catalog, but they can still be ``passively'' connected via other sources in the group. 
This leads to the \textit{confusion} of multiple catalog sources, where only one (or rarely, none) of these sources can reasonably represent the measured properties of the group.
When such confusion occurs, we need to select the right source (or ``representative'' source) for the properties of the object measured by this survey.
The procedure is described in Appendix \ref{appendix:groupproperties}.
It is worth noting that even we distinguish stars from galaxies here, we still allow possible stellar objects to be identified as host galaxies, given the inevitable confusion of star/galaxy separation in external catalogs (Section \ref{sec:newhost}).

\begin{deluxetable}{lc|lc}
\tablecaption{Astrometric tolerances for external catalogs \label{tab:astromtol}}
\tablenum{3}
\tablehead{
\colhead{Catalog} & \colhead{Tolerance} & \colhead{Catalog} & \colhead{Tolerance}
}
\startdata
NED                 & $0.937''$     & 2MASS XSC         & $0.968''$ \\
SIMBAD              & $1.230''$     & 2MASS PSC         & $0.752''$ \\
HyperLEDA           & $1.227''$     & UKIDSS LAS        & $0.474''$ \\
{\it GALEX} MIS     & $1.641''$     & VHS               & $0.904''$ \\
{\it GALEX} AIS     & $1.824''$     & {\it AllWISE}     & $1.211''$ \\
SDSS                & $0.456''$     & unWISE            & $1.336''$ \\
PS1                 & $1.421''$     & NSA               & $1.433''$ \\
DES                 & $0.227''$     & SuperCOSMOS       & $0.813''$ \\
DESI LS             & $1.196''$     & {\it Gaia}        & $0.010''$ \\
SkyMapper           & $1.077''$     & MPA-JHU           & $0.010''$ \\
VST ATLAS           & $0.293''$     &                   &           \\
\enddata
\tablecomments{The per-catalog astrometric tolerances are estimated by maximizing the overall connectivity score, as described in Appendix \ref{appendix:astromtol}. \textit{Gaia} and MPA-JHU catalogs have fixed astrometric tolerances of $0.01''$. These catalogs are matched passively with other catalogs. }
\end{deluxetable}


\subsection{Finding sources matching known hosts}
\label{sec:knownhosts}

For events with known hosts, we only need to identify their counterparts in external catalogs, or more accurately, the cross-matched group near each transient that contains these catalog sources. We put the name-resolved or as-provided host coordinates into the field as a virtual source and cross-match it with other catalog sources using zero per-catalog astrometric tolerance. The cross-matched group that contains this virtual source is considered as the counterpart of the known host.

When using name-resolved host coordinates, the coordinate always coincides with (and thus matches) a NED or SIMBAD source with the same object name. We then mark the group as a ``confirmed-by-name'' host. When using as-provided host coordinates, and the coordinate matches a group that has \textit{not} been identified as a star, we consider the group a ``confirmed-by-coordinate'' host. If any cross-matched source in the group is believable a star rather than a galaxy, we only consider the group a primary host candidate instead of a confirmed host.

Finally, if we cannot identify a  ``confirmed-by-name'' or ``confirmed-by-coordinate'' host using host coordinates, or if only transient coordinates are used for source searching, we proceed to rank cross-matched groups by their likelihood to be the true host using the procedure described in the following subsection.
This includes the case in which the as-reported host coordinates coincide with a star or even match nothing in the search radius.

{To ensure that these known host coordinates have successfully matched the indicated galaxies, we perform a comprehensive visual inspection of these fields (Appendix \ref{appendix:inspectionknown}).
Occasionally, if the host coordinate failed to match the center of the indicated galaxy or the most prominent component of an irregular host, we manually reassign the host to the correct cross-matched group.
We focus on the results of cross-matching. Generally, we do not judge the correctness of transient-host association in our upstream data sources.
However, if there is \textit{clearly} a better choice of host galaxy than the indicated one, we mark the galaxy as an ``alternative host'' and flag the case \textit{without} reassigning the host.
To summarize, in 91 cases, the input host coordinate missed the indicated galaxy; in 257 cases, we noticed better hosts than the indicated ones.
For various reasons, like image quality issues, we are unable to inspect 106 cases. The majority of known and cross-matched hosts (17556) passed the inspection without any issues.}

\subsection{Finding new host candidates} \label{sec:newhost}

For events without known \textit{and} confirmed hosts, we identify their best host candidates among cross-matched groups.
Host galaxies reported in previous surveys are often visually identified by experienced observers in discovery or archival images.
{Such a manual or semi-automatic workflow may remain effective if transients, mainly supernovae, are classified and reported at the current rate; however, this would become infeasible for future high-efficiency follow-up programs or large transient samples, as we discussed here.}
Meanwhile, for state-of-the-art and future time-domain sky surveys, the growing involvement of galaxy properties in real-time alert processing and follow-up scheduling also requires efficient and reliable methods for automated host identification.

{Finding host galaxies is usually straightforward for human observers using images. However, it may not be an easy problem for automated algorithms when only cataloged source properties are available.}
Transients may occur anywhere within and even far outside the optical radii of their host galaxies. Therefore, the right host is not always obvious, and confusion of multiple possible hosts is sometimes inevitable, even for human observers.
Only one (or even none) could be the true host among those cross-matched groups in each field, assuming that sources are properly matched. We need an effective and robust method to \textit{rank} these cross-matched groups by their possibility to be the host.
We formulate the problem as follows: each cross-matched group could be characterized with a set of numeric or binary-valued parameters. We aim to construct a continuous scalar function of these parameters, whose value indicates the possibility that a cross-matched group can be the true host.
The function need not be a probability estimate in nature, as we only use it as a numeric score to rank groups and identify the best host candidate of each event. Also, the function should differentiate multiple possible hosts effectively to avoid confusion, i.e., being reasonably sensitive and ``monotonic.''
We noticed that the decision functions inside conventional binary classifiers might satisfy these requirements if the classifier is trained to separate cross-matched groups into true hosts and other non-host objects.
We, therefore, train binary classifiers to distinguish \textit{known} hosts from other nearby non-host objects and then use the trained decision function to rank cross-matched groups of events \textit{without} known hosts.

{There are several important aspects of this machine learning-based host ranking method: the construction of the training dataset, the parameterization of cross-matched groups, the evaluation of performance, and finally, the choice of classifiers. Here we outline the overall procedure to construct ranking functions. Relevant details are discussed in the Appendix \ref{appendix:rankingfunction}.}

{The training set lays the groundwork of our ranking functions.
Aiming for a clean training set, we choose known and properly cross-matched hosts that passed our visual inspection without quality issues or alternative hosts (Section \ref{sec:knownhosts}, Appendix \ref{appendix:inspection}).}
We access external catalogs and cross-match sources again, supposing that neither host name nor coordinate is available so that the training set best resembles the actual situation of host ranking.
Cross-matched groups near these transients are labeled as either ``true hosts'' or ``other objects'' based on our existing knowledge about their true hosts.
We include both ``true hosts'' and ``other objects'' in the training set, where the latter enable the ranking function to reject non-host objects near transient coordinates. The training set, as a result, is much larger than the actual number of transients used for training.

There is a wide range of options for the input variables of ranking functions, such as transient-host offsets, results of cross-matching, or detailed source properties.
{We call a particular combination of input variables a \textit{feature set}.}
As the starting point, we construct a basic feature set using only those universally available parameters of each group, regardless of which catalogs have been cross-matched. We also expand the basic feature set into its redshift-dependent version using a few transient redshift-relevant parameters.
{Some more detailed (but not universally-available) properties in external catalogs, including optical-infrared magnitude, angular size, photometric redshift, and other derived parameters, may further improve the performance of ranking functions.
Assuming that some particular wide-field surveys have the required sensitivity to detect the true hosts, we further expand the basic feature set with similar source properties measured in these catalogs.
However, these feature sets are only applicable in the coverage of these surveys, and the choice of host is also limited to groups with required parameters available.}

Since the ranking functions are trained using events with known hosts, we can evaluate their performance by checking if a ranking function can recover those hosts.
We define the \textit{accuracy} of a ranking function as the chance that the group of a known host ranked the highest among all cross-matched groups in its field.
In other words, the accuracy here is the fraction of known hosts that have been successfully recovered by the ranking function, which differs from the accuracy commonly used in statistics, machine learning, or other research disciplines.
{We choose this particular performance metric because the goal of training differs from the actual way the trained model was applied.
The classifiers are trained to distinguish individual ``true hosts'' from ``other objects,'' without the contextual information about other groups in the field; while the trained decision functions are used to rank (instead of to classify) all cross-matched groups in the same field and identify the best one. A result-oriented metric may better reflect the expected outcome of the training process.}

The decision functions of trained binary classifiers are the actual ranking function we use. To choose the best classifier for the purpose, we compare several conventional algorithms in the training and testing process, including Logistic Regression, Support Vector Machine \citep[SVM;][]{Cortes95}, Random Forest \citep[RF;][]{Ho95, breiman01}, AdaBoost \citep{Freund97}, Stochastic Gradient Descent (SGD), and Multilayer Perceptron (MLP).
We test their performance under different input feature sets, including the basic feature set and its redshift-dependent version, as well as other feature sets that rely on certain detailed source properties in external catalogs.
{These classifiers all reach above 95\% accuracy.
Logistic Regression classifier, in particular, achieves $97.3\pm0.3\%$ and $97.5\pm0.6\%$ accuracy using the basic feature set and its redshift-dependent version. It also stably outperforms other classifiers when using those derived feature sets with detailed source properties.}
{Therefore, we use Logistic Regression as the default classifier. When transient redshift is available, we use the redshift-dependent version of the basic feature set; otherwise, we use the default, redshift-independent version.
When applicable, ranking scores from other classifier and feature set combinations are also provided, but we do not use them to rank candidates directly.}

As a final note, the highest-ranking group in a field is referred to as the primary candidate, and other groups are collectively considered secondary candidates.
However, if the primary candidate is already assigned in a field, other groups are all secondary candidates.
We list the primary candidate first in the dataset, followed by secondary candidates in descending order of their ranking scores. The structure of the database is described in Section \ref{sec:datastructure}.

\subsection{Quality Control of Newly-identified hosts}
\label{sec:qualitycontrolmetrics}

{We use trained ranking functions to identify the best host candidates for events without reported hosts. Although performing reasonably well in cross-validation, the ranking functions have not been tested for performance for newly identified hosts.
Meanwhile, quality control flags and metrics are also desired for individual events to facilitate the use of the database.
We provide two sets of quality control flags and metrics based on visual inspection and mock sample tests.}

\subsubsection{Visual Inspection}
\label{sec:qualitycontrolvisualinsp}

{We perform a comprehensive visual inspection of new hosts identified by us. The inspection is aimed to 1) evaluate the accuracy of ranking functions on these new hosts; 2) understand the failure modes of ranking functions; 3) provide quality flags for the use of their host properties; and 4) make corrections to misidentified hosts when possible.}

{We rely on visual inspection because the true hosts remain unknown at this point. Other transient host catalogs may serve as the ``ground truth'' here, but these catalogs may also have misidentified hosts, leading to inaccurate estimates of accuracy.
Indeed, visual inspection is a \textit{subjective} evaluation of quality and performance, which depends on the designed workflow, image quality, and our discernment; while being an independent test, it reveals true performance and failure modes that are otherwise hard to characterize.}

{We have previously inspected the cross-matching of known hosts (Section \ref{sec:knownhosts}, Appendix \ref{appendix:inspectionknown}). For the sample of new hosts identified in this work, we focus on whether the default ranking functions have identified the most likely host we noticed in image cutouts. We make corrections when hosts are likely misidentified.
Also, when there are multiple possible hosts, or when we cannot identify the most likely host in image cutouts, we mark these indecisive cases with appropriate flags.
The procedure and criteria of visual inspection are discussed in Appendix \ref{appendix:inspectionnew}.}

{As a summary, without considering cases in which we are unsure about the most likely hosts, the default ranking functions achieved $97.0\%$ overall accuracy, comparable to the results from cross-validation.
Taking those indecisive cases into consideration, the accuracy tops at $97.3\%$ under the optimistic assumption that indecisive cases are always correctly identified; but this accuracy may also drop to $89.1\%$ in an improbable pessimistic situation that the ranking function failed in all indecisive cases.}

\subsubsection{Confidence Scores}
\label{sec:qualitycontrolconfidencescore}

{As a complementary method for quality control, we introduce a numeric metric for the reliability of new hosts, namely the confidence score.
The confidence score characterizes the degree to which a host candidate stands out among nearby non-host objects.
True hosts are, in fact, outliers among those more abundant non-host objects in the field. They should have significantly higher ranking scores than nearby non-host objects, assuming properly trained ranking functions.
If the ranking score of a candidate is comparable to the best scores that non-host objects can reach by chance, then the candidate is, to a certain extent, indistinguishable from non-host objects. This makes the candidate a less reliable one.
We use randomized mock transient samples and metrics for outlier detection to carry out the comparison here.}

{We feed a randomized mock transient sample into the workflow to construct baseline distributions, i.e., the distribution for the best-ranking score that non-host objects can get by chance. With the corresponding cumulative distributions, we can map the ranking score of a new candidate into a percentage score, which we refer to as the confidence score.
For a well-behaved ranking function, reliable candidates should have confidence scores close to $100\%$, because their ranking scores are significantly higher than non-host field objects; transients with multiple possible hosts may have more than one group with high confidence scores; finally, if no candidate reached a high confidence score, then all its candidates are likely indistinguishable from non-host field objects, and the event itself is likely hostless.
Indeed, the baseline distributions depend on transient redshift and catalog coverage and should be constructed separately for each ranking function. We discuss the details of implementation in Appendix \ref{appendix:confidencescore}.}

{Nevertheless, the confidence scores only serve as a reference. They are still based on the trained ranking functions themselves, which do \textit{not} directly indicate the correctness of transient-host association.
Misidentified hosts could have high confidence scores, while genuine hosts may also have low confidence scores. Cross-checking with visual inspection results, or performing some basic quality assessment, is encouraged when using confidence scores.}

\subsection{Accuracy of Ranking Functions}

{We use trained ranking functions to identify host candidates for events without reported hosts. The default ranking functions achieve above $97\%$ accuracy from the cross-validation using known hosts, and our visual inspection reveals a similar empirical accuracy in new hosts identified in this work.
To further understand the behavior of ranking functions and to characterize the quality of the dataset in detail, here we analyze the dependence of accuracy on some key transient or host parameters, including transient redshift, transient type, transient-host angular offset, and host optical-near infrared (NIR) magnitude.}

{For the comparison of accuracy across types, we group transients into broad classes using the hierarchy in Table \ref{tab:eventstatistics}.
Non-supernova transients, mainly GRBs, are grouped into ``Other.'' Some unclassified transients, which enter our sample because either redshift or host galaxy is known, are grouped into ``Unknown.''
For optical-NIR magnitudes of hosts, we use $r$, $i$, or $z$-band magnitudes from DESI Legacy Surveys (LS) and PS1. Outside their coverage, SDSS, VST ATLAS, or SkyMapper, are also used.
For continuous variables, we divide events into bins of equal sample size along the axis of interest to characterize the dependency of accuracy.}

\subsubsection{Accuracy from Cross-validation}
\label{sec:accuracyknown}

{We do not use trained classifiers to rank cross-matched groups when the host is already known.
Instead, we trust the reported host and directly use host coordinate to match the right group.
However, as the training set, we can use these known hosts to test the accuracy and analyze the behavior of our ranking functions.}

{We estimate the accuracy with standard 10-fold cross-validation.
We divide the sample into 10 subsamples, where each subsample contains about the same number of fields. During each test, we use one subsample for testing and the rest for training. We repeat the test 10 times and use each subsample once in turn for testing. The results are averaged over 10 tests to represent the performance.
Besides estimating the overall accuracy, we also derive the dependence of accuracy on transient redshift, transient type, transient-host angular offset, and host optical-NIR magnitude by binning the test sample, estimating the per-bin accuracy, and taking the average of 10 separate tests.
For each test, to estimate the per-bin accuracy to percent level, each bin must contain at least a few hundred events, which limits the bin number along the axis.}

{We summarize the dependency of accuracy in the upper sub-panels of Figure \ref{fig:ridgeaccuracy}.
We only show the accuracy curves for the two default ranking functions for clarity. Accuracy curves using other feature sets are summarized in Appendix \ref{appendix:rankingfunctionaccuracy}.
The change of accuracy over the redshift range is at the level of a few percent, indicating the stable performance of the ranking functions.
The accuracy is lower than average for low-redshift or high-redshift events. Also, at larger transient-host offsets or for fainter hosts, the accuracy is lower than average.
The degradation of accuracy for low-redshift events is likely attributable to larger transient-host angular offsets.
Towards higher redshifts, the properties of true hosts become less distinguishable from those ubiquitous faint galaxies, possibly leading to higher failure rates.
Notably, even at relatively small transient-host angular offsets, the accuracy is only at $98\%$. Failure happens here when a more distant object is chosen rather than the true host in close proximity.
We do not see a significant variation of accuracy across transient types. The accuracy in ``SLSN'' and ``Other'' (mainly TDEs and GRBs) is noisy due to their limited sample size compared to the number of folds.
Using transient redshift-relevant parameters marginally boosts the overall accuracy, and the improvement is clear at large transient-host offsets.}

{For the accuracy of other feature sets (Appendix \ref{appendix:rankingfunctionaccuracy}), we notice that the inclusion of some source properties can further improve the accuracy above our basic feature sets.
Even only at the percent level, the increase in accuracy indicates a significant reduction of failure rates.
However, this comes with the cost that some host galaxies may not have the required features and are thus ignored by these ranking functions.}

\subsubsection{Empirical Accuracy from Visual Inspection}
\label{sec:empiricalaccuracy}

{We also estimate the accuracy of new hosts from the results of visual inspection.
Because the inspection relies on our discernment instead of some ``ground truth,'' we refer to this as extit{empirical} accuracy.
In the subsample of events with unambiguous and clearly visible hosts, we use the fraction of successfully identified cases as the empirical accuracy.
There are, however, many cases in which the most likely host cannot be determined visually, usually due to image quality issues, confusion of multiple nearby galaxies, and even the complete absence of any likely hosts. The uncertainty in accuracy due to these cases should be taken into consideration.
Therefore, we further estimate the optimistic and pessimistic limits of empirical accuracy supposing that \textit{all} these indecisive cases are either successfully identified or missed by the ranking functions (Section \ref{sec:qualitycontrolvisualinsp}).}

{We focus on the dependence of empirical accuracy on key transient and host parameters, which is outlined in the lower sub-panels in Figure \ref{fig:ridgeaccuracy}.
We find similar patterns of dependence as we have seen in known hosts.
The accuracy does not change significantly along the axes.
We find lower than average accuracy in low-redshift or high-redshift bins, larger transient-host angular offsets, and fainter hosts.
Considering the loss of accuracy due to indecisive cases, even in the pessimistic situation, the accuracy of low- to mid-redshift events remains around $95\%$.
However, for high redshift events, the large fraction of indecisive cases implies a likely lower accuracy than outlined here -- the pessimistic lower limit of the last redshift bin is under $70\%$.
For this reason, ``SLSN'' and ``Other'' (mainly GRBs), usually high-redshift events, have pessimistic lower limits around $80\%$ and $50\%$, respectively.
Using redshift as input may improve the results for high-redshift or fainter hosts among cases with identifiable hosts, although the improvement may not be significant considering the broad range of accuracy due to indecisive cases.}

{Finally, it shall be noted that for indecisive cases, their redshift and type are intrinsic properties of the events, so that we can quote the optimistic and pessimistic limits for these two curves. However, transient-host angular offset and host optical-NIR magnitude depend on the host and are unknown for indecisive cases. Therefore, we are unable to quote the optimistic and pessimistic limits for these two panels.}

\subsubsection{Typical Failure Modes}

{To understand the reason for which the ranking functions miss the true hosts, we rerun our redshift-independent default ranking function on the entire training set, i.e., known and properly cross-matched hosts that passed our visual inspection, and then select cases in which the top-ranking groups are not the true hosts.
On the other hand, from the visual inspection of new hosts, we also summarize cases where the ranking functions missed the most likely hosts as we noticed in the images.
We analyze the failed cases on either side and find some common failure modes.}

\begin{figure*}
\centering
\includegraphics[width=0.04571428571428572\textwidth]{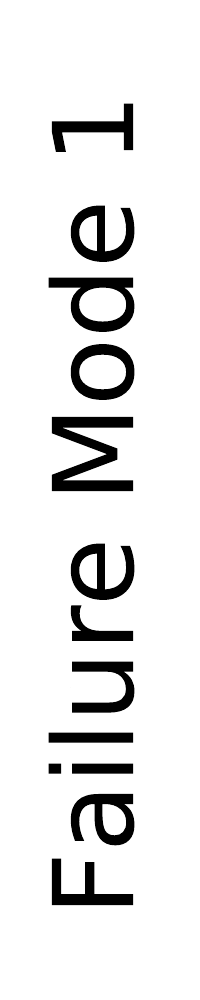}
\includegraphics[width=0.22857142857142858\textwidth]{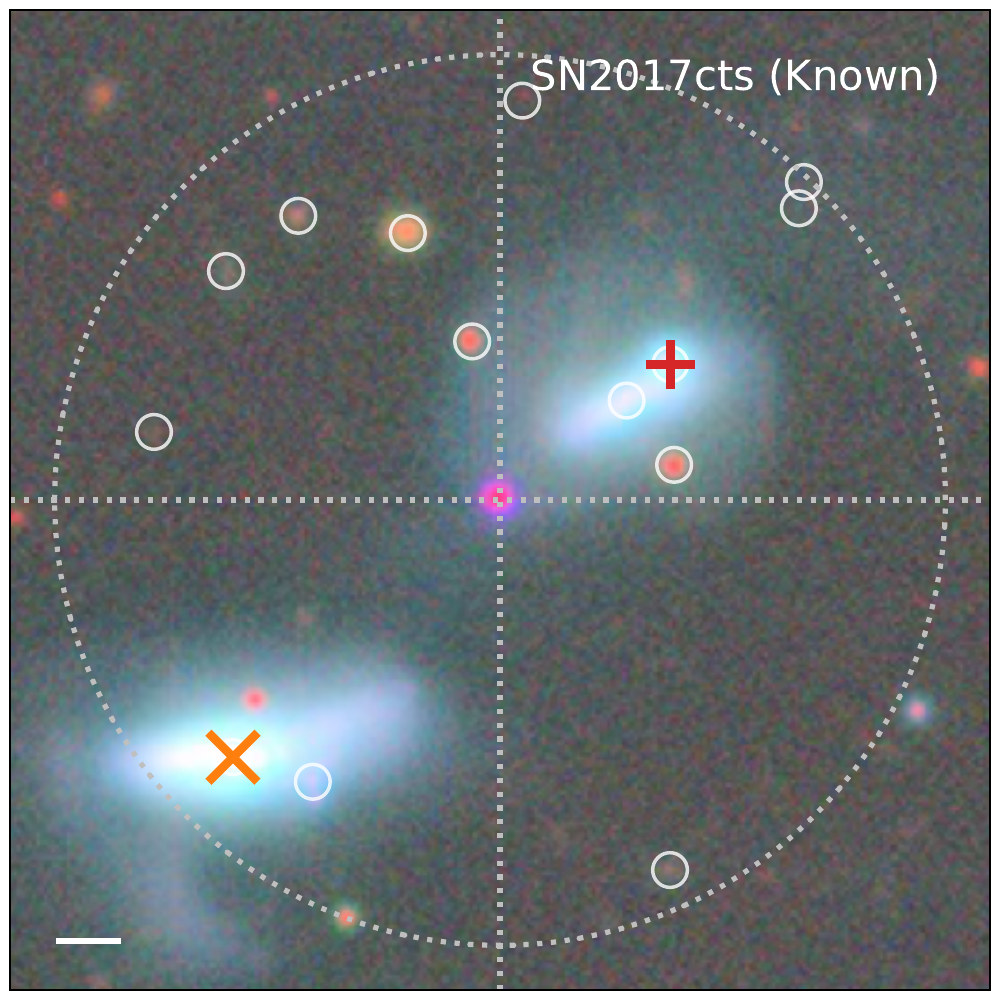}
\includegraphics[width=0.22857142857142858\textwidth]{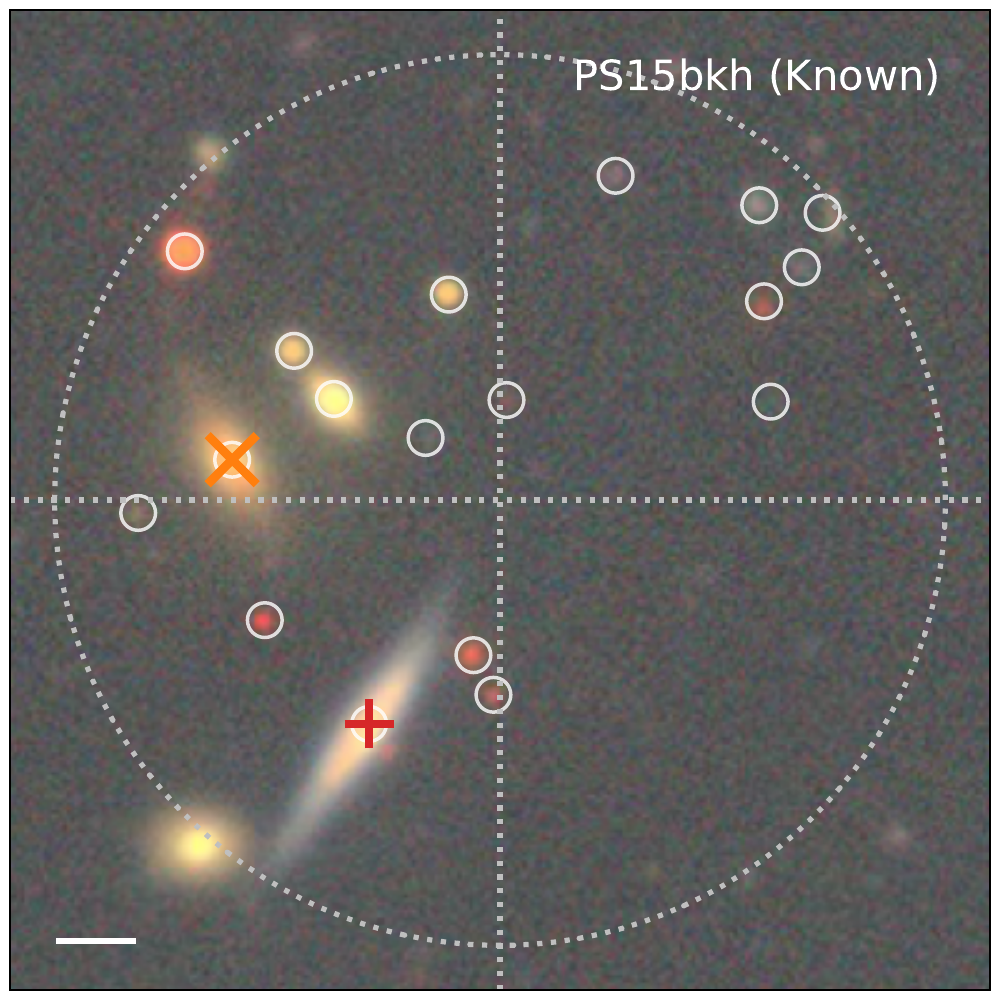}
\includegraphics[width=0.22857142857142858\textwidth]{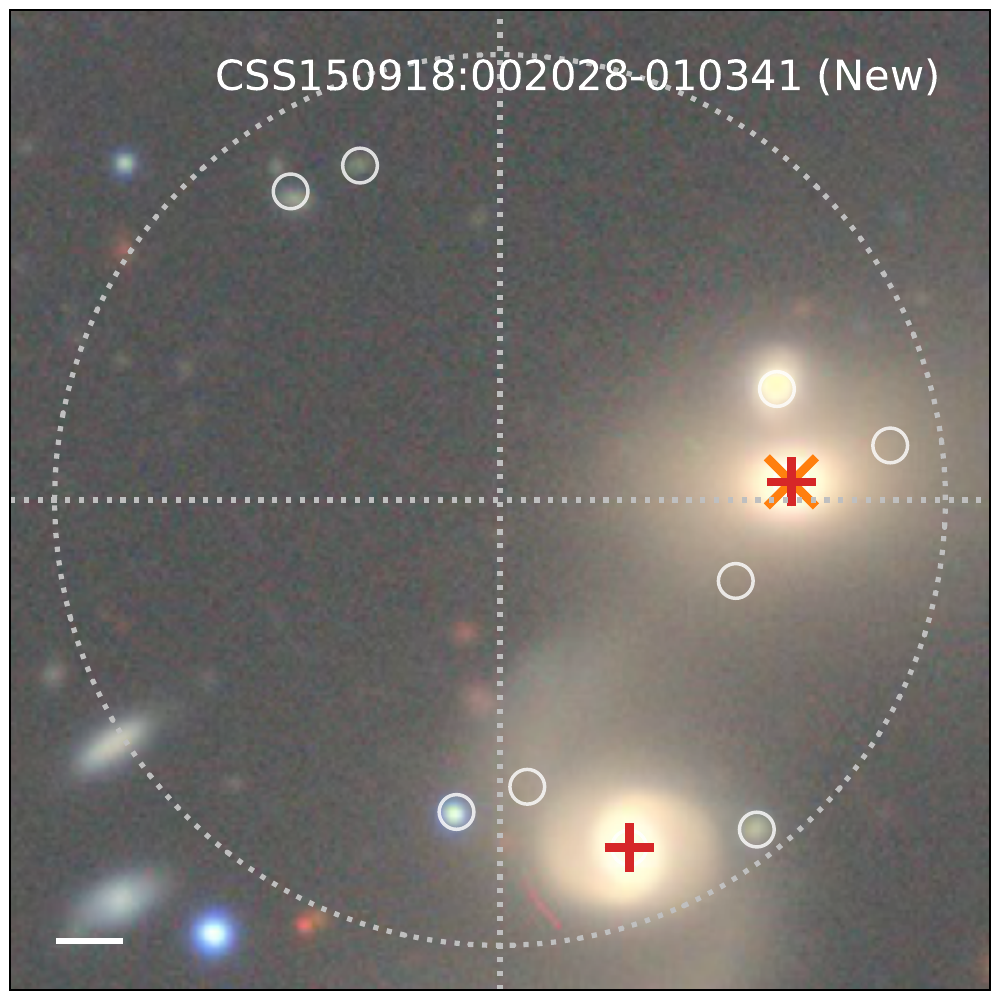}
\includegraphics[width=0.22857142857142858\textwidth]{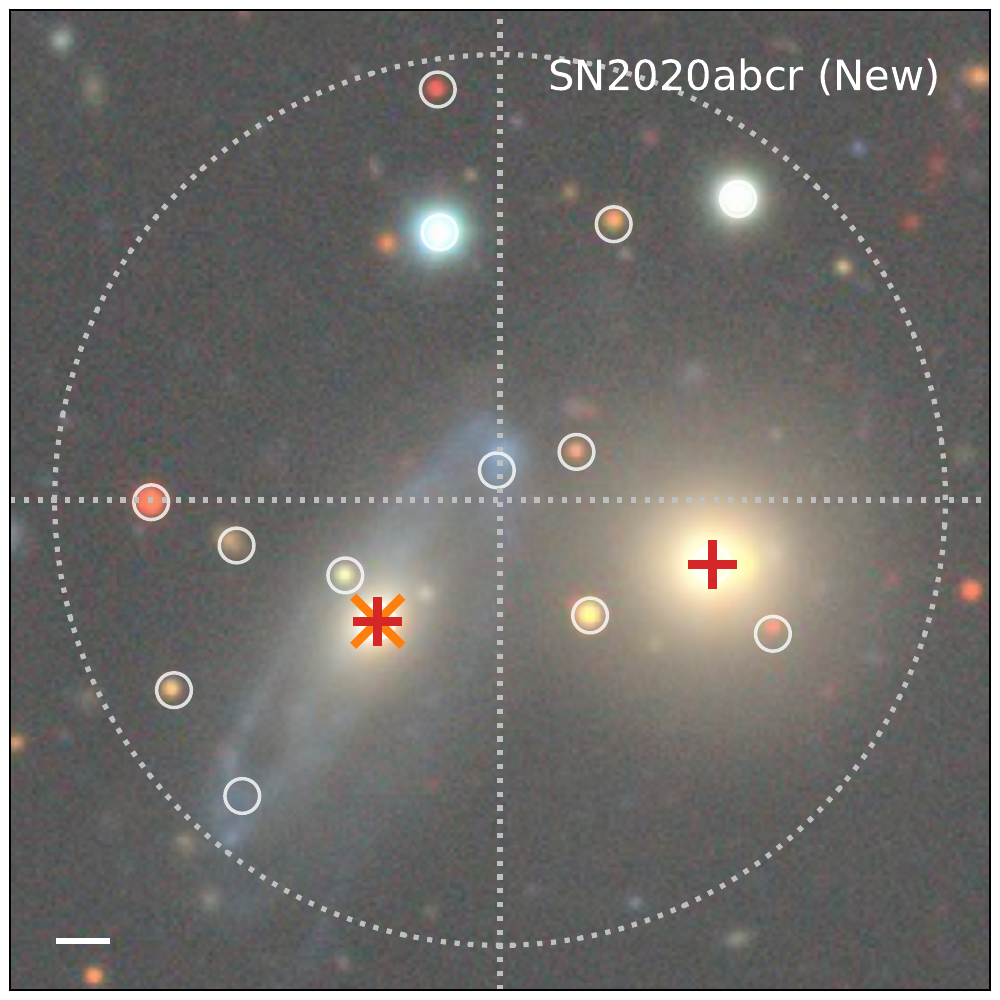} \\
\vskip 0.2cm \hrule \vskip 0.2cm
\includegraphics[width=0.04571428571428572\textwidth]{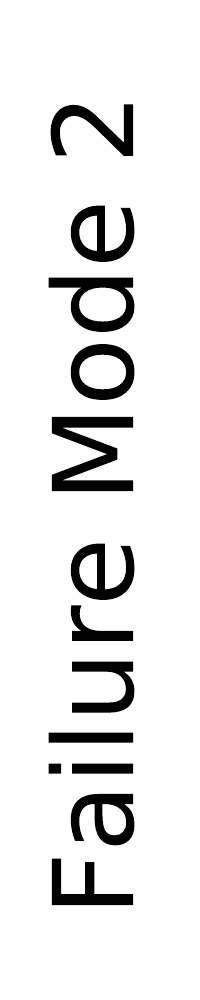}
\includegraphics[width=0.22857142857142858\textwidth]{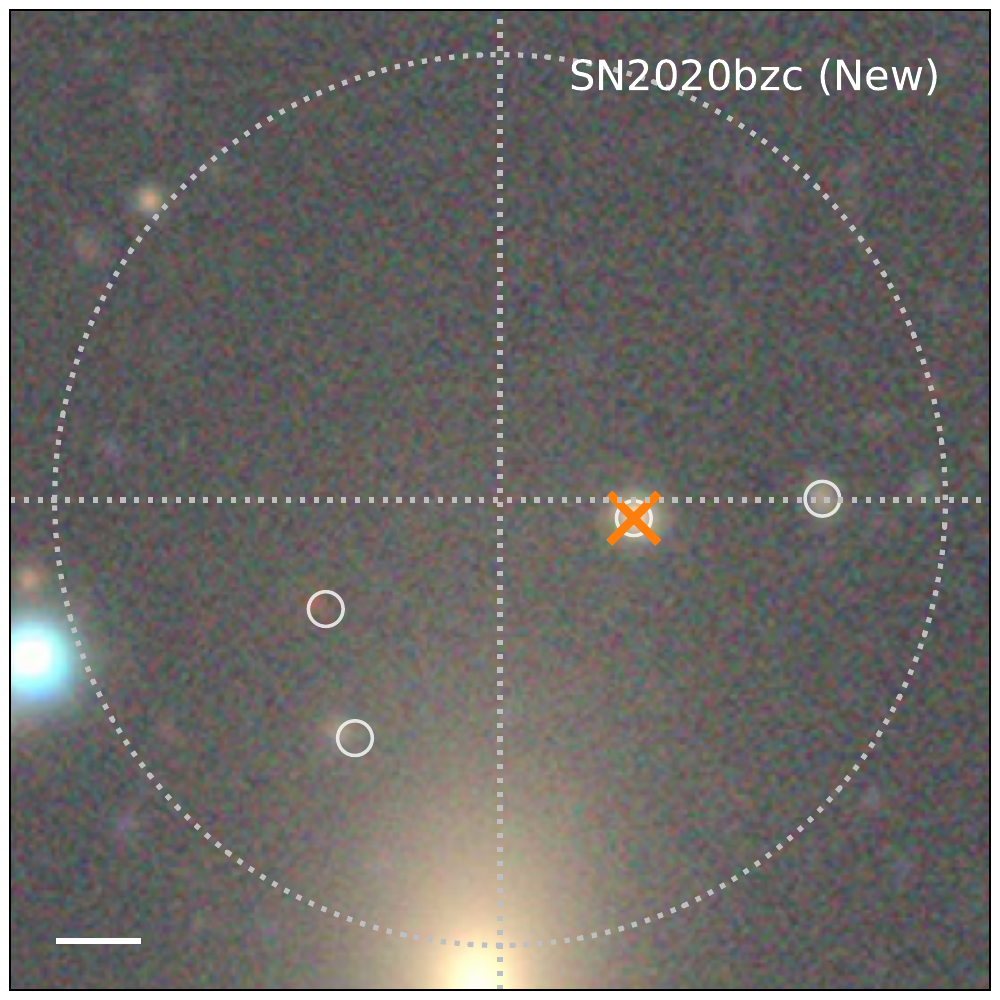}
\includegraphics[width=0.22857142857142858\textwidth]{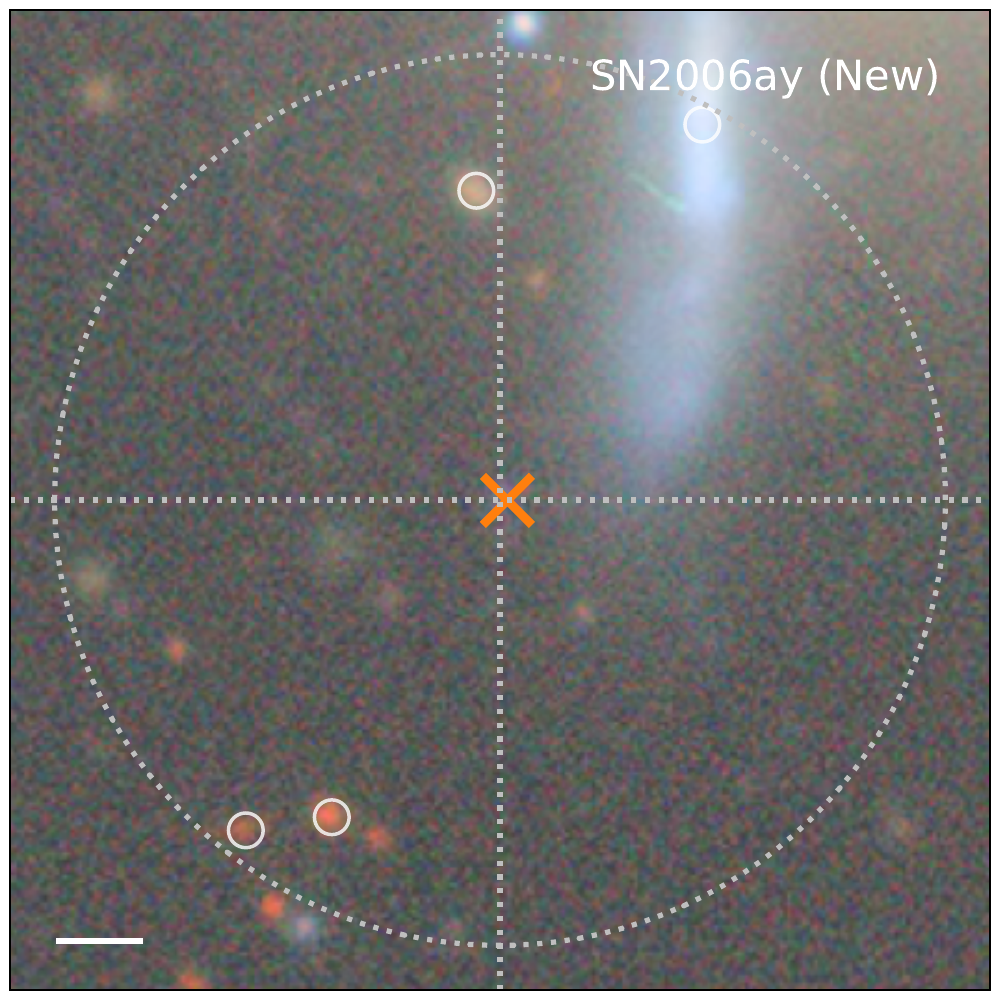}
\includegraphics[width=0.22857142857142858\textwidth]{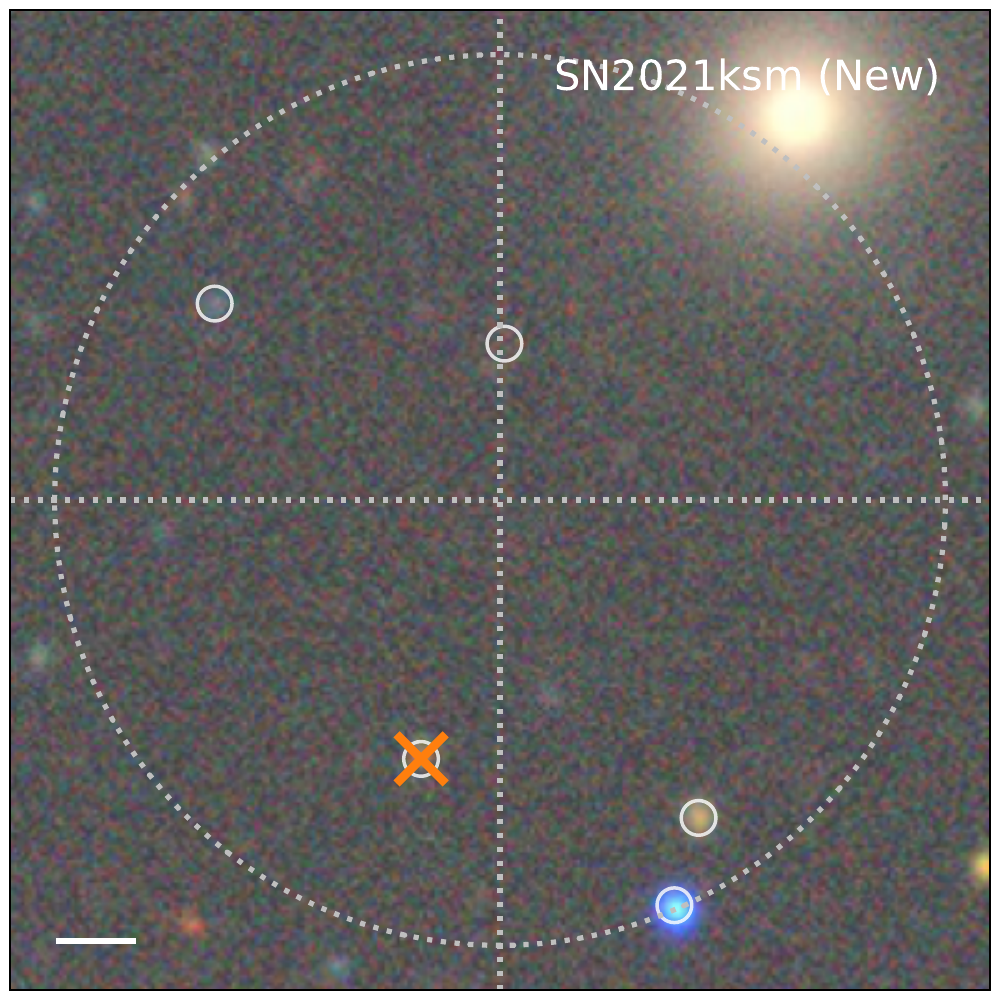}
\includegraphics[width=0.22857142857142858\textwidth]{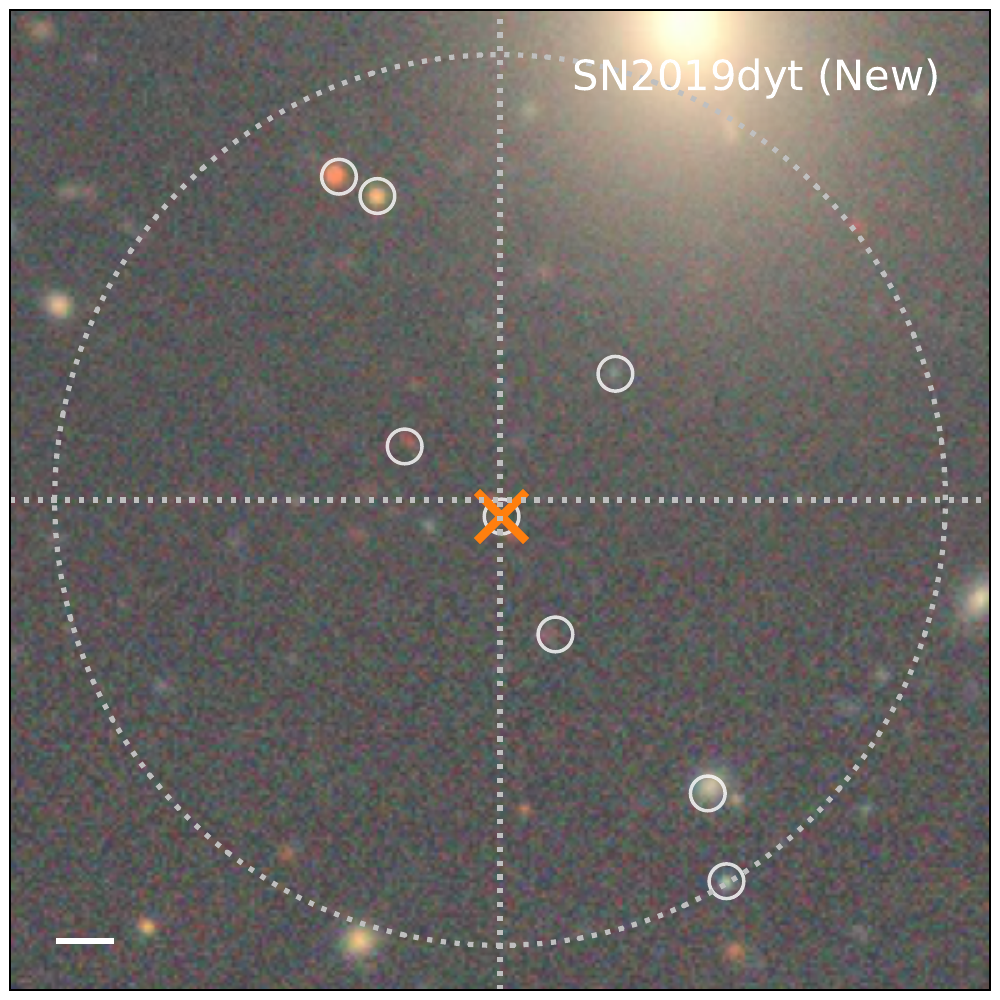} \\
\vskip 0.2cm \hrule \vskip 0.2cm
\includegraphics[width=0.04571428571428572\textwidth]{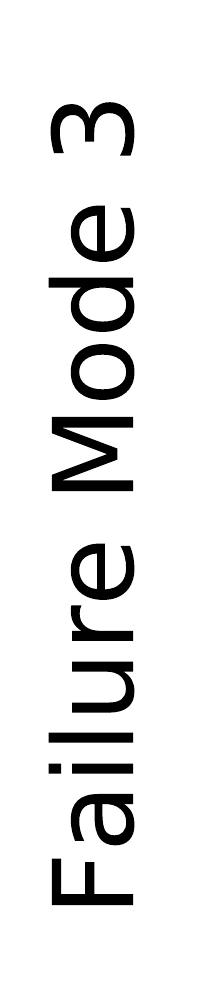}
\includegraphics[width=0.22857142857142858\textwidth]{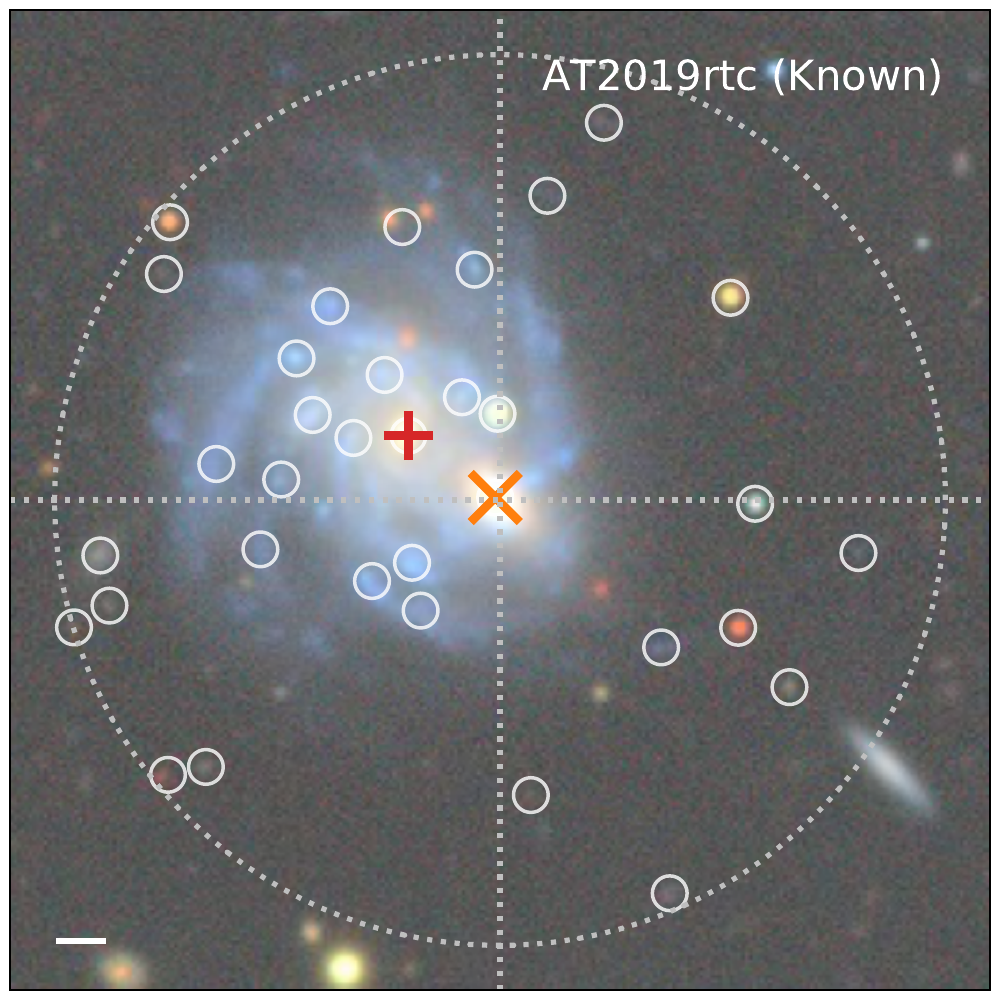}
\includegraphics[width=0.22857142857142858\textwidth]{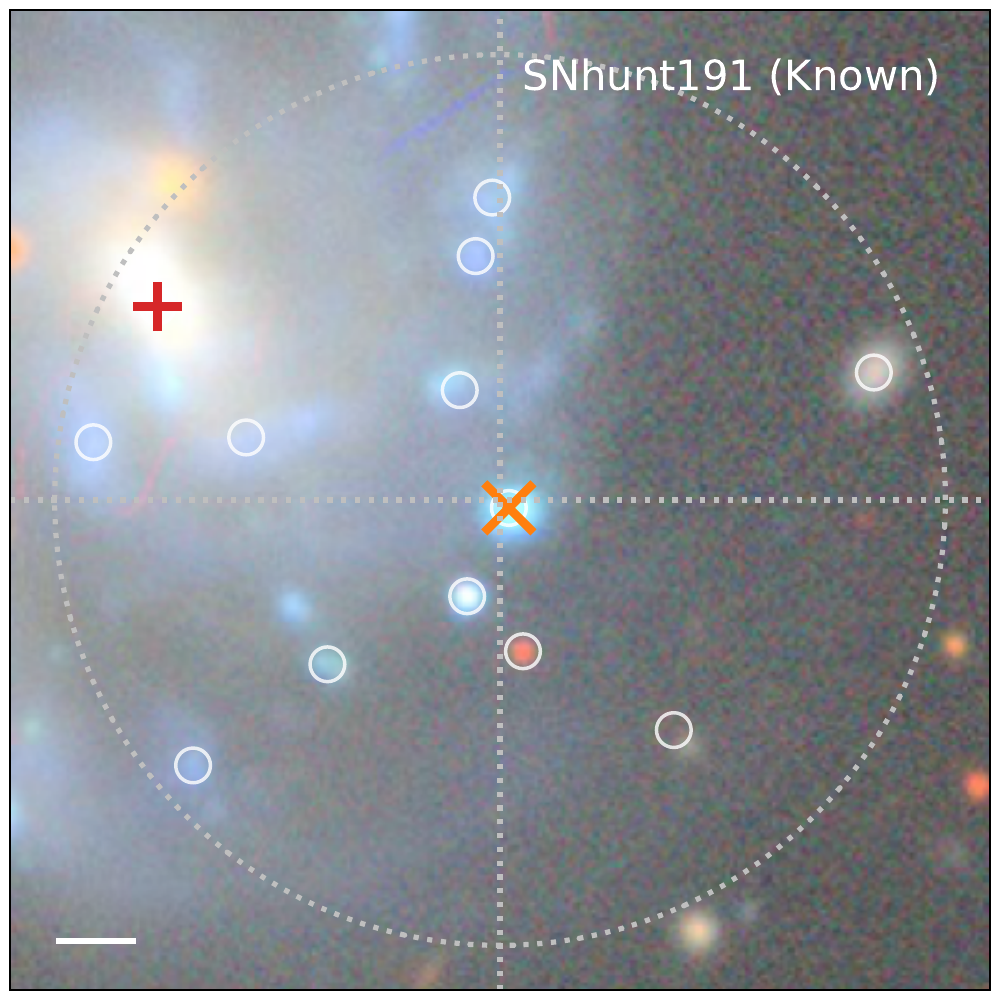}
\includegraphics[width=0.22857142857142858\textwidth]{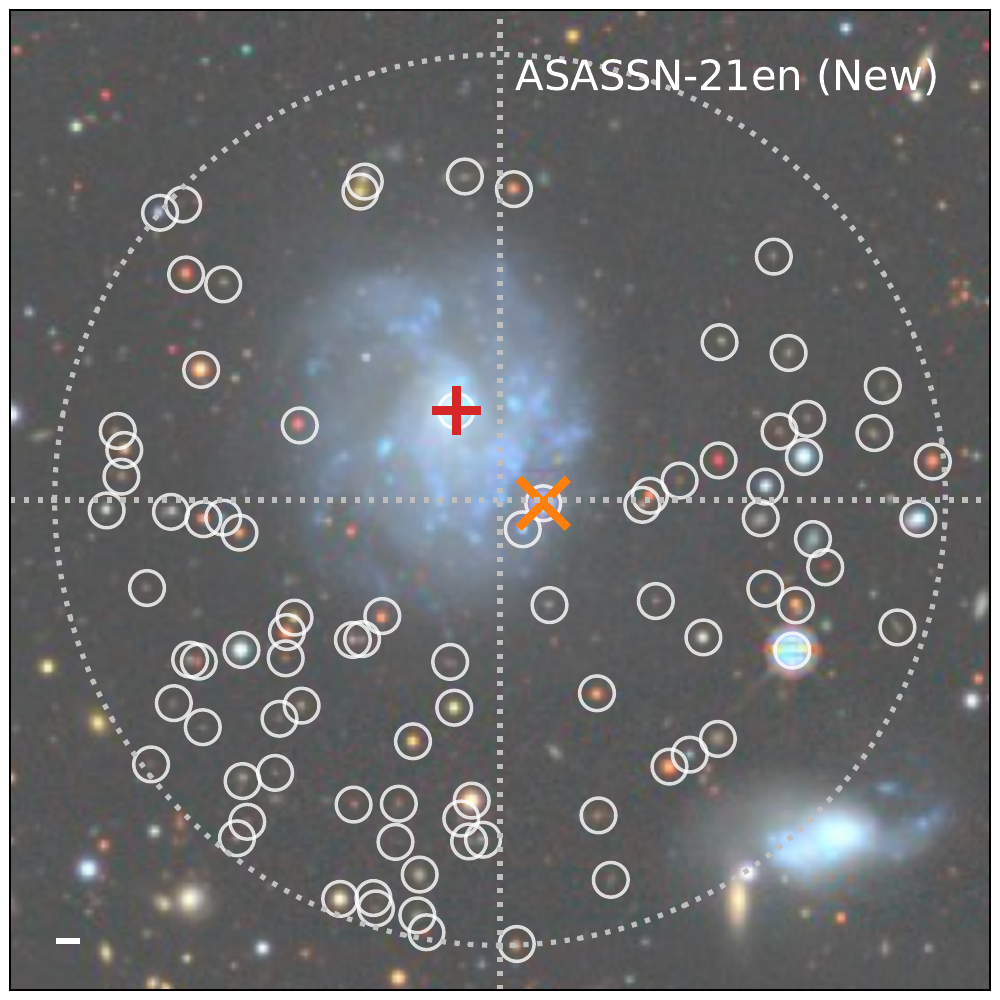}
\includegraphics[width=0.22857142857142858\textwidth]{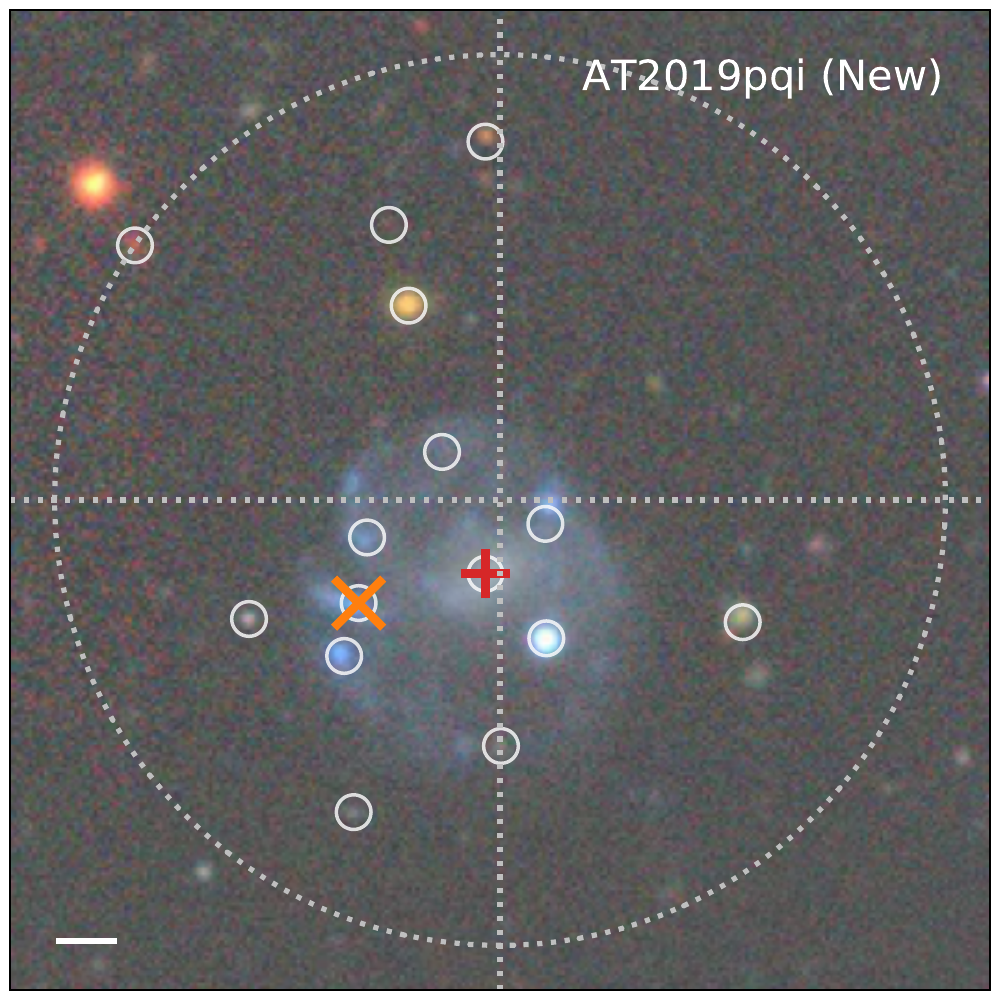}
\vskip 0.2cm \hrule \vskip 0.2cm
\includegraphics[width=0.04571428571428572\textwidth]{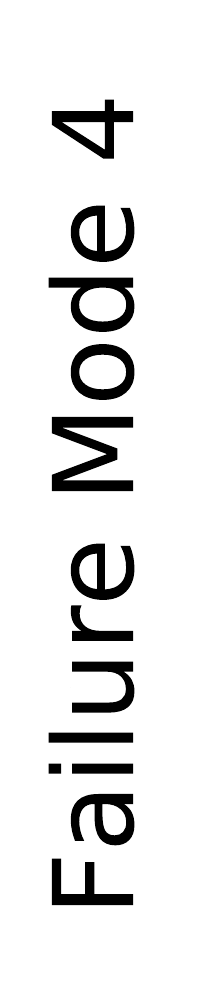}
\includegraphics[width=0.22857142857142858\textwidth]{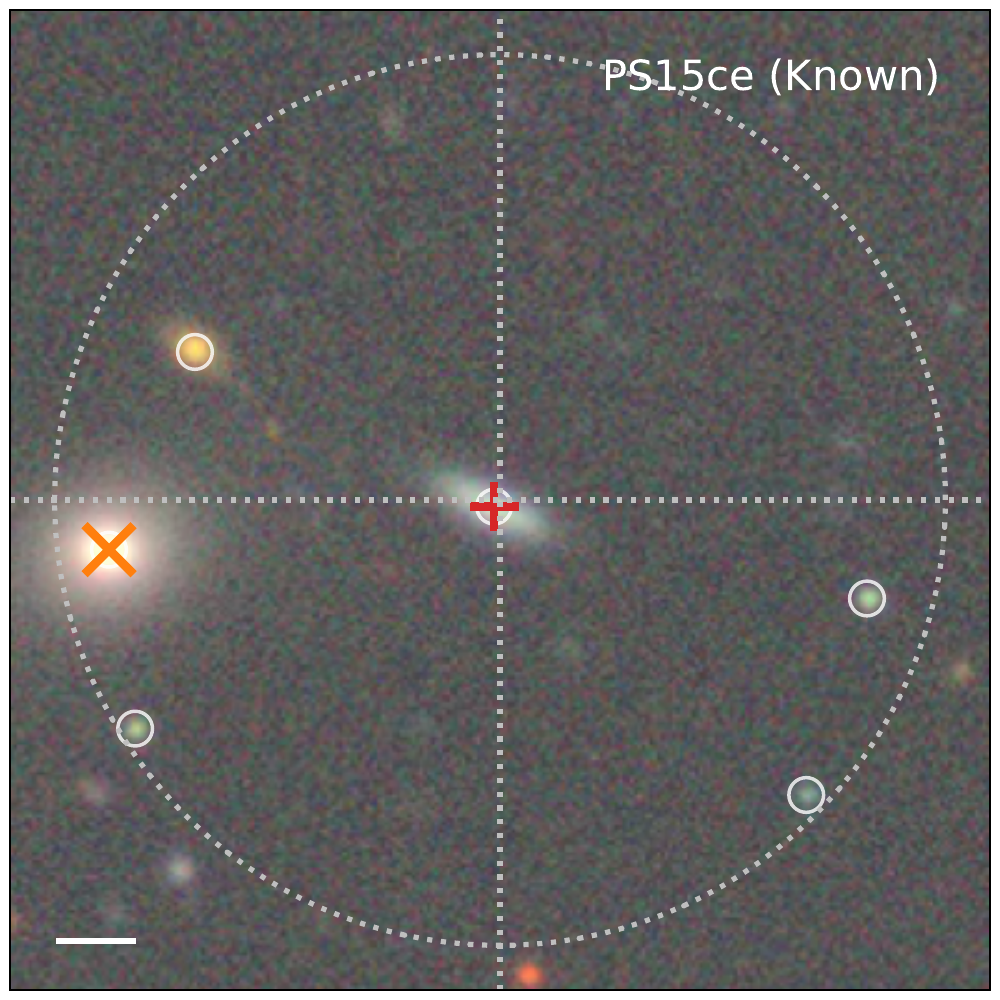}
\includegraphics[width=0.22857142857142858\textwidth]{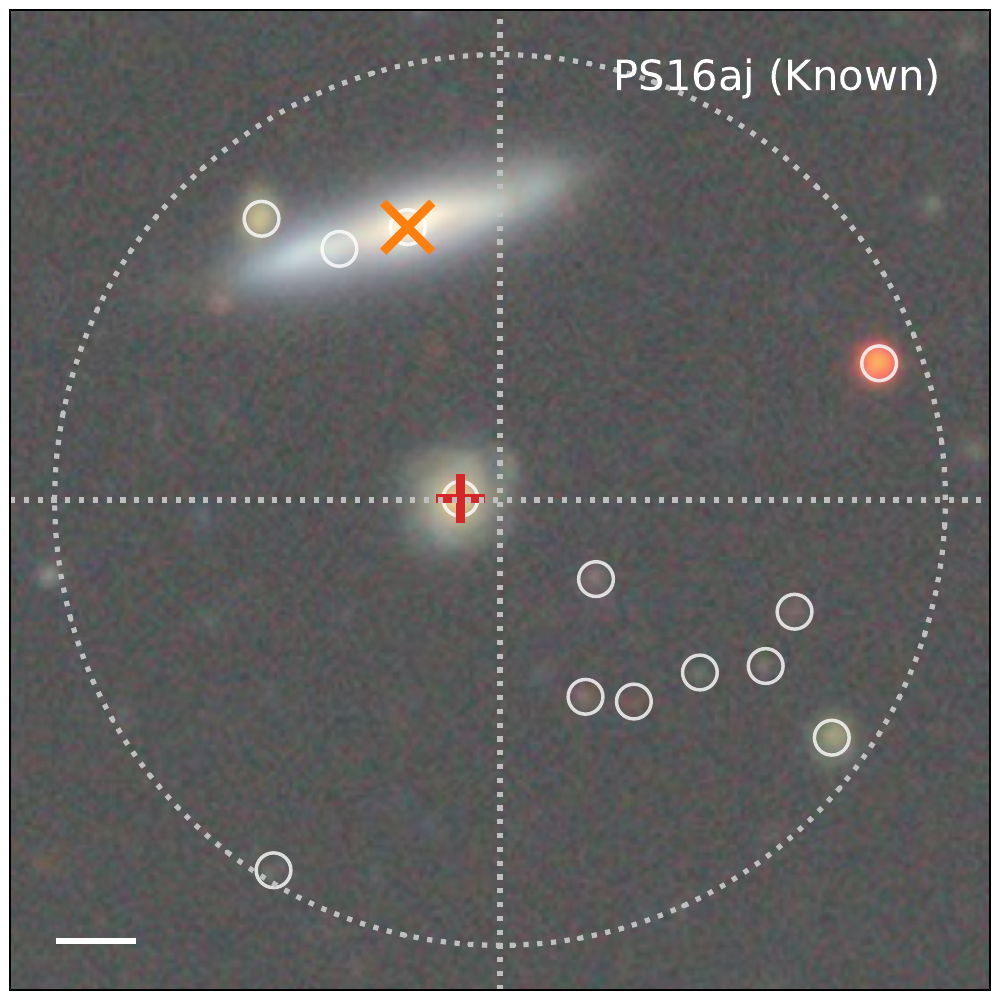}
\includegraphics[width=0.22857142857142858\textwidth]{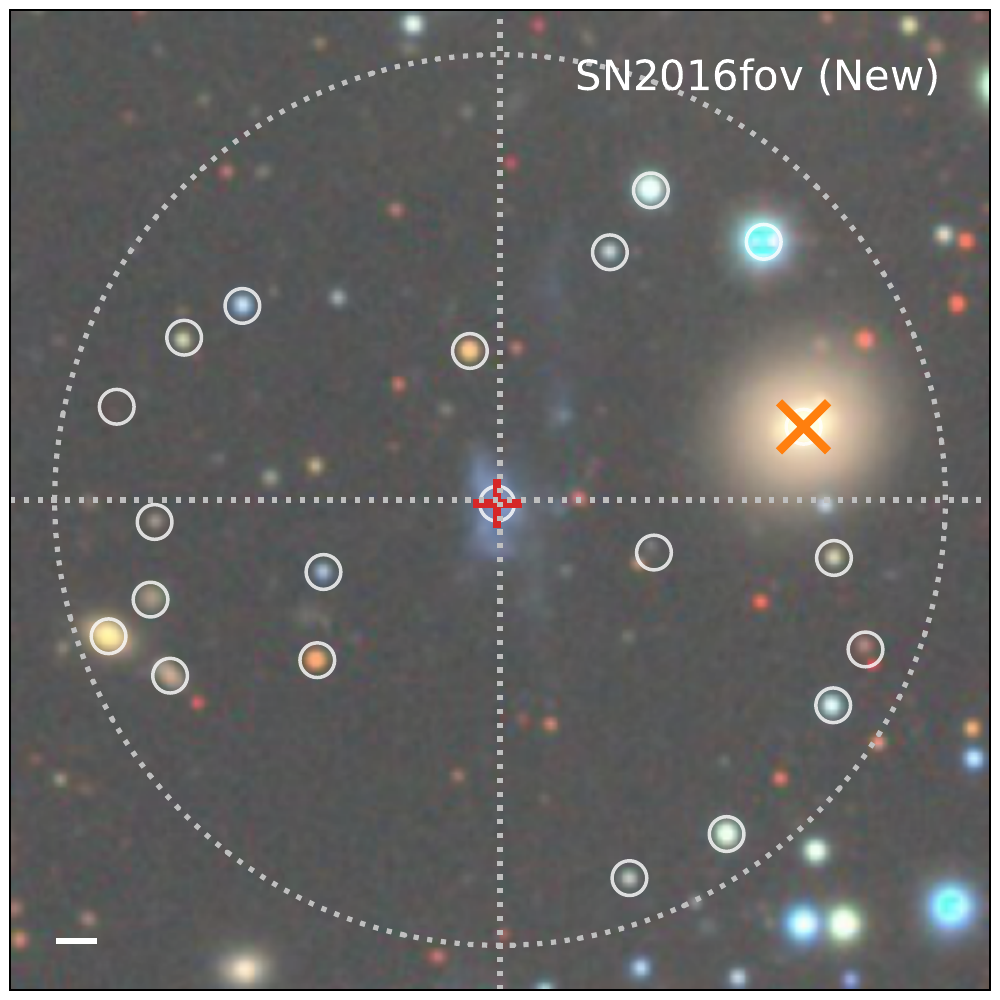}
\includegraphics[width=0.22857142857142858\textwidth]{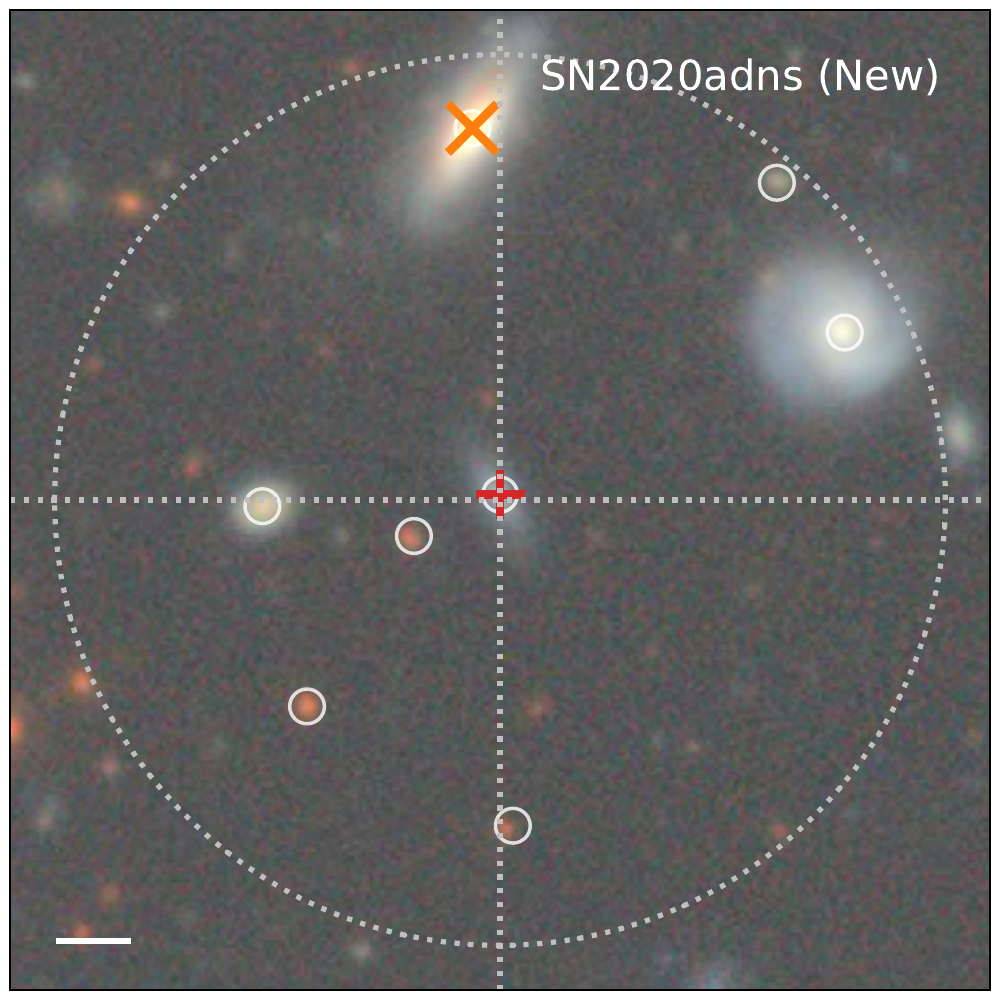}
\caption{
{Examples of incorrectly identified hosts from cross-validation of known hosts and visual inspection of new hosts.}
Each panel shows a field centered at the transient event, with the event name and angular scale of 5'' indicated at the corners.
{Inside the search radius of each event (dotted circle), the best host (either previously reported or visually identified) and algorithm-identified host are highlighted with red and orange crosses, and other cross-matched sources are marked with white circles.}
Each row here represents a common failure mode:
1) inevitable confusion of multiple likely host candidates in the field, in which the true host could be ambiguous even for visual identification;
2) the true host is located beyond the search radius due to an exceptionally large transient-host offset;
3) part of a well-resolved, nearby large galaxy is identified as the host, instead of its main component;
and 4) another cross-matched object, either nearby or far in the background, is erroneously identified as the host at higher ranks.
Events at lower redshifts are more susceptible to the second and third cases.
\label{fig:failedhostxid}}
\end{figure*}

Figure \ref{fig:failedhostxid} shows examples of failed cases, grouped into four common failure modes.
First, in some fields, the true host could be ambiguous given the existing input parameters. Some of these cases can be resolved using more detailed source properties like shape parameters or galaxy redshifts, but some remain challenging even for human observers.
{Second, in some cases, the true host lies beyond the search radius, so another group inside the radius is mistakenly chosen as the host. These are usually low-redshift events with large transient-host angular offsets than typical. Since we exclude such cases for the training set, the examples here are selected from new hosts. Failure like this can be reduced with larger search radii at the cost of higher computational and storage overloads.}
Third, some well-resolved nearby galaxies are split into multiple sources in certain survey catalogs. These parts or substructures of galaxies, when cross-matched into groups, become interlopers that confuse our ranking functions and reduce the accuracy (see also Appendix \ref{appendix:xmatchqc}).
Fourth, the ranking scores can be undesirably boosted or penalized in certain situations, leading to the ignorance of a clear and unambiguous nearby host that other methods may not easily miss.
Here the second and third failure modes are closely related to low-redshift events, which are less of a concern at higher redshifts. However, the other two failure modes are limits of the method itself and can only be resolved using better-tailored ranking functions.

We estimate the accuracy based on the assumption that the true host is always inside the search radius.
In reality, there are occasional cases where the true host is beyond the radius, as shown in Figure \ref{fig:failedhostxid}. Most of the time, the true host is just outside the search radius.
For the training set, we excluded 262 events due to the absence of the true host in the radius. This represents a small fraction compared to the resulted training set (17421). For new hosts, at least 91 events are affected by this issue (Appendix \ref{appendix:inspectionnew}).

\subsection{Notes on undetected hosts}
\label{sec:undetectedhost}

{Our training and testing process implicitly assumes that every transient should have a host. In reality, it is not uncommon to see transients without apparent host galaxies in discovery or archival images.
We adhere to public, wide-field surveys to identify host galaxies, while many hosts, especially those of high-redshift transients (e.g., SLSNe, GRBs), may drop below the sensitivity limits of these wide-field surveys and are thus undetected.
Intergalactic stars may also contribute to such hostless transients.
Their ``host galaxies'' are certainly misidentified and should be excluded from any analysis.}

{We use visual inspection to flag possible hostless transients (Appendix \ref{appendix:inspectionnew}).
The criteria we apply are rather conservative.
We only rely on DESI LS images, or occasionally, PS1 images without stacking artifacts, to identify them.
We avoid labeling an event as hostless if its vicinity (angular separation $\lesssim15''$) contains any possible faint galaxies. These faint and sometimes barely-resolved galaxies are ubiquitous in survey images, and they are generally distant background galaxies per the photometric redshift catalog of \cite{Zhou21}. However, we cannot eliminate the possibility that some of them are actually real host galaxies at large projected distances compared to their sizes.
We check image cutouts of larger field-of-view for low-redshift cases to exclude possible hosts at extreme angular distances.
Finally, if there is neither object in the image nor sources in survey catalogs, but the name of a nearby VAC source clearly indicates the host, we do not label the event as hostless.
We assume that their hosts, if exist, are low-mass galaxies with small sizes. If marginally detectable, they should be excluded by the criteria above. Our inspection leaves $408$ likely hostless events at the sensitivity limits of our background images and external catalogs.}

{These $408$ possible hostless transients include $176$ GRBs, $158$ SNe Ia, $30$ SLSNe, $43$ CC SNe that are not SLSNe, and a few more unclassified events. Since one event can have multiple types, the numbers above do not match the sample size.
More than half of these possible hostless events have redshifts reported, with a median value of $0.4$. GRBs and SLSNe dominate the high redshift side of the population, and the low redshift side ($z<0.1$) consists exclusively of supernovae ($42$ in total).
Therefore, both low-redshift and high-redshift events may contribute to the population of hostless transients here.}

{However, the actual number of hostless transients could be underestimated here. Taking the actually detected GRB hosts in \citet{Hjorth12} for instance. At a limiting magnitude of $R\simeq 24$, typical of modern wide-field sky surveys, only $24\%$ of their hosts are detectable; at a deeper limiting magnitude of $R\simeq24.5$, more than half of their hosts remain undetected. Yet, this does not include the upper limits reported in \citet{Hjorth12}. Our visually identified sample of hostless transient is certainly non-exhaustive.}

{Finally, the confidence score is an alternative, quantitative approach to identify hostless events. These events do not have hosts; as a result, their best-ranking candidates are nearby non-host objects, which are expected to follow the baseline distributions as we discussed in Section \ref{sec:qualitycontrolconfidencescore}.
We show the distribution of confidence scores for possible hostless transients in Appendix \ref{appendix:confidencescore}.}

\subsection{Notes on the workflow}

We discussed the cross-identification of host galaxies under two different circumstances: events with known hosts and events without hosts reported. The procedure to access external catalogs and cross-match catalog sources are similar, as outlined in Figure \ref{fig:fullworkflow}.
In fact, we combine these two parallel workflows into a single pipeline. Accessing and compiling properties of known hosts is just a trivial case of the entire procedure, where the part of ranking host candidates can be skipped.
The entire pipeline is summarized as pseudo-code in Appendix \ref{appendix:xidcode}.

In Figure \ref{fig:xidexample}, we show some typical host galaxies that went through our pipeline, including both known hosts and new hosts. These hosts are randomly selected near a few redshift values as examples.
Even having more than 20 catalogs with non-negligible catalog-to-catalog offsets of sky coordinates, our cross-matching algorithm works reasonably well for hosts across the entire redshift range.

Similar to the transient collection, we store the compiled properties of host galaxies in a separate data collection, indexed by the unique identifier of each event.
For events with confirmed hosts, either by name or by coordinate, we store the properties of the cross-matched host and other cross-matched groups.
When there is no confirmed host besides the properties of the primary candidate, we also store the properties of other cross-matched groups (``secondary candidates'') in descending order by their ranking score in the field.
Note that when the as-reported host coordinates match a stellar object, the group is marked as a primary candidate instead of a ``confirmed-by-coordinate'' host (Section \ref{sec:knownhosts}). Otherwise, the group with the highest-ranking score becomes the default primary candidate.

We retrieve and compile host properties separately for each event, and consequently, there could be duplicates of hosts if several transients have been detected in the same galaxy. 
However, this is a relatively uncommon situation, as only 1886 known hosts in our database have more than one transient. Therefore, we do not index these cross-matched hosts uniquely to eliminate duplicates at this moment.
Our statistics and analyses are also based on the idea of transient-host pairs rather than individual hosts.

\begin{figure*}[h]
\centering
\includegraphics[width=0.49\linewidth]{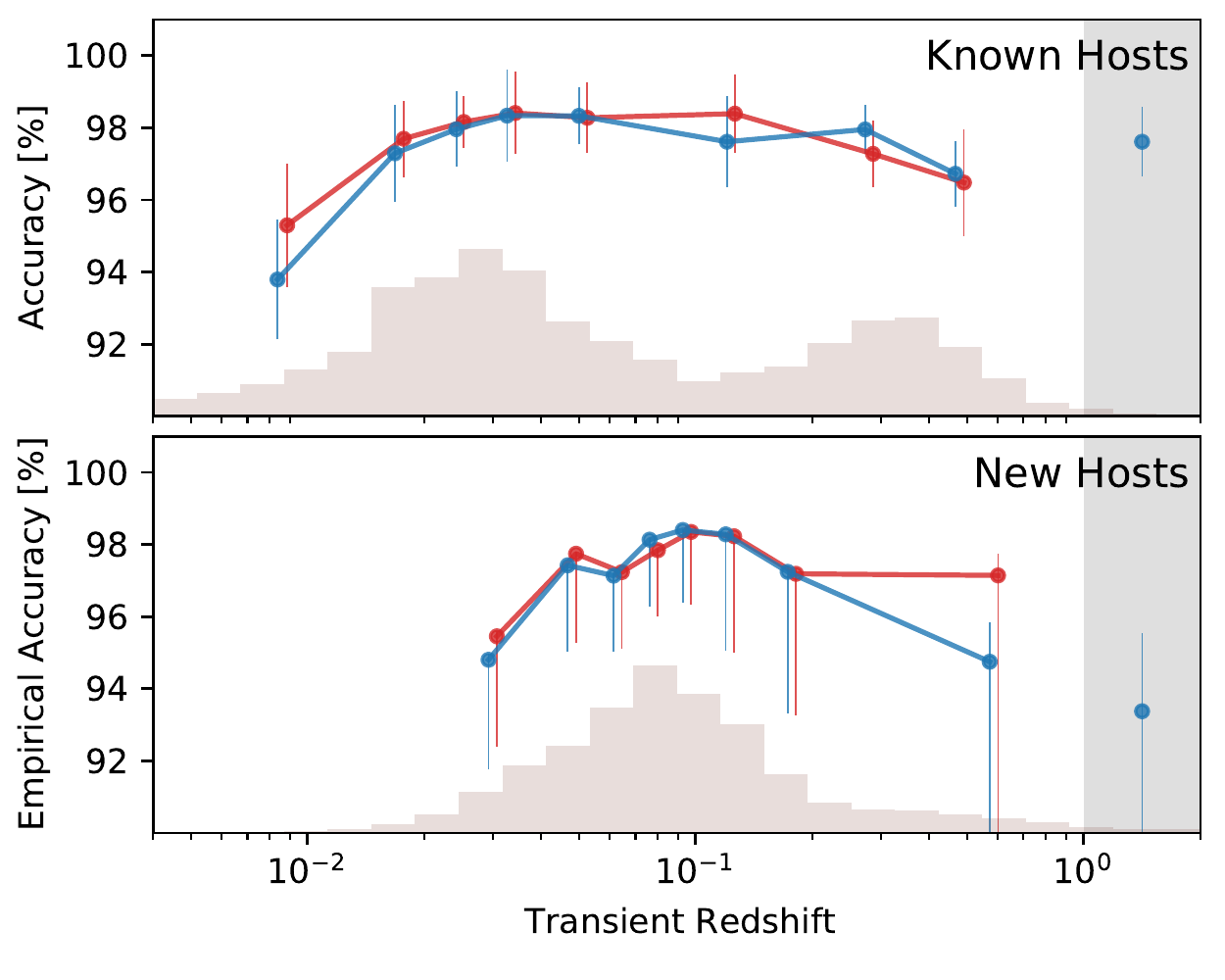}
\includegraphics[width=0.49\linewidth]{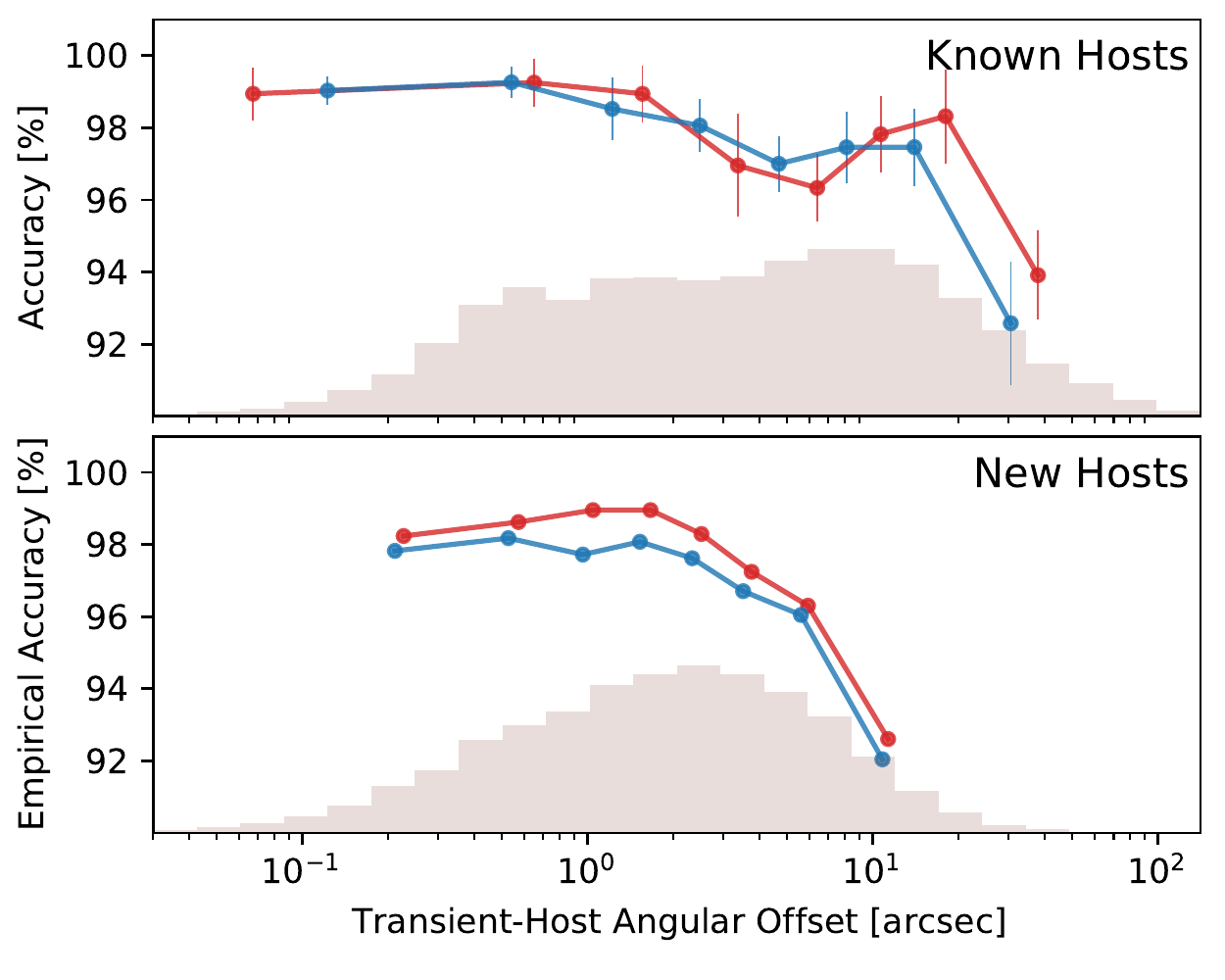} \\
\includegraphics[width=0.49\linewidth]{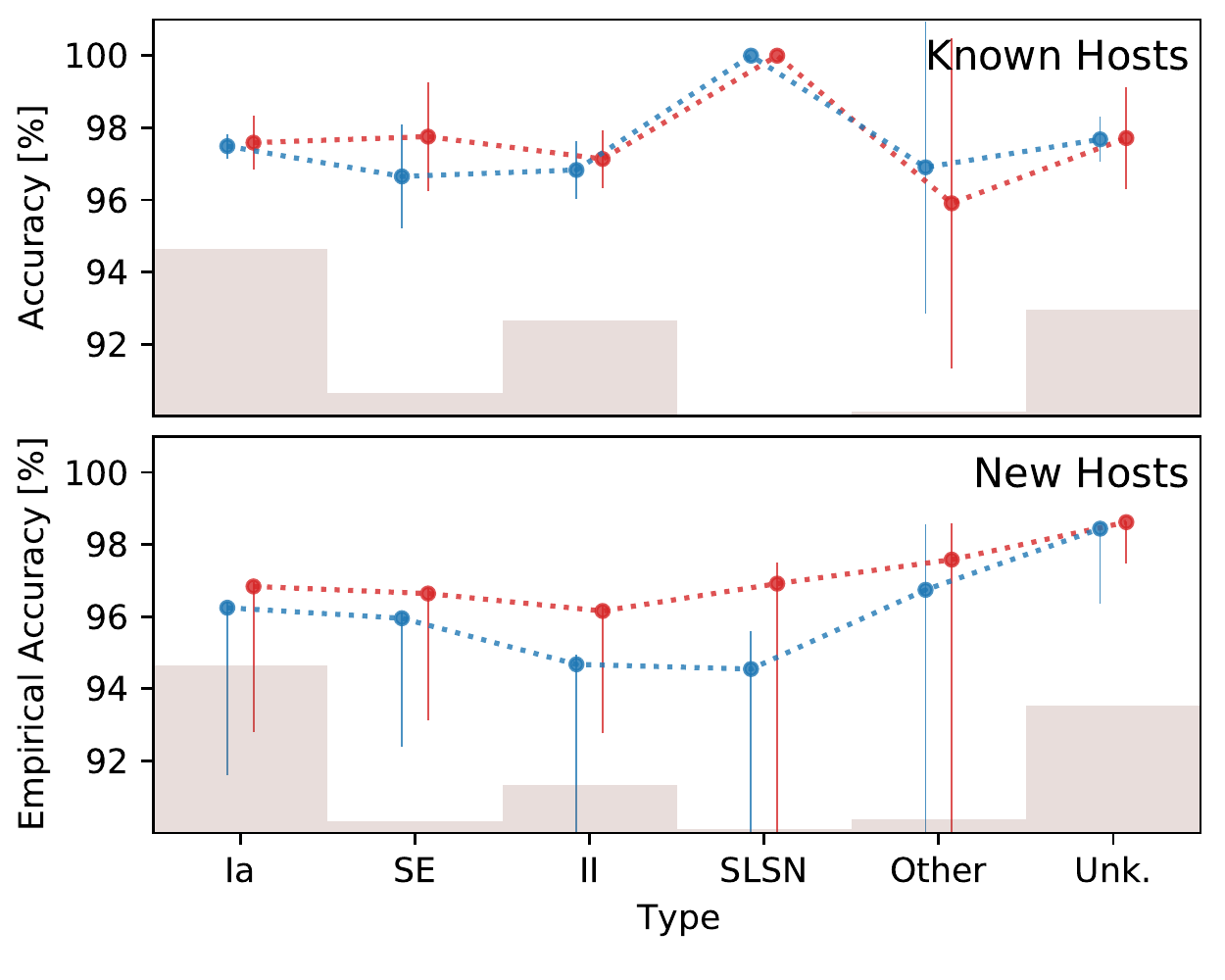}
\includegraphics[width=0.49\linewidth]{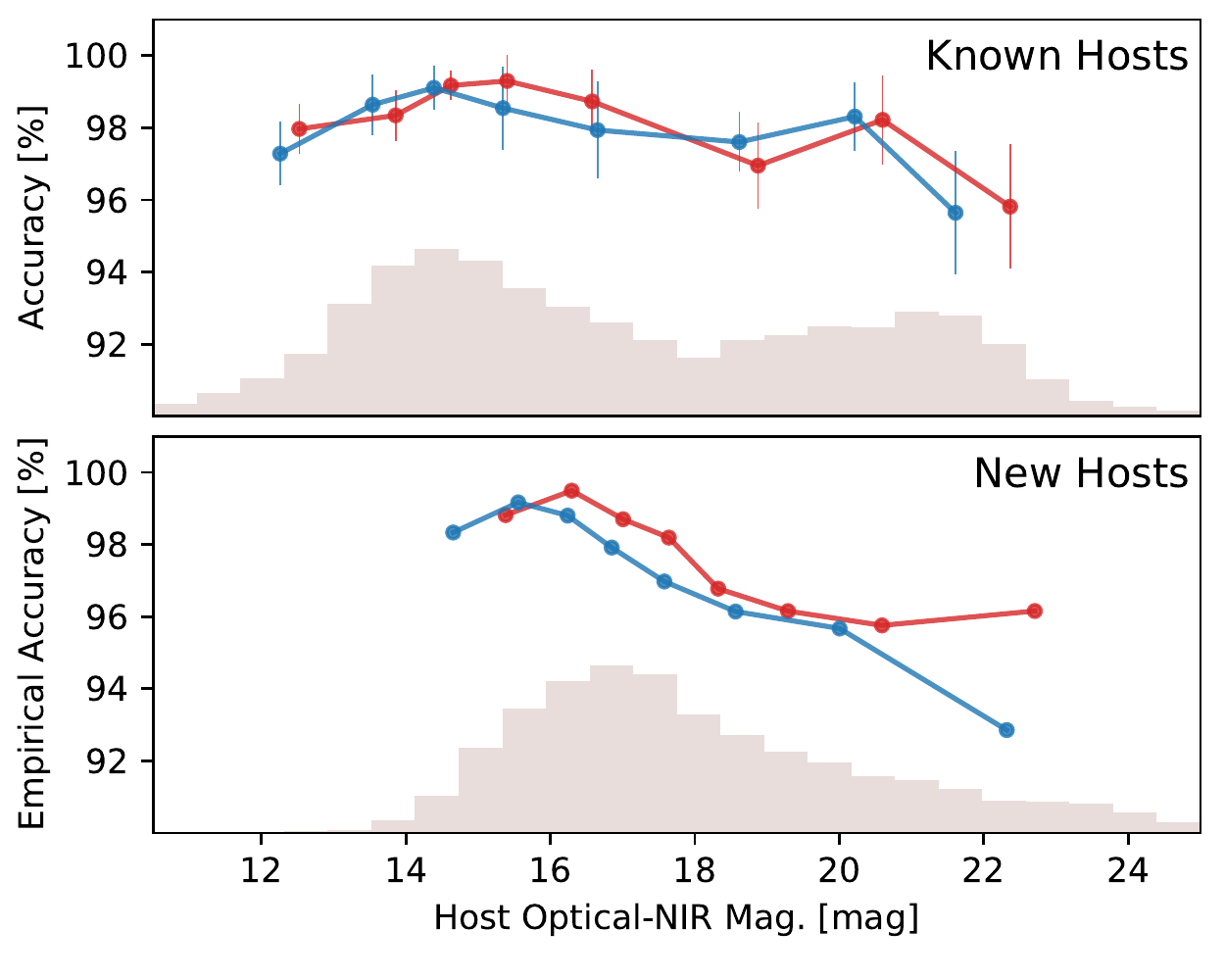} \\
\includegraphics[width=0.98\linewidth]{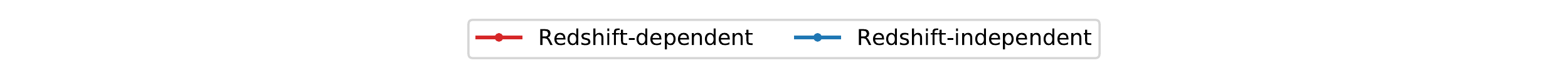}
\caption{
{The accuracy of host cross-identification with our default ranking functions, using redshift-dependent variables (red) or not (blue). The four panels here show the dependence of accuracy on transient redshift, transient-host angular offset, transient type, and host optical-NIR magnitude.
Except for the panel of transient type, we split the sample into bins of equal number counts along the axis of interest and plot the per-bin accuracy at the median value of the bin.
We estimate the accuracy in two separate ways.
With the sample of known hosts (i.e., the training set), we calculated the average accuracy using 10-fold cross-validation, where the standard deviations are indicated as error bars.
For the sample of new hosts, we estimated an \textit{empirical} accuracy  with a visual inspection, defined as the fraction of visually identified hosts that are also chosen by our algorithm. Vertical ``error bars'' outline the range of accuracy due to the uncertain cases (Appendix \ref{appendix:inspectionnew}).
We plot the distribution of both samples as shaded areas in each sub-panel. The densities are over the logarithmic axis for redshift and transient-host angular offset, so the histogram area corresponds to the sample size.
For the dependence on redshift, we also plot the subset of points without redshift in the gray shaded area of the first panel.} 
\label{fig:ridgeaccuracy}}
\end{figure*}

\subsection{Comparison with similar works} \label{sec:hostremark}

The technique to cross-identify host galaxies and the data products presented in this work is unique in many aspects.
Here we compare our method and dataset with similar works in the literature and discuss the advantages and drawbacks of different approaches.

\subsubsection{The Directional Light Radius Method}

The Directional Light Radius (DLR) method \citep[e.g.,][]{Sullivan06, Gupta16, Sako18}, which relies on galaxy shapes and sizes to identify hosts, is probably the most widely used technique in previous works.
The angular distances of potential host galaxies to a transient are normalized by their elliptical radii in the direction of the transient (i.e., the DLR).
The resulted parameter ($d_{\mathrm{DLR}}$) is a comparison of transient-galaxy angular distance and galaxy angular size, taking axis ratio and position angle (inclination or projection effects) into consideration. Nearby galaxies with smaller $d_{\mathrm{DLR}}$ are more likely to be the host.
\footnote{In this work and in \citet{Gupta16}, $d_{\mathrm{DLR}}$ refers to the normalized transient-galaxy angular distance by DLR. In \citet{Gagliano20}, $d_{\mathrm{DLR}}$ refers to the DLR itself, and $\theta/d_{\mathrm{DLR}}$ is the normalized angular distance (Eq. 9).} 
The DLR method implicitly assumes that transients in galaxies have a similar radial extent as the stellar light, which is reasonable for most stellar explosions.
Calculating $d_{\mathrm{DLR}}$ requires no more than the basic shape and size parameters of galaxies, making the method easy to implement and thus popular for similar studies.

The method also has some clear drawbacks.
First, shape and size parameters, such as moments of light, are method- and data-dependent. The resulted $d_{\mathrm{DLR}}$ are thus comparable only within the same survey. Such parameters are also noisy for low signal-to-noise detections or marginally-resolved galaxies, leading to significant errors in the estimated $d_{\mathrm{DLR}}$.
Second, even assuming that transients follow stellar light and galaxy isophotes are elliptical, the dramatic differences in radial light profiles (or concentration of light) remains a source of bias.
For example, the effective radius ($r_{50}$) is commonly used to calculate $d_{\text{DLR}}$, but the fraction of enclosed light in units of $r_{50}$ depends on the light profile, even for the simplest models.
At four times the effective radius, the fraction of enclosed stellar light is $99.1\%$ for an exponential disk, but for de Vaucouleurs' profile, the fraction is only $84.7\%$. 
The same $d_{\text{DLR}}$ value, therefore, corresponds to a different fraction of enclosed light and hence the probability of being (or not being) the true host.
{Third, the comparison of $d_{\mathrm{DLR}}$ implicitly assumes uniform transient rate and detection efficiency across galaxies, which may become an issue when there are multiple candidates with comparable $d_{\text{DLR}}$.}
Finally, for many photometric catalogs, there is a specific issue with nearby large galaxies. Most photometric pipelines are not optimized for galaxies of large angular sizes. Their resolved sub-structures, such as clumps, spiral arms, and even massive star clusters can be broken into individual sources by profile-fitting or de-blending algorithms. Such substructures may impede host identifying, where the source with the smallest $d_{\text{DLR}}$ could be a substructure instead of the galaxy core. %

{The DLR method is not immediately applicable to our entire transient sample due to the heterogeneity in transient types, redshift range, and catalog coverage. For example, shape and size parameters required by the DLR method are not always provided in external catalogs. Even when available, they are in diverse representations and are survey-dependent.}
{However, for some specific catalogs, we implement the method. We test the accuracy when only $d_{\text{DLR}}$ is used to rank host galaxies. We also include this $d_{\text{DLR}}$ parameter in some feature sets and analyze the performance.
Only using $d_{\text{DLR}}$ to rank nearby groups, we can recover nearly $90\%$ of known hosts. The implied $10\%$ misassociation rates is higher than the estimate of $5\%$ in \citealt{Gagliano20}, while close to the mock sample performance in \citep{Gupta16}.
On the other hand, feature sets including $d_{\text{DLR}}$ achieve higher accuracy than the default ranking functions. However, the improved performance may not be fully attributed to this single parameter.}

\subsubsection{The Gradient Ascent Method}

The recent work of \citet{Gagliano20} represents a more sophisticated genre of technique, where the gradient of surface brightness extracted from image cutouts, instead of cataloged source properties, is used to associate transients with potential hosts.
They locate the peak of surface brightness (i.e., the core of the potential host) in the image cutout and use the cataloged source at the peak position for host properties.
Finding the peak of light in an image is a non-trivial task in the presence of foreground stars and resolved galaxy substructures.
They remove point sources, smooth out the fluctuations over the extended light component, and construct the gradient field for the light. After then, starting from the transient location, the algorithm ascends to the core of a nearby galaxy following this gradient field.
This method works remarkably well for low-redshift host galaxies with resolved substructures where the conventional DLR method could easily fail.

However, the technique requires direct analysis of image cutouts, potentially with higher computational costs. The quality and availability of image data could also be a constraint of its applicability.
Also, the hyper-parameters in the entire workflow, such as the smoothing scale and the star/galaxy separation criteria, must be tuned for each survey.
Finally, in analogous to an optimization problem, the chance remains that the current solution (i.e., the peak of surface brightness) is only a ``local minimum'' instead of the global best solution, or the solution converges in the wrong direction.

\subsubsection{The Ranking Function Method}

In this work, we introduce a machine learning-based host candidate ranking technique.
We characterize potential hosts by various parameters, including basic parameters that are universally available for every cross-matched group, and optionally, detailed source properties in external catalogs.
These parameters are then fed into a ranking function, which sorts these potential hosts by their estimated possibility to be the right host.
The ranking function is trained using known, properly cross-matched, and visually inspected host galaxies in our dataset, essentially turning the problem into a regular classification or regression problem.

Using a \textit{trained} algorithm instead of a designed one based on certain assumptions is the primary difference of our approach compared to other methods.
Those already identified hosts, including non-host objects in the same field, are used to train ranking functions and objectively evaluate their performance.
A trained algorithm generally uses fewer assumptions and is thus less vulnerable to biases inherited from the imposed assumptions.
The method is also extensible and flexible. Any characteristics of potential hosts can be used as input, and many conventional classification and regression algorithms can act as the ranking function. Actually, this is a framework or a family of methods.
However, the method relies on a well-constructed and sufficiently large training set, which may not be available for other similar problems.
{One may create a training set using mock transient-host pairs as did in \citet{Gupta16}, but the assumptions made when generating mock transient-host pairs could implicitly bias the results.}
We also assume that the cases to which the trained ranking function is applied are similar to those in the training dataset. {This is less of a concern since the new hosts have empirical accuracy comparable to the accuracy from cross-validation, but the transferability of the trained model should be tested in other similar applications.}

We provide a matrix of trained ranking functions, but the default ranking functions we used deserve further discussion.
To ensure the completeness of input variables, the default ranking functions only take the basic characteristics of cross-matched groups that are independent of which catalogs have been matched. Detailed source properties in external catalogs, which proved to improve the accuracy, are not used. Therefore, the behavior of the default ranking functions might be partly driven by non-physical factors.
We use a simple Logistic Regression classifier here. Such a simple linear model may suffer from underfitting, but the model is also insensitive to outliers in the training set and systematic variations in some variables. In other words, a simple linear algorithm guarantees the robustness of the trained model, but it may also lack the capability to differentiate multiple likely candidates.

\subsubsection{Single-catalog vs. multiple-catalog techniques}

{We choose hosts from cross-matched groups, where each group may contain sources from multiple external catalogs. This multi-catalog approach, driven by our motivation to provide rich host properties across multiple surveys, clearly differs from previous works.}
Our method and dataset naturally provide better coverage of host properties over a wide range of wavelengths, robustly cross-matched across various catalogs, but the downsides are also clear.
To begin with, cross-matching multiple catalogs, which itself is a challenging problem for galaxies with a wide range of angular sizes and redshifts, becomes an extra step.
{More importantly, due to the sensitivity limits and partial sky coverage of surveys, cross-identified hosts cannot have complete source properties across these 21 external catalogs.
Consequently, there is a trade-off between the accuracy and applicability of training functions if one plans to use detailed source properties as input parameters. Adhering to those universally available basic features avoids the problem, but the performance could be driven by non-physical factors.}
Also, the compiled dataset could be sparse, especially for high redshift hosts or hosts beyond the coverage of major sky surveys.
On the contrary, single-catalog techniques can fully utilize the existing source properties when finding host galaxies. Host properties compiled in this way are also consistently measured and are only subject to the selection function of one survey. However, the data provided are always limited without further cross-matching with other catalogs.

\subsubsection{Comparison with Gagliano et al. (2020) dataset}

Finally, besides the different methods to identify host galaxies as we discussed above, it is also worth comparing the publicly available dataset presented here with the \texttt{GHOST} database maintained by \cite{Gagliano20}.
First, our work includes more transient types, such as TDEs, GRBs, and other rare events, with fine-grained taxonomy, while \citet{Gagliano20} focus mainly on spectroscopically-confirmed supernovae.
Second, our transient sample includes \NStatConfirmed events with known and cross-matched hosts and \NStatPrimaryCand new hosts. The larger sample size compared to the \texttt{GHOST} database is mainly due to our different sample selection criteria and the inclusion of other transient types.
Third, we choose to trust the reported hosts when available and only use the algorithm when there is no usable host information, while \citet{Gagliano20} perform consistent catalog-based and image-based host matching for all their supernovae.
Fourth, the dataset presented here includes cross-matched properties in 21 catalogs, while \citet{Gagliano20} provides source properties mainly in the Pan-STARRS catalog, with limited data in other catalogs. We also include properties of secondary host candidates.
Finally, besides the static version of the database in various formats, at the moment, we do not provide visualization and data access tools as \citet{Gagliano20} did. This could be a part of our future updates.

 \onecolumngrid \clearpage
\begin{sidewaysfigure*}[t]
\centering
\includegraphics[width=0.2784\textwidth]{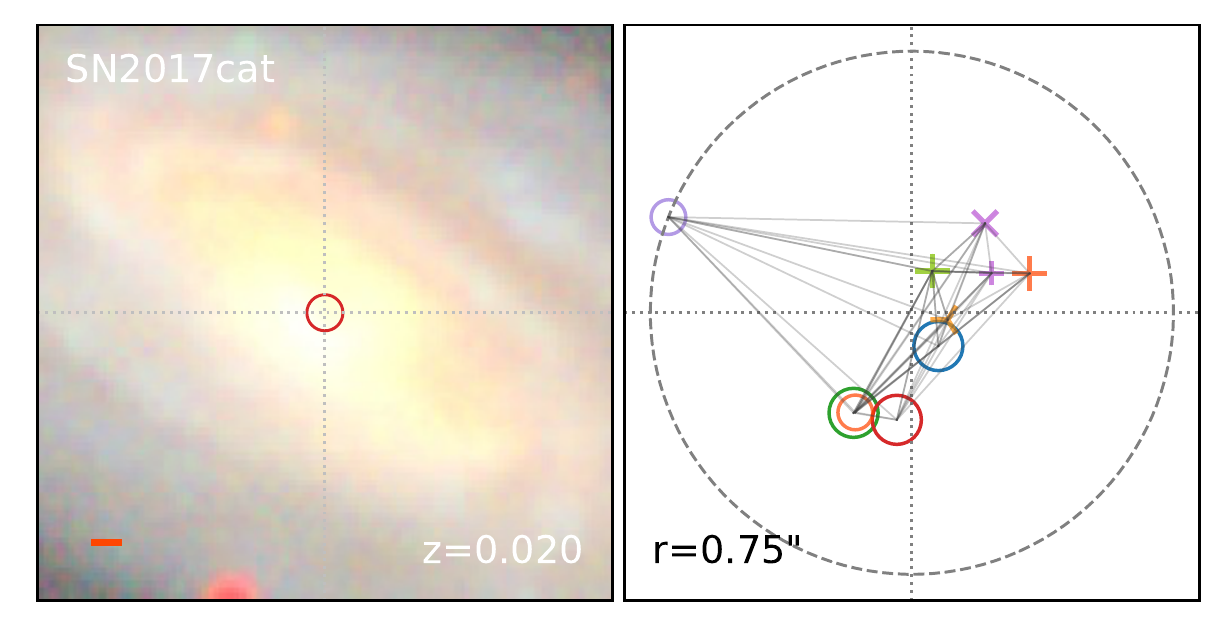}
\includegraphics[width=0.2784\textwidth]{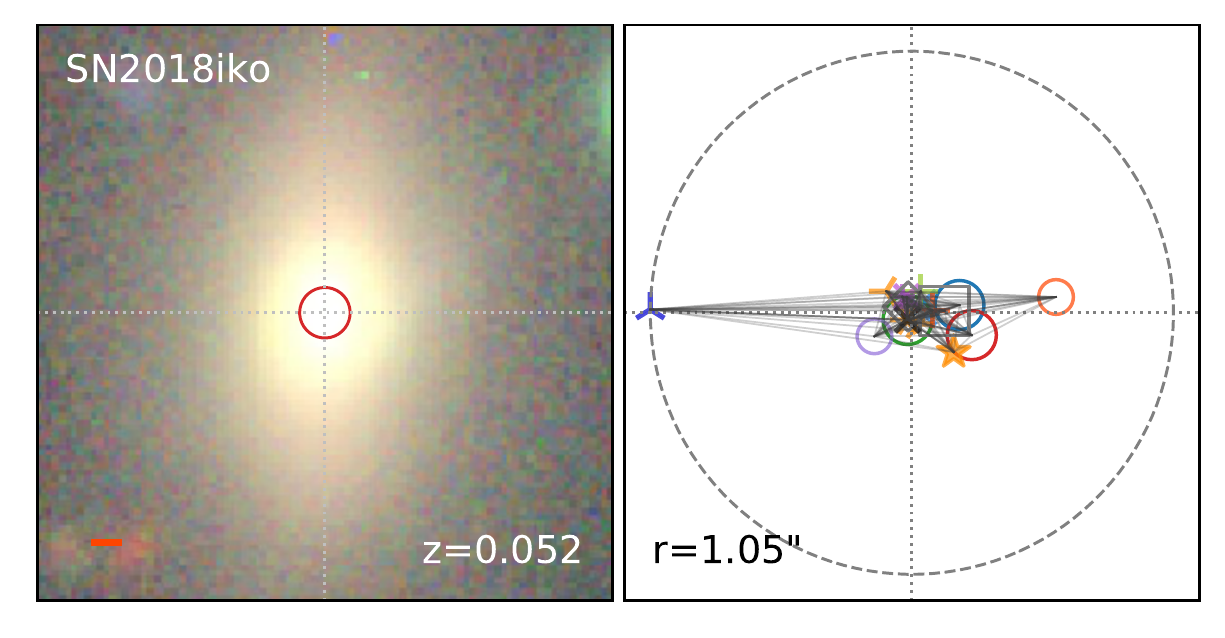}
\includegraphics[width=0.2784\textwidth]{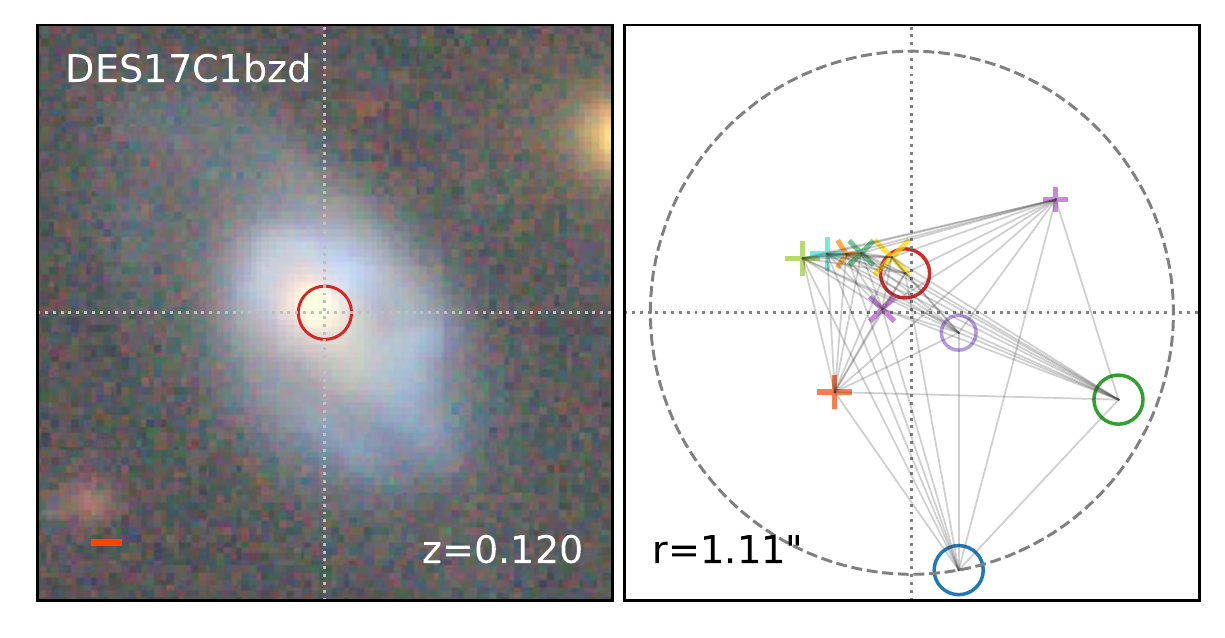}
\includegraphics[width=0.1449\textwidth]{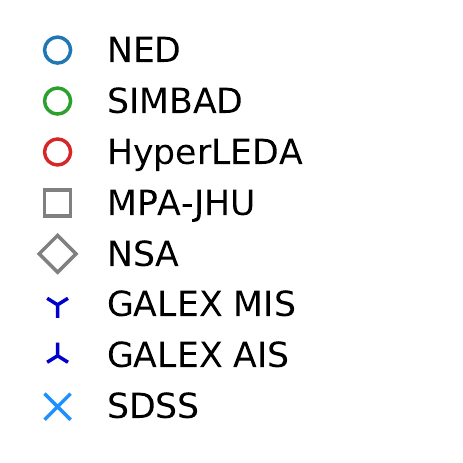} \\ 
\includegraphics[width=0.2784\textwidth]{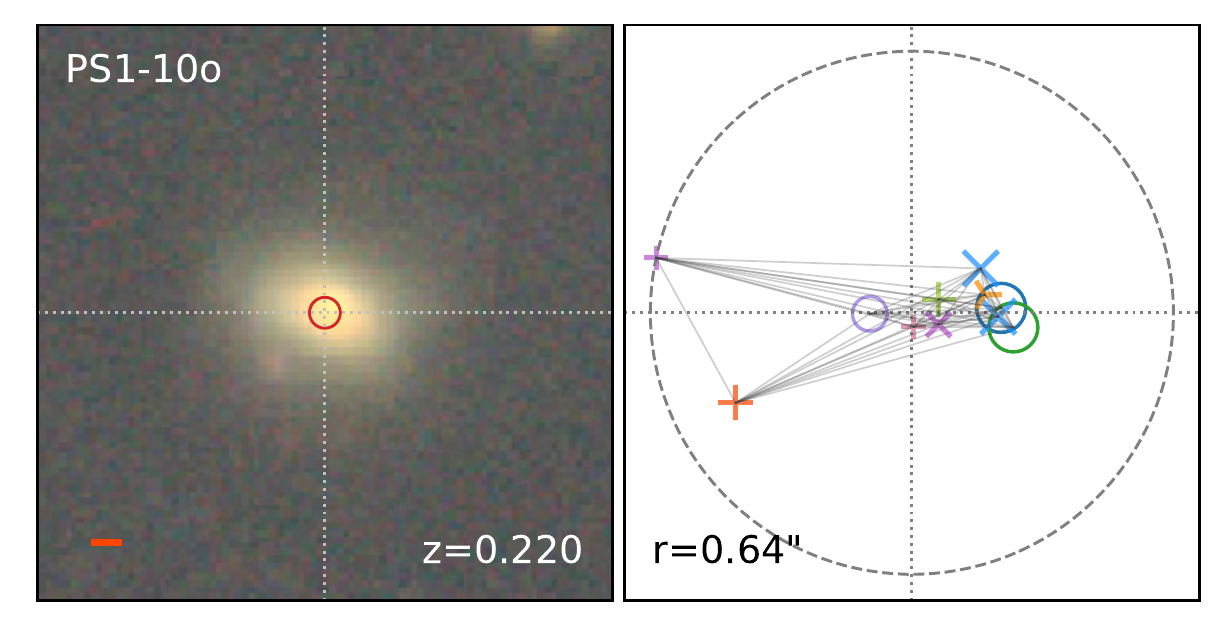} 
\includegraphics[width=0.2784\textwidth]{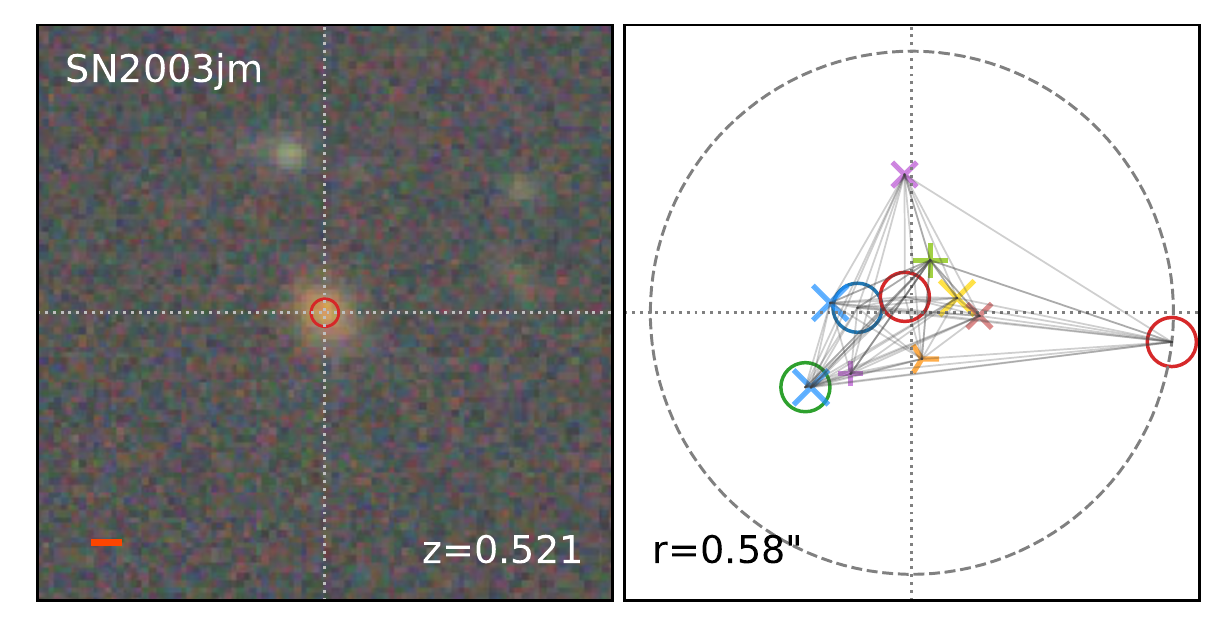}
\includegraphics[width=0.2784\textwidth]{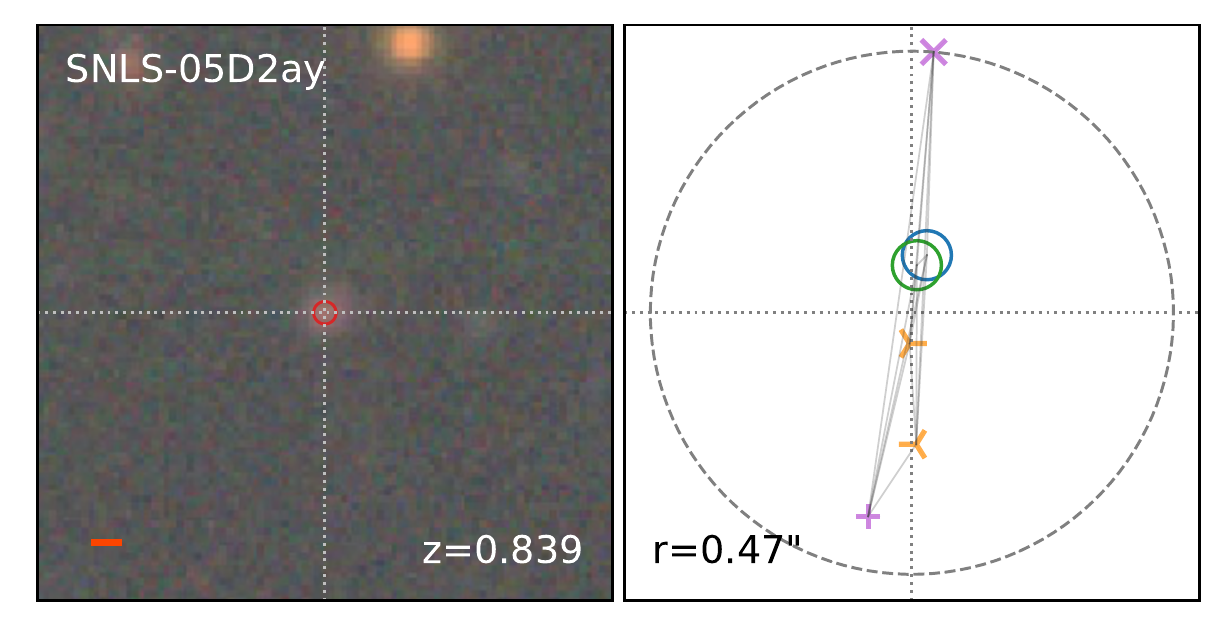} 
\includegraphics[width=0.1449\textwidth]{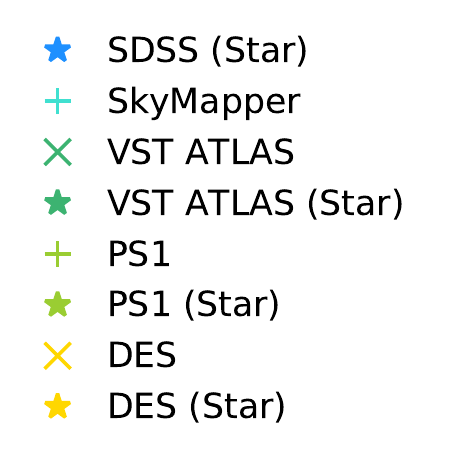} \\ 
\vskip 0.1cm \hrule width19.9cm \vskip 0.1cm
\includegraphics[width=0.2784\textwidth]{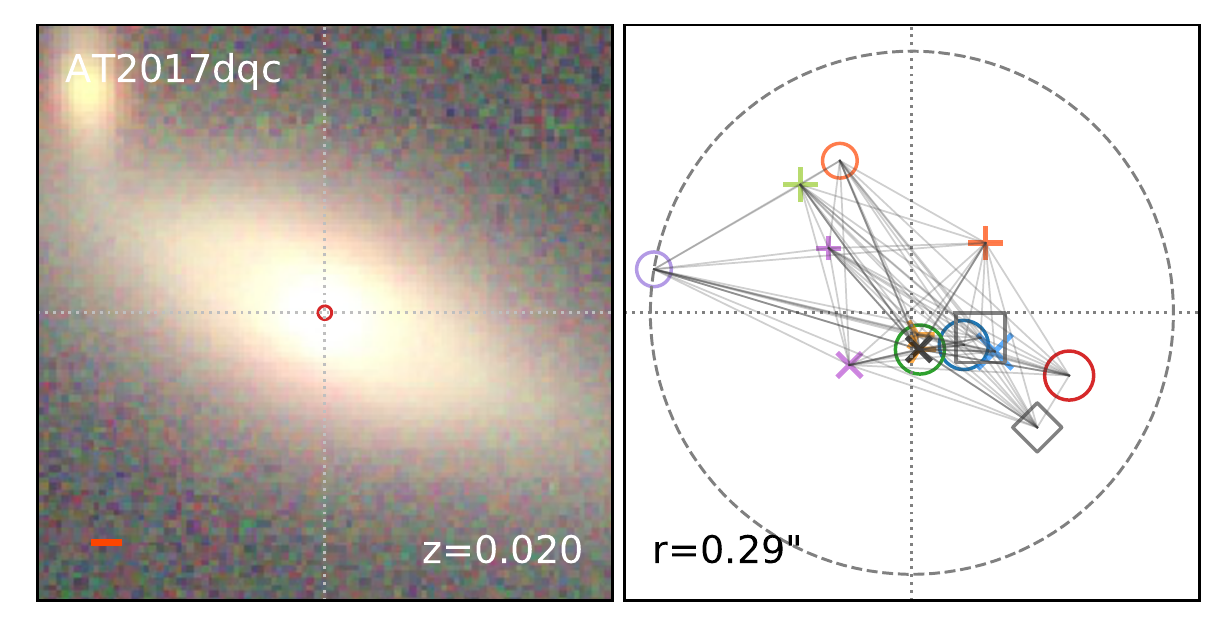} 
\includegraphics[width=0.2784\textwidth]{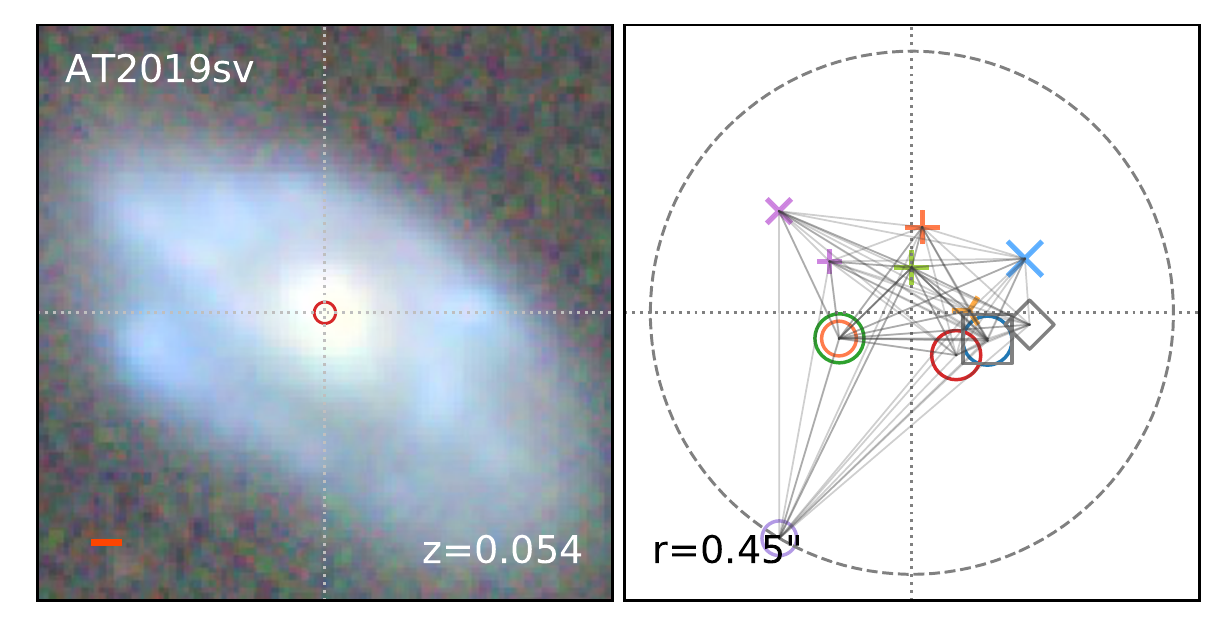} 
\includegraphics[width=0.2784\textwidth]{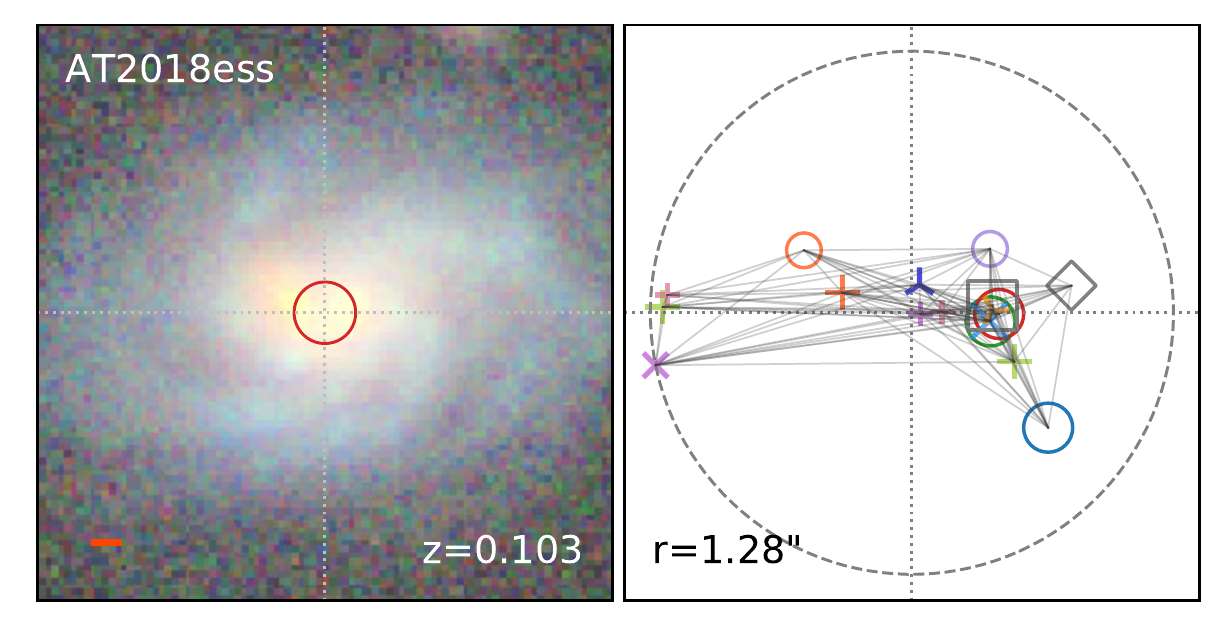}
\includegraphics[width=0.1449\textwidth]{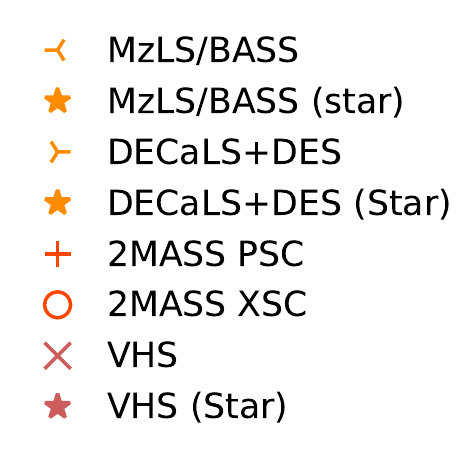} \\ 
\includegraphics[width=0.2784\textwidth]{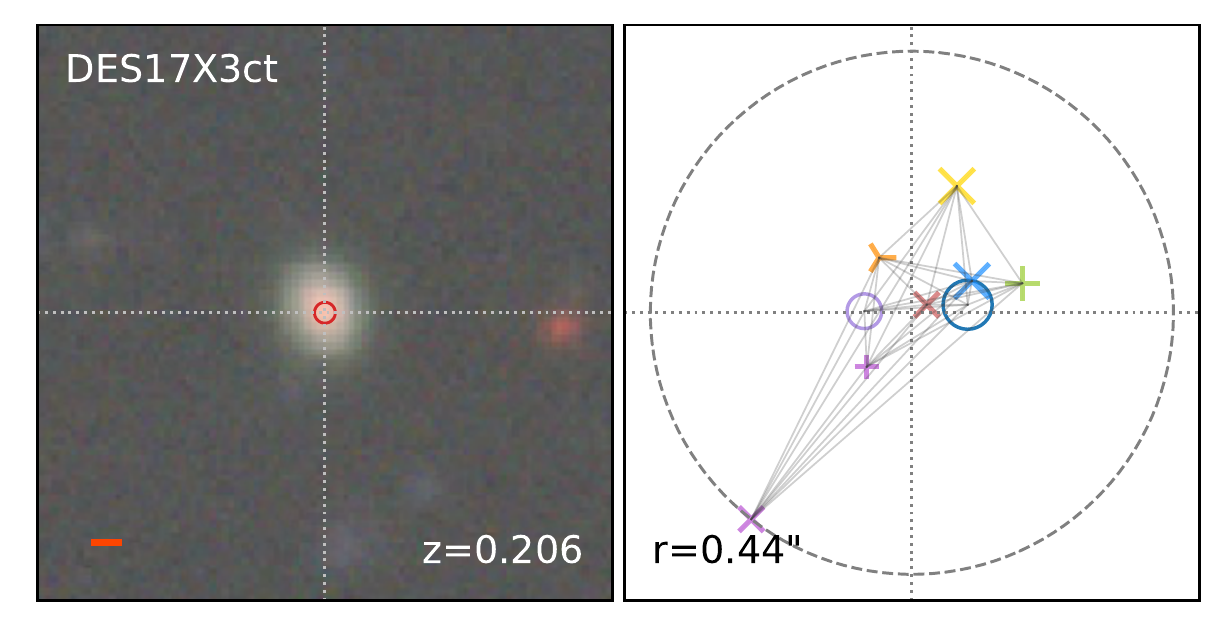} 
\includegraphics[width=0.2784\textwidth]{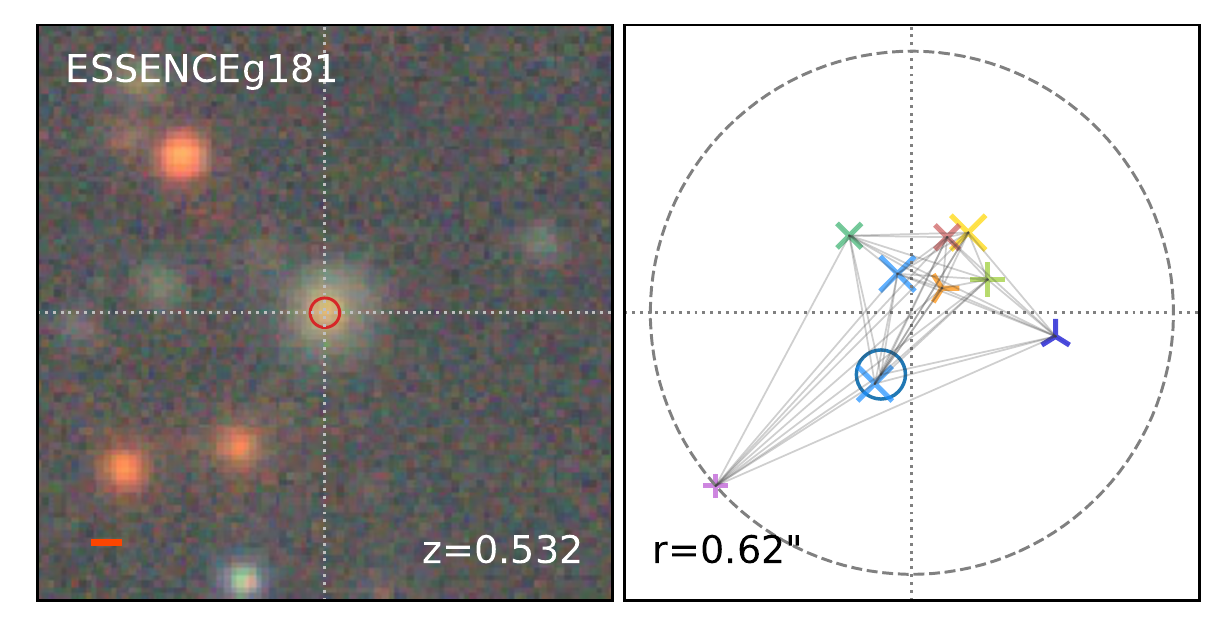} 
\includegraphics[width=0.2784\textwidth]{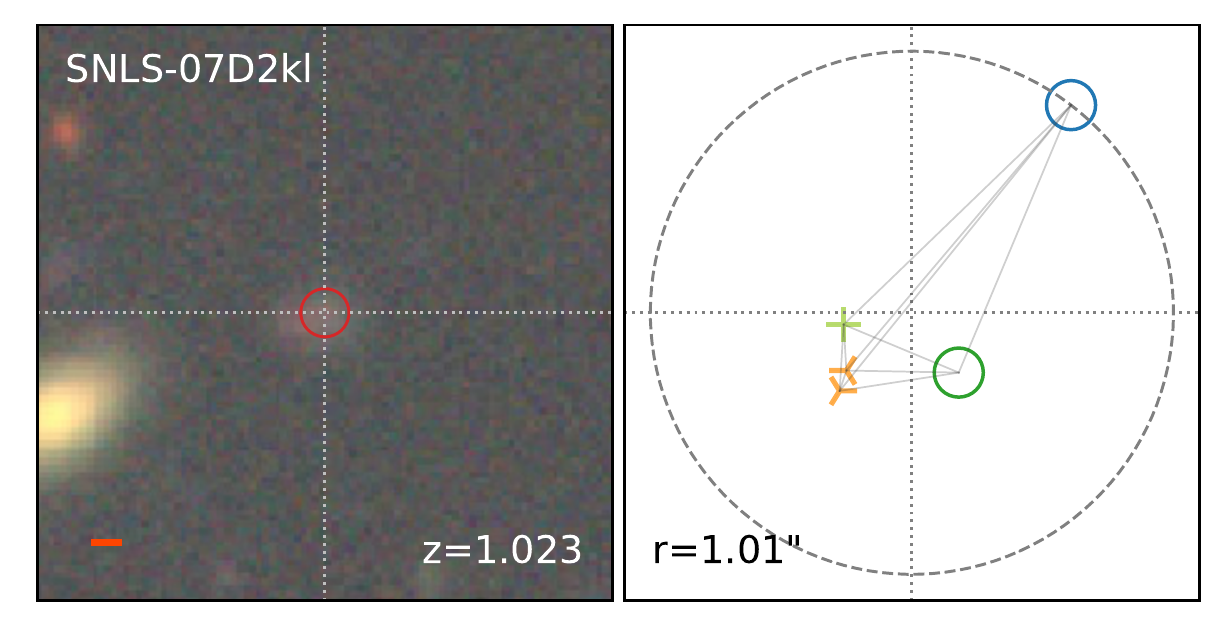} 
\includegraphics[width=0.1449\textwidth]{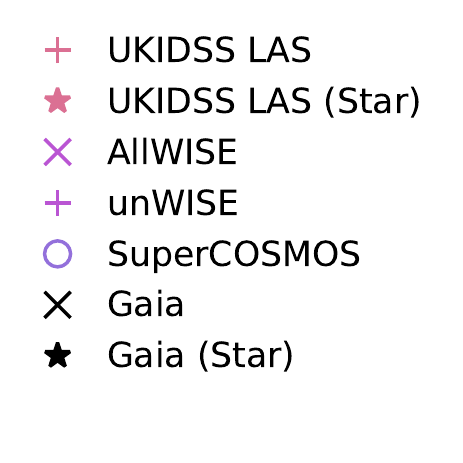} \\ 
\caption{
Examples of cross-matching catalog sources using astrometric tolerances and connectivity of objects.
Each panel here shows the color composite image of a host galaxy on the left side, with transient name, redshift, and 1" angular scale (red bar) indicated.
The cross-matched catalog sources are plotted with various symbols on the right side, a zoom-in view of the red circle (radius indicated) centered at the host.
To cross-match objects, the distances of catalog sources are normalized by their pairwise astrometric tolerances and then compared to their thresholds of connectivity (Section \ref{sec:xmatch}).
Connected pairs under this criteria are indicated as gray lines, and groups of inter-connected catalog sources are further labeled as cross-matched objects.
For clarity, we only zoom into the host here, while other catalog sources in the search radius are also cross-matched into groups in the same way.
These panels are randomly selected near redshifts $0.02$, $0.05$, $0.1$, $0.2$, $0.5$, and $1$, where the upper six panels are known hosts, and the lower six panels are new hosts with our algorithm (Section \ref{sec:newhost}).
The cross-matching algorithm we developed is effective yet flexible for most hosts, while more complex cases are discussed in Section \ref{appendix:xmatchqc}.
\label{fig:xidexample}}
\end{sidewaysfigure*}
\twocolumngrid

\section{Host Data Compilation} \label{sec:extcatalogs}

The general guideline to choosing external catalogs is that the data should be available for a large fraction of our hosts, and in combination, these catalogs should cover the entire UV-optical-IR wavelength range.
Following this principle, we select a wide variety of catalogs (Tables \ref{tab:photocatalogs}, \ref{tab:othercatalogs}).
Here, we discuss the basic characteristics, source selection, and star-galaxy separation criteria used for our external catalogs. We also briefly summarize the coverage of host properties in these catalogs.
Note that the column names mentioned are as appeared in our data sources.

\subsection{GALEX}

We choose the All-sky Imaging Survey \citep[AIS GR6/7;][]{Bianchi11} and the Medium-depth Imaging Survey \citep[MIS GR5;][]{Bianchi17} catalogs of the \textit{GALEX} mission for UV photometry of host galaxies.
They provide fixed circular apertures (8 and 17 pixels) and elliptical Kron magnitudes in FUV and NUV bands, with related shape parameters for apertures and standard \texttt{SExtractor} flags in each band.
The AIS has point source limiting magnitudes ($5\sigma$, AB) of $20.8$ (NUV) and $19.9$ (FUV), covering $2.6\times 10^4\sqdeg$ ($63\%$ of the sky), while the MIS covers only 1000 $\sqdeg$ but reaches deeper limiting magnitudes of $22.7$ in NUV and $22.6$ in FUV.

We do not impose source selection criteria for \textit{GALEX} catalogs. Sources in both catalogs are always assumed to be valid objects.
There are sources that are affected by dichronic, detector window, or detector edge reflection artifacts. These sources, although shall be excluded for photometric analysis as recommended by the references above, usually match true objects in deeper optical surveys.
We, therefore, preserve these sources in our cross-matching process, and these potentially corrupted photometric measurements can be excluded later using corresponding artifact bits in \texttt{Nafl} and \texttt{Fafl} columns.
Given the relatively low angular resolution of the \textit{GALEX} survey, we do not use their star-galaxy separation parameters as given in the catalogs.
Taking both AIS and MIS into account, \NStatGALEXanybandConf of our known hosts have \textit{GALEX} magnitudes in at least one band (usually NUV), while \NStatGALEXallbandConf have magnitudes in both \textit{GALEX} bands. Meanwhile, \NStatGALEXanybandCand new hosts have at least one band available, and \NStatGALEXallbandCand have both bands available.

\subsection{SDSS photometric and spectroscopic catalogs}

{We also use the SDSS DR16 source catalog \citep{Ahumada20} for optical photometry of host galaxies.}
This data release covers $1.4\times 10^4\sqdeg$ ($34\%$ the sky) in \textit{ugriz} bands, reaching point source sensitivity down to $22.2$ mag in \textit{g} and \textit{r} (95\% completeness, asinh magnitude).
{We accessed the photometric object table at SDSS SkyServer \footnote{\url{skyserver.sdss.org/dr16/en/home.aspx}}. The table provides complete PSF, profile-fitting (de Vaucouleurs, exponential and composite), fixed aperture and Petrosian photometry of detected objects.}
We exclude duplicate sources by selecting primary survey objects (\texttt{mode=1}).
Stars are identified using their pipeline morphological classification (\texttt{class}) when cross-matching objects. This morphological indicator is based on the comparison of model-fitting magnitudes with PSF-fitting magnitudes.
However, we only perform star-galaxy separation for sources brighter than $i=20$ (asinh magnitude). Fainter than this magnitude, the purity of stars begins to drop\footnote{\url{classic.sdss.org/dr7/products/general/stargalsep.html}}.

We also search for spectroscopic targets in MPA-JHU catalogs \citep{Kauffmann03, Brinchmann04, Tremonti04}.
These catalogs are based on the spectra of SDSS DR8 \citep{Aihara11}, which provide spectral properties (spectral indices, line equivalent widths, and fluxes) and derived physical parameters (star formation rates, stellar masses and gas-phase metallicities) based on SDSS spectroscopy.
We select objects that are classified as galaxies (\texttt{SPECTROTYPE=`GALAXY'}) with reliable measurements (\texttt{RELIABLE=1}). Spectra are also deduplicated using the list provided by the authors.

Within our transient-host pairs, \NStatSDSSConf known hosts and \NStatSDSSCand new hosts have SDSS photometry.
Given that SDSS uses asinh magnitudes \citep{Lupton99}, reported magnitudes close to the detection limit may have a very low signal-to-noise ratio.
Requiring a minimal SNR of 5 in $g$, $r$, and $i$ bands, \NStatSDSSHighSNRgriConf known hosts and \NStatSDSSHighSNRgriCand new hosts have reliable SDSS magnitudes. Further including the two shallower bands ($z$, $u$), the number count drops to \NStatSDSSHighSNRallConf and \NStatSDSSHighSNRallCand for known hosts and new hosts. Meanwhile, \NStatMPAJHUConf known hosts and \NStatMPAJHUCand new hosts have reliable spectroscopic measurements in the MPA-JHU catalogs.

\subsection{DESI Legacy Imaging Survey}\label{sec:desilsdata}

The DESI Legacy Imaging Surveys \citep[LS;][]{Dey19} includes three photometric surveys that are complementary in sky coverage and filter set: the Mayall z-Band Legacy Survey (MzLS), which maps the high galactic altitude region in the northern hemisphere above Dec=$30\degree$ in $z$-band; the Beijing-Arizona Sky Survey (BASS), which has a similar footprint above Dec=$30\degree$ in $g$ and $r$-bands; and the Dark Energy Camera Legacy Survey (DECaLS), which covers the equatorial region.
They jointly cover $1.4\times 10^4\sqdeg$ in $g$, $r$ and $z$ bands.

While conducted using different facilities, the reduced images of these surveys are processed using the same photometric pipeline \citep[\texttt{tractor};][]{Lang16}, which creates catalogs by measuring source properties jointly over multiple images.
This allows the incorporation of data from other surveys.
Besides the existing unWISE stacked images (see Section \ref{sec:wisedata}), the latest LS DR8 also processed DES DR1 images, further extending the total sky coverage to $1.94\times 10^4$ $\sqdeg$ ($grz$; $W1$, $W2$), reaching Dec=$-60\degree$ in the southern Galactic cap region.

The \texttt{tractor} catalogs provide profile-fitting measurements of sources, where we take the geometric parameters related to de Vaucouleurs and exponential components, the fraction of de Vaucouleurs components, and the pipeline-chosen types and magnitudes.
We search for primary objects within these three surveys (\texttt{brick\_primary=1}), so that duplicates can be excluded.
When cross-matching sources, stellar sources are identified using the morphological classification flag (\texttt{type}), where moderately bright sources ($g<22$) with ``\texttt{PSF}'' type are considered as stellar sources.
At $g=22\,\text{mag}$, about half of extended sources are classified as ``\texttt{REX},'' a special type for \textit{possibly} extended sources that cannot be robustly classified due to their low signal-to-noise ratios.
There are also about an equal number of point-like (``\texttt{PSF}'') sources and ``\texttt{REX}'' sources near this magnitude.
We conclude that the signal-to-noise of detection fainter than this limit is insufficient to make definitive source morphology classification, so we do not use source morphology classification beyond this limit.
{Besides source characteristics in \texttt{tractor} catalogs, we further includes the photometric redshift estimated of \citet{Zhou21}}

Within our database, \NStatLSConf known hosts and \NStatLSCand new hosts have photometric measurements from these three surveys.

\subsection{Pan-STARRS}

We also retrieve photometric measurements of our host galaxies and candidates in the Pan-STARRS DR2 (PS1 DR2) stacked object table, which mainly includes the 3$\pi$ Steradian Survey \citep{Chambers16}.
The 3$\pi$ Steradian Survey imaged the entire northern sky above Dec$=-30\degree$ in $grizy$, reaching multi-epoch stacked limiting magnitudes ($5\sigma$, AB) of $21.4$ (\textit{y}) to $23.3$ (\textit{g}).
The catalog provides band-wise Kron and PSF magnitudes with related geometric parameters.  
Given its wide coverage and high sensitivity, PS1 DR2 photometry is vastly available and relatively complete for our known hosts and new hosts.

We select primary detections (\texttt{primaryDetection=1}) of objects within \texttt{StackObjectThin} table, with valid g-band detection (\texttt{gKronMag > -999}).
When cross-matching objects, bright objects that have significant difference in point-source and Kron magnitudes ($i_{\mathrm{PSF}}<20\,\text{mag}$, $i_{\mathrm{PSF}} - i_{\mathrm{Kron}} > 0.05\,\text{mag}$) are considered as point-like sources.
There are \NStatPSConf known hosts and \NStatPSCand new hosts that match Pan-STARRS DR2 sources.

\subsection{2MASS Point and Extended Source Catalogs}

We use the 2MASS Point Source Catalog \citep[PSC;][]{Skrutskie06} and the Extended Source Catalog \citep[XSC;][]{Jarrett00} as our primary data source for NIR photometry.
They both cover the entire sky in \textit{J}, \textit{H} and \textit{Ks} bands.
The PSC measured fixed aperture magnitudes of over $5\times 10^8$ presumably point sources, reaching limiting magnitudes ($10\sigma$, Vega) of $J$=15.8 and $K_s$=14.3.
The XSC contains $1.6\times 10^6$ sources, measured with isophotal, Kron and extrapolated ``total radius'' apertures, reaching limiting magnitudes ($10\sigma$, Vega) of $J\simeq14.7$ to $K_s\simeq13.1$, depending on the surface brightness profile.
We accessed both catalogs as a few percent for objects in the PSC are actually unresolved extragalactic sources \citep{Rahman16}, and the PSC also has better depth.

We neither imposed selection criteria when searching for objects nor did we use point source or extended source classifications to identify stellar and galaxies when cross-matching objects.
Within our hosts, \NStatPSCConf known hosts and \NStatPSCCand new hosts have matched PSC sources. Meanwhile, \NStatXSCConf known hosts and \NStatXSCCand new hosts have matched XSC sources.
Most XSC objects also have detection in PSC, while \NStatPSConlyConf known hosts and \NStatPSConlyCand new hosts only cross-matched with PSC objects.
However, as we have discussed, PSC is optimized for point-source photometry, and well-resolved galaxies might be split into multiple point sources in PSC. Therefore, we always use XSC photometry when available.

\subsection{Mid-IR photometry based on \textit{WISE} data}\label{sec:wisedata}

Several different catalogs based on the images of the \textit{WISE} survey \citep{Wright10} contribute to mid-infrared host magnitudes.
\textit{WISE} surveyed the entire sky in four mid-infrared passbands ($W1$--$W4$; 3.4, 4.6, 12, and 22 $\micron$).
Primarily we use the \textit{AllWISE} Source Catalog \citep{Cutri14}, which measures source magnitudes using co-added images of the original \textit{WISE} mission and the extended, post-cryogenic \textit{NEOWISE} phase \citep{Mainzer11}.
The sensitivity is higher at shorter wavelengths with more episodes of stacked imaging, reaching limiting magnitudes (95\% completeness, point source, Vega) of $17.1$ in $W1$ and $15.7$ in $W2$.
In $W3$ and $W4$ bands, the limiting magnitudes (95\% completeness, point source, Vega) are $11.5$ and $7.7$, respectively.

We also included measurements in a few other \textit{WISE}-based catalogs.
Notably, \cite{Lang14} produced resolution-optimized co-adds of \textit{WISE} images (``unWISE''), allowing more sources to be detected than the original \textit{AllWISE} catalog.
These co-added images are already used for model-fitting photometry in DESI LS \citep{Dey19} (see Section \ref{sec:desilsdata}).
Beyond that, we also use the All-sky unWISE Catalog \citep{Schlafly19} as a complement to the official \textit{AllWISE} catalog and \texttt{tractor}-measured mid-IR magnitudes in LS DR8.
Primary sources (\texttt{primary>0}) are selected to avoid duplicates.
Note that both \textit{AllWISE} and unWISE catalogs measure objects as point-like sources, and consequently, the magnitudes of extended sources would be inaccurate compared to model-fitting magnitudes.
Well-resolved nearby galaxies could also be split into multiple sources in \textit{AllWISE} and unWISE catalogs.

Of our host galaxies, \NStatAllWISEConf known hosts and \NStatAllWISECand new hosts have measurements in the \textit{AllWISE} catalog.
Meanwhile \NStatunWISEConf known hosts and \NStatunWISECand new hosts matched objects in the unWISE catalog.
Finally, DESI LS catalogs perform forced photometry over unWISE images, and magnitudes are thus available for all optically detected objects, although the signal-to-noise ratio of detection may vary.

\subsection{DES}

{For hosts in the southern hemisphere, the DES DR2 catalog is our primary source for deep optical photometry.}
The survey imaged 5000 deg$^2$ of the southern Galactic cap region in \textit{grizY} bands, reaching co-added, point source limiting magnitudes ($10\sigma$, AB) of $g$=24.33 and $r$=24.08 in its first major data release \citep{Abbott18}.
The catalog provides band-wise fixed circular aperture magnitudes and elliptical Kron magnitudes, with aperture geometric parameters, other \texttt{SExtractor} flags, and star-galaxy separation parameters in each band.
{We note that the catalog of DESI LS already contains source properties in the footprint of DES. However, the ``official'' DES catalog contains full five-band data with source properties measured using an alternative photometry method. We, therefore, include DES in our database.}

We searched objects in the co-added source catalog of DES DR2. Stars are identified using the criteria in \citep[Eq. 4,][]{SevillaNoarbe18}, with an additional magnitude limit of $i<21.5$ for better purity of stars.
Within our transient-host pairs, \NStatDESConf known hosts and \NStatDESCand new hosts have DES sources matched. All of them have complete photometry in all five DES bands.

\subsection{UKIDSS LAS}

The Large Area Survey (LAS) of UKIDSS \citep{Lawrence07} is a near-infrared (\textit{YJHK}) survey covering two continuous regions in the northern Galactic cap region and the southern side of SDSS Stripe 82.
The 4000 $\sqdeg$ survey area has been intensively imaged by several major optical surveys, making it an important data source for NIR photometry.
Sources in the catalog have fixed aperture and Petrosian magnitudes (with aperture parameters) reported in each band, with designed photometric depth ($5\sigma$ point source, Vega) of 18.2 mag in $K$ and 20.3 mag in $J$.
Some sources have two separate epochs of $J$-band measurements.

We search primary objects (\texttt{m=1}) in the main source table, where artifacts (\texttt{cl=0} or \texttt{pn>0.1}) are rejected. Sources with high probability of being stars (\texttt{p*}$>$0.95) are marked as stars when cross-matching with other catalogs.
We have valid UKIDSS LAS photometry in at least one band for \NStatLASConf known hosts and \NStatLASCand new hosts. Meanwhile, \NStatLASfullConf known hosts and \NStatLASfullCand new hosts have complete photometric detection in all four bands.

\subsection{VHS}

The VHS DR4 catalog \citep{McMahon13} is our primary source for deep NIR photometry in the southern hemisphere.
This hemispheric survey imaged most of the southern sky in \textit{YJHKs} bands, with comparable sensitivity to the UKIDSS LAS, reaching point-source limiting magnitudes ($5\,\sigma$, Vega) of 20.2 in $J$ and 18.1 in $Ks$.
Fixed aperture magnitudes of 3 diameters (2'', 2.8'', 5.7'') and Petrosian magnitudes are reported for each band.

We searched for primary objects (\texttt{PriOrSec=1} or \texttt{priOrSec=FrameId}) in the source table, where artifacts (\texttt{Mclass=0} or \texttt{pNoise>0.1}) are excluded. When cross-matching with other catalogs, objects with high probabilities of being stars (\texttt{p*}$>$0.95) are marked as stellar objects.
Within our hosts, \NStatVHSConf known hosts and \NStatVHSCand new hosts have valid photometry in VHS.

\subsection{VST ATLAS}

The survey footprint of the VST ATLAS consists of two separate fields in the southern hemisphere that substantially overlap with DECaLS and DES, covering a total area of $4700\sqdeg$ in \textit{ugriz} bands.
The sensitivities are comparable to the SDSS, reaching 23.2 mag in \textit{g} and 22.6 mag in \textit{r} ($5\sigma$ median depth, AB).
Sources have fixed aperture magnitudes in 3 diameters (2'', 2.8'', 5.7'') and Petrosian magnitudes reported in each band.
We use the VST ATLAS as a complementary data source for optical photometry, especially in the $u$ band that is not imaged by DECaLS or DES.

When accessing the catalog, we select primary objects and exclude duplicates of known sources (\texttt{PriOrSec=1} or \texttt{priOrSec=FrameId}). Possible noise or artifacts (\texttt{Mclass=0} or \texttt{pNoise>0.1}) are also excluded.
When cross-matching with other catalogs, objects that are very likely to be stars by probability (\texttt{p*}$>$0.95) are considered as stars.
Within our hosts, \NStatATLASConf known hosts and \NStatATLASCand new hosts have valid photometry in the VST ATLAS.

\subsection{SkyMapper}

The SkyMapper Southern Survey \citep{Keller07} aims to image the entire southern hemisphere in \textit{uvgriz} bands with multi-epoch sensitivity similar as SDSS.
{The latest publicly available data release \citep[DR2][]{Onken19} we used here contains images of the snapshot-style shallow survey and the deeper main survey. The local sensitivity ranges from about 18 mag to 22 mag in $g$ and $r$ ($10\sigma$, point source, AB), depending on the completeness of the survey.}
We accessed the main table in which PSF and Petrosian magnitudes of sources are provided in each band, along with source geometric properties.
Though not as deep as other optical catalogs here, its hemispheric coverage and \textit{u}-band filter remains a great advantage over other surveys.

We searched for sources in the master table without a selection cut.
Given that the angular resolution of SkyMapper is slightly lower than other optical surveys we use, we also do not use their star-galaxy separation parameters when cross-matching with other catalogs.
In our catalog, \NStatSkyMapperConf known hosts and \NStatSkyMapperCand new hosts have valid photometric magnitudes in the SkyMapper catalog.

\subsection{SuperCOSMOS}

We include the all-sky galaxy catalog \citep{Peacock16} of SuperCOSMOS surveys \citep{Hambly01} as an auxiliary data source for optical photometry.
The catalog is based on the digitized photographic plates with UK Schmidt Camera, Palomar Schmidt Telescope, and ESO Schmidt Camera, providing calibrated \textit{B}, \textit{R} and \textit{I}-band photometry down to limiting magnitudes (AB) of $B_J\lesssim 21$, $R_F\lesssim 19.5$ and $I_N\lesssim 18.5$.
Unlike other optical catalogs we use here, the magnitudes in this catalog are measured in the traditional Johnson--Cousins system, but a significant advantage is its complete coverage of the entire sky and its good sensitivity in the \textit{B}-band.

We obtained the static version of this catalog at the SuperCOSMOS Science Archive\footnote{\url{ssa.roe.ac.uk}, accessed on Jun 20, 2018}.
No specific selection cut was made when searching for objects.
For our hosts, \NStatSCOSConf known hosts and \NStatSCOSCand new hosts have matched photometry in the SuperCOSMOS all-sky galaxy catalog.

\subsection{NSA}

The NASA-Sloan Atlas (NSA) is a catalog of local galaxies ($z\lesssim 0.15$) based on SDSS and \textit{GALEX} datasets.
With improved image reduction \cite{Blanton11} and source extraction optimized for nearby large galaxies, the catalog measured the detailed properties of about 0.6 million galaxies.
Notably, Petrosian photometry with elliptical aperture, best-fitting Sersic profile parameters with variable index, derived asymmetry and concentration indices, and K-corrected source characteristics are provided for objects, besides their classical SDSS-style circular Petrosian photometry and azimuthally-averaged radial light profile.

We do not make specific selection cuts when searching for objects in this catalog, and all objects are considered to be galaxies. Within our database, \NStatNSAConf known hosts and \NStatNSACand new hosts matched objects in this catalog.

\subsection{HyperLEDA}

HyperLEDA \citep{Makarov14} is another value-added catalog that compiles properties of galaxies across various surveys and reference sources. While HyperLEDA is not used in resolving host names, we accessed their table of homogenized astrophysical parameters to further enrich our available host properties.
These properties include but are not limited to, morphological T-type and detailed signatures (bars, rings, etc.), geometric properties, stellar and gas-phase kinematics, HI 21 cm and FIR fluxes, Mg II spectral indices, etc.

We only select objects that are classified as galaxies in HyperLEDA. Multiplicities of galaxies, like pairs, triplets, and groups, are not used as their sky coordinates could be quite uncertain.
We have \NStatHyperLEDAConf known hosts and \NStatHyperLEDACand new hosts matched with HyperLEDA objects, but the availability of galaxy properties varies significantly from one object to another. 

\begin{figure}[t]
\centering
\includegraphics[width=\linewidth]{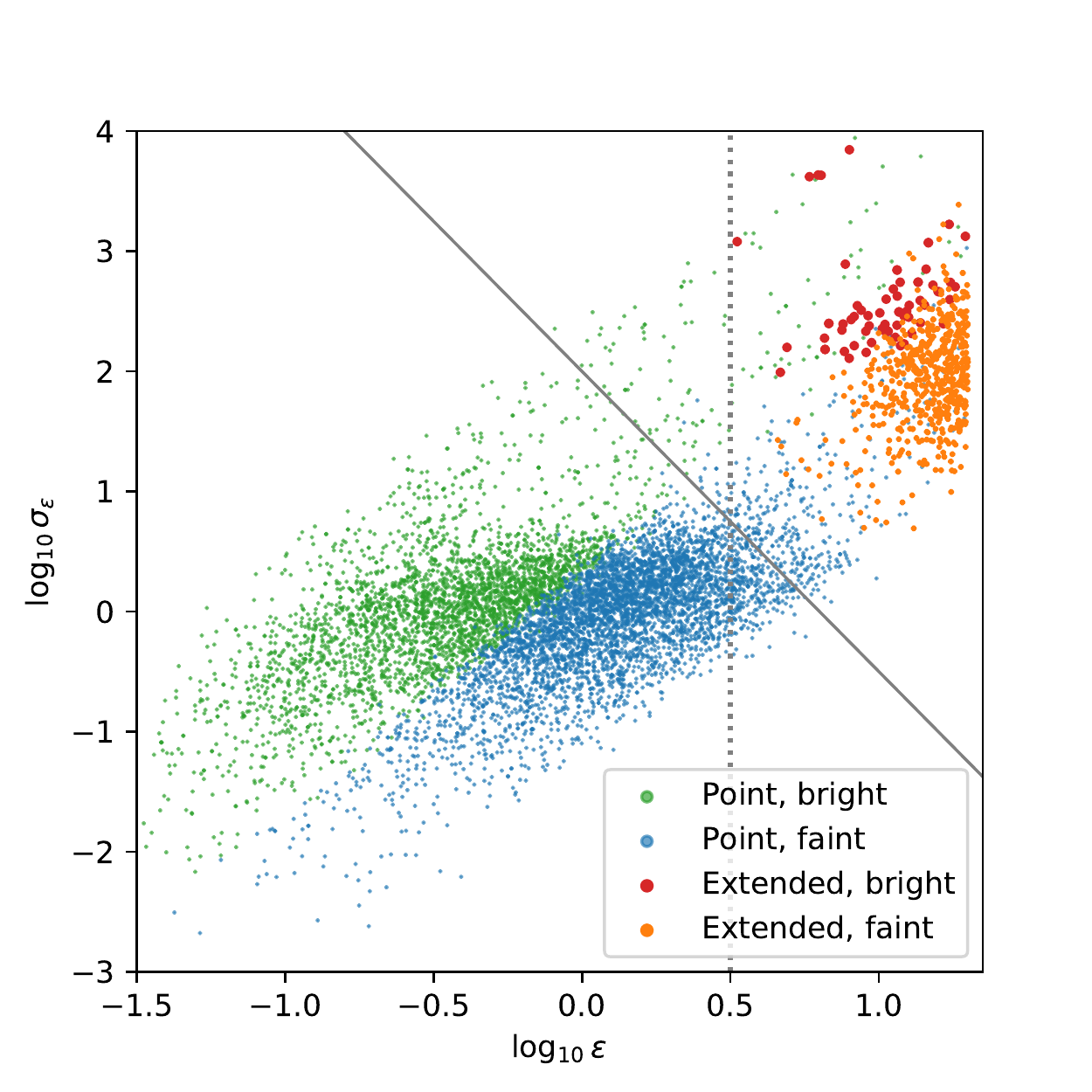}
\caption{Star-galaxy separation using \textit{Gaia} astrometric excess noise ($\epsilon$) and its statistical significance ($\sigma_{\epsilon}$), illustrated with a cross-matched example of \textit{Gaia} and MzLS/BASS.
Sources are classified into ``point'' (\texttt{PSF}) or ``extended'' (\texttt{EXP}, \texttt{DEV}, \texttt{COMP}, \texttt{REX}) by DESI LS pipeline morphological types, and further separated into ``bright'' or ``faint'' groups by \textit{Gaia} G-band magnitude of 19.
DESI LS use G$<$19 and $\log_{10}\epsilon<0.5$ (vertical dashed line) as their criterion for astrometric point sources, which is independent of their morphological classification.
To improve the effectiveness at fainter magnitudes while ensuring the purity of point sources, we use a magnitude-independent selection cut (solid line) that aligns with the gap that separates point-like and extended sources in this plane.
\label{fig:gaiastarcut}}
\end{figure}

\subsection{Gaia}

We also searched for objects in the \textit{Gaia} DR2 catalog \citep{Gaia16, Gaia18}. \textit{Gaia} is a specialized astrometric mission for Galactic stars, which we do not use for host properties here.
Instead, we use the astrometric properties of cross-matched \textit{Gaia} sources to identify foreground stars against distant host candidates.
Most galaxies, even in the magnitude limit of the \textit{Gaia} instrument, are absent in the catalog, while bright and compact galaxy nuclei are sometimes detected and cataloged.

Here we identify stars with the astrometric excess noise parameter ($\epsilon$).
This parameter, measured in the unit of milliarcseconds, characterizes the goodness-of-fit with their astrometric model, where values closer to zero indicate astrometrically well-behaved sources.
Stars usually have lower $\epsilon$ compared to galaxies at similar magnitudes.
This parameter has been used in previous works to separate stars and galaxies, while the detailed criteria vary.
DESI LS Pipeline labels sources above $G$=19 mag with $\epsilon<\sqrt{10}$ as point sources\footnote{See \url{github.com/legacysurvey/legacypipe}, in \texttt{ py/legacypipe/reference.py}, commit \texttt{a4a2fcd}}. This is independent of their source morphological classification.
Considering the degradation of astrometric fits at fainter magnitudes, \citet{Koposov17} select stars with a magnitude-dependent cut (Eq. 1) of $\log_{10}\epsilon<0.15(G-15)+0.25$.
Though known to be effective, the purity and accuracy of similar methods are yet to be tested and compared.

We use the following empirical criterion for stellar sources:
\[\log_{10}\epsilon < -0.4\log_{10}\sigma_{\epsilon} + 0.8.\]
Rather than doing a simple magnitude-dependent cut on $\epsilon$, we use the significance level of astrometric excess noise ($\sigma_{\epsilon}$) in our criterion to optimize the efficiency at fainter magnitudes.
As illustrated in Figure \ref{fig:gaiastarcut}, there is a clear gap that separates point-like and extended sources in $\log_{10}\epsilon$--$\log_{10}\sigma_{\epsilon}$ plane, which extends to fainter magnitudes.
To ensure that our selection cut yields a complete sample of extended sources, or equivalently, a pure sample of point sources, we cut near the cloud of point sources.
Note that objects with non-negligible $\epsilon$ are not guaranteed to be galaxies, as unresolved binaries may also have positive $\epsilon$. This is not against our goal to keep a complete sample of galaxies.

\subsection{Extinction corrections}

Host properties like photometric magnitudes and intrinsic colors need correction of foreground Galactic extinction to be fairly interpreted and compared.
Accurate correction of Milky Way extinction requires high-resolution reddening maps, extinction curves, filter transmission profiles, and source spectral characteristics, which are beyond the scope of this paper.
Without introducing extra complexities, we provide empirical extinction corrections of host magnitudes in each band.

The foreground reddening at the positions of host galaxies is estimated using the map of \cite{Schlegel98}, as implemented in \texttt{mwdust}\footnote{\url{github.com/jobovy/mwdust}} \citep{Bovy16}.
Extinction correction in \textit{GALEX}, SDSS, 2MASS, and \textit{WISE} (\textit{W1}, \textit{W2}) bands are estimated using the empirical coefficients in \cite{Yuan13}.
We also calculate another set of corrections for \textit{GALEX} bands using \cite{Peek13}, given the vast range of estimated extinction coefficients for \textit{GALEX} bands in the literature.
These extinction corrections are provided separately in the database and are not directly applied to the measured magnitudes in each catalog.

\begin{longrotatetable}
\begin{deluxetable}{llllll}
\tablecaption{Photometric Catalogs\label{tab:photocatalogs}}
\tablenum{4}
\tablehead{
\colhead{Catalog}                                           & \colhead{Version}             & \colhead{Filters}                         & \colhead{Hosted at}       & \colhead{References}
}
\startdata
{\it GALEX} Medium-depth Imaging Survey (MIS)               & GR5                           & FUV, NUV                                  & Vizier                    & \citet{Bianchi11} \\
{\it GALEX} All-sky Imaging Survey (AIS)                    & GR6/7                         & FUV, NUV                                  & Local                     & \citet{Bianchi17} \\
SDSS primary survey objects (\texttt{photoPrimary})         & DR16                          & u, g, r, i, z, y                          & SkyServer                 & \citet{Ahumada20} \\
Pan-STARRS 3$\pi$ Survey, stacked object                    & DR2                           & g, r, i, z, y                             & MAST                      & \citet{Chambers16} \\
DECaLS, tractor catalog                                     & DR8                           & g, r, z; W1--W4                           & Datalab                   & \citet{Dey19} \\
MzLS/BASS, tractor catalog                                  & DR8                           & g, r, z; W1--W4                           & Datalab                   & \citet{Zou17} \\
2MASS, Extended Source Catalog (XSC)                        & --                            & J, H, Ks                                  & Vizier                    & \citet{Skrutskie06} \\
2MASS, Point Source Catalog (PSC)                           & --                            & J, H, Ks                                  & Vizier                    & \citet{Jarrett00} \\
\textit{AllWISE} source catalog                             & --                            & W1--W4                                    & Vizier                    & \citet{Cutri14} \\
unWISE source catalog                                       & --                            & W1, W2                                    & Datalab                   & \citet{Schlafly19} \\
UKIDSS Large Area Survey (LAS)                              & DR9                           & Y, J, H, K                                & Vizier                    & \citet{Lawrence07} \\
DES, co-added source catalog                                & DR2                           & g, r, i, z, y                             & Datalab                   & \citet{Abbott18} \\
SkyMapper, main table (``\texttt{master}'')                 & DR2                           & u, v, g, r, i, z                          & Local                     & \citet{Wolf18} \\
VHS band-merged multi-waveband catalog                      & DR4.1                         & Y, J, Ks                                  & Vizier                    & \citet{McMahon13} \\
SuperCOSMOS all-sky galaxy catalog                          & --                            & B, R, I                                   & Local                     & \citet{Peacock16} \\
NSA                                                         & 1.0.1                         & FUV, NUV;                                 & Local                     & -- \\
                                                            &                               & u, g, r, i, z                             &                           &    \\
\enddata
\end{deluxetable}
\end{longrotatetable}


\begin{longrotatetable}
\begin{deluxetable}{lllll}
\tablecaption{Spectroscopic, Value-added and Astrometric Catalogs\label{tab:othercatalogs}}
\tablenum{5}
\tablehead{
\colhead{Catalog}                                   & \colhead{Type}    & \colhead{Version}             & \colhead{Hosted at}           & \colhead{References}
}
\startdata
SIMBAD basic data                                   & VAC               & --                            & \url{simbad.u-strasbg.fr}     & -- \\
NED basic data                                      & VAC               & --                            & \url{ned.ipac.caltech.edu}    & -- \\ 
HyperLEDA astrophysical parameters                  & VAC               & --                            & Local                         & -- \\
SDSS MPA-JHU catalogs                               & spectroscopic     & DR7                           & Local                         & \citet{Kauffmann03}; \\
                                                    &                   &                               &                               & \citet{Brinchmann04};\\
                                                    &                   &                               &                               & \citet{Tremonti04} \\
\textit{Gaia} sources (``\texttt{gaia\_source}'')   & astrometric       & DR2                           & Vizier                        & \citet{Gaia18} \\
\enddata
\end{deluxetable}
\end{longrotatetable}

\section{Statistics of Transient-Host Pairs} \label{sec:hostpropstat}

In this section, we summarize the basic statistics of our transient-host pairs, including both confirmed host galaxies and primary host candidates. Secondary host candidates in each field are not included in the statistics of host properties here.

\subsection{Angular and physical distances}

\label{sec:angulardistance}
\begin{figure}
\centering
\includegraphics[width=\linewidth]{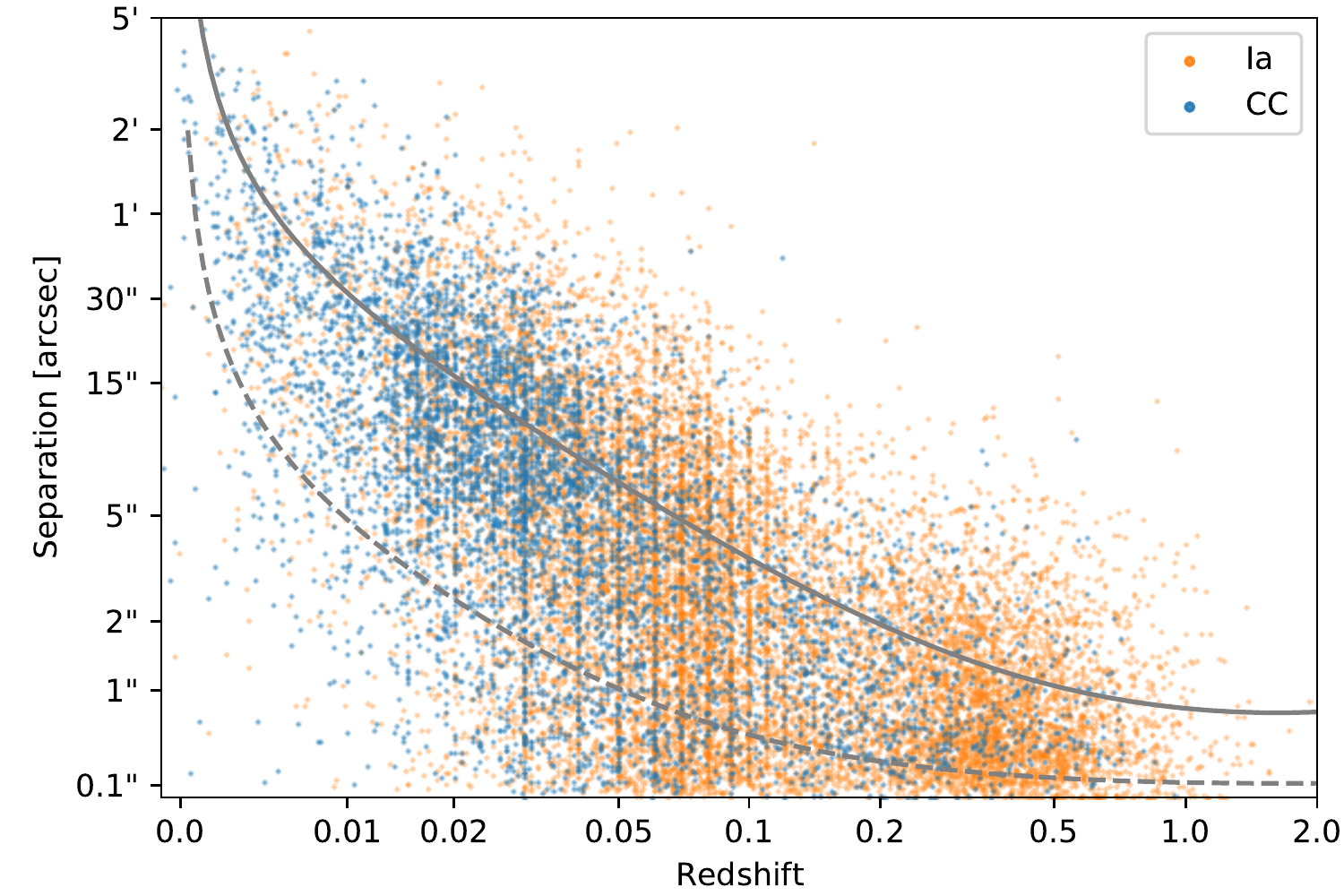}
\includegraphics[width=\linewidth]{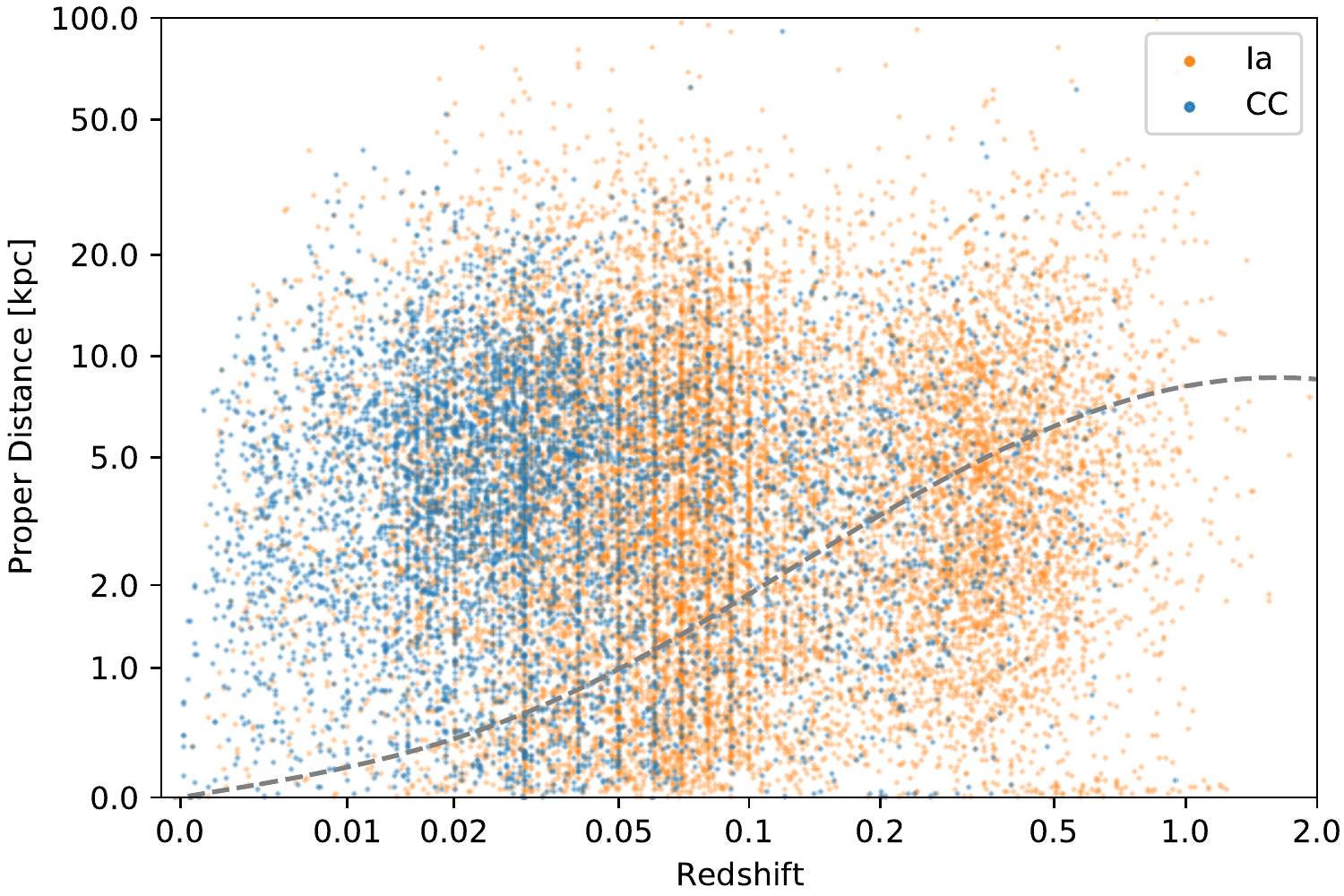}
\caption{(Top) The angular transient-host distances in our database, including both known host galaxies and newly identified hosts.
Thermonuclear (Ia) and core-collapse (CC) supernovae are plotted separately with orange and blue dots. Angular distances matching 1 kpc and 6.5 kpc (average deprojected physical distance) are outlined in dashed and solid gray lines, respectively.
(Bottom) The deprojected physical distance of transient-host pairs (in kiloparsecs), with similar symbols as in (a). Physical distance matching 1'' angular scale (typical position accuracy of galaxy coordinates) is indicated in gray dashed line.
The angular distance (a) mainly follows the cosmological angular diameter distance, where most events happen within 1 kpc and 10 kpc. Converted to the physical distances, we find little or no redshift dependence of transient-host distance, with an average value of 6.5 kpc.
\label{fig:zredoffset}}
\end{figure}

The transient-host angular distance is expected to be comparable to the optical sizes of galaxies, except for nuclear events such as TDEs, and outlying events like Ca-rich gap transients \citep{Kasliwal12, Lunnan17}.
Therefore, without considering the size evolution of galaxies, the angular transient-host distance should scale with the cosmological angular size-redshift relation.
The distances of our transient-host pairs agree with this general trend (Figure \ref{fig:zredoffset}).
Towards higher redshifts, the angular distance decreases following the angular diameter distance until $z\sim2$, where the angular scale peaks.
Converted to the proper distance at their redshifts, the deprojected physical distance is similar across the entire redshift range, with very little or no redshift evolution.
The average physical distance of transient-host pairs is 6.50 kpc, with 25th, 50th, and 75th percentiles of 1.66 kpc, 4.01 kpc, and 8.29 kpc, respectively.
This also justifies our redshift-dependent search radius of 30 kpc when searching for objects in external catalogs for events without any host information reported.

As a side note, at intermediate-to-high redshift, an angular distance of 1'' projects to a few kiloparsecs, as indicated by the dashed line in the right panel of Figure \ref{fig:zredoffset}.
The deprojected physical distance here may have larger errors than for lower-redshift events.
Also, our deprojected distance does not consider the ellipticity of host galaxies. A more comprehensive analysis of transient-host distances will be performed in our future work.

\subsection{Redshift coverage of host properties}

\begin{figure}
\centering
\includegraphics[width=\linewidth]{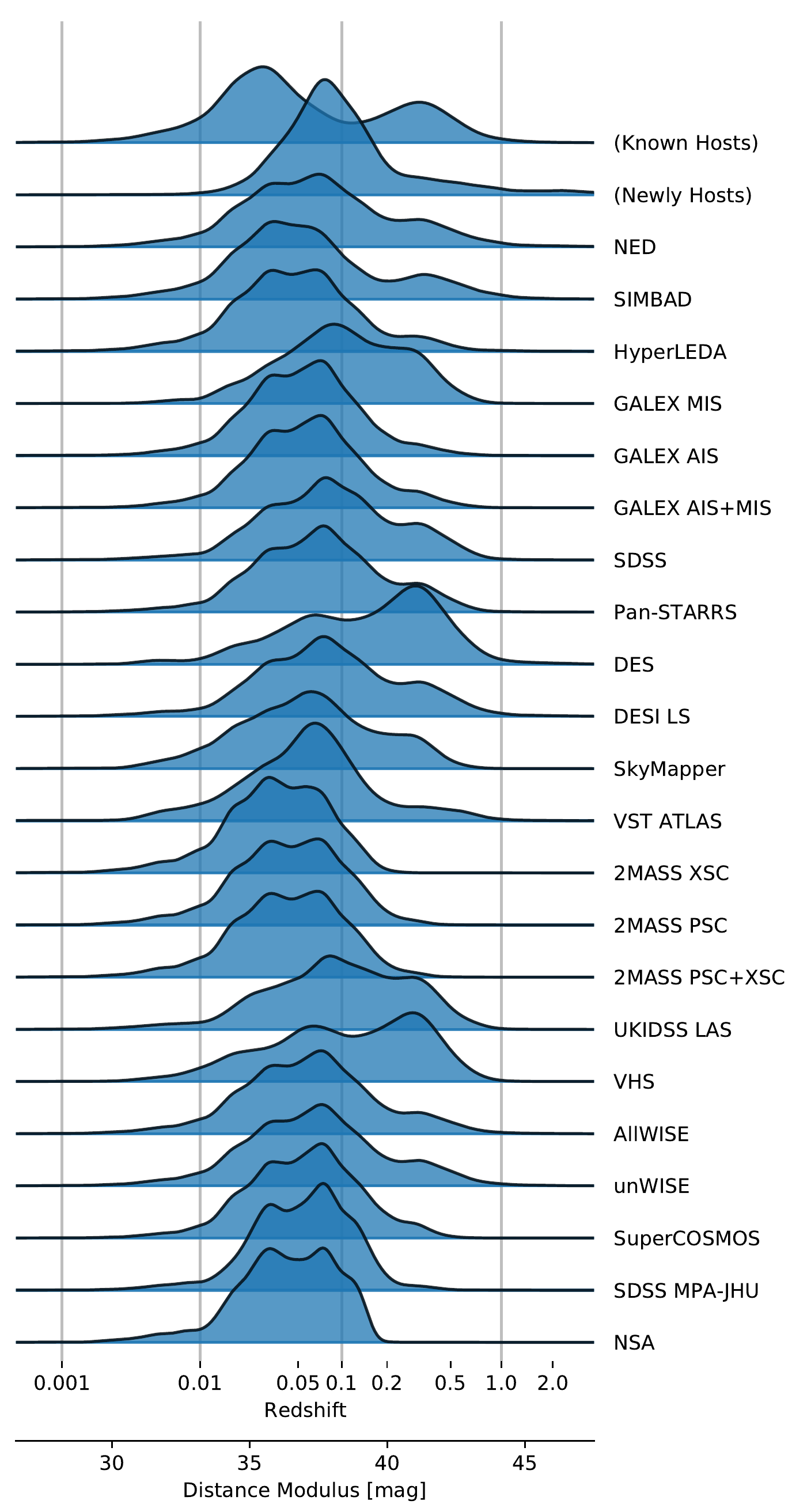}
\caption{Redshift and luminosity distance distributions of transients with host properties available in each external catalog, shown as normalized densities with respect to distance modulus.
The top two rows show the distribution for events with confirmed known hosts and newly identified candidates, while other rows show subsets of all hosts (confirmed and newly identified) with data in the external catalog available, as indicated at the left side.
Here the distributions mainly reflect the depth of external catalogs, where deeper surveys like {\it GALEX} MIS and DES have more higher redshift events than those shallow surveys (e.g., SkyMapper).
The double-peaked redshift distribution of transients is clear in the known hosts.
\label{fig:zredcatalog}}
\end{figure}

For transients with host properties available in a certain external catalog, we inspect their redshift distribution (Figure \ref{fig:zredcatalog}).
We convert redshifts to cosmological luminosity distance moduli for better illustration and comparison with host photometric magnitudes.
The aforementioned double-peaked transient redshift distribution remains clear in the known hosts (top row, Figure \ref{fig:zredcatalog}), but newly identified hosts do not have such a double-peaked distribution.
The double-peaked transient redshift distribution is also imprinted onto other external catalogs.

For external catalogs, besides being shaped by the underlying transient sample, their redshift distributions also reflect their relative sensitivity or depth.
External catalogs with better sensitivity, such as \textit{GALEX} MIS and DES, tend to have a higher fraction of high-redshift events. Conversely, external catalogs with relatively lower sensitivity, like 2MASS, \textit{GALEX} AIS, and SkyMapper, have more events at lower redshifts.
As a result of such vastly different catalog-wise redshift distributions, it could be non-trivial to find a proper set of external catalogs that provide adequate coverage in both redshift and photometric wavelength.

\subsection{Coverage of host properties by types}

\begin{figure*}
\centering
\includegraphics[width=\linewidth]{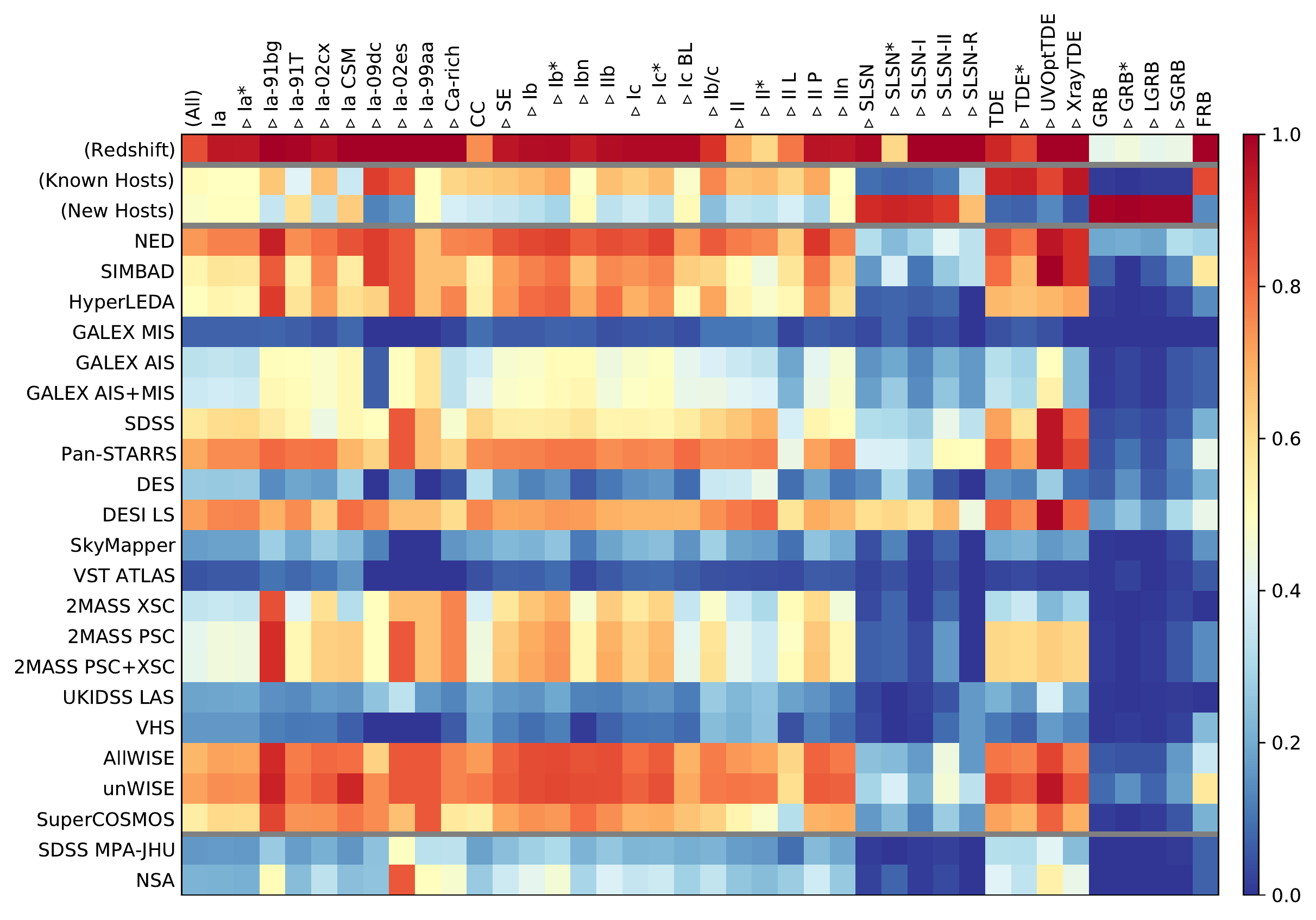}
\caption{Fractional converge of host properties within each dataset, for each major transient type.
The fractions of events with valid redshift, known hosts, or newly identified hosts are also indicated in three separate columns.
The fractional coverage accounted here primarily reflects the coverage of a catalog in sky area and depth.
Wide-area sky surveys, including Pan-STARRS, \textit{WISE}, SDSS, and SuperCOSMOS, have generally better coverage over almost all transient types.
The depth or sensitivity of a catalog may also have a non-negligible effect on the availability.
Furthermore, transient sub-types that require detailed follow-up to identify have relatively complete host properties as they are usually biased to lower redshifts. 
\label{fig:typecatalog}}
\end{figure*}

Here we summarize the fraction of hosts (both known and newly identified) that have valid catalog objects cross-matched in an external catalog, under each transient type label.
Results are listed in Figure \ref{fig:typecatalog}.
In the first three rows, we also show the fraction of transients with redshift, with confirmed hosts or newly identified primary host candidates.
Note that only types with at least 5 events are listed here.
The coverage of host properties here also reflects the sky coverage and depth (or sensitivity) of these external catalogs.
Full-sky and hemispheric surveys, including \textit{WISE}, \textit{GALEX}, SkyMapper, Pan-STARRS, and SuperCOSMOS, have better coverage of host properties in general.
Meanwhile, some transient types have slightly better coverage of host properties than others, such as Ia-91bg and Ia-02cx.
This could be a result of their redshift distributions, as events at lower redshifts may have more complete properties compiled in external catalogs. Secondary factors, like systematic differences of their host luminosities or SEDs, may also have a non-negligible effect.
Such non-uniform coverage of host properties is naturally expected due to the heterogeneity of our transient sample and external catalogs, where the cosmic volume is not coherently sampled on both sides and events are classified in different ways.

\subsection{Host photometric magnitudes}

\begin{figure*}
\centering
\includegraphics[width=0.85\linewidth]{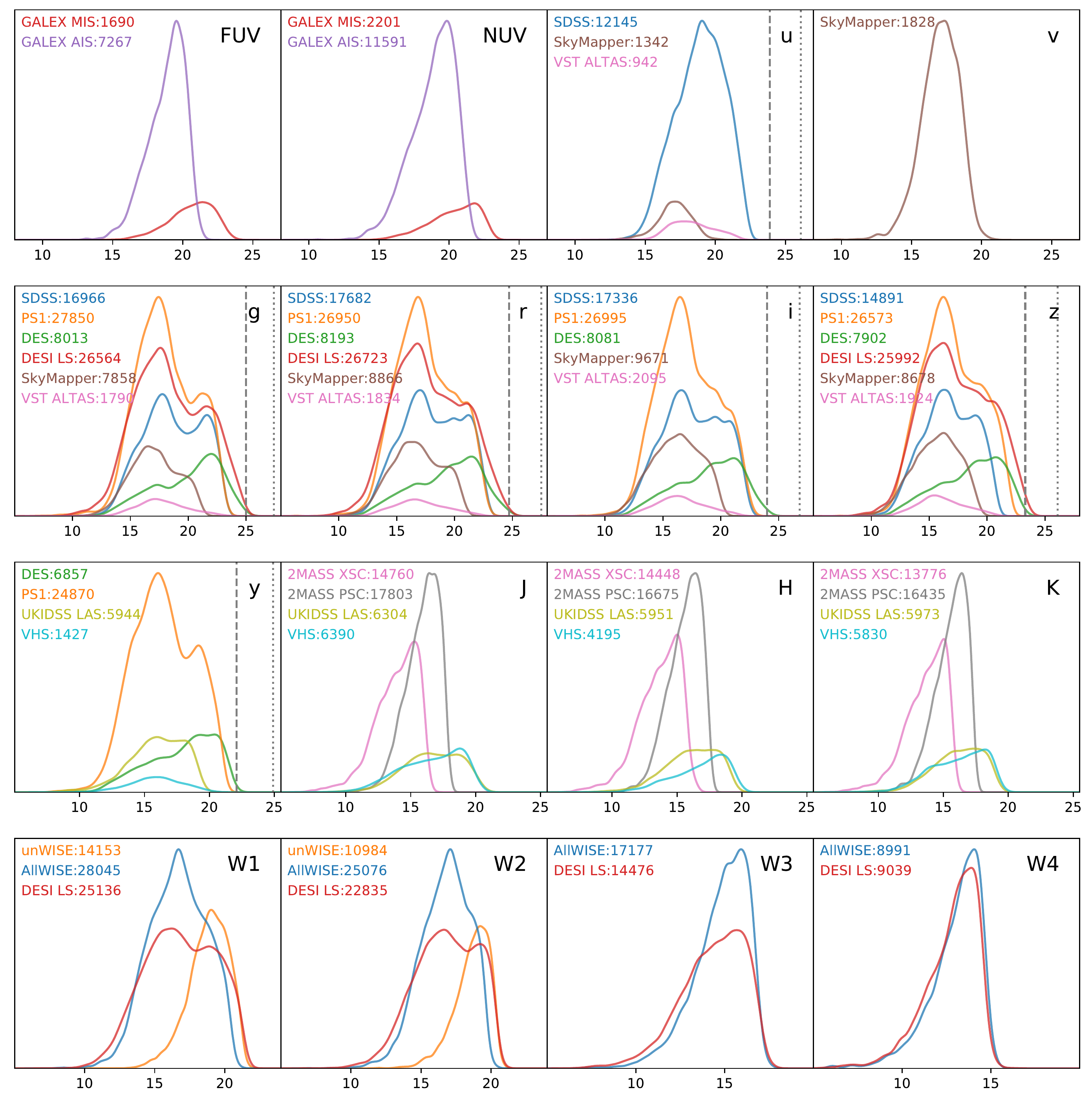}
\caption{The photometric magnitude (AB) distribution of known hosts and newly identified hosts in our external catalogs, grouped by their nominal photometric bands.
In each panel, the photometric band is indicated at the upper right corner, while the catalog and number count of valid values are listed in the legend. Histogram heights are scaled to the highest peak in each photometric band.
We applied a signal-to-noise cut of 5 for a fair comparison.
Vega magnitudes are converted to the AB system using the zero points in \citet{Jarrett11}.
Note that deep optical catalogs with substantial coverage in the northern hemisphere have prominent double-peaked magnitude distributions.
These events are mostly at higher redshifts, rather than in fainter hosts at lower redshifts. 
\label{fig:magdistr}}
\end{figure*}

The external catalogs we choose have substantial overlap in their nominal photometric bands. Here we compare photometric magnitude distributions of known hosts and newly identified primary candidates, grouped by their photometric bands (Figure \ref{fig:magdistr}).
Even for the same photometric band, the filter profiles, photometeric techniques, and calibrations could be quite dissimilar across these surveys.
We could compare the reported magnitudes in different surveys if they measure the same region of SED and the results are in the same magnitude system.
For a fair comparison, we cut magnitudes to a signal-to-noise ratio of 5.

Besides the clearly different sample sizes and limiting magnitudes as we have discussed in Section 4, we also noticed that the double-peaked redshift distribution also imprints on the photometric magnitudes of a few bands, including \textit{grizy} and W1, W2.
More specifically, optical surveys with substantial coverage in the northern hemisphere, like SDSS, MzLS/BASS and Pan-STARRS, tend to have such features. Southern hemisphere surveys, such as DES, VST ATLAS, and SkyMapper, do not have a secondary peak.
Even excluding supernovae in SDSS Stripe 82, which contributed a significant fraction of the medium-to-high redshift events (Figure \ref{fig:disctime}, \ref{fig:skydistribution}), the second peak remains in \textit{grz} bands.
However, when excluding events above $z\simeq0.1$, the second peak in \textit{grz} bands disappears.
Therefore, we expect that the double-peaked magnitude distribution is related to the double-peaked transient redshift distribution, where high redshift events in the northern hemisphere (or northern Galactic cap) contribute to the fainter, secondary peak.
The fainter hosts of low-redshift events do not lead to such double-peaked distributions.

\subsection{Photometry calibration across catalogs}

As we have discussed above, the external catalogs we choose overlap significantly in sky coverage and filter set.
Many hosts, as a result, could have multiple magnitudes reported for the same (yet not exactly identical) band, contributed by different surveys, reductions, or measurements.
These magnitudes might be inconsistent, i.e., their offsets are greater than their claimed random errors.
There could be catalog-wise systematic errors, which are not properly accounted for in their reported magnitudes.

Many factors, like detailed filter profiles and photometric calibration, may contribute to the offsets here. Beyond that, perhaps the most important factor is the choice of photometric techniques.
Modern surveys measure extended objects using either aperture photometry \citep[e.g.,][]{Petrosian76, Kron80} or profile-fitting photometry \citep[e.g.,][]{Lupton99, Lang16}.
Considering factors like galaxy light profile, morphological uniformity, source angular size, and sky background level, none of these methods is fully immune to potential biases.
Moreover, de-blending (or not) can also be a source of inconsistencies in magnitudes, especially for crowded fields or well-resolved galaxies.
Therefore, careful selection and cross-calibration of external catalogs might be required to reduce biases when combining or comparing photometric measurements for each band.
Or at least, the systematic uncertainties should be estimated and taken into account for these measured magnitudes. This remains a part of our future work.

\section{Data Format and Availability} \label{sec:datastructure}

This database is designed to be maintainable and updated regularly to incorporate recently detected and classified transient events, as well as newly released survey datasets for host galaxies. The changes in upstream data sources and the updates to this catalog should be traceable using a version control solution.

The data is publicly available as a document-based database and can be imported and accessed with standard database protocols.
Finally, static, ``clean'' versions of the database are provided on a regular basis, in standard machine-readable formats like \texttt{FITS} and \texttt{NumPy}.

\subsection{Database structure}

Most modern astronomical databases, particularly those used in sky surveys, are relational databases, or sometimes loosely called ``\texttt{SQL}-like'' databases.
Records in a relational database are organized into separate tables with strict pre-defined formats (``data model'' or ``schema''), and records are interlinked by reference keys within, and across tables.
However, the heterogeneity and complexity of our data bring challenges to conventional table-based relational database systems.
Creating separate tables for every data source would be inevitable in this case, let alone accessing or updating tables via enormous reference keys and status flags.

We use \texttt{Mongodb}\footnote{\url{www.mongodb.com}, \url{github.com/mongodb/mongo}}, a document-oriented database solution, to store our data.
Its design philosophy grants great flexibility in data structure and naturally fits our purpose.
Such a database keeps collections of \textit{documents}, where every document is an independent, dictionary-like (key-value style) data container with versatile, even nested or hierarchical structures.
For a single event, the query results and cross-identified hosts are independent of other events.
Rather than distributing these records across multiple tables, it is more convenient to group them into one self-contained document per event.
This is also easier to access and maintain compared to joining multiple tables by reference keys in \texttt{SQL}-like languages.

Our database contains multiple data collections: (1) a master collection of transient events (``\texttt{events}``), including their sky coordinates, redshifts, classification, and reported host galaxies, along with other basic characteristics;
(2) a collection for query results in NED and SIMBAD (``\texttt{vacs}''), as the center coordinates and searching radii used in other catalogs relies on results returned from VACs;
(3) multiple data collections for the query results in survey catalogs, grouped by their data sources (``\texttt{vizier}'', ``\texttt{local}'' for local catalogs, ``\texttt{mast}'', ``\texttt{datalab}'', ``\texttt{sdss}'' for SDSS SkyServer);
(4) a collection for cross-identified hosts (``\texttt{host\_summary}''), including the list of cross-matched objects, their properties in various catalogs, and their rank when multiple candidates are present.
Every unique event points to a document in each collection that shares a unique index.

Maintaining multiple collections is mainly for easier implementation and better maintainability.
Finally, as an end-user product, we also created a cleaned-up version, where transient characteristics and properties of hosts or candidates are combined into a stand-alone collection, with user-friendly structures.
Tables in standard \texttt{FITS} and \texttt{NumPy} formats are also released on a regular basis, but only the properties of confirmed hosts and primary candidates are included.

\subsection{Maintenance of the database}

We plan to update the database on a regular basis to incorporate recent transient events in our data sources and the latest survey catalogs for host properties.
This requires: (1) a mechanism to trace the change in upstream data sources, (2) bookkeeping of historical versions after a document is updated or removed, (3) extendable catalog searching and object cross-matching code to easily incorporate new survey catalogs.

To trace the change in transient event data sources, we generate ``version control codes'' (VCCs) for each event, with one VCC per data source.
This is the \texttt{MD5} hash value of the original record in formatted, plain text representation.
Any change in the data source, even a very minor one, would result in a different VCC than the previous one, indicating that catalog queries and host cross-matching must be performed again for this event.
Besides VCCs, we also set human-readable date stamps for each update.
Moreover, events that no longer exist in any data source would be marked for removal.

To trace the change within our database, when a document is updated or removed, we keep a snapshot of the previous version.
These snapshots are stored in separate collections, indexed with the unique identifiers of transient events, and organized in chronological order for each event.
This is done only for our transient, VAC, and host collections, as the contents of survey catalogs are static, and their versions are indicated by the generation of the data release.

Finally, to ensure that our database is extendable, we have developed standardized, general-purpose query scripts for external survey catalogs.
New catalogs on \texttt{Vizier}, Data Lab and \texttt{MAST} can be included using configuration files, without adding new code.
We also create separate task scripts for our transient data sources yet share those commonly-used routines (e.g., determining search radius) to retain the modularity of the program and the consistency of their behavior.

\subsection{Data Access}

We provide the cleaned-up, reorganized version of the host property collection in three formats for users with different purposes:

\begin{enumerate}
    \item[-] Human-readable, self-contained, plain text \texttt{JSON} format, one file per event. \texttt{JSON} is a conventional data exchange format that stores data in nested dictionary-like (``key-value'' style) or array-like structures.
    This is suitable for those interested in individual, or a smaller sample of events.
    Also, when performance is not a critical factor of consideration and query operations are not required, \texttt{JSON} files can also be used instead of other formats.
    The format is also used by the Open Supernova Catalog.

    \item[-] Full snapshot of the host property collection in \texttt{BSON} format.
    \texttt{BSON} is a binary analogue and extension of \texttt{JSON} format that is used internally in \texttt{MongoDB} for data storage.
    The snapshot contains one single large file that can be directly imported into a local \texttt{MongoDB} database.
    This version is suitable for those using a significant portion of the dataset, or those who need query and aggregation capabilities.
    
    \item[-] Compiled tables of transient and host properties in standard \texttt{NumPy} and \texttt{FITS} formats.
    This format only includes properties of known and newly identified hosts, without detailed data quality control flags and properties of other cross-matched objects that may contain the right host when our host identification fails.    
    This is a convenient format when only properties of robustly identified and cross-matched hosts are concerned.
\end{enumerate}

These data are available at \url{https://sandbox.zenodo.org/record/743438}. The data schema for the host collection is illustrated using an example in Appendix \ref{appendix:dataschema}.

\section{Summary}\label{sec:summary}

We assemble a database for extragalactic transient events and the properties of their host galaxies.
Transients including supernovae, tidal disruption events, gamma-ray bursts, fast radio bursts, and gravitational wave events are compiled from various data sources.
Based on the availability of host names or coordinates in those data sources, we search value-added catalogs and survey catalogs for photometric and spectroscopic properties of potential host galaxies.
The returned objects and sources in those catalogs are then cross-matched to find host properties.
For events without known hosts, we also rank cross-matched objects to identify their best host candidates and compile their properties.

We design detailed strategies for source queries in external catalogs to maximize the chance of finding the right host.
We search sky coordinates of host galaxies whenever available, and we always obtain host coordinates using value-added catalogs if their names are provided.
For transient events without host coordinates, we search their event coordinates.
The search radius is then determined with a fixed projected distance at the transient redshift, or a default radius when transient redshift is unknown.
The search radii of gamma-ray bursts, as an exception, are partly based on their localization accuracy.

We also design custom procedures to improve the accuracy of object cross-matching and host candidate ranking when necessary.
When cross-matching sources across catalogs, we identify groups of spatially associated sources by their pairwise angular distances and the empirically-calibrated per-catalog astrometric tolerances and per-field matching thresholds.
When required, we rank cross-matched groups to find new host candidates as per their angular distances to reported transient coordinates, results of source matching in external catalogs, the geometric properties of the cross-matched groups, their possibility of being foreground stars, and optionally, detailed source properties in cross-matched catalogs.
Here stars are identified and excluded using relatively conservative photometric, morphological, or astrometric criteria.
Primary host candidates are identified using a Logistic Regression-based ranking algorithm, trained with transient-host pairs with known, properly cross-matched and visually-inspected host galaxies.
The ranking algorithm reaches above $97\%$ overall accuracy in our cross-validation using known hosts.
{Also, visual inspection of newly identified hosts reveals a similar accuracy.}

Aiming to optimize the coverage of host properties, we select source catalogs of wide-area sky surveys with complementary survey regions and filter sets.
We search catalogs of \textit{GALEX} and \textit{WISE} missions for UV and mid-infrared photometry.
Meanwhile, we use catalogs of various ground-based surveys for optical-IR properties, including 2MASS, VHS, and UKIDSS for near-infrared, as well as SDSS, DES, DECaLS, MzLS/BASS, Pan-STARRS, VST ATLAS, and SkyMapper for optical photometry.
Extra properties are supplied by NASA-Sloan Atlas, HyperLEDA, and MPA-JHU catalogs.

We provide our data in human-readable, plain text \texttt{JSON} files, \texttt{MongoDB} database snapshots, and standard \texttt{NumPy}/\texttt{FITS} tables. The data, including documentation and examples, can be accessed online at \url{https//sandbox.zenodo.org/record/743438}. Source code for this project is also available at \url{https://github.com/shiaki/THEx-database}.

By the sample size of transients, the coverage of transient types, the number of newly identified host candidates, and the availability of host galaxy properties, our database is the largest publicly-available dataset for transient host galaxies.

\acknowledgments{
I.A. is a CIFAR Azrieli Global Scholar in the Gravity and the Extreme Universe Program and acknowledges support from that program, from the Israel Science Foundation (grant numbers 2108/18 and 2752/19), from the United States - Israel Binational Science Foundation (BSF), and from the Israeli Council for Higher Education Alon Fellowship.

K.D.F. is supported by Hubble Fellowship grant HST-HF2-51391.001-A, provided by NASA through a grant from the Space Telescope Science Institute, which is operated by the Association of Universities for Research in Astronomy, Incorporated, under NASA contract NAS5-26555.

This research is based on observations made with the Galaxy Evolution Explorer, obtained from the MAST data archive at the Space Telescope Science Institute, which is operated by the Association of Universities for Research in Astronomy, Inc., under NASA contract NAS 5–26555.

Funding for SDSS-III has been provided by the Alfred P. Sloan Foundation, the Participating Institutions, the National Science Foundation, and the U.S. Department of Energy Office of Science. The SDSS-III web site is \url{http://www.sdss3.org/}.

SDSS-III is managed by the Astrophysical Research Consortium for the Participating Institutions of the SDSS-III Collaboration including the University of Arizona, the Brazilian Participation Group, Brookhaven National Laboratory, Carnegie Mellon University, University of Florida, the French Participation Group, the German Participation Group, Harvard University, the Instituto de Astrofisica de Canarias, the Michigan State/Notre Dame/JINA Participation Group, Johns Hopkins University, Lawrence Berkeley National Laboratory, Max Planck Institute for Astrophysics, Max Planck Institute for Extraterrestrial Physics, New Mexico State University, New York University, Ohio State University, Pennsylvania State University, University of Portsmouth, Princeton University, the Spanish Participation Group, University of Tokyo, University of Utah, Vanderbilt University, University of Virginia, University of Washington, and Yale University. 

The Pan-STARRS1 Surveys (PS1) and the PS1 public science archive have been made possible through contributions by the Institute for Astronomy, the University of Hawaii, the Pan-STARRS Project Office, the Max-Planck Society and its participating institutes, the Max Planck Institute for Astronomy, Heidelberg and the Max Planck Institute for Extraterrestrial Physics, Garching, The Johns Hopkins University, Durham University, the University of Edinburgh, the Queen's University Belfast, the Harvard-Smithsonian Center for Astrophysics, the Las Cumbres Observatory Global Telescope Network Incorporated, the National Central University of Taiwan, the Space Telescope Science Institute, the National Aeronautics and Space Administration under Grant No. NNX08AR22G issued through the Planetary Science Division of the NASA Science Mission Directorate, the National Science Foundation Grant No. AST–1238877, the University of Maryland, Eotvos Lorand University (ELTE), the Los Alamos National Laboratory, and the Gordon and Betty Moore Foundation.

The Legacy Surveys consist of three individual and complementary projects: the Dark Energy Camera Legacy Survey (DECaLS; Proposal ID 2014B-0404; PIs: David Schlegel and Arjun Dey), the Beijing-Arizona Sky Survey (BASS; NOAO Prop. ID 2015A-0801; PIs: Zhou Xu and Xiaohui Fan), and the Mayall z-band Legacy Survey (MzLS; Prop. ID 2016A-0453; PI: Arjun Dey). DECaLS, BASS and MzLS together include data obtained, respectively, at the Blanco telescope, Cerro Tololo Inter-American Observatory, NSF’s NOIRLab; the Bok telescope, Steward Observatory, University of Arizona; and the Mayall telescope, Kitt Peak National Observatory, NOIRLab. The Legacy Surveys project is honored to be permitted to conduct astronomical research on Iolkam Du\'ag (Kitt Peak), a mountain with particular significance to the Tohono O\'odham Nation.

NOIRLab is operated by the Association of Universities for Research in Astronomy (AURA) under a cooperative agreement with the National Science Foundation.

This project used data obtained with the Dark Energy Camera (DECam), which was constructed by the Dark Energy Survey (DES) collaboration. Funding for the DES Projects has been provided by the U.S. Department of Energy, the U.S. National Science Foundation, the Ministry of Science and Education of Spain, the Science and Technology Facilities Council of the United Kingdom, the Higher Education Funding Council for England, the National Center for Supercomputing Applications at the University of Illinois at Urbana-Champaign, the Kavli Institute of Cosmological Physics at the University of Chicago, Center for Cosmology and Astro-Particle Physics at the Ohio State University, the Mitchell Institute for Fundamental Physics and Astronomy at Texas A\&M University, Financiadora de Estudos e Projetos, Fundacao Carlos Chagas Filho de Amparo, Financiadora de Estudos e Projetos, Fundacao Carlos Chagas Filho de Amparo a Pesquisa do Estado do Rio de Janeiro, Conselho Nacional de Desenvolvimento Cientifico e Tecnologico and the Ministerio da Ciencia, Tecnologia e Inovacao, the Deutsche Forschungsgemeinschaft and the Collaborating Institutions in the Dark Energy Survey. The Collaborating Institutions are Argonne National Laboratory, the University of California at Santa Cruz, the University of Cambridge, Centro de Investigaciones Energeticas, Medioambientales y Tecnologicas-Madrid, the University of Chicago, University College London, the DES-Brazil Consortium, the University of Edinburgh, the Eidgenossische Technische Hochschule (ETH) Zurich, Fermi National Accelerator Laboratory, the University of Illinois at Urbana-Champaign, the Institut de Ciencies de l’Espai (IEEC/CSIC), the Institut de Fisica d’Altes Energies, Lawrence Berkeley National Laboratory, the Ludwig Maximilians Universitat Munchen and the associated Excellence Cluster Universe, the University of Michigan, NSF’s NOIRLab, the University of Nottingham, the Ohio State University, the University of Pennsylvania, the University of Portsmouth, SLAC National Accelerator Laboratory, Stanford University, the University of Sussex, and Texas A\&M University.

BASS is a key project of the Telescope Access Program (TAP), which has been funded by the National Astronomical Observatories of China, the Chinese Academy of Sciences (the Strategic Priority Research Program ``The Emergence of Cosmological Structures'' Grant XDB09000000), and the Special Fund for Astronomy from the Ministry of Finance. The BASS is also supported by the External Cooperation Program of Chinese Academy of Sciences (Grant 114A11KYSB20160057), and Chinese National Natural Science Foundation (Grant 11433005).

The Legacy Survey team makes use of data products from the Near-Earth Object Wide-field Infrared Survey Explorer (NEOWISE), which is a project of the Jet Propulsion Laboratory/California Institute of Technology. NEOWISE is funded by the National Aeronautics and Space Administration.

The Legacy Surveys imaging of the DESI footprint is supported by the Director, Office of Science, Office of High Energy Physics of the U.S. Department of Energy under Contract No. DE-AC02-05CH1123, by the National Energy Research Scientific Computing Center, a DOE Office of Science User Facility under the same contract; and by the U.S. National Science Foundation, Division of Astronomical Sciences under Contract No. AST-0950945 to NOAO.

This publication makes use of data products from the Two Micron All Sky Survey, which is a joint project of the University of Massachusetts and the Infrared Processing and Analysis Center/California Institute of Technology, funded by the National Aeronautics and Space Administration and the National Science Foundation.

This publication makes use of data products from the Wide-field Infrared Survey Explorer, which is a joint project of the University of California, Los Angeles, and the Jet Propulsion Laboratory/California Institute of Technology, funded by the National Aeronautics and Space Administration.
}

\appendix
\twocolumngrid

\section{Acronyms and references for galaxy data sources} \label{appendix:acronyms}

We list the acronyms of survey programs and data sources in Table \ref{tab:datasourceacronyms}, along with their references.

\begin{deluxetable*}{lll}
\tablecaption{Acronyms and references for missions, surveys and online services\label{tab:datasourceacronyms}}
\tablenum{6}
\tablehead{\colhead{Name} & \colhead{Acronym} & \colhead{References}}
\startdata
Galaxy Evolution Explorer & \textit{GALEX}  & \citet{Martin05} \\
Sloan Digital Sky Survey & SDSS             & \citet{York00} \\
Panoramic Survey Telescope and & Pan-STARRS & \citet{Chambers16} \\
\quad\quad Rapid Response System &          &    \\
Mayall z-band Legacy Survey & MzLS          & \citet{Dey19} \\
Beijing-Arizona Sky Survey & BASS           & \citet{Zou17} \\
Dark Energy Camera Legacy Survey & DECaLS   & \citet{Dey19} \\
Dark Energy Survey & DES                    & \citet{DES16} \\
VLT Survey Telescope ATLAS & VST ATLAS      & \citet{Shanks15} \\
SkyMapper Southern Sky Survey & SkyMapper   & \citet{Keller07} \\
Two-Micron All-sky Survey & 2MASS           & \citet{Skrutskie06} \\
Vista Hemisphere Survey & VHS               & \citet{Arnaboldi07} \\
UKIRT Infrared Deep Sky Survey & UKIDSS     & \citet{Lawrence07} \\
Wide-field Infrared Survey Explorer & \textit{WISE} & \citet{Wright10} \\
SuperCOSMOS Sky Surveys & SuperCOSMOS       & \citet{Hambly01} \\
NASA/IPAC Extragalactic Database & NED      & \citet{Helou91} \\
SIMBAD Astronomical Database & SIMBAD       & \citet{Wenger00} \\
HyperLeda Extragalactic database & HyperLEDA & \citet{Paturel03} \\
NASA-Sloan Atlas & NSA                      & -- \\
\textit{Gaia} & --                          & \citet{Gaia16}
\enddata
\tablecomments{References listed are project or mission description papers.}
\end{deluxetable*}

\section{Standardization of Type Labels} \label{appendix:typelabels}

Below we discuss our detailed procedure to standardize transient type labels. Here typewriter fonts indicate the actual standardized labels used in our database.

\textit{Homogenizing type labels.}
When more than one type label refers to essentially the same group of transients, we choose a single, unique label to designate all of them.
To decouple type labels from yet-unclear physical scenarios or progenitor properties, we follow the rule that rare subtypes are named either after their commonly-referred prototype events or by their most distinctive observational signatures.
For example, ``Iax'' and ``Ia-02cx''  are used interchangeably in the literature for a subgroup of faint-and-slow thermonuclear supernovae, so we combine events under ``Iax'' label into ``\texttt{Ia-02cx}.''
Similarly, ``Ia-02ic,''``Ia-11kx'' and ``Ian'' labels are merged into the more descriptive and commonly used ``\texttt{Ia-CSM}.''
Events labeled as ``Ia-SC'' (super-Chandrasekhar) and  ``Ia-06gz'' are grouped into ``\texttt{Ia-09dc}'' because ``Ia-SC'' makes assumptions about the progenitor properties and ``Ia-06gz'' is less commonly used compared to ``\texttt{Ia-09dc}.''
This is already partly done within OSC, and we further extended their list of pre-defined synonyms\footnote{\url{github.com/astrocatalogs/supernovae/blob/master/input/type-synonyms.json}}.

\textit{Handling hybrid, intermediate or transitional types.}
A very small fraction of events are labeled as such composite types. The rule to relabel these composite types depends on the reason for the classification.
For composite labels due to the mixed usage of criteria or techniques, we split the labels and set them separately.
As an example, supernovae classified as ``IIn P'' are labeled as \textit{both} ``\texttt{IIn}'' and ``\texttt{II P}'' simultaneously, because ``n'' indicates spectral signature while ``P'' indicates light curve shape.
For unstandardized notations of later-defined subtypes in this style, we use the proper subtype labels instead.
Events classified as  ``Ia/IIn,'' for example, will be labeled as ``\texttt{Ia CSM}'' following this rule, rather than \texttt{Ia} \textit{and} \texttt{IIn} as in the previous case.
Finally, for composite types due to their ambiguous photometric or spectroscopic signatures, we also split the labels. Meanwhile, we add an additional ``\texttt{AMBIGUOUS}'' flag to these events (see later).
Events classified as ``Ia-CSM/IIn'' (not Ia/IIn) will receive both ``\texttt{Ia-CSM}'' and ``\texttt{IIn}'' labels, with ``\texttt{AMBIGUOUS}'' flags.
An exception to this rule is that Type Ib/c supernovae are \textit{not} labeled as both ``\texttt{Ib}'' and ``\texttt{Ic}.'' Instead, we reserve a ``\texttt{Ib/c}'' label for these events.

\textit{Rebuilding the hierarchy of type labels.}
The type labels of each event are aggregated from multiple reference sources, often with variations in the granularity of their label system.
To fit these type labels into a consistent and maximally-compatible classification scheme, we assign each event its best-refined subtype (there can be multiple) \textit{and} all their \textit{physical} parent labels, so that each label in our database points to a complete sample of events, including its subtypes.
The hierarchy of transient types is listed in Table \ref{tab:eventstatistics}.
For example, we always add ``\texttt{Ia}'' labels to events with ``\texttt{Ia-91T}'', ``\texttt{Ia-91bg},'' and other similar SN Ia subtype labels.
Events classified as ``\texttt{Ic BL}'' are also labeled as ``\texttt{Ic}'', ``\texttt{SE}'' (stripped envelope), and ``\texttt{CC}'' (core collapse).
Note that SNe IIb are considered as a subtype of Ib, instead of Type-II. These events are more related to stripped-envelope SNe (Ib, Ic) \citep{Filippenko93}.
Similarly, earlier works classify Ca-rich transients as a subtype of SN Ib, but here we group these events under SN Ia, as they are likely from thermonuclear events \citep{Perets10}.
Furthermore, we set a separate branch for superluminous supernovae (``\texttt{SLSN}''), due to their non-homogeneous observational signatures and possibly diversity of physical scenarios \citep{GalYam19}.

\textit{Flagging ambiguous and conflicting cases.}
Besides curated type labels, we also set special quality control flags to indicate the status of label homogenization.
For a single event, when its original list of type labels contains an ambiguous label that has been split into two labels with physically incompatible types, we mark the event as an ambiguous case (``\texttt{AMBIGUOUS}'').
Meanwhile, when its original list of type labels contains any two labels with incompatible types (i.e., inconsistent classification in different reference sources), we label the event as a conflicting case (``\texttt{CONFLICT}'').
Here any two labels become incompatible when they are not on the same major branch of the classification tree (\texttt{Ia}, stripped-envelope ``\texttt{SE}'', \texttt{II} and non-SN labels), as listed in Table \ref{tab:eventstatistics}.

\textit{Flagging peculiar events.}
Events of relatively rare subtypes are always labeled as peculiar (``\texttt{Pec}''), and ``Pec'' labels in original data sources are also preserved.
Here we do not consider ``\texttt{Pec}'' as a subtype or an intermediate level in the hierarchy of classification. This special label is used as a flag, rather than a subtype.
Therefore, ``Ia Pec'' events are always labeled as ``\texttt{Ia}'' and ``\texttt{Pec}'' simultaneously. Meanwhile, other rare and more specified subtypes of SNe Ia, such as Ia-91T and Ia-91bg, are also labeled as ``\texttt{Pec}''.
There could be known sub-types that are only classified as peculiar events, without detailed subtype labels. We further consult the review paper of \citet{Taubenberger17} to assign proper subtype labels for 60 peculiar SNe Ia in our sample.

\textit{Assigning type labels for GRBs, TDEs, and rare events.}
As discussed in Section \ref{sec:grbdatasources}, we always assign ``\texttt{GRB}'' labels for both long and short GRBs. Whenever possible, we use their $T_{90}$ value (the time when 90\% of photons reach the detector) to further classify them into short GRBs (``\texttt{SGRB},'' $T_{90}<2\,\text{s}$) or long GRBs (``\texttt{LGRB},'' $T_{90}\geq 2\,\text{s}$).
When there are associated optical supernova detection (rather than the afterglow), we also preserve other type labels assigned to this event.
Similarly, we assign ``\texttt{TDE}'' labels for all TDEs in our combined list.
For records in \citet{French20} review, we further assign ``\texttt{UVOptTDE}'', ``\texttt{XrayTDE}'', ``\texttt{LikelyXrayTDE}'' and ``\texttt{PossibleXrayTDE}'' per their wavelength and quality of detection.
Fast radio bursts and gravitational wave events are labeled as ``\texttt{FRB}'' and ``\texttt{GW},'' respectively.
Except for GW170817, other events in the Open Kilonova Catalog are only labeled as candidates (``\texttt{PossibleKilonova}'').

\section{Host galaxy reference sources} \label{appendix:hostinfoquality}

{The host galaxy names and coordinates in the Open Supernova Catalog (OSC) are collected from a wide variety of reference sources, including astronomical catalogs, individual papers, as well as supernova discoveries reported to the Transient Name Server (TNS), The Astronomer’s Telegram (ATel), and historically, to the Central Bureau of Astronomical Telegrams (CBAT) which issues International Astronomical Union Circulars (IAUC) and Central Bureau Electronic Telegrams (CBET).
We use their host names and coordinates to match sources in external catalogs and to train our ranking functions. The contribution and data quality of these reference sources, therefore, must be assessed beforehand. Here we focus on whether the provided host names or coordinates point to any possible host galaxies, while the reliability of transient-host association here (against other nearby galaxies) will be addressed later in Appendix \ref{appendix:inspectionknown}.}

{We collect the host names and coordinates reported in OSC along with their reference sources (catalogs, individual papers, TNS/ATel reports, IAUC/CBET issues). For supernova reports in TNS/ATel/IAUC/CBAT, we consider the entire platform or series, instead of individual indexed reports therein, as the data source. We resolve host names in NED and SIMBAD to find their corresponding sky coordinates. Coordinates in NED are preferred when names are recognized at both sides.
To examine if these name-resolved or as-provided coordinates are valid coordinates of galaxies, we obtained DESI LS image cutouts using the Sky Viewer\footnote{\url{https://www.legacysurvey.org/viewer}}. These image stamps are centered at transient coordinates, with box size covering four times the largest transient-host angular offset among all known host coordinates of the event.
Name-resolved and as-reported host coordinates of this event are plotted onto the image, with anonymized names of reference sources. Images are also randomly shuffled to minimize consistency issues in this quick inspection.}

{We classify these host coordinates into four cases: 1) “Good,” host coordinate is clearly centered on a possible host galaxy in the image; 2) “Off,” host coordinate points to a possible host galaxy in the image but deviates from its center (or the most prominent component) by more than $2''$, which \textit{may} challenge the cross-matching algorithm; 3) “Missed,” host coordinate points to a galaxy whose association with the transient is dubious, or no galaxy is visible at the indicated position due to erroneous coordinate or very faint host, and 4) ``Unknown,'' inspection cannot be performed due to image issues.
When the host is ambiguous in the image, we consider the coordinate ``Good'' or ``Off'' when it points to any possible galaxy, without determining if the galaxy appears to be the most likely host or not.}

{In Table \ref{tab:hostreference}, we summarize the number of host names and coordinates provided in each reference source (``Total'') and the number of host names that are recognized in either NED or SIMBAD (``Resolved''). Due to the limited sky coverage of DESI LS, we are only able to inspect a subset of host names and coordinates (``Inspected''). We evaluate the quality using the ``Good'' fraction and ``Missed'' fraction, with the total number of ``Good'', ``Off'', and ``Missed'' as the denominator.
Most reference sources have “Good” fractions above $90\%$. Notably, the host coordinates in the Asiago Supernova Catalogue have a relatively lower “Good” fraction due to the truncation of right ascension (in ``hh:mm:ss'' format) at seconds, leading to an average position shift of several arcseconds. The issue is also propagated into the dataset of \citet{Lennarz12}. Therefore, we ignore the host coordinates provided in these two reference sources. Furthermore, host coordinates from individual papers are also ignored for their low overall ``Good'' fraction.}

{Finally, it should be emphasized that the results shall not be considered as an examination of their original host galaxy assignment. The resolution and sensitivity of image cutouts, the procedure of data compilation in OSC and its upstream data sources, the accuracy of coordinates in NED and SIMBAD, and fundamentally, the empirical and non-exhaustive nature of this inspection can all affect the “Good” fraction of these reference sources.}

\begin{longrotatetable}
\begin{deluxetable}{lrrrrrrrrr}
\tablecaption{Major Contributors of Host Information in the Open Supernova Catalog \label{tab:hostreference}}
\tablenum{7}
\tablehead{
\colhead{Reference Source} & \colhead{Total} & \colhead{Resolved} & \colhead{Inspected} & \colhead{Good} & \colhead{Off} & \colhead{Missed} & \colhead{Unknown} & \multicolumn{2}{c}{Fraction} \\
\cline{9-10}
\colhead{} & \colhead{} & \colhead{} & \colhead{} & \colhead{} & \colhead{} & \colhead{} & \colhead{} & \colhead{Good} & \colhead{Missed}
}
\startdata
Latest Supernovae\TNM{a}                             & 8716     & 8654     & 6482     & 6020     &  186     &   68     &  208  & 0.960 & 0.011 \\
Asiago Supernova Catalogue \citep{Barbon99}          & 6746     & 3461     & 2625     & 2437     &   60     &   13     &  115  & 0.971 & 0.005 \\
Transient Name Server\TNM{b}                         & 4822     & 4785     & 3731     & 3475     &  104     &   38     &  114  & 0.961 & 0.011 \\
SDSS-II Supernova Survey Data Release \citep{Sako18}
                                                     & 4300     &  487     &  491     &  463     &    1     &   14     &   13  & 0.969 & 0.029 \\
``A Unified Supernova Catalogue'' \citep{Lennarz12}
                                                     & 3860     & 3393     & 2706     & 2519     &   47     &   15     &  125  & 0.976 & 0.006 \\
The Pan-STARRS Survey for Transients\TNM{c} \citep{Huber15}
                                                     & 2770     & 2755     & 2236     & 2109     &   47     &   40     &   40  & 0.960 & 0.018 \\
The Astronomer's Telegram\TNM{d}                     & 2689     & 2552     & 1636     & 1515     &   48     &   27     &   46  & 0.953 & 0.017 \\
International Astronomical Union Circulars           & 1785     & 1783     & 1406     & 1300     &   34     &   10     &   62  & 0.967 & 0.007 \\
Redshift-independent Distances in NED \citep[NED-D v13.1.0;][]{Steer17}
                                                     & 1566     & 1564     & 1432     & 1315     &   24     &   20     &   73  & 0.968 & 0.015 \\
Central Bureau Electronic Telegrams\TNM{e}           & 1417     & 1409     & 1013     &  931     &   24     &    6     &   52  & 0.969 & 0.006 \\
ASAS-SN Supernovae\TNM{f}                            & 1110     & 1060     &  779     &  726     &   22     &   10     &   21  & 0.958 & 0.013 \\
Gaia Photometric Science Alerts\TNM{g} \citep{Hodgkin21}
                                                     &  845     &  841     &  525     &  492     &   18     &    7     &    8  & 0.952 & 0.014 \\
UC Berkeley Filippenko Group's Supernova Database\TNM{h}
                                                     &  656     &  655     &  520     &  484     &    9     &    2     &   25  & 0.978 & 0.004 \\
Berkeley SN Ia Program Low-redshift SN Ia Sample \citep{Silverman12}
                                                     &  578     &  577     &  460     &  424     &   18     &    5     &   13  & 0.949 & 0.011 \\
Catalina Real Time Transient Survey, Supernova Hunt\TNM{i}
                                                     &  393     &  367     &  296     &  279     &    6     &    6     &    5  & 0.959 & 0.021 \\
\ldots                                               &  266     &  255     &  200     &  187     &    3     &    4     &    6  & 0.964 & 0.021 \\
CBAT Transient Objects Confirmation Page\TNM{j}      &  335     &  334     &  259     &  243     &   10     &    0     &    6  & 0.960 & 0.000 \\
SDSS-II Photometrically-classified SN Ia \citep{Campbell13}
                                                     &  317     &  317     &  320     &  303     &    1     &   11     &    5  & 0.962 & 0.035 \\
Cosmicflows-2 \citep{Tully13}                        &  288     &  288     &  228     &  212     &    7     &    0     &    9  & 0.968 & 0.000 \\
LOSS Stripped-Envelope SN Sample \citep{Shivvers19}
                                                     &  218     &  218     &  167     &  155     &    2     &    0     &   10  & 0.987 & 0.000 \\
LOSS SN Ia Light Curve Dataset \citep{Ganeshalingam10}
                                                     &  162     &  162     &  139     &  127     &    6     &    0     &    6  & 0.955 & 0.000 \\
SDSS-II, SN Ia Host Properties \citet{Gupta11}
                                                     &  156     &  156     &  159     &  150     &    0     &    4     &    5  & 0.974 & 0.026 \\
\citet{Jha07}                                        &  131     &  131     &  104     &   95     &    6     &    1     &    2  & 0.931 & 0.010 \\
\citet{Reindl05}                                     &  111     &  111     &   92     &   83     &    6     &    1     &    2  & 0.922 & 0.011 \\
\citet{Weyant14}                                     &  100     &  100     &   73     &   63     &    2     &    0     &    8  & 0.969 & 0.000 \\
\citet{Wang06}                                       &   98     &   98     &   73     &   66     &    5     &    0     &    2  & 0.930 & 0.000 \\
CfAIR2 SN Ia Near-infrared Light Curve Sample \citep{Friedman15}        
                                                     &   95     &   93     &   71     &   64     &    2     &    1     &    4  & 0.955 & 0.015 \\
CfA4 SN Ia Light Curve Sample \citep{Hicken12}       &   89     &   89     &   74     &   67     &    2     &    3     &    2  & 0.931 & 0.042 \\
LOSS SN Ia Photometric Data Release \citet{Stahl19}  &   86     &   86     &   69     &   63     &    3     &    0     &    3  & 0.955 & 0.000 \\
\citet{Parodi00}                                     &   67     &   67     &   56     &   50     &    4     &    1     &    1  & 0.909 & 0.018 \\
\citet{Rodriguez14}                                  &   52     &   52     &   40     &   38     &    1     &    0     &    1  & 0.974 & 0.000 \\
\citet{Galbany16.2}                                  &   50     &   50     &   35     &   34     &    0     &    0     &    1  & 1.000 & 0.000 \\
Other Reference Sources (five or more SNe)           & 1429     & 1422     & 1236     & 1072     &   37     &   49     &   78  & 0.926 & 0.042 \\
Other Reference Sources (fewer than five SNe)        &  389     &  362     &  328     &  270     &   14     &    2     &   42  & 0.944 & 0.007 \\
\hline
Asiago Supernova Catalogue \citep{Barbon99}          & 6491     &      --  & 5654     & 1712     & 1637     & 1901     &  404  & 0.326 & 0.362 \\
``A Unified Supernova Catalogue'' \citep{Lennarz12}
                                                     & 5509     &      --  & 4892     & 2448     &  346     & 1732     &  366  & 0.541 & 0.383 \\
SDSS-II Supernova Survey Data Release \citep{Sako18}
                                                     & 3813     &      --  & 3757     & 3619     &   48     &   34     &   56  & 0.978 & 0.009 \\
Pan-STARRS Photometrically Classified SNe \citet{Jones18}
                                                     & 1163     &      --  & 1198     & 1152     &    2     &    8     &   36  & 0.991 & 0.007 \\
SDSS-II SN Ia Rates Sample, \citep{Smith12}          &  341     &      --  &  334     &  321     &    1     &    9     &    3  & 0.970 & 0.027 \\
\citet{Prieto08}                                     &  254     &      --  &  258     &  249     &    3     &    0     &    6  & 0.988 & 0.000 \\
Nearby Supernova Factory SN Ia Host Sample \citet{Childress13.2}
                                                     &  253     &      --  &  229     &  200     &   17     &    6     &    6  & 0.897 & 0.027 \\
CANDELS SN Ia Rate Sample \citep{Rodney14}           &   63     &      --  &   78     &   58     &    1     &    0     &   19  & 0.983 & 0.000 \\
Other Reference Sources                              &   15     &      --  &   12     &    5     &    4     &    2     &    1  & 0.455 & 0.182 \\
\enddata
\tablecomments{Table shows the statistics of reference sources for host names (before the horizontal line) and for host coordinates (after the horizontal line). Columns are as defined in Appendix Z.}
\TNT{a}{\url{https://www.rochesterastronomy.org/supernova.html}}
\TNT{b}{\url{https://www.wis-tns.org/}}
\TNT{c}{\url{https://star.pst.qub.ac.uk/ps1threepi/psdb/}}
\TNT{d}{\url{https://astronomerstelegram.org/}}
\TNT{e}{\url{http://www.cbat.eps.harvard.edu/cbet/RecentCBETs.html}}
\TNT{f}{\url{http://www.astronomy.ohio-state.edu/asassn/sn_list.html}}
\TNT{g}{\url{http://gsaweb.ast.cam.ac.uk/alerts/home}}
\TNT{h}{\url{http://heracles.astro.berkeley.edu/sndb/}}
\TNT{i}{\url{http://nesssi.cacr.caltech.edu/SNhunt/}, which appeared as two separate reference sources in OSC (see the following line).}
\TNT{j}{\url{http://www.cbat.eps.harvard.edu/unconf/tocp.html}}
\end{deluxetable}
\end{longrotatetable}

\section{Accessing External Catalogs} \label{appendix:externalcatalogs}

\subsection{Resolving host names in value-added catalogs} \label{appendix:resolving}

{Here we discuss the procedure to resolve existing host names and choose the best name-resolved coordinate for the host.}

We choose NED and SIMBAD to resolve existing host names\footnote{accessed via \texttt{astroquery} \citep{Ginsburg19}, \url{github.com/astropy/astroquery}}, as they are the most commonly used astronomical databases and presumably have the best-available compilation of object names, sky coordinates, redshifts, and other basic data.
These VACs indeed provide some measured and derived properties of hosts, but the data are usually incomplete or not up-to-date. Accessing other survey catalogs for host properties remains necessary to compile their properties.
There are other online services like HyperLEDA that are capable of resolving object names, while scripted access is not supported yet.

{We collect host names of each event from reliable reference sources (Appendix \ref{appendix:hostinfoquality}). These host names are resolved individually and separately in NED and SIMBAD to obtain their corresponding coordinates and other basic data.
We exclude resolved coordinates with large uncertainties (error circle above $2.5''$ in NED, precision grade of $5$ or below in SIMBAD), and then choose the best resolved coordinate of each host name from the remaining ones.
For each name, we use coordinate in NED, unless the name is \textit{only} resolved in SIMBAD. Host names without resolved and precise coordinates in either VAC are considered as unresolved.}

{When there are more than one resolved host names at this moment, we check their consistency.
Pairs of name-resolved coordinates with separation above $5''$ are considered inconsistent, which implies possible disagreement of transient-host association among reference sources.
In the presence of any inconsistent pair, we skip the remaining steps and conclude without a best name-resolved coordinate.}

{Finally, when no host name is successfully resolved, we leave without a best name-resolved coordinate. When there is only one resolved name, we use it for the best name-resolved coordinate. When multiple names are resolved, we choose one for the best name-resolved coordinate. We prefer common galaxy names (e.g., NGC, IC, UGC, and PGC galaxies) over IAU-style source names (catalog name + sexagesimal coordinate), and anonymous galaxies (``Anon'' + sexagesimal coordinate) have the lowest priority. We also prefer names resolved simultaneously in NED and SIMBAD if the rule above cannot decide the best host.}

\subsection{Searching nearby sources in external catalogs}

\label{appendix:searchingsources}
\begin{figure*}[]
\centering
\includegraphics[width=\textwidth]{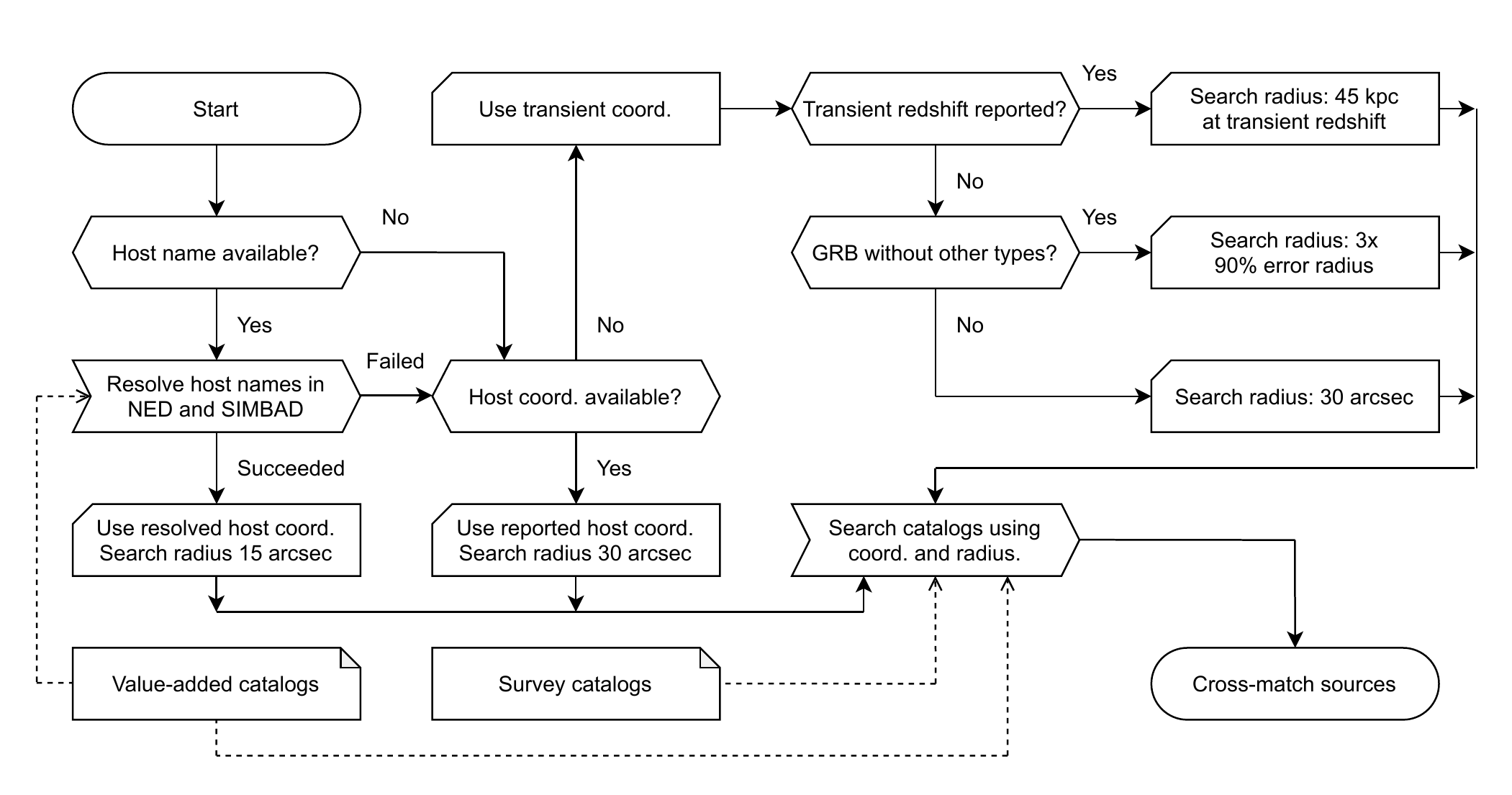}
\caption{The procedure to access value-added catalogs and survey catalogs for host properties.
For each event, we determine the coordinate to search (host or transient coordinate) and the search radius based on the availability of information in our transient data sources.
Whenever possible, we search host coordinates resolved using VACs or as provided in the transient dataset, instead of the transient coordinate.
When searching a transient coordinate, to optimize the chance of covering the true host, we determine the search radius using transient redshift if available.
In the absence of transient redshift, the error circle is also used for GRBs.
\label{fig:xidproc}}
\end{figure*}

When searching host (or transient) coordinates in external catalogs for nearby sources, to optimize the chance of finding the right host in external catalogs, we set sky coordinate and radius to search based on the following approach.
{If there is a best name-resolved host coordinate, we use this coordinate to search other catalogs with a search radius of $15''$. When resolving host names, we already have their basic properties in NED or SIMBAD, but we also perform a coordinate-based search again in the same catalog to access other nearby sources.}
If there is no name-resolved coordinate, but a host coordinate is directly provided in any of those upstream data sources, we use this as-reported host coordinate with a search radius of $30''$.
Finally, when neither name-resolved nor as-reported host coordinate is available, we use transient coordinates to search for nearby sources. The search radius in this situation is set based on the redshift of the event.
{For events with known redshifts, we set the search radius to match a projected distance of $45$ kpc at the transient redshift. Otherwise, we use a default search radius of $30''$.
Specially, for GRBs with neither redshift estimate nor later supernova detection, we use three times the 90\% error radius as the search radius.}
When searching transient coordinates, we clip the search radius within $15''$ and $2'$ to increase the chance of enclosing the right host, without including too many irrelevant non-host objects.
{For GRBs with search radii set by error circles, we clip the search radii within $5''$ and $15''$.}
The workflow is also summarized in Figure \ref{fig:xidproc}.

When coordinates and search radii are determined, we search for nearby sources of transient events in external catalogs.
Depending on their availability, our external catalogs can be classified into two categories, those accessible at online services (via web APIs, client-side packages or other protocols), and those provided as static tables in text or binary formats.

For online catalogs, as per the service where they are hosted, we use the proper interface or tool to access their data.
Catalogs hosted by NSF’s OIR Lab Astro Data Lab (formerly NOAO Data Lab) are accessed using their official \texttt{Python} client\footnote{\url{github.com/noaodatalab/datalab}}.
{Meanwhile, catalogs hosted at VizieR\footnote{\url{vizier.u-strasbg.fr}} and SDSS SkyServer\footnote{\url{skyserver.sdss.org/dr16/en/home.aspx}} are accessed via the third-party \texttt{astroquery} package \citep{Ginsburg19}.}
Finally, one catalog hosted at MAST\footnote{\url{archive.stsci.edu}} is accessed directly using their web API.
We record the detailed return status of queries when accessing online catalogs.
Failed requests due to any error are flagged to be made again to ensure that our database includes everything available in those external catalogs.

Local catalogs, on the contrary, are provided as static binary or plain text tables and thus cannot be accessed using coordinates and search radii like the online services.
We spatially index their catalog sources so that they can be queried with the same interface as we used for online services.
Depending on the size of the catalog, we use two different solutions.
Sources in smaller catalogs (below $10^7$ records) are indexed with kd-tree (as implemented in \texttt{SciPy}) using their 3D Cartesian coordinates on the unit celestial sphere.
Regarding algorithm complexity, efficiency and scalability, this is not the optimal solution, but it is quick and easy to deploy.
Sources in larger catalogs (above $10^7$ records) are indexed with \texttt{healpix} (via \texttt{healpy}), where sky coordinates are converted to 1D pixel indices in which they reside.
Searching for sources near a position becomes finding sources with certain pixel indices, which significantly improves performance.
We index sources with different resolution levels, and when searching sources, the best spatial resolution is chosen to match the search radius so that the region is fully covered and speed is optimized at the same time.
Sources inside the pixels of interest but beyond the search radius are excluded.
To further improve performance, before indexing spatial coordinates using either method, we make quality control cuts and drop unused columns to reduce the size of the catalog.
Catalogs published in text format are also converted to binary formats for better speed.

Having queried results in various external catalogs, we create local data collections in our database to store them. We do this for each service or interface via which the data are accessed (VizieR, Data Lab, local, etc.).
In each data collection, besides lists of returned sources in external catalogs, we also add auxiliary information including search radius and the source of searched coordinates (host coordinates obtained by resolving names, host coordinates as provided in data sources, or transient coordinates).

\section{Algorithm for cross-matching} \label{appendix:xidcode}

We list our algorithm for cross-matching as pseudo-code below.

\onecolumngrid \clearpage
\begin{algorithm}[H]
\caption{Cross-identification of host galaxies}
\begin{algorithmic}[]

    \item[]
    \Procedure{cross\_identify}{$s,k,\mathtt{is\_star},\mathtt{catalog}$}
    \State \textbf{input:}  \tabto{2.0cm} $s$: length-$n$ array of real number pairs, projected Cartesian coordinates of sources;
    \State                  \tabto{2.0cm} $k$: length-$n$ array of real numbers, per-catalog astrometric tolerances for sources in $s$;
    \State                  \tabto{2.0cm} \texttt{is\_star}: length-$n$ array of binary flags, whether the source is star-like (\texttt{true}) or not (\texttt{false});
    \State                  \tabto{2.0cm} \texttt{catalog}: length-$n$ integer array, external catalog identifiers of sources.
    \State \textbf{output:} \tabto{2.0cm} \texttt{is\_confirmed}: binary flag, if \texttt{primary\_host} is a confirmed host (\texttt{true}) or a primary candidate (\texttt{false});
    \State                  \tabto{2.0cm} \texttt{primary\_host}: integer set, indices of sources for the confirmed host or primary candidate;
    \State                  \tabto{2.0cm} \texttt{secondary\_hosts}: list of integer sets, indices of sources in each secondary host candidate.

    \item[]
    \State \texttt{is\_confirmed}    \tabto{3.5cm} $\gets$ \texttt{false}
    \State \texttt{primary\_host}    \tabto{3.5cm} $\gets$ \texttt{null}
    \State \texttt{secondary\_hosts} \tabto{3.5cm} $\gets$ \texttt{empty list}

    \item[]
    \ForAll {$s_i$, $s_j$ $\in S$ where $i<j$}
        \State $d_{ij}$, $d_{ji}$ $\gets \|s_i - s_j\| \,/\, (k_i^2 + k_j^2)^{1/2}$ \Comment{$d$: ($n$+1, $n$+1) matrix, normalized distance;}
    \EndFor
    \For {$s_i \in S$}
        \State $d_{i,n+1}$, $d_{n+1,i}$ $\gets \|s_i\| \,/\, k_i$ \Comment{normalized distance to the queried coordinate.}
    \EndFor

    \item[]
    \For {$t_i \in \mathtt{thres\_axis}$} \Comment{$\mathtt{thres\_axis}$: array of real numbers, trial matching thresholds;}
        \State $\mathcal{W}_i$ \tabto{1.75cm}$\gets$ \textit{connected\_components}$(d, t_i)$ \Comment{$\mathcal{W}$: array of integer sets, possible matching configurations;}
        \State $\mathcal{V}_i$ \tabto{1.75cm}$\gets$ \textit{count\_valid\_pairs}$(\mathcal{W}_i, d, t_i,  \mathtt{catalog})$ \Comment{$\mathcal{V}$: array of integers, number of valid pairs.}
    \EndFor
    
    \item[]
    \State $W_{\mathrm{opt}}$ \tabto{1.375cm}$:=$ the unique element in $\mathcal{W}$, which maximizes $\mathcal{V}$ at minimal $|\log(\mathtt{thres\_axis})|$  \Comment{the optimal configuration;}
    \State $t_{\mathrm{opt}}$ \tabto{1.375cm}$:=$ the unique element in \texttt{thres\_axis} corresponding to $W_{\mathrm{opt}}$ \Comment{the optimal matching threshold;}
    \State $w_0$ \tabto{1.375cm}$:=$ the unique element in $W_{\mathrm{opt}}$, where $(n + 1) \in w_0$ \Comment{group matching the queried coordinate.}
    \State $Q_i$ \tabto{1.375cm}$\gets$ \textit{quality\_metrics}$(w_i, s, d, t_{\mathrm{opt}})$ for $w_i$ in $W_{\mathrm{opt}}$ \Comment{$Q$: array of tuples, quality metrics of groups.}

    \item[]
    \If {queried host coordinate \textbf{and} \textit{length}$(w_0)>1$} \Comment{host position is queried and properly matched.}
        \State \texttt{primary\_host} $\gets w_0$
        \If {for all $p \in$ \textit{representative\_source}$(w_0)$, $\mathtt{is\_star}[p]=\mathtt{false}$} \Comment{host position matched a non-stellar object.}
            \State \texttt{is\_confirmed} $\gets$ \texttt{true} 
        \EndIf
        \If {using name-resolved coordinate}
            \State \texttt{is\_confirmed} $\gets$ \texttt{true}
        \EndIf
    \EndIf

    \item[]
    \If {$\mathtt{is\_confirmed}=\mathtt{false}$}
        \For {$w_i \in W_{\mathrm{opt}}$}
            \State $v_i \gets$ \textit{representative\_sources}$(w_i)$
            \State $\theta_i \gets \overline{\|s_p\|}$ where $p \in w_i$ \Comment{$\theta_i$: mean angular distance to the queried coordinate.}
            \State $f_i \gets$ \textit{ranking\_function}$(\theta_i, v_i, \mathtt{catalog}, \mathtt{is\_star}, Q_i, \ldots)$
        \EndFor
        \For {$w_i$ in $W$, sorted by $f$ in descending order,}
            \If{$\mathtt{primary\_host}=\mathtt{null}$} \Comment{transient coordinate is queried, or host coordinate unmatched.}
                \State \texttt{primary\_host} $\gets w_i$
            \Else
                \State insert $w_i$ to the end of \texttt{secondary\_hosts}
            \EndIf
       \EndFor
    \EndIf
    \EndProcedure
    
    \item[]
    \Function{count\_valid\_pairs}{$W$, $d$, $t$, \texttt{catalogs}}{$\;\to$ $V$}
        \State \textbf{input:}  \tabto{2.0cm} $W$:                  \tabto{3.75cm} list of integer sets, indices of sources in cross-matched groups;
        \State                  \tabto{2.0cm} $d$:                  \tabto{3.75cm} $(n+1, n+1)$ matrix, normalized pairwise distances of sources;
        \State                  \tabto{2.0cm} $t$:                  \tabto{3.75cm} positive real number, matching threshold.
        \State                  \tabto{2.0cm} \texttt{catalogs}:    \tabto{3.75cm} external catalog identifiers of sources;
        \State \textbf{output:} \tabto{2.0cm} $V$:                  \tabto{3.75cm} integer, the number of ``valid pairs'' under this configuration.

        \item[]
        \State $V$ $\gets$ $0$
        \For {$w_i$ $\in$ $W$}
            \ForAll{$p$, $q$ $\in$ $w_i$ where $p < q$}
                \If {$d_{pq}$ $<$ $t$ \textbf{and} \texttt{catalog}$[p]$ $\neq$ \texttt{catalog}$[q]$}
                    \State $V$ $\gets$ $V + 1$
                \EndIf
            \EndFor
        \EndFor
    \EndFunction
    
\end{algorithmic}
\end{algorithm}

\begin{algorithm}[H]
\caption{Cross-identification of host galaxies, continued}
\begin{algorithmic}[]
    \item[]
    \Function{connected\_components}{$d$, $t$}{$\;\to$ $W$}
        \State \textbf{input:}  \tabto{2.0cm} $d$:      \tabto{2.75cm} $(n+1, n+1)$ matrix, normalized pairwise distances of sources;
        \State                  \tabto{2.0cm} $t$:      \tabto{2.75cm} positive real number, matching threshold.
        \State \textbf{output:} \tabto{2.0cm} $W$:      \tabto{2.75cm} list of integer sets, a unique partition of integer set $\{1, \ldots, n+1\}$ that cannot be further refined, \\
                                                        \tabto{2.75cm} where for any $w_i \in W$ and for any $i\in w_i$, $\{j; j\in Q$, $j\neq i$ and $d_{ij}<t\} \subseteq w_i$
        \item[] \tabto{0.45cm}(abridged)
    \EndFunction

    \item[]
    \Function{representative\_sources}{$w$}{$\;\to$ $v$}
        \State \textbf{input:}  \tabto{2.0cm} $w$:      \tabto{2.75cm} set of integers, indices of sources in a cross-matched group.
        \State \textbf{output:} \tabto{2.0cm} $v$:      \tabto{2.75cm} set of integers, a subset of $w$, in which each external catalog contributes at most one source.
        \item[] \tabto{0.45cm}(abridged)
    \EndFunction
    
    \item[]
    \Function{ranking\_function}{$\theta$, $v$, $\mathtt{catalogs}$, $\mathtt{is\_star}$, $Q$, \ldots}{$\;\to$ $f$}
        \State \textbf{input:}  \tabto{2.0cm} $\theta$:            \tabto{3.75cm} real number, average angular distance to the origin, i.e., the queried coordinate;
        \State                  \tabto{2.0cm} $v$:                 \tabto{3.75cm} set of integers, indices of representative sources in a cross-matched group;
        \State                  \tabto{2.0cm} $\mathtt{catalogs}$: \tabto{3.75cm} array of integers, external catalog identifiers of sources;
        \State                  \tabto{2.0cm} $\mathtt{is\_star}$: \tabto{3.75cm} array of binary flags, indicator for stellar sources.
        \State                  \tabto{2.0cm} \ldots:              \tabto{3.75cm} other source properties.
        \State \textbf{output:} \tabto{2.0cm} $f$:                 \tabto{3.75cm} real number, the ranking score.
        \item[] \tabto{0.45cm}(abridged)
    \EndFunction

    \item[]
    \Function{quality\_metrics}{$w$, $s$, $d$, $t$}{$\;\to$ $Q$}
        \State \textbf{input:}  \tabto{2.0cm} $w$:                 \tabto{2.75cm} set of integers, indices of representative sources in a cross-matched group;
        \State                  \tabto{2.0cm} $s$:                 \tabto{2.75cm} array of real number pairs, projected Cartesian coordinates of sources;
        \State                  \tabto{2.0cm} $d$:                 \tabto{2.75cm} $(n+1, n+1)$ matrix, normalized pairwise distances of sources;
        \State                  \tabto{2.0cm} $t$:                 \tabto{2.75cm} real number, a cross-matching threshold.
        \State \textbf{output:} \tabto{2.0cm} $Q$:                 \tabto{2.75cm} 3-tuple, quality metrics of the cross-matched group.
        \item[] \tabto{0.45cm}(abridged)
    \EndFunction
    
\end{algorithmic}
\end{algorithm}
\twocolumngrid

\section{Finding astrometric tolerances and matching thresholds} \label{appendix:astromtol}

Every unique object in the sky, in ideal situations, corresponds to a cross-matched group of catalog sources, in which each catalog contributes at most one source.
Given a constant matching threshold, when per-catalog astrometric tolerances are excessively large, irrelevant nearby sources can be included in the cross-matched groups, and even nearby cross-matched groups can be ``glued'' together into one group;
conversely, insufficient per-catalog astrometric tolerances, in this case, may leave some catalog sources unmatched, and even split a properly-matched group into multiple smaller groups, leading to the incompleteness of compiled properties and non-unique correspondence of cross-matched groups and true objects in the sky.
The astrometric tolerance of each catalog should depend on its astrometry calibration and image resolution, as well as the angular size and signal-to-noise ratio of individual detected sources. Such information, however, is not always provided in external catalogs.
We, therefore, determine the per-catalog astrometric tolerances in an empirical approach. %

Finding the best per-catalog astrometric tolerances requires a performance metric, i.e., a numeric indicator for the goodness of cross-matching. 
We use the number of ``valid pairs'' cross-matched under constant, unity per-field matching threshold as the performance metric (or ``objective function'') to be optimized. Here two catalog sources form a valid pair if 1) they are \textit{directly} connected under the existing per-catalog astrometric tolerances and constant, unity per-field matching threshold, and 2) both sources are uniquely matched in this group, without the confusion of multiple sources in the same catalog.
Such a combinatorial objective function is tailored towards large cross-matched groups with the least confusion. We use the first criterion to increase the chance of having sources matched, while the second criterion penalizes confusion of multiple sources -- inside a group, we only count valid pairs of uniquely-matched sources.
The number of sources and their maximal number of possible connections are always limited in a field. When properly matched, the same source can participate in more valid pairs and thus have a higher per-source contribution; meanwhile, a higher fraction of those possible connections would become real.
From either perspective, counting the number of connections is a reasonable choice.

Maximizing the objective function is a computationally intensive optimization problem, where the value for each catalog is a free parameter to tune, and evaluating the objective function requires cross-matching all catalog sources in all existing fields for once.
Even more complicated is that the objective function is a non-differentiable and non-continuous one, which may have a plethora of local extrema that challenges conventional optimization algorithms.
Given the unknown properties of this objective function, instead of using any local or global optimization algorithm, we search the best astrometric tolerances with a Monte-Carlo approach.
We replace the log-likelihood function in Markov chain Monte Carlo algorithm with our objective function so that astrometric tolerances with better overall performance (i.e., a higher number of valid pairs) are preferred during the sampling process.
Given enough time, we could find a set of best global astrometric tolerances.
It must be emphasized that the Monte Carlo sample, in this case, should not be interpreted as the ``posterior'' of the best-fitting parameters, as the objective function itself is not a likelihood or any other probability density in nature.

Converting the objective function to a ``log-likelihood'' function needs some scaling.
Any monotonically increasing function, in principle, can be used here as a ``scaling function.''
However, if the gradient of the scaling function is too steep, then minor changes in the number of valid pairs may lead to a significant change in the density of the Monte Carlo sample. 
Given that the Monte Carlo sample traces linear density while the value of the ``log-likelihood'' function is supposed to be logarithmic, the sampling process can be thus extremely sensitive to local maxima, leading to ``trapped'' chains, slow convergence, and over-constrained, ``spiky'' Monte Carlo sample.
On the contrary, if the gradient of the scaling function is too shallow, then the ``log-likelihood'' function cannot effectively perceive the improvement of the objective function, leading to an overly smooth Monte Carlo sample with low contrast of density.
We tested several possible scaling functions, and a simple power function with an index of $0.618$ leads to the best balance of sampling efficiency and the contrast of sample density.

\begin{figure*}
\centering
\includegraphics[width=\linewidth]{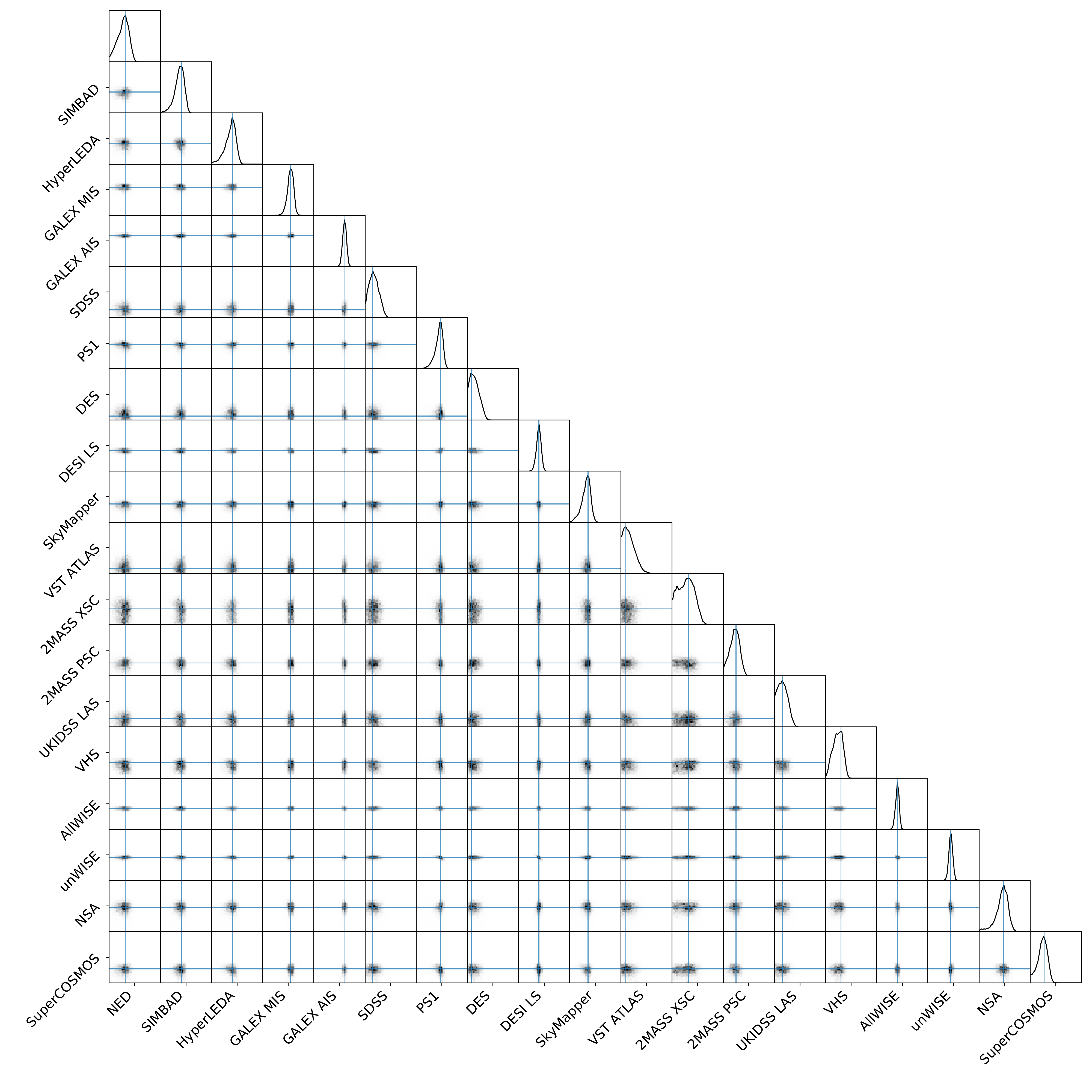}
\caption{
Finding the best astrometric tolerances for external catalogs using Monte-Carlo sampling.
The results here show a Monte Carlo sample of 200,000 points, optimized for an objective function with a constant power index of 0.618 for the scaling function, as discussed in Appendix \ref{appendix:astromtol}.
We use a fixed axis range of $(0'', 3'')$ for all sub-panels, including marginalized distributions and two-dimensional projections of the sample.
The best-overall astrometric tolerances, determined by searching the peak of marginalized distribution using a Gaussian kernel of $0.125''$, are indicated with blue lines. The detailed values are listed in Table \ref{tab:astromtol}. 
\label{fig:astromtol}}
\end{figure*}

There are 30 thousand fields in our sample. To evaluate the objective function, all these fields must be cross-matched again using the newly proposed astrometric tolerances.
{To make the problem computationally feasible, we pre-calculate pairwise distance matrices and combine them into larger block diagonal matrices so that each evaluation of the objective function takes only a few seconds on a single core of a conventional desktop computer (Intel Core i7-9700K, 3.6 GHz), and sampling 200,000 steps using \texttt{emcee} \citep{ForemanMackey13} takes less than a day.}
We locate the best astrometric tolerances by finding the maximum of the marginalized Monte-Carlo sample density using a Gaussian smoothing kernel of 0.125'', after throwing away the first 20,000 points. The final sampled parameter space is illustrated in Figure \ref{fig:astromtol}, and the best-overall parameters are listed in Table \ref{tab:astromtol}.

When finding the astrometric tolerance using Monte Carlo sampling, we used a fixed, unity matching threshold for all our fields.
We consider the astrometric tolerance as a globally-averaged property of a catalog, reflecting the typical scatter of source coordinates averaged over all existing fields.
However, the globally-averaged astrometric tolerances are not necessarily the optimal values for individual fields. There could be field-to-field variations of source density and catalog coverage. Meanwhile, the scatter of source coordinates for hosts may also depend on their redshifts.
The same set of globally-averaged astrometric tolerances could be overly large in certain fields but too small for some other fields, combining or splitting properly-matched groups and leading to mismatching or incompleteness of host properties.

To enable some local fine-tuning of the cross-matching criteria, we allow the matching threshold to vary in each field. Tuning the matching threshold is equivalent to scaling the astrometric tolerances by a constant factor in this field.
When cross-matching, we use the matching threshold that maximizes the number of valid pairs matched in this field, and the matching threshold is chosen from 31 logarithmically-spaced values from $0.316$ to $3.16$.
When several matching thresholds produce the same maximal number of valid pairs, we use the one that is closest to unity.

\begin{figure}
\centering
\includegraphics[width=\linewidth]{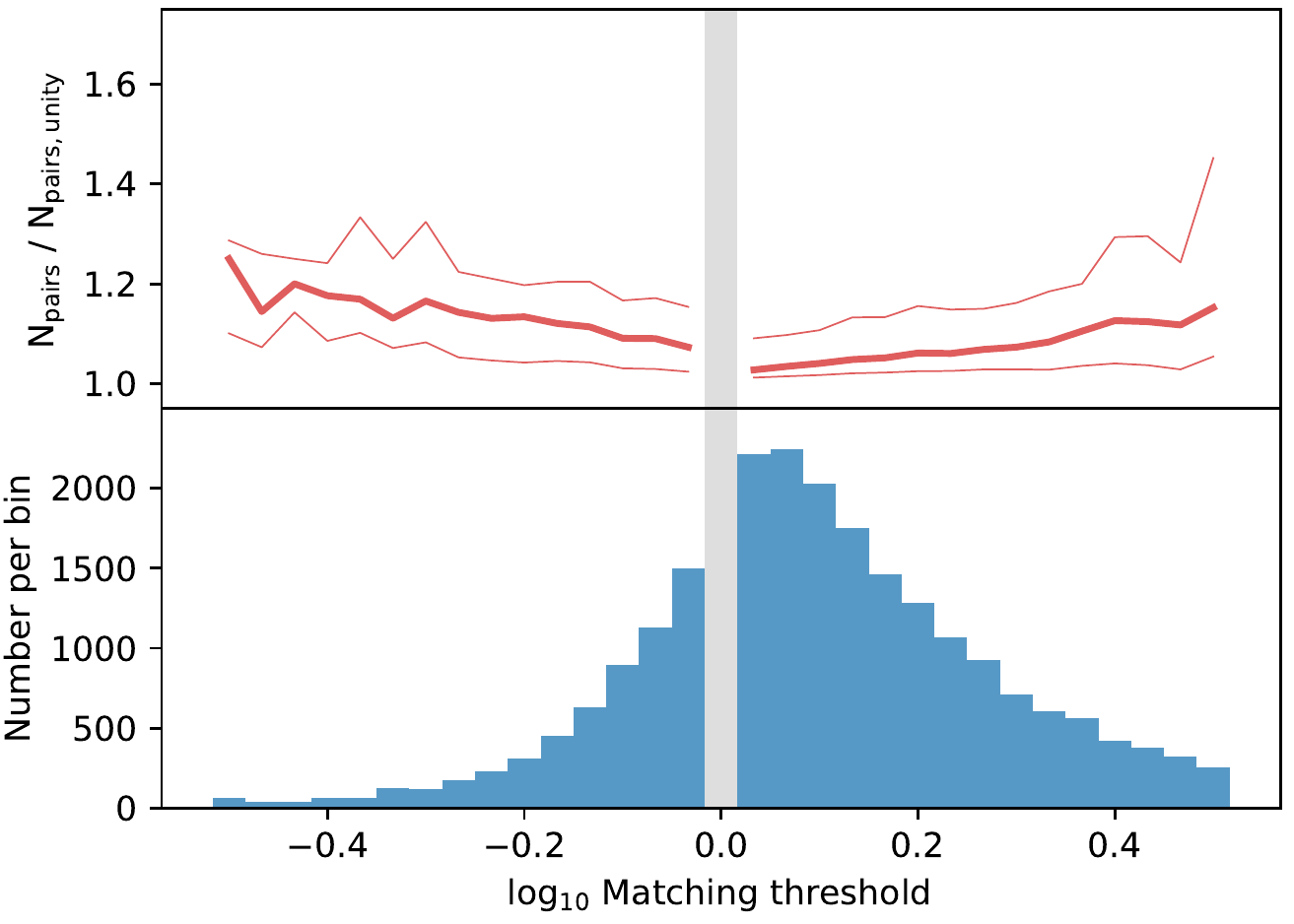}
\caption{(Top) The improvement of connectivity after tuning the matching threshold. We count the number of valid pairs ($N_{\mathrm{pair}}$) using the adjusted matching threshold and compare it to the number of valid pairs matched under unity threshold ($N_{\mathrm{pair,\,unity}}$). Their ratio represents the improvement of cross-matching after per-field fine-tuning of the matching threshold. The solid line show the moving median of $N_{\mathrm{pair}}/N_{\mathrm{pair,\,unity}}$, while the thin lines show the corresponding 25\%-75\%iles. 
(Bottom) Histogram of the locally-adjusted matching threshold. More than half of the fields are cross-matched using a default unity matching threshold (gray zone), so we only show the distribution for those with different matching thresholds.
\label{fig:matchingthreshold}}
\end{figure}

In Figure \ref{fig:matchingthreshold}, we show the distribution of the fine-tuned per-field cross-matching threshold, along with the improvements in the number of valid pairs, compared to the number of pairs matched under a constant threshold of unity.
For confirmed hosts and primary candidates, after the cross-matching threshold is tuned, there is about 10\% to 20\% increase in the number of valid pairs.
We note that even we allow the matching threshold to vary in each field, the default matching threshold of unity is still the best value in most fields.

\section{Quality Metrics of Cross-matching} \label{appendix:xmatchqc}

Cross-matching sources across multiple external catalogs is a non-trivial problem in many aspects. We use per-catalog astrometric tolerances and per-field matching thresholds to determine the connectivity of sources, which we have already discussed in Appendix \ref{appendix:groupproperties}.
To further ensure that our cross-matching is valid and the compiled host properties are accurate, we introduce three quality control metrics for cross-matched groups:
\begin{enumerate}
    \item[-] Mean offset ($d$), which measures the average angular distance of sources to the group centroid, i.e., the equal-weighted mean position of sources in this group. This metric measures the size of a cross-matched group, where compact groups should have a lower mean offset.
    \item[-] Axis ratio ($\sigma_a/\sigma_b$), which measures the ``roundness'' of a cross-matched group, calculated using the eigenvalues of a second-order moment matrix. This is the axis ratio (semi-major to semi-minor) of sky coordinate distribution, where lower values (closer to 1) indicate a more ``isotropic'' distribution.
    Along with this axis ratio, we also provide similar indicators, such as ellipticity and the absolute value of Pearson's r in this case.
    \item[-] Degree of connectivity ($F$). This is the number of direct connections in a group divided by the maximal possible number of connections in a group with $N$ sources.
    This parameter measures the compactness of a group, where a compact group is expected to have most sources inter-connected, except for source pairs from the same catalog; while a loose, wide-spanning group should have many source pairs unconnected, resulting in a lower value for this number.
\end{enumerate}
For all cross-matched groups, including known hosts, newly identified host candidates, and other cross-matched groups, we always provide these three quality control metrics.
We also make sure that these parameters behave well in the limiting case of only a few sources matched, when some parameters may have singular values. For the connectivity, we use the number of direct matches divided by the number of maximal possible matches plus one to avoid division by zero for single-source groups. For ellipticity, we assume that single-source groups are close to groups with only two sources. Their axis ratio equals $0$, ellipticity equals $1$, and Pearson's r is also $1$.

Besides these quality metrics, we also provide lists of catalogs that cover the field and that have been matched by the confirmed host or primary candidate of the field.
For catalogs covering the field but not matched by the confirmed host or primary candidate, we also provide the nearest source in these catalogs to our confirmed host or primary candidate.
We note that such kind of situation does not necessarily indicate quality issues in cross-matching, but this can be used as a diagnostic for potentially under-matched (``splitted'') groups and to locate missed sources.
Detailed statistics of cross-matching are listed in Table \ref{tab:crossmatchingsummary}.

\begin{deluxetable}{lrrrr}
\tablecaption{Summary of Cross-matching \label{tab:crossmatchingsummary}}
\tablenum{8}
\tablehead{
\colhead{Catalog} & \colhead{Matched} & \multicolumn{2}{c}{Nearby} & \colhead{Confusion} \\
\cline{3-4}
\colhead{}        & \colhead{}        & \colhead{$<2''$} & \colhead{$<5''$} & \colhead{}}
\startdata
NED                     &    29070 &       18 &      239 &      733 \\
SIMBAD                  &    21005 &       38 &      275 &      184 \\
HyperLEDA               &    21123 &       25 &      237 &       27 \\
\textit{GALEX} MIS      &     2991 &       18 &      251 &        2 \\
\textit{GALEX} AIS      &    16181 &       93 &     1170 &       48 \\
SDSS                    &    22945 &       33 &      306 &      504 \\
PS1                     &    28145 &       44 &      450 &      302 \\
DES                     &     8559 &       19 &      116 &       72 \\
DESI LS                 &    28439 &       34 &      313 &     1538 \\
SkyMapper               &    11070 &       31 &      249 &      101 \\
VST ATLAS               &     3045 &        5 &       32 &      508 \\
2MASS XSC               &    14900 &       11 &       63 &        3 \\
2MASS PSC               &    18745 &       31 &      211 &      389 \\
UKIDSS LAS              &     8183 &       16 &      181 &      304 \\
VHS                     &     7968 &       24 &      185 &      122 \\
\textit{AllWISE}        &    28009 &      118 &     1175 &      102 \\
unWISE                  &    31530 &      150 &     1126 &      966 \\
NSA                     &     9634 &       14 &      161 &        1 \\
SCOS                    &    24265 &       87 &      597 &      375 \\
\textit{Gaia}           &     9608 &       23 &      750 &      174 \\
MPA-JHU                 &     7558 &        5 &       81 &        6 \\
\enddata
\tablecomments{\textit{Matched} indicates the number of confirmed hosts and primary candidates with the catalog matched. \textit{Nearby} shows the number of cases when a confirmed host or primary candidate did not match source in a certain catalog, but there exists sources in this catalog within some angular distance. This need not to be a problem of cross-matching. \textit{Confusion} shows the number of confirmed hosts and primary candidates with multiple sources in this catalog matched.}
\end{deluxetable}

\begin{figure*}
\centering
\includegraphics[width=0.45\linewidth]{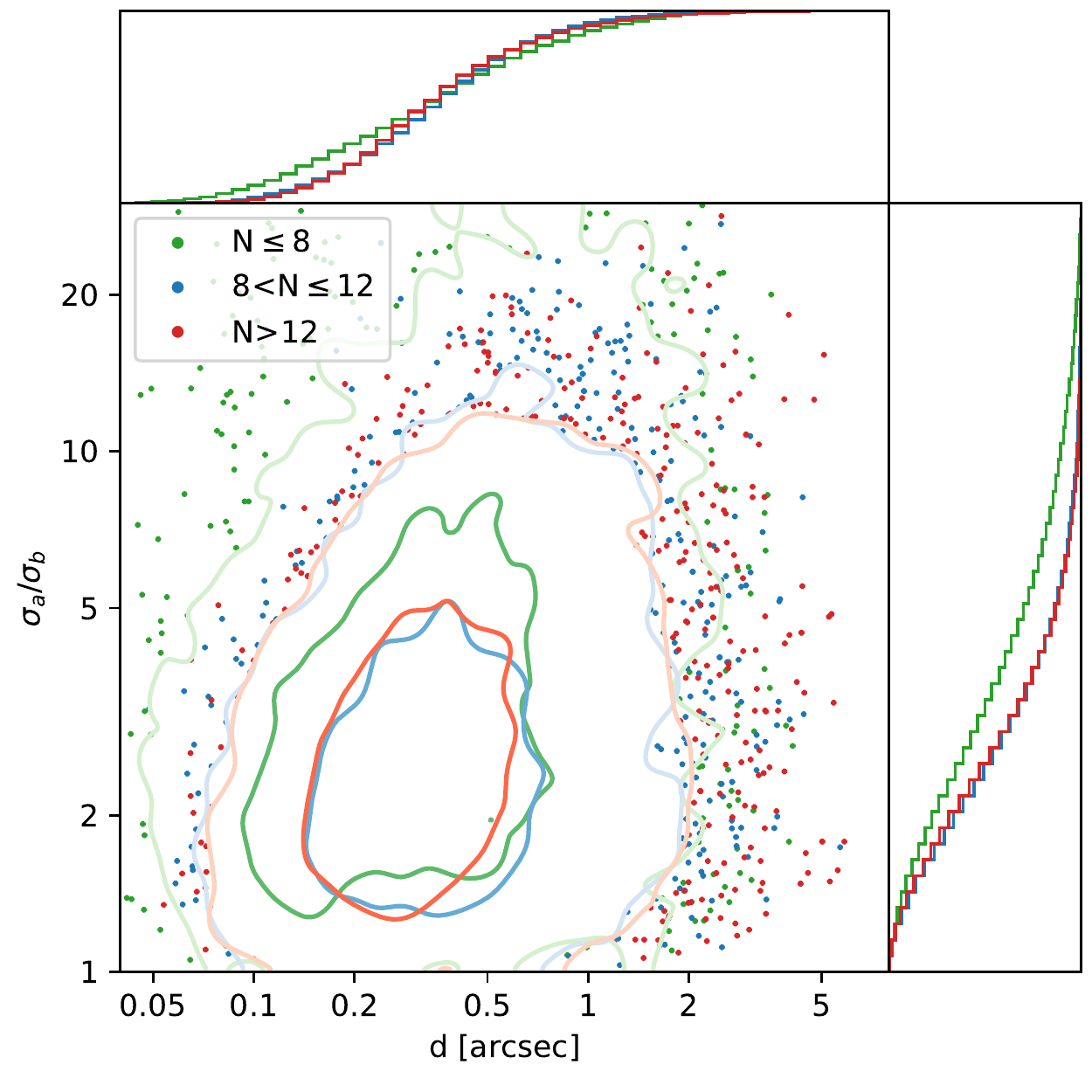}
\includegraphics[width=0.45\linewidth]{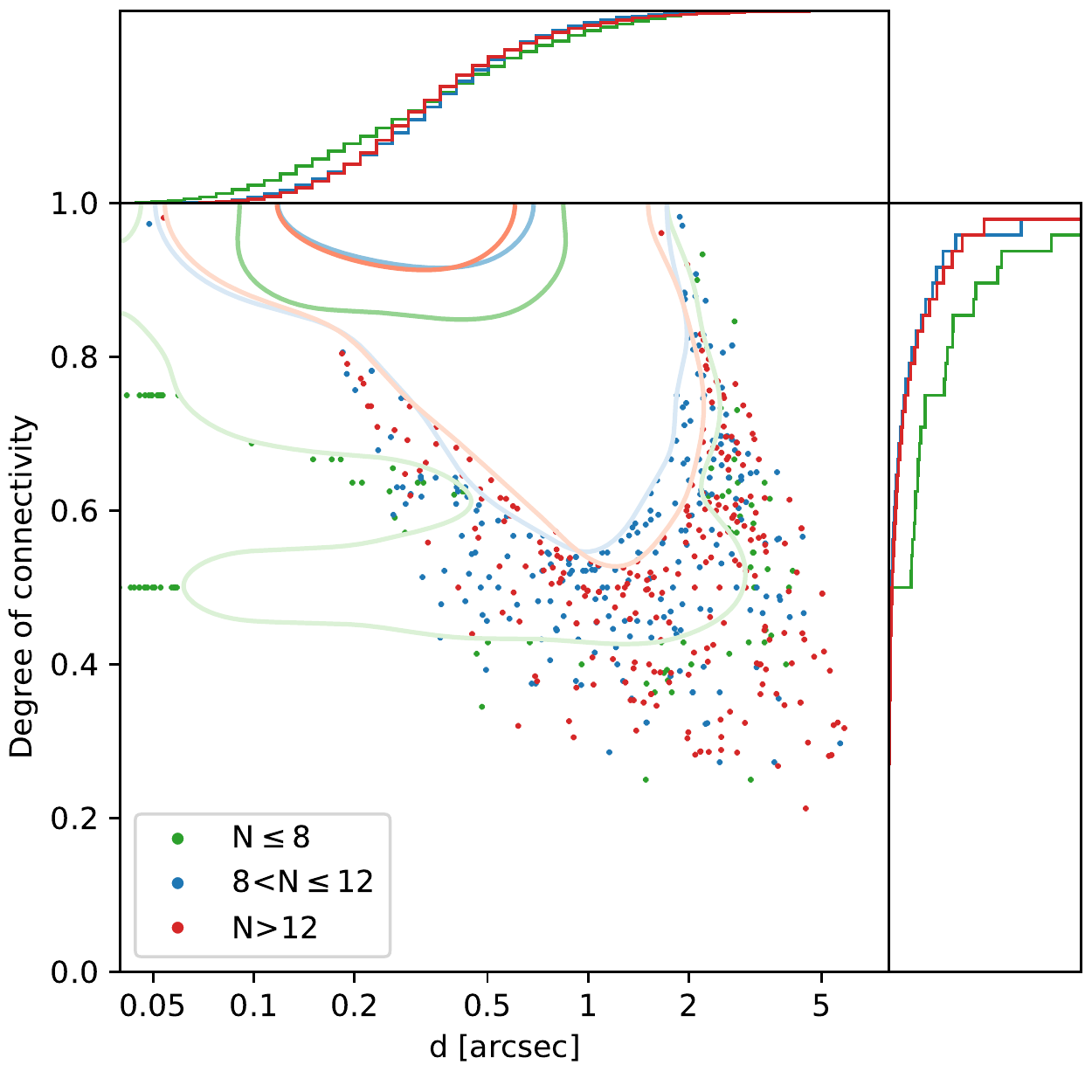}
\caption{
Three quality control metrics of cross-matching for known and newly identified hosts.
For cross-matched objects in a group, we calculate their average distance to the group centroid (``mean offset,'' $d$), the axis ratio (semi-major to minor) of covariance ellipse ($\sigma_a/\sigma_b$), and the number of directly connected pairs to the maximal number of allowed connections (i.e., ``degree of connectivity'').
We split our hosts into 3 sub-groups of nearly equal sizes per their number of cross-matched catalog objects ($N$). The dark and light contours in each panel enclose 50\% and 95\% hosts in each sub-group, while hosts outside 95\% contours are plotted as individual points.
The top and right sides of each panel show the cumulative histograms of each metric for each sub-group.
Properly matched groups are expected to be compact (low $d$) and isotropic ($\sigma_a/\sigma_b$ near unity), in which catalog objects are adequately connected (high degree of connectivity).
Cross-matched groups with extreme values, like large $d$ and $\sigma_a/\sigma_b$, or very low degree of connectivity, should be used with care.
\label{fig:xidqc}}
\end{figure*}

\begin{figure*}
\centering
\includegraphics[width=0.45\linewidth]{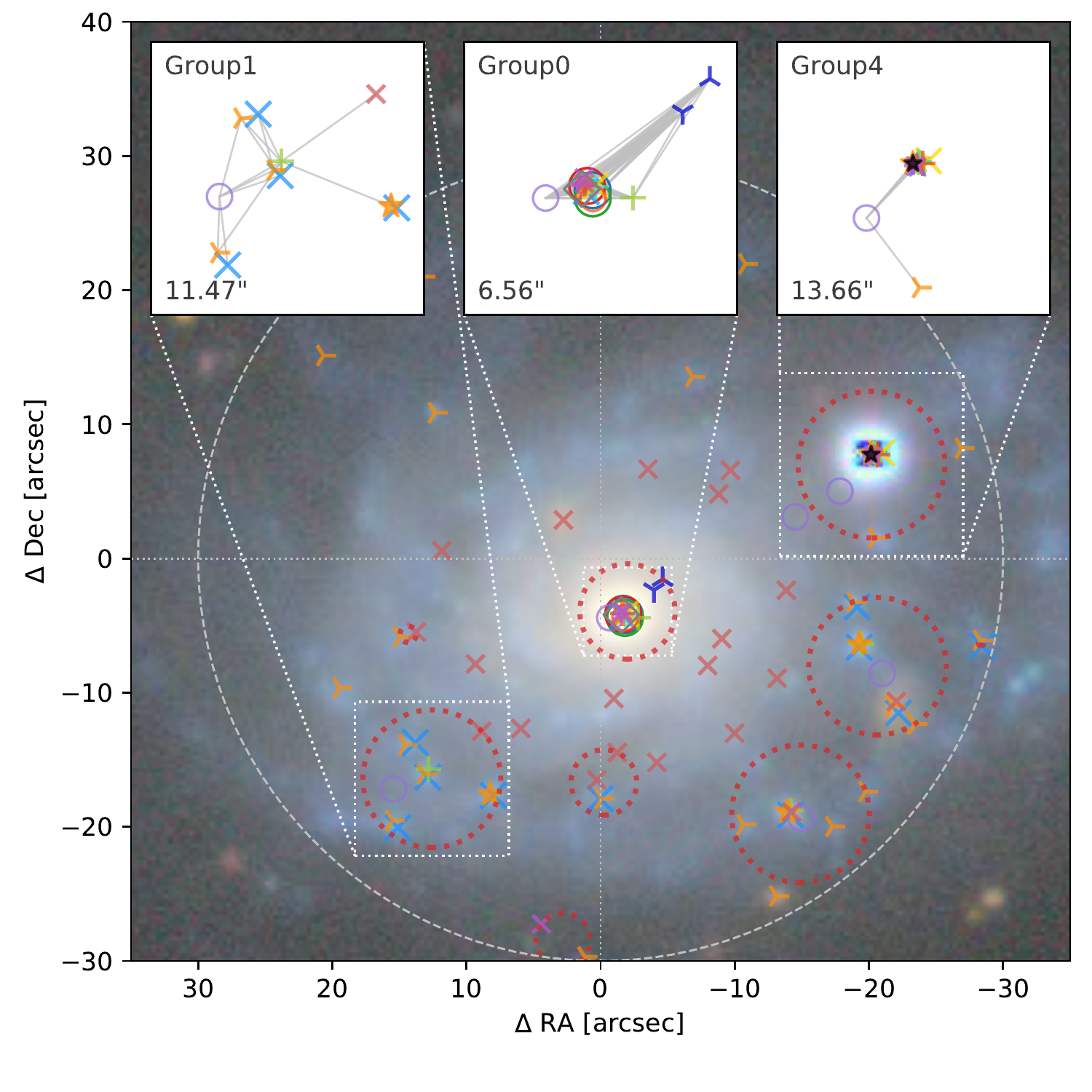}
\includegraphics[width=0.45\linewidth]{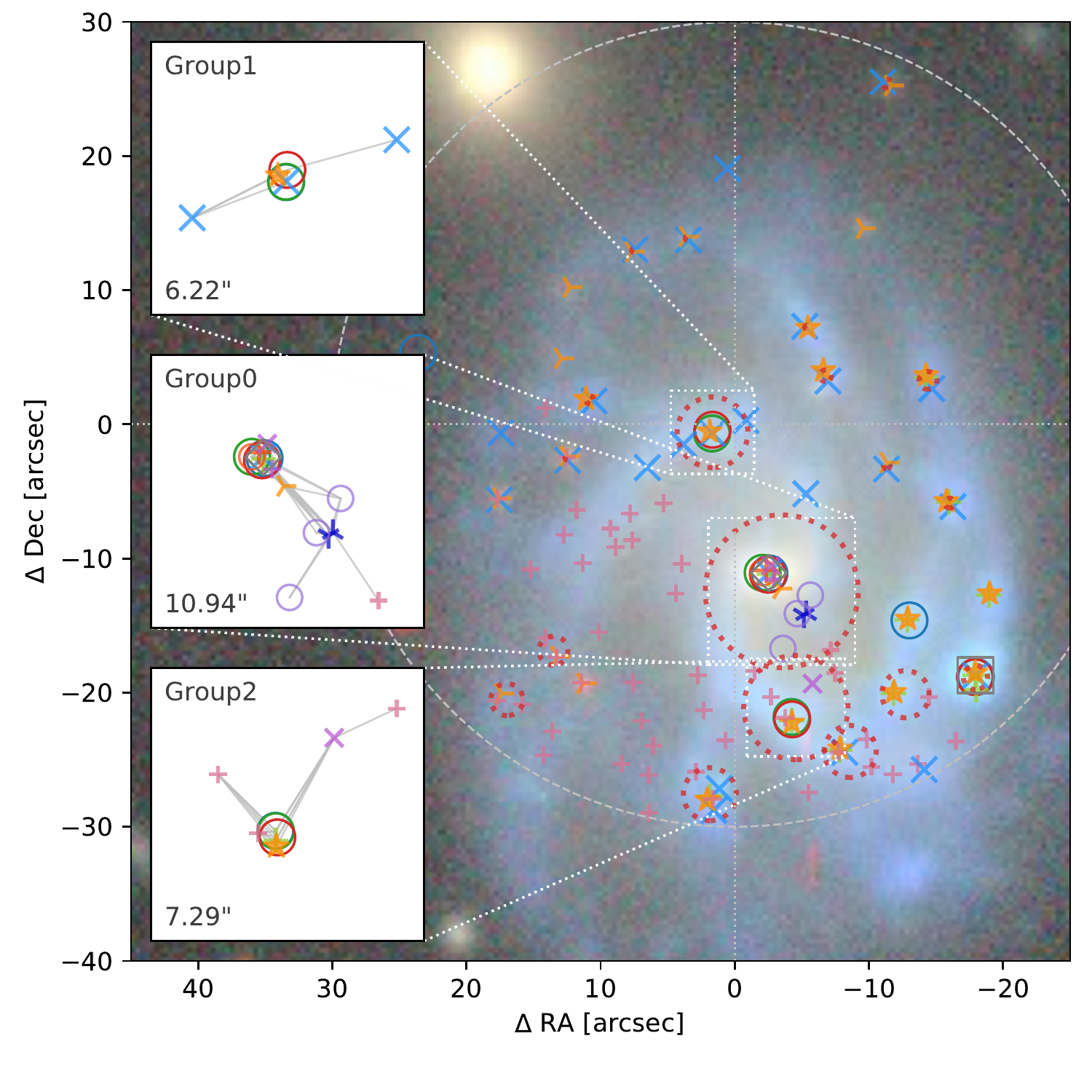} \\
\includegraphics[width=0.45\linewidth]{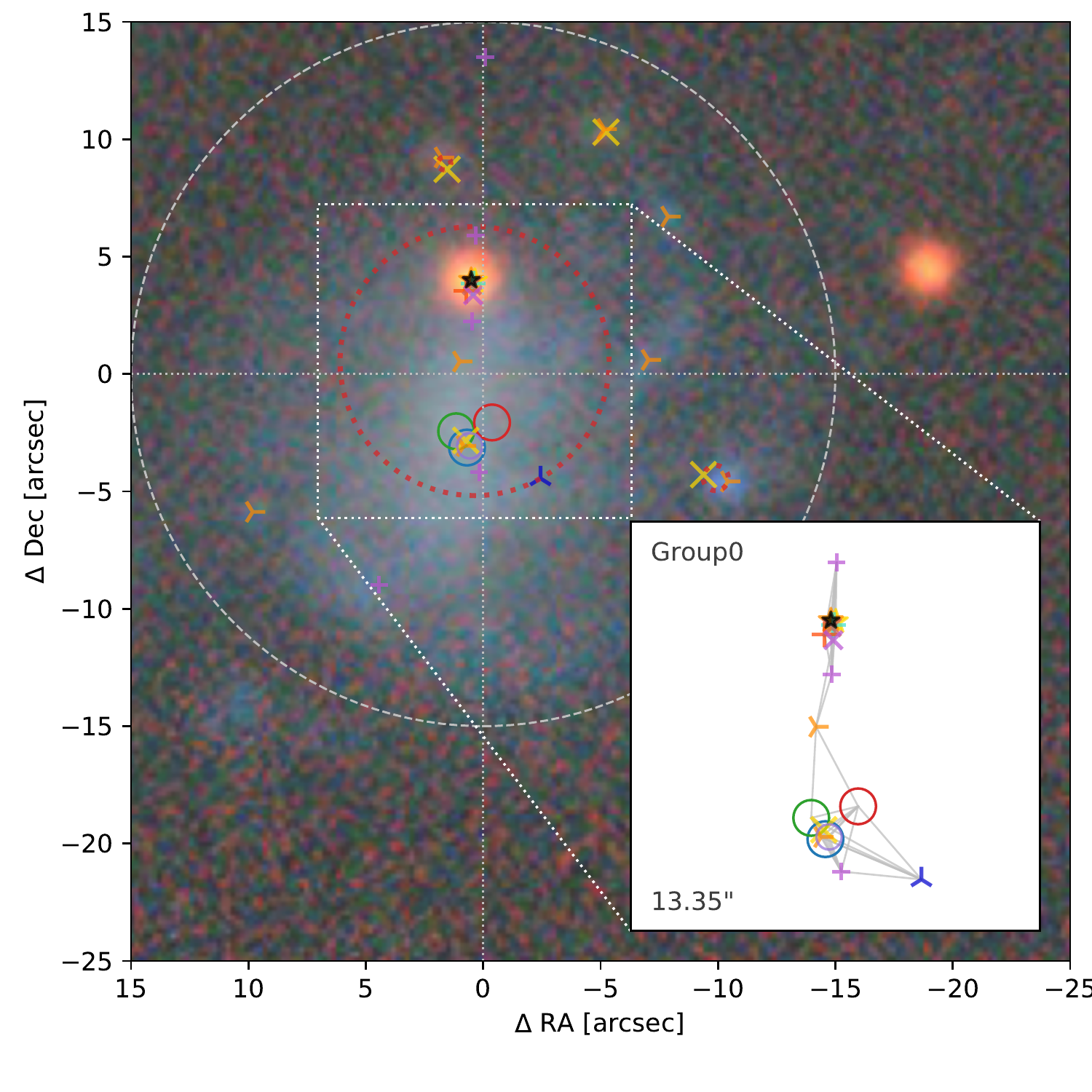} 
\includegraphics[width=0.45\linewidth]{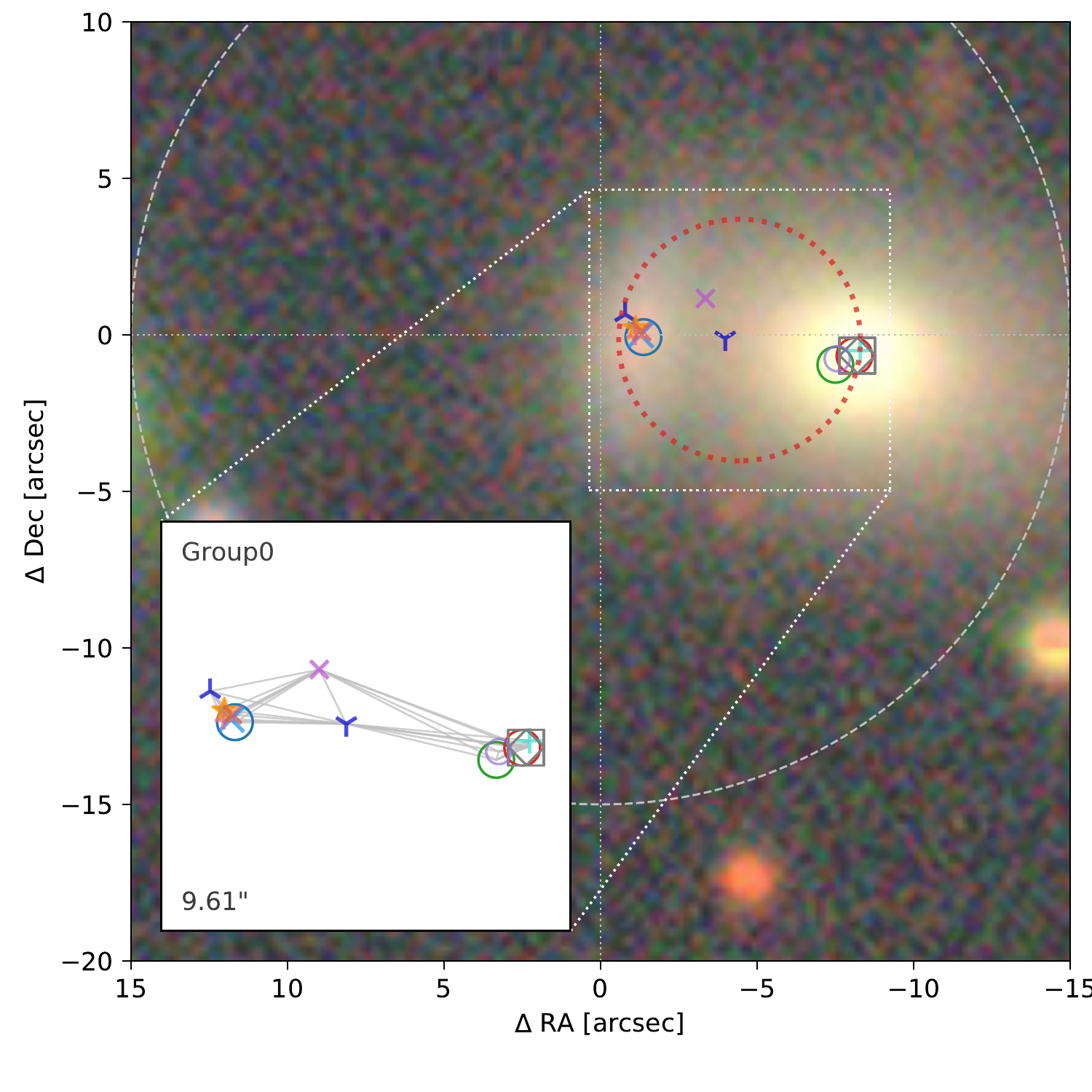}
\includegraphics[width=0.90\linewidth]{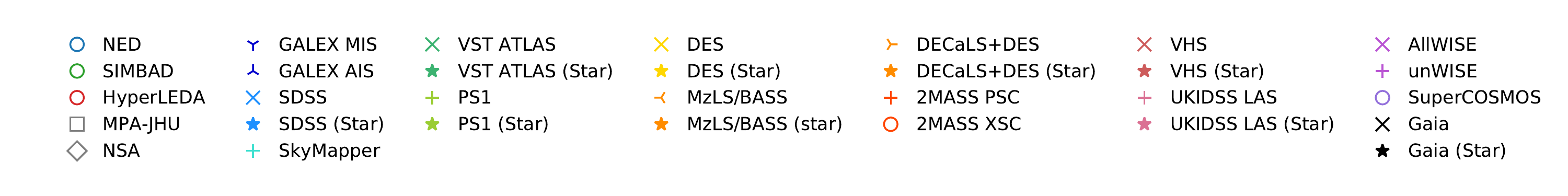}
\caption{
(Top) Cross-matching catalog sources in fields with nearby large galaxies. Two examples here are PS16fdq (left) and SNhunt124 (right).
The spiral arms, bright clumps, and extended outskirts of nearby large galaxies are often broken into individual sources when the object detection algorithm is not tailored towards such situations.
Nearby large galaxies with low surface brightness, flat light profile, or irregular morphology may also have significant catalog-to-catalog offsets in source coordinates.
Consequently, their compiled host properties are subject to incompleteness or mismatching. Even the cross-matched substructures may interfere with the identification of new hosts.
(Bottom) Cross-matching catalog sources with non-optimal astrometric tolerances or matching thresholds. Examples here are SN2017hdn (left) and SDSS-II SN 15822 (right).
The cross-matched groups in both cases contain at least two distinct objects, which are ``glued'' into a single group either due to large scatter of source coordinates or by a few ``bridging sources.''
Such groups can be excluded by their extreme values of connectivity and mean offset.
We illustrate the connectivity of catalog sources in selected groups with zoom-in panels, where panel sizes are indicated at lower left corners. Red dashed circles indicate cross-matched groups with spanning distance above 1''.
\label{fig:largegalaxyxmatch}}
\end{figure*}

It is worth noting that, nearby large galaxies pose a significant challenge to our cross-matching algorithm and their properties should be used with care.
These well-resolved galaxies could have large catalog-to-catalog offsets in sky coordinates, and sometimes these large galaxies are even split into multiple sources in certain catalogs.
We demonstrate relatively successful cases and failed cases in Figure \ref{fig:largegalaxyxmatch}.

\section{Star/galaxy separation and resolution of confusion} \label{appendix:groupproperties}

Before locating the groups corresponding to our known hosts or ranking cross-matched groups to identify the best host candidates, we need to assign source properties to cross-matched groups.
The procedure here includes a simple star/galaxy separation, as foreground stars should not be considered as host candidates, as well as the selection of source properties when a group includes multiple sources from the same catalog, i.e., resolving confusion of multiple matched sources.

To accurately identify foreground stars in each field, upon cross-matching catalog sources, we utilize source morphology indicators or star/galaxy separation parameters provided in these catalogs to flag potential stellar sources.
More accurately, these indicators or parameters only separate point and extended sources. Some extragalactic objects, such as quasars and cores of extremely compact galaxies, could be classified as point sources. Conversely, some extended sources could be foreground sources like stellar multiplicities or nebulae. 
Here ``stars'' and ``galaxies'' are only used loosely to refer to these two types of sources.
We choose rather strict criteria for stellar sources so that the identified subset of sources is pure and genuine galaxies would not be easily flagged as stellar sources.
After flagging individual stellar sources in external catalogs, we check if our cross-matched groups contain stellar sources. Without the confusion of multiple sources, when any catalog source in a cross-matched group satisfies the criteria for stellar sources, we flag the entire group as a star.

Even with optimized astrometric tolerances and matching thresholds, confusion or missing of sources could still occur in some fields, i.e., a certain catalog may contribute more than one valid source to a cross-matched group, or a catalog source that corresponds to the true host is not included in the group.
Both cases are penalized by our target function of optimization as described in Appendix \ref{appendix:astromtol} and should have already been minimized. However, the first case, if present, could still lead to mismatching of host properties.

Whenever a cross-matched group contains more than one source from a certain catalog, we keep all sources but select only one of them to represent the measured properties of the cross-matched group within that catalog.
For photometric catalogs, the brightest source is selected as the representative source, and the deepest photometric band of each survey is used to rank confusing sources.
For other catalogs, the one closest to the group centroid, i.e., the average position of sources in this group, is selected to represent the properties of this group.
In any case, we prefer non-stellar sources over stellar ones when finding the representatives within confusing sources.

Finally, in case a cross-matched group contains confusing sources in certain catalogs, and the group is not yet labeled as a star under stellar criteria for other catalogs without confusion, we select representative sources for catalogs with confusion first, and then label the entire group as a star only if any of these representative sources satisfies our stellar criteria.
In other words, when there is a confusion of multiple sources in a catalog with stellar source criteria, and the group is not labeled as a star by other catalogs without confusion, it would only be labeled as a star when all confusing sources satisfy stellar criteria in at least one of the catalogs that have multiple confusing sources.

\section{The Ranking Functions} \label{appendix:rankingfunction}

For events without known hosts, we rank cross-matched groups to identify their best host candidates. A ranking function, which takes the properties of cross-matched groups as input and returns scores that reflect their likelihood of being the right host, is thus necessary.
Here we use the decision functions of conventional binary classifiers as the ranking function. Constructing a ranking function can be done by training a binary classifier using existing host galaxies.

\subsection{Constructing the training set}

{To construct a training set for ranking functions, we choose transients that 1) have name-resolved or as-reported host coordinates in our trusted reference sources, 2) have valid sources in external catalogs cross-matched using host coordinates, and 3) have been visually inspected by us to have acceptable quality, i.e., ``OK'' cases without further flags, comments, or alternative hosts (Appendix \ref{appendix:inspectionknown}). 
We search their transient coordinates again in external catalogs, with search radius set as if neither host names nor host coordinates are known.
We then cross-match the retrieved catalog sources in these fields with the same astrometric tolerances. Query results and cross-matched groups are stored in a separate ``training'' database, with an identical structure as the main database.
Cross-matched groups closest to the known, visually-confirmed hosts in the main database are labeled as ``true hosts,'' while other groups in the field are labeled as ``other objects.''
In the absence of a ``true host,'' we exclude the field from the training set. About $0.3\%$ ``true hosts'' have large angular offsets ($\geq 3''$) to their counterparts in the main database, which are usually caused by the change of cross-matching thresholds (and hence group members) under different search radii and field centers (Appendix \ref{appendix:astromtol}). We do not use these fields for the training set.}

{When assembling the training set, we do not include fields in which the true hosts are beyond the search radii. However, we keep track of these excluded fields to estimate their impact on the accuracy afterward.}

\subsection{Constructing the feature sets}

To characterize cross-matched groups in each field, we start from the following parameters:
\begin{enumerate}

    \item [-] The angular distances of group centroids to the transient coordinate ($\delta \theta$, in arcseconds). Besides linear angular distances, we also include $\mathrm{arcsinh}(\delta\theta/\text{arcsec})$ and $1/(1 + \delta \theta/\text{arcsec})$ in the list of parameters.

    \item [-] The projected physical distances ($\delta r$, in kiloparsecs), for events with known redshifts only. Similarly, besides linear distances, we also include $\mathrm{arcsinh}(\delta r/15 \, \text{kpc})$ and $1/(1 + \delta r / 15\,\text{kpc})$ in the list of parameters.

    \item [-] Binary flags indicating if the cross-matched group contains source(s) from a specific catalog. Every survey or value-added catalog corresponds to one flag.

    \item [-] Geometric signatures and quality metrics of cross-matched groups, including the mean offset ($d$), axes ratio ($\sigma_b/\sigma_a$) and the degree of connectivity ($F$) of each group. These parameters are discussed in detail in Appendix \ref{appendix:xmatchqc}.

    \item [-] {``Stellarity'' parameters, including a binary flag indicating if any representative source satisfies the criteria for stellar sources and the fraction of representative sources from catalogs with star-galaxy separation criteria that are likely stars.}

    \item [-] Field-aware contextual parameters, including the quantile ranks within the field for 1) the number of catalogs matched, 2) the linear distance to transient coordinates, and 3) the connectivity parameter ($F$).

\end{enumerate}

The list above includes 32 or 35 parameters (3 angular distances, 3 optional projected physical distances, 21 binary flags for external catalogs, 3 geometric and quality metrics, 2 parameters for ``stellarity,'' 3 contextual parameters).
We choose these parameters because this is a \textit{complete} set of parameters, which are universally available for any cross-matched groups, even in sky areas that are not covered by certain surveys or for hosts that are beyond the sensitivity limits of certain surveys. Also, they do not rely on detailed galaxy properties.
{These two feature sets, which we name as \texttt{Basic8} (32 parameters) and \texttt{Basic8\_z} (35 parameters, including redshift-dependent projected distances), lay the groundwork for other input feature sets.}

{We also extend the ``basic'' feature sets above using other more detailed source properties. Since optical-near infrared (NIR) magnitudes are the most available kind of source properties in external catalogs, we create the \texttt{AnyMag}/\texttt{AnyMag\_z} feature sets, which the include existing parameters in \texttt{Basic8}/\texttt{Basic8\_z} feature sets, with one more parameter, the optical-NIR magnitude of the group. This optical-NIR magnitude is selected in DESI LS, PS1, DES, SDSS, VST ATLAS, and SkyMapper catalogs, in the preferred order here. We use $r$, $i$, or $z$-band extended source magnitude in these catalogs, whichever is available first.
Due to their dependence on source optical-NIR magnitudes, the feature sets also limit the choice of possible hosts to groups with any of these catalogs matched. Other groups would not be ranked and selected as the host candidate.}

{Furthermore, since PS1 and DESI LS are the two most widely available external catalogs to our host candidates, we construct feature sets based on their measured source properties. The \texttt{PS1}/\texttt{PS1\_z} feature sets, which are also based on the existing \texttt{Basic8}/\texttt{Basic8\_z} feature sets, have the following additional parameters:}
\begin{enumerate}
\item[-] {The Kron magnitudes \cite{Kron80} of PS1 sources, in $r$, $i$ or $z$-band, whichever is properly measured first in this preferred order;}
\item[-] {The Kron radii in the same band as the magnitude above, which measure the angular sizes of objects using the first radial moments of their surface brightness profiles;}
\item[-] {The ratio of $\delta \theta$ to the Kron radius above, which is analogous to $d_{\mathrm{DLR}}$ used in the Directional Light Radius (DLR) method, but without correcting for the axis ratio or inclination of galaxies.}
\end{enumerate}
{Since the Kron radius here includes seeing contribution, the sources sizes are overestimated, especially for compact sources. This may reduce the distinguishing power of the ranking function in some situations.}

{We also constructed \texttt{LS}/\texttt{LS\_z} feature sets, which include parameters in the existing \texttt{Basic8}/\texttt{Basic8\_z} feature sets, with the following additional ones:}
\begin{enumerate}
    \item [-] {The profile-fitting magnitude of DESI LS sources, preferably in $r$-band, but $z$ or $g$-band are also used in the absence of $r$-band data;}
    \item [-] {The measured half-light radius $R_{50}$ and the calculated $R_{90}$ (radius enclosing 90\% of total flux) under the best-fitting light profile. For point sources, both radii are set to 0.1'';}
    \item [-] {$d_{\mathrm{DLR}}$ calculated using $R_{50}$ and $R_{90}$, considering the position angle and axis ratio of galaxies, following the definition in \citet{Sullivan06};}
    \item [-] {The photometric redshifts $z_{\mathrm{ph}}$ of LS sources from \citep{Zhou21}, the squared difference of transient redshift and $z_{\mathrm{ph}}$, and finally,
    $$\sqrt{(z - z_{ph})^2 / \sigma_{z_{ph}}^2 + 1} - 1$$
    where $\sigma_{z_{\mathrm{ph}}}$ is the uncertainty of $z_{\mathrm{ph}}$ estimate.}
\end{enumerate}

{Note that $R_{90}$ of galaxies are derived from $R_{50}$ using the ratio of $R_{90}$ to $R_{50}$ under the best-fitting light profile. For exponential profile, we use $R_{90}/R_{50} \simeq 3.52$; while for de Vaucouleurs' profile, we use $R_{90}/R_{50} \simeq 5.58$. The position angles and axis ratios are converted from ellipticities in complex representation.
When fitted with composite models (exponential plus de Vaucouleurs'), the complex ellipticity, $R_{50}$, and $R_{90}$ are the linear combination of the two components, weighted by their fractions in the total flux.
The last three $z_{\mathrm{ph}}$-related parameters are only used for transients with known redshift (i.e., only in \texttt{LS\_z} feature set).
The function forms are chosen to separate true hosts (with consistent redshift) and other objects (with possible inconsistent redshift) easily using simple linear classifiers.}

{For testing and comparison, we also assemble feature sets that only use information relevant to PS1 or LS catalogs. We derived \texttt{PS1sub}/\texttt{PS1sub\_z} feature sets from the existing \texttt{PS1}/\texttt{PS1\_z} feature sets, in which parameters related to multi-catalog cross-matching and other cross-matched catalogs are ignored. Similarly, we assemble \texttt{LSsub}/\texttt{LSsub\_z} feature sets using only parameters available from LS sources alone, including $z_{\mathrm{ph}}$-related parameters.}

{Feature sets using more detailed source properties in external catalogs are only available for some host candidates. We plot the fraction of availability in the bottom sub-panels of Figure \ref{fig:logisticaccuracy}.}

\subsection{Testing the classifiers}
\label{appendix:rankingfunctionaccuracy}

{Having the training set ready, we trained several commonly-used binary classifiers, including Logistic Regression, Support Vector Machine (SVM; with regularization parameter $C=0.25$), Random Forest (RF; with a maximal depth of $5$), AdaBoost, Stochastic Gradient Descent (SGD), and Multilayer Perceptron (MLP), using the default hyper-parameters in \texttt{scikit-learn} (v0.22.1) unless noted.}

\begin{figure*}
\centering
\includegraphics[width=0.45\linewidth]{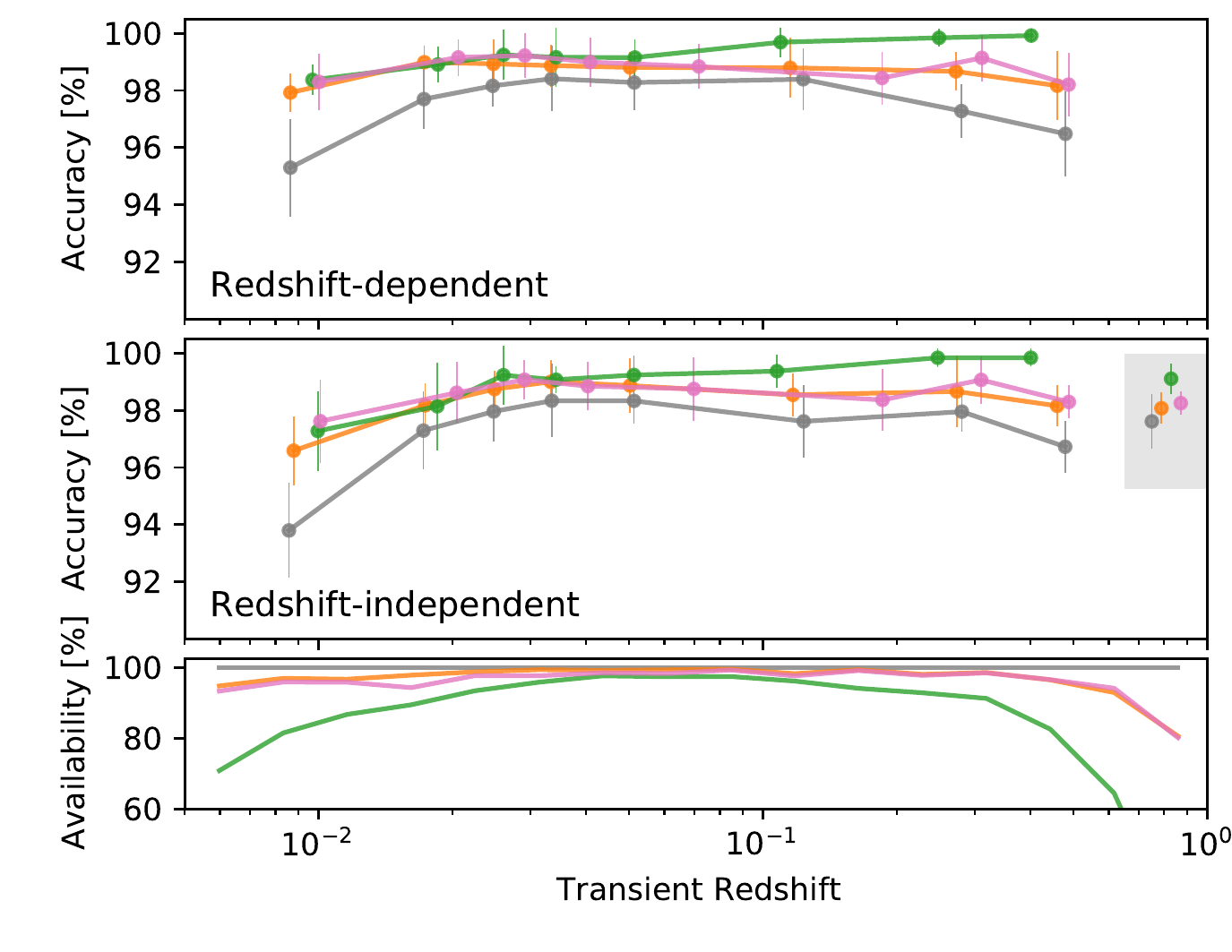}
\includegraphics[width=0.45\linewidth]{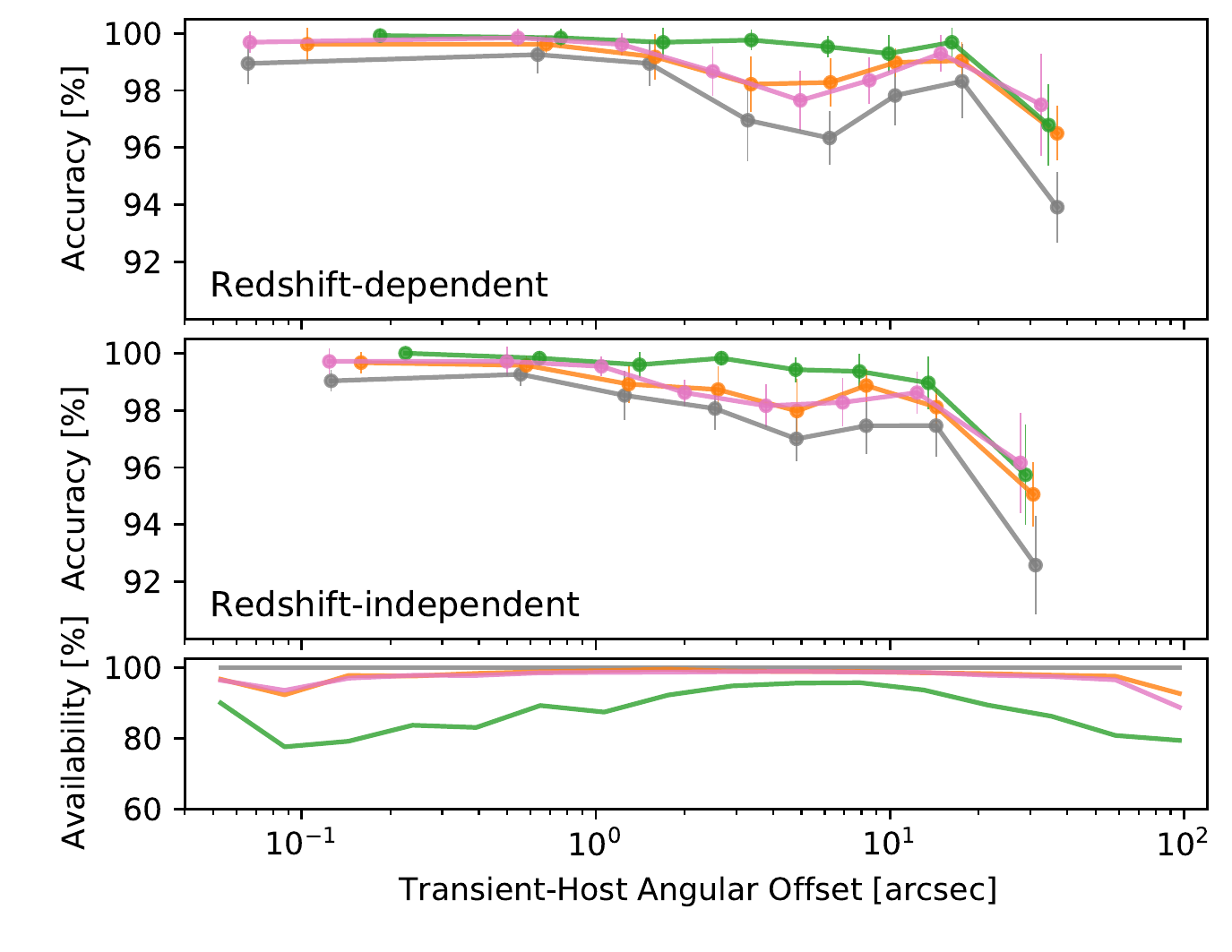} \\
\includegraphics[width=0.45\linewidth]{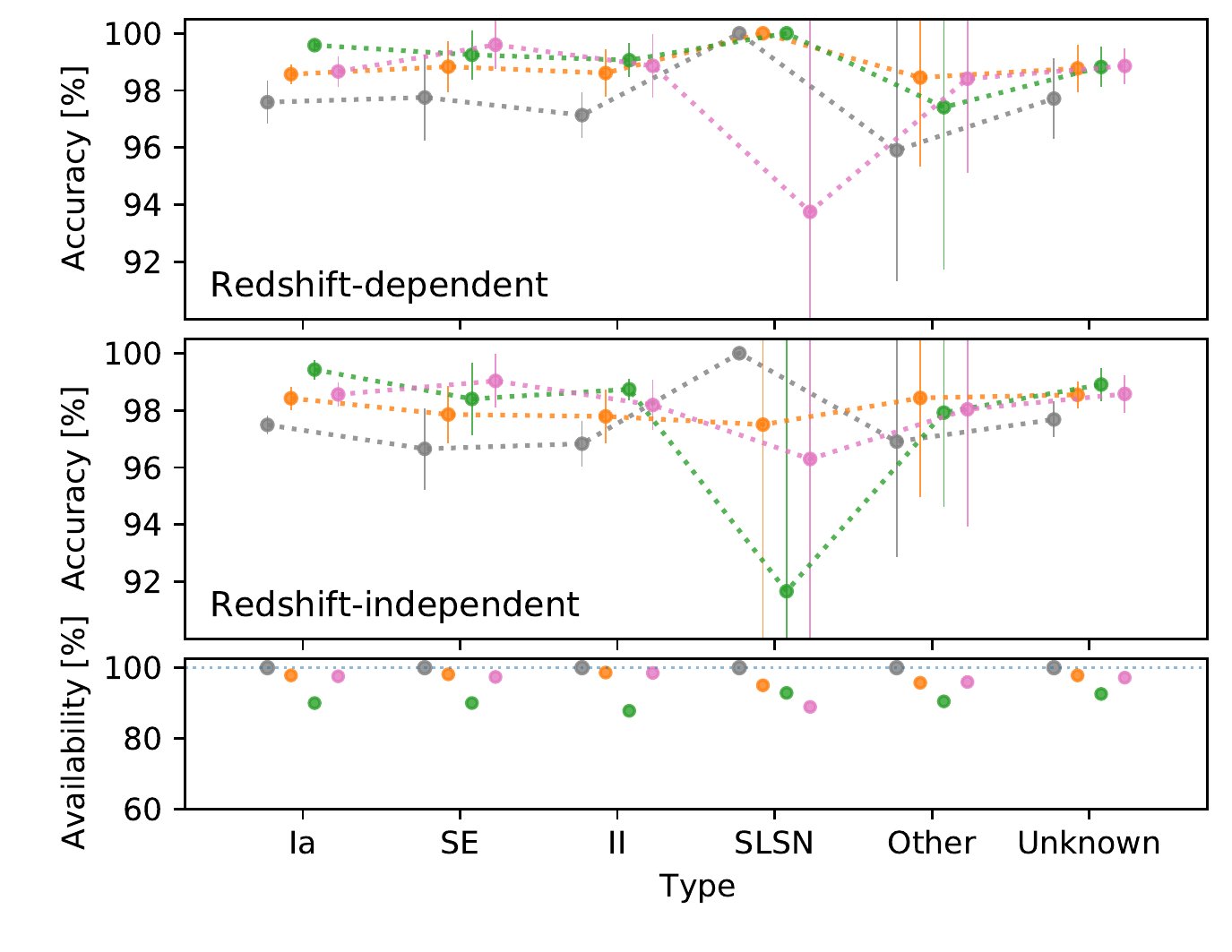} 
\includegraphics[width=0.45\linewidth]{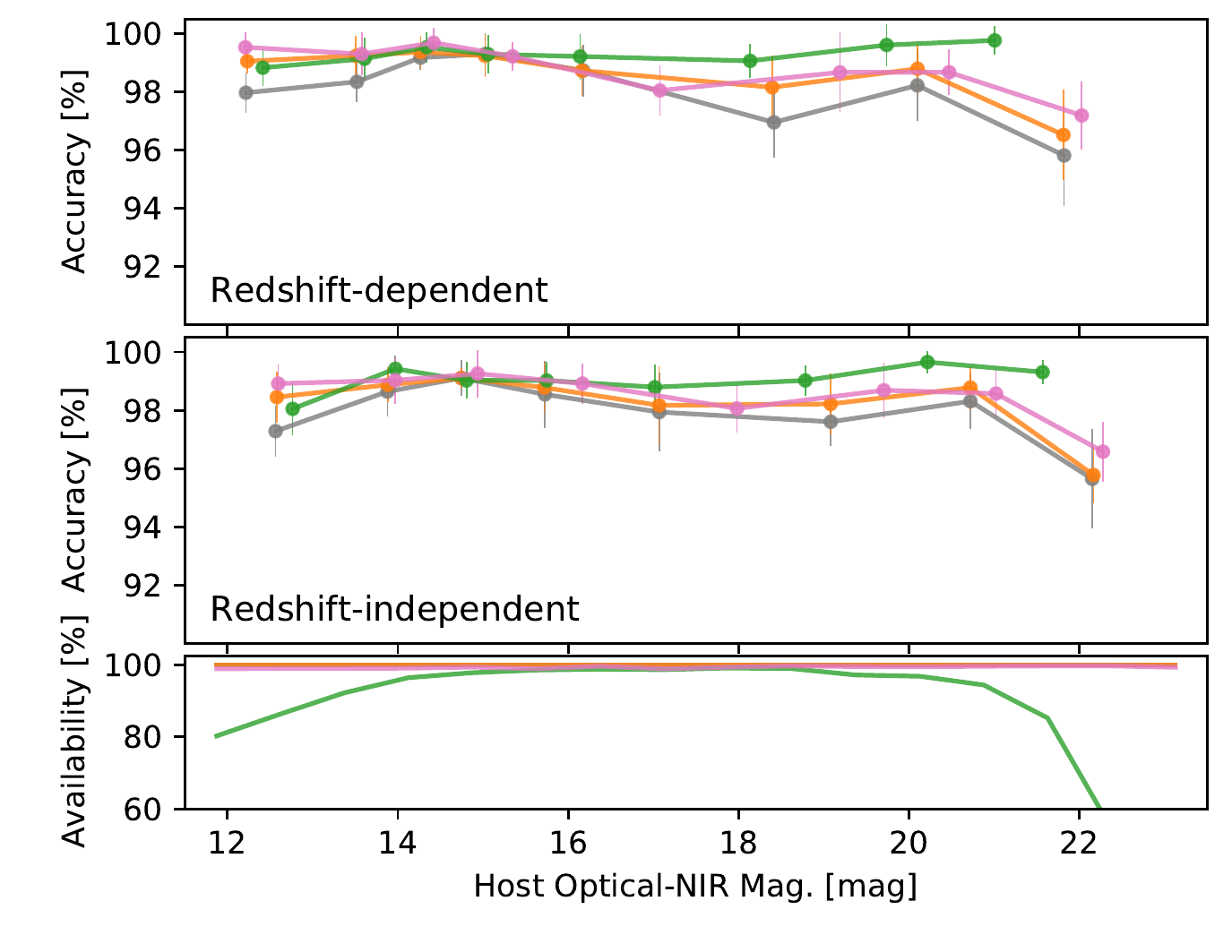} \\
\includegraphics[width=0.90\linewidth]{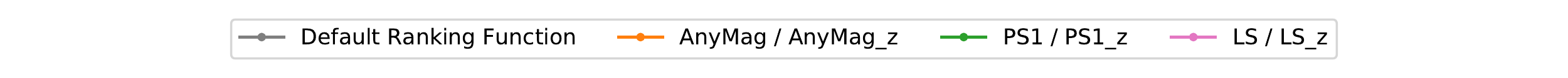}
\caption{{Accuracy of the ranking function estimated using the training sample, averaged over standard 10-fold cross-validation.
We show the dependence of accuracy on transient redshift, transient-host angular offset, transient type, and host optical-NIR magnitudes in four separate panels. Except for the panel of transient type, we divide the sample into equal-sized bins over the axis of interest and plot the per-bin accuracy at the median value of each bin.
Each curve shows the accuracy of a feature set. For clarity, the accuracy curves of redshift-dependent feature sets (with ``\texttt{\_z}'' suffix) and redshift-independent feature sets are grouped into two separate sub-panels.
Some feature sets rely on certain source properties that are not always available for those known hosts. We estimate the accuracy using the subset of known hosts with the relevant features available and outline the fraction of true hosts with relevant parameters available in the bottom sub-panels.
Specifically, in the panel of transient redshift, we plot the accuracy of events without redshift in the gray shaded area, if applicable.}
\label{fig:logisticaccuracy}}
\end{figure*}

{We train ranking functions with every possible combination of classifiers and input feature sets. When using feature sets with transient redshift-dependent parameters (i.e., feature sets with ``\texttt{\_z}'' suffix), we limit our training and testing to transients with known redshifts; otherwise, the entire training set is used.
If the feature set relies on source properties in a particular external catalog, we further limit the training set to transients in the coverage of that survey, whose ``true hosts'' have a source in that catalog matched.
Even for transients in the coverage of the survey, the ``true hosts'' may not have the required parameter available. We show the fraction of availability in Figure \ref{fig:logisticaccuracy}.}

{Furthermore, we create two special ranking functions for comparison, one selects the nearest group to the transient coordinate (``Naive Nearest''), and the other selects the group with the minimal $d_{\mathrm{DLR}}$ (based on $R_{50}$) in the field (``Simple DLR'').
They directly use $-\delta\theta$ or $-d_{\mathrm{DLR}}$ as the ranking scores, where the negative signs select groups with the lowest $\delta\theta$ or $d_{\mathrm{DLR}}$.
The latter ranking function is only limited to transients and host candidates with DESI LS parameters.}

We assess the performance of these trained ranking functions by the chance that the true host has the highest ranking score in its field, which we loosely refer to as the accuracy.
The accuracy is averaged over a standard 10-fold cross-validation (Section \ref{sec:accuracyknown}), where training and test datasets are split based on fields rather than individual cross-matched groups.
{Besides the overall accuracy, we also analyzed the dependence of accuracy on transient redshift, transient-host angular offset, host optical-NIR magnitude, and transient type. The results are estimated in bins of equal sample size (except for transient type), and the accuracy in each bin is averaged over 10-fold tests. The results are summarized in Figure \ref{fig:logisticaccuracy}.}

{We summarize the overall performance of ranking functions in Table \ref{tab:accuracymatrix}.
With the ``basic'' feature sets, the accuracy is above $96\%$ for all classifiers, in which the Logistic Regression classifier achieves $97.5\pm0.6\%$ accuracy using transient redshift-related parameters, or $97.3\pm3\%$ without using redshift-related parameters.
Therefore, we use Logistic Regression and ``basic'' feature sets as our default ranking functions. When transient redshift is available, we use the redshift-dependent feature set (\texttt{Basic\_z}); otherwise, we use the redshift-independent version (\texttt{Basic}).}

{We calculate the ranking scores of other feature sets in our released dataset, but we do not use the scores to rank cross-matched groups. However, it is worth comparing their performance with the default ranking functions.
We see a clear increase in accuracy when source properties are included (Table \ref{tab:accuracymatrix}, Figure \ref{fig:logisticaccuracy}).
For example, adding any optical-NIR magnitude to the existing redshift-dependent ``basic'' feature set improves the accuracy to $98.6\%$. The inclusion of source properties in PS1 and LS can even push the accuracy further to $99.3\%$ and $98.8\%$, respectively.
For their redshift-independent versions, the accuracy is similar, if not marginally lower.
Even a percent-level increase of accuracy in this range indicates a significant reduction of failure rates compared to the default ranking function. Therefore, the inclusion of detailed source properties brings a substantial performance improvement.}

{The improvements in accuracy, however, comes with the cost that true hosts without the required extra parameters measured are ignored. 
For example, many true hosts do not have PS1-relevant parameters measured, especially for fainter or brighter hosts close to sensitivity and saturation limits (Figure \ref{fig:logisticaccuracy}, ``Availability'' sub-panels). Even PS1-based feature sets outperform our default ranking functions (``basic'' feature sets), this may not benefit host matching in reality.
On the contrary, feature sets with DESI LS-relevant parameters are available for most of the true hosts. \textit{Assuming} those true hosts, if above the sensitivity of DESI LS, will be detected and cataloged, the improved accuracy would facilitate host matching of transient surveys.}

{We also notice that when using DESI LS-relevant parameters alone (\texttt{LSsub}/\texttt{LSsub\_z}), one can reach comparable accuracy to our default ranking functions (i.e., ``basic'' feature sets). Also, \texttt{LSsub}/\texttt{LSsub\_z} feature sets have a reasonable fraction of availability in the survey footprint (Figure \ref{fig:logisticaccuracy}), clearly higher than the coverage of PS1-relevant parameters.
This implies that the training framework would also work for single-catalog host matching if the catalog and input parameters are properly chosen.
On the contrary, PS1-relevant parameters improve the accuracy of the default feature sets, but using these parameters alone leads to low accuracy of around $90\%$.}

{Finally, ranking cross-matched groups by $d_{\text{DLR}}$, the accuracy is about $90\%$, lower than the DLR-only accuracy in \citet{Gagliano20}, and close to the DLR-only mock sample performance of \citep{Gupta16}.
Choosing the nearest group to the transient position as the host, only about half of the true hosts are successfully recovered. The ranking function method we present here significantly outperforms these methods.}

\begin{deluxetable*}{lrrrrrr}
\tablecaption{Average Accuracy of Ranking functions from Cross-validation \label{tab:accuracymatrix}}
\tablenum{9}
\tablehead{\colhead{Name} & \colhead{\texttt{Basic}(\texttt{\_z})} & \colhead{\texttt{AnyMag}(\texttt{\_z})} & \colhead{\texttt{PS1}(\texttt{\_z})} & \colhead{\texttt{LS}(\texttt{\_z})} & \colhead{\texttt{PS1sub}(\texttt{\_z})} & \colhead{\texttt{LSsub}(\texttt{\_z})}}
\startdata
\rule{0pt}{4ex}    
         & \multicolumn{6}{c}{With redshift-dependent parameters} \\
\cline{2-7}
Logistic        & $97.5\pm0.6$ & $98.6\pm0.3$ & $99.3\pm0.1$ & $98.8\pm0.4$ & $90.7\pm1.1$ & $97.6\pm0.4$  \\ 
SVM             & $97.4\pm0.7$ & $98.6\pm0.3$ & $99.3\pm0.1$ & $98.7\pm0.4$ & $90.3\pm1.1$ & $97.5\pm0.4$  \\ 
RF              & $97.3\pm0.6$ & $98.4\pm0.3$ & $99.3\pm0.1$ & $98.6\pm0.5$ & $87.6\pm1.5$ & $96.1\pm0.5$  \\ 
AdaBoost        & $97.0\pm0.4$ & $98.4\pm0.2$ & $99.0\pm0.4$ & $98.6\pm0.3$ & $89.8\pm1.0$ & $97.1\pm0.4$  \\ 
SGD             & $96.9\pm0.6$ & $98.1\pm0.2$ & $99.2\pm0.2$ & $98.5\pm0.4$ & $91.0\pm1.6$ & $96.6\pm0.4$  \\ 
MLP             & $96.8\pm0.6$ & $97.9\pm0.3$ & $99.1\pm0.2$ & $98.3\pm0.6$ & $89.9\pm1.3$ & $94.5\pm0.8$  \\ 
\rule{0pt}{4ex}    
         & \multicolumn{6}{c}{Only redshift-independent parameters} \\
\cline{2-7}
Logistic        & $97.3\pm0.3$ & $98.3\pm0.3$ & $99.0\pm0.3$ & $98.5\pm0.3$ & $89.0\pm0.8$ & $95.7\pm0.5$  \\ 
SVM             & $97.4\pm0.4$ & $98.2\pm0.3$ & $99.0\pm0.3$ & $98.5\pm0.3$ & $88.2\pm0.7$ & $95.6\pm0.6$  \\ 
RF              & $97.1\pm0.3$ & $98.1\pm0.4$ & $99.0\pm0.3$ & $98.3\pm0.3$ & $88.5\pm0.8$ & $95.0\pm0.5$  \\ 
AdaBoost        & $96.7\pm0.4$ & $98.0\pm0.3$ & $98.9\pm0.3$ & $98.3\pm0.3$ & $90.4\pm0.8$ & $95.6\pm0.4$  \\ 
SGD             & $96.9\pm0.5$ & $97.8\pm0.4$ & $99.0\pm0.3$ & $98.2\pm0.3$ & $88.7\pm0.7$ & $94.8\pm0.6$  \\ 
MLP             & $96.4\pm0.4$ & $97.6\pm0.4$ & $99.0\pm0.3$ & $97.9\pm0.5$ & $89.1\pm0.8$ & $93.6\pm0.8$  \\ 
\rule{0pt}{4ex}    
         & \multicolumn{6}{c}{Other methods} \\
\cline{2-7}
Naive Nearest  & $53.2\pm1.0$ & --           & --           & --           & --           & --            \\ 
Simple DLR     & --           & --           & --           & --           & --           & $89.8\pm0.4$  \\ 
\enddata
\tablecomments{{Each row represents the accuracy of a classifier using the input feature set indicated in each column. The upper and lower parts shows the results for redshift-dependent feature sets (with ``\texttt{\_z}'' suffix) and redshift-independent feature sets. The last two rows shows the accuracy of two special ranking functions for comparison.}}
\end{deluxetable*}

\section{Visual Inspection of Cross-identified Host Galaxies}
\label{appendix:inspection}

{We rely on existing name-resolved or as-reported host coordinates to create training sets for ranking functions, where the quality of input data may affect the training results. Meanwhile, the trained ranking functions, although achieving good accuracy in known transient-host pairs, have not been tested for performance on newly identified hosts. A comprehensive visual inspection of cross-identified hosts, in either case, can serve as a subjective yet independent test of data quality.}

{To inspect the cross-identified hosts, we obtained image cutouts of their fields, including DESI LS $grz$ color composite images from the Sky Viewer\footnote{\url{www.legacysurvey.org/viewer}}, PS1 $gri$ color composite and $g$-band \texttt{FITS} images from the Hierarchical Progressive Survey (HiPS) datasets hosted at Centre de Donn\'{e}es astronomiques de Strasbourg (CDS)\footnote{\url{alasky.u-strasbg.fr/hips-image-services/hips2fits}}, as well as DSS2 color composite and red-band \texttt{FITS} images from the HiPS datasets at CDS.
Images cutouts are centered at the queried coordinates (i.e., known host coordinates or transient coordinates). The box size is set to 2.5 times the queried radius, rounded up to the nearest multiplicity of $15”$.
Preferably we use DESI LS images for visual inspection for their better sensitivity over PS1 images. DSS2 images, on the contrary, are only used when neither DESI LS  images nor PS1 images are available due to their lower sensitivity and resolution compared to modern digital sky surveys.}

{We developed specialized software to facilitate the inspection, which overlays symbols and markers of transient coordinates, cross-matched groups, possible stellar objects, and projected distance scales ($5$, $10$, and $20$ kpc) onto the background image.
We indicate the cross-matched catalogs of each group using color bands on these symbols. Furthermore, for groups with DESI LS sources cross-matched, we also use upward or downward arrows to indicate if transient redshift is below or above the $95\%$ confidence interval of their photometric redshift estimates in \citet{Zhou21}.}

\begin{sidewaysfigure*}[t]
\centering
\includegraphics[width=\linewidth]{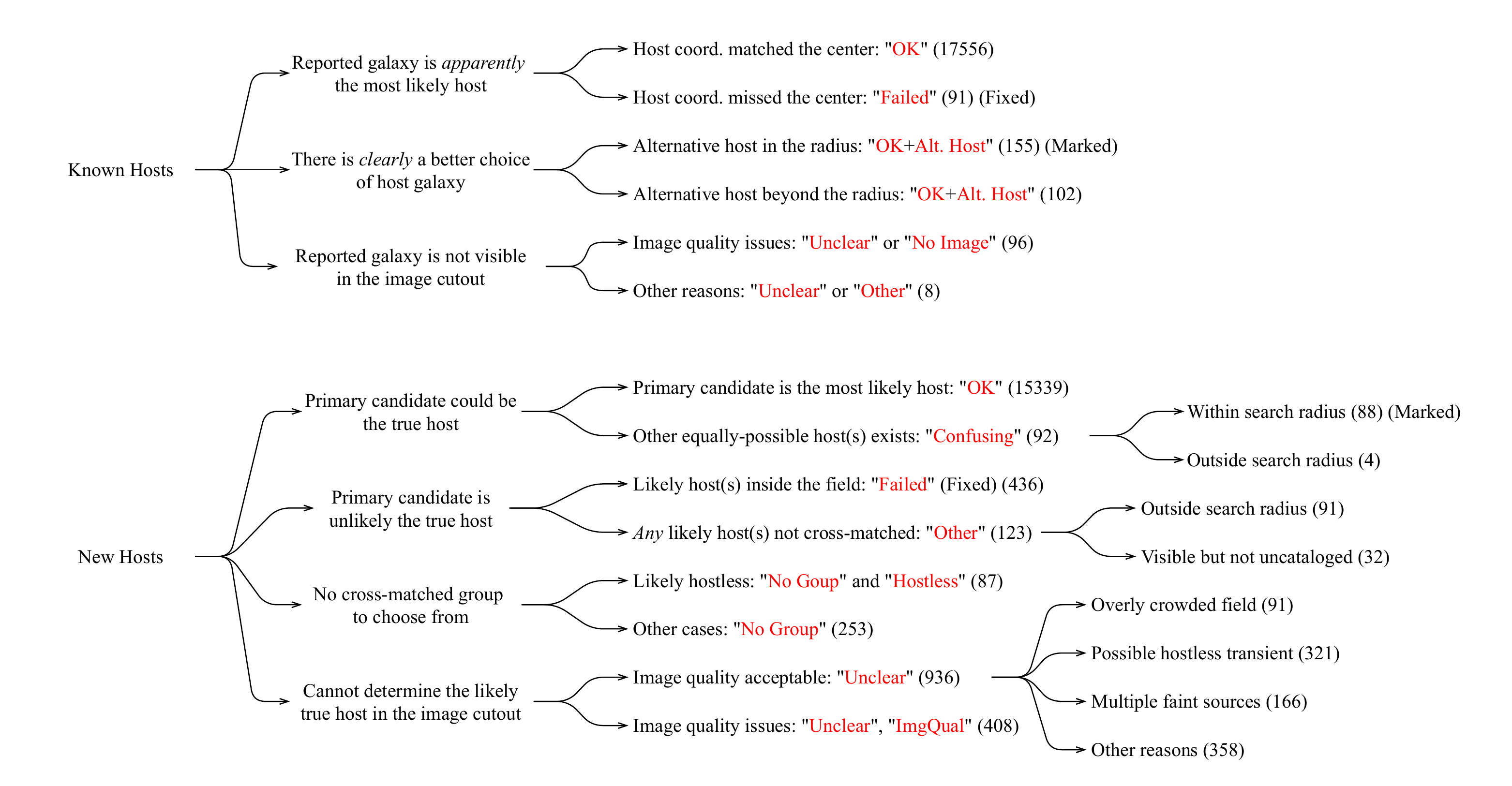}
\caption{{The decision tree of visual inspection for transients with known and properly cross-matched hosts (Appendix \ref{appendix:inspectionknown}) and transients with newly identified hosts (Appendix \ref{appendix:inspectionnew}).
For transients with known hosts, we assign ``OK'' or ``Failed'' based on the results of cross-matching.
If there is clearly a better host than the reported one, we also flag the case and mark the ``alternative host'' in the image.
If we cannot determine if the indicated host is successfully cross-matched and if any better host exists, we consider this an ``Unclear'' case and set flags for the detailed reasons.
For events with new hosts, if the algorithm chose the unique best host in the images, we consider the case ``OK.'' If the algorithm chose one likely host while there are other equally possible hosts, we consider this a ``Confusing'' case.
If the algorithm missed the best host in the radius, we consider the case a ``Failed'' one and then manually reassign the host if we can. However, if the best host is outside the radius, we group the case under ``Other.''
If no group has been cross-matched near the event, the case falls into ``No Group.'' Finally, similar to the case of known hosts, if we cannot determine the best host from the image, we consider the case ``Unclear'' with appropriate flags indicating the detailed reason.}
\label{fig:visinspworkflow}}
\end{sidewaysfigure*}

{We conducted the visual inspection separately for the samples of known hosts and new hosts. We summarize the workflow in Figure \ref{fig:visinspworkflow}. Below we describe the detailed procedure and the results.}

\subsection{Inspection of known hosts}
\label{appendix:inspectionknown}

{We group cross-identified hosts with known coordinates into three cases: ``OK,'' ``Failed,'' or ``Unclear.'' We focus on whether the known host coordinate successfully matches the galaxy it indicates.}

{First, we check the image cutout and cross-matched groups to find the galaxy indicated by the host coordinate. We consider the case ``OK'' if the known host coordinate matched the core of the galaxy or the most prominent ``main component'' of an irregular galaxy.
If the known host coordinate failed to match the core or the main component of the galaxy, we consider this a ``Failed'' case. Such a situation occurs when the reported coordinate has large offsets from the actual central component cross-matched from multiple external catalogs (e.g., irregular galaxies, nearby large galaxies).
We then mark the correct group manually. Host properties, in this case, will be updated with our manual correction.}

{Generally, we do not judge the correctness of reported hosts during this inspection. When there are more than one equally possible host as we see and the input host coordinate points to any of them, we accept the galaxy as the true host.
Occasionally, the known host coordinate points to a galaxy, but there is an \textit{obviously} better choice of the host, not just equally possible, from our visual inspection. We consider the case ``OK'' if the known host coordinate matched the galaxy it indicates, but at the same time, we set an ``alternative host'' flag to indicate that we prefer another host.
We manually mark the preferred host when it has been cross-matched in external catalogs. If the preferred host lies beyond the search radius and is therefore not cross-matched, we set an ``alternative host beyond radius'' flag to indicate the situation and then mark the position of the alternative host (RA, Dec offsets from mouse click) if its core is visible.}

{In some cases, we cannot distinguish the indicated galaxy in the image, and hence the robustness of cross-matching and the existence (or not) of any better host. We consider these cases as ``Unclear.'' Fields can be marked ``Unclear'' for various reasons. Still, image sensitivity or resolution issues, for which we set the ``image quality issue'' flag, is a primary cause. High-redshift hosts outside the coverage of DESI LS and PS1 sometimes fall in this category.}

{Furthermore, if the transient position is likely inaccurate, due to either equinox conversion issue of historical coordinates, or possible mistake in upstream data sources, we set an additional ``possible error in metadata'' flag.
If matched by name-resolved coordinates, their host properties remain usable, but transient-host offsets could be wrong.
Other uncommon cases are grouped under ``Other'' with appropriate descriptive flags.}

\subsection{Inspection of newly identified hosts}
\label{appendix:inspectionnew}

{Newly identified hosts are grouped into four major cases: “OK,” “Failed,” “Confusing,” or “Unclear.” Cases including ``Other'' and ``No Group'' are also used when applicable.}

{First, we visually identify the galaxy that appears to be the most likely host in the field. Then we reveal the host identified by the default ranking function. Revealing the choice of the algorithm later may reduce the confirmation bias in this inspection. If the default ranking function chose the most likely host as we see in the image, we consider the case “OK.” Similar to the inspection of known hosts, we require the center or the ``main component'' of the galaxy to be selected. Otherwise, this would be considered as a ``Failed'' case, with manual reassignment of the host.}

{Occasionally, there could be more than one visually identified, equally-possible host (for example, the transient is in a group or cluster environment but not clearly associated with a specific member), and the default ranking function chose \textit{any} of them, we consider the case a “Confusing” one and then manually mark other possible hosts.
If any possible host is beyond the search radius and is therefore not cross-matched into a group, we set a ``host beyond radius'' flag and then mark the galaxy (RA, Dec offsets) in the image when possible.}

{If the ranking function missed the most likely host, or rarely, all possible hosts when there are multiple, we consider this a ``Failed'' case and then manually mark the most likely host(s). This also includes the situation in which the ranking function chose some substructure of the most likely host rather than the main component.
However, if the most likely host lies beyond the search radius, or if it is marginally visible but not cataloged and cross-matched, we group the case into ``Other'' instead of ``Failed,'' because this is a failure in the source accessing and cross-matching process, rather than the ranking function itself.}

{When there is no cross-matched group for the ranking function to choose from, a possible situation for high-redshift transients such as GRBs, we mark the case with ``No Group.'' If the sensitivity of the background image is fairly good (DESI LS image, or PS1 image without stacking artifacts) and there is no possible host near the transient position, we further flag the case as ``hostless.''}

{For any reason, if the most likely host is not clearly visible in the image, we consider the case as “Unclear.” This includes, but is not limited to the following situations as we flagged: 1) the transient is likely hostless (``hostless'') in images of fairly good sensitivity (see also \ref{sec:undetectedhost}); 2) the field is overly crowded (``crowded,'' e.g., low-galactic latitude regions), hindering the visual identification of the host; 3) there are multiple faint sources, likely background galaxies, in the vicinity (``multiple faint sources''), in which no one appears to be a better host than others, yet possible association with the transient cannot be fully excluded; 4) low-resolution, low-sensitivity, or corrupted images (``image quality issue''). When possible, we also comment with the detailed reason if the case does not fit in these typical reasons.}

\subsection{Empirical accuracy from visual inspection}

\begin{figure*}
\centering
\includegraphics[width=0.45\linewidth]{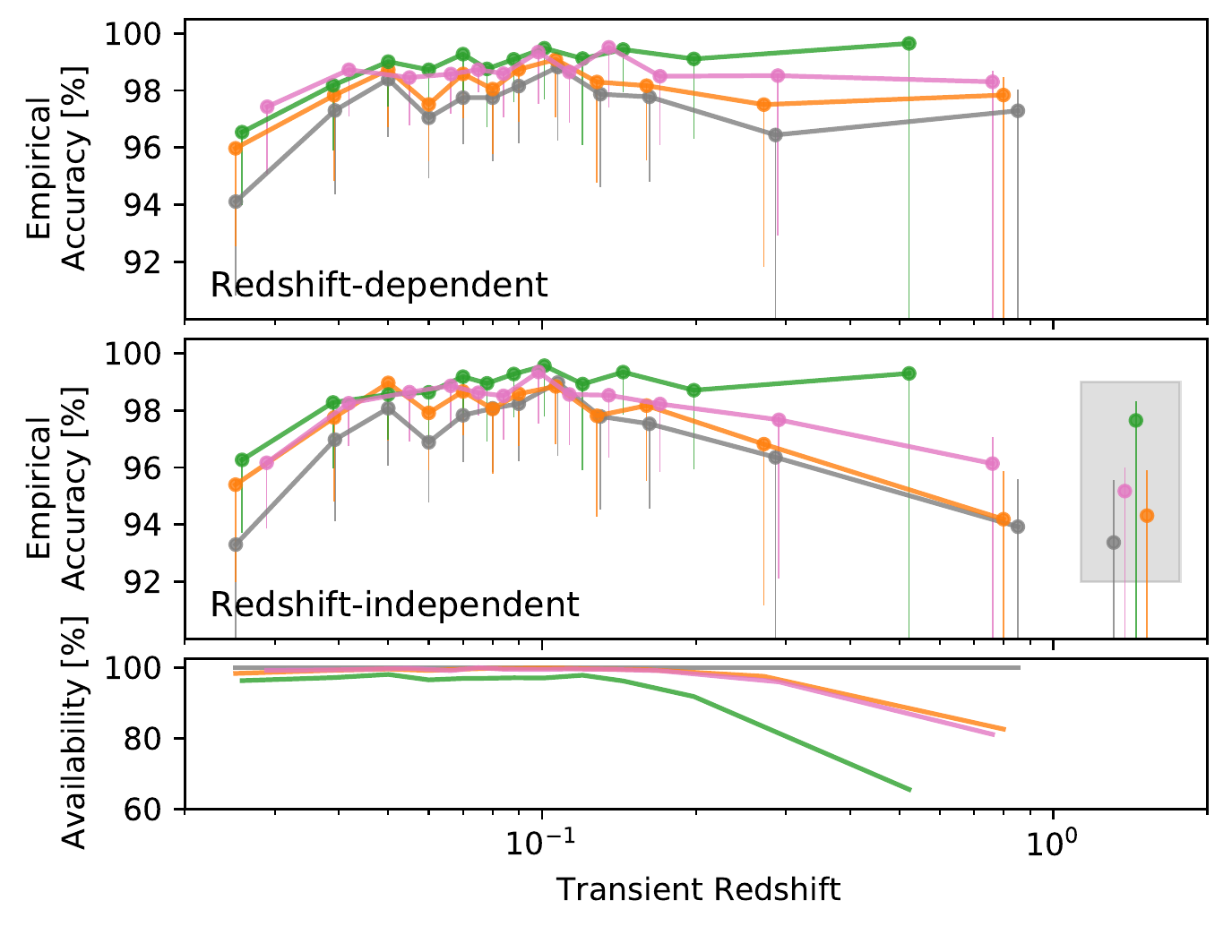}
\includegraphics[width=0.45\linewidth]{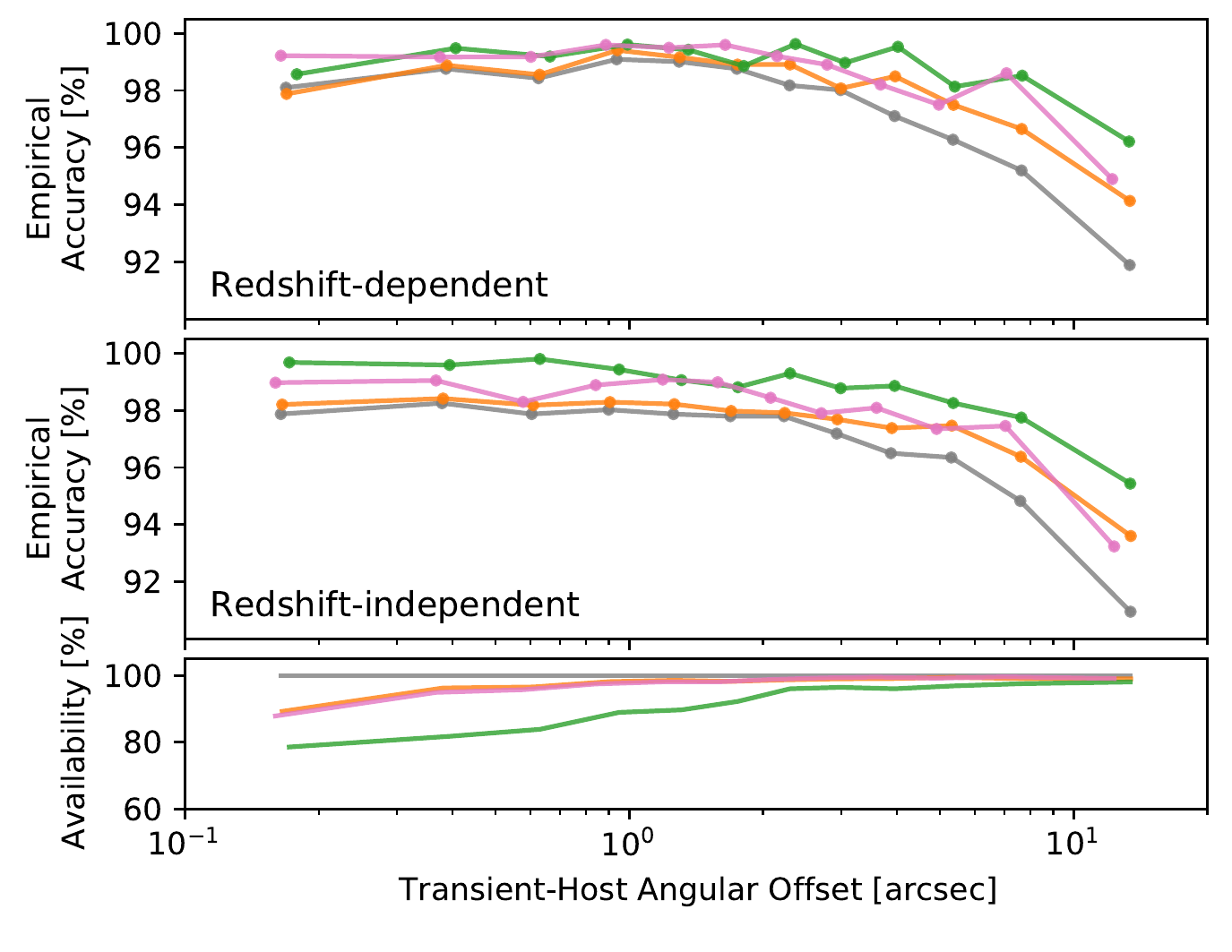} \\
\includegraphics[width=0.45\linewidth]{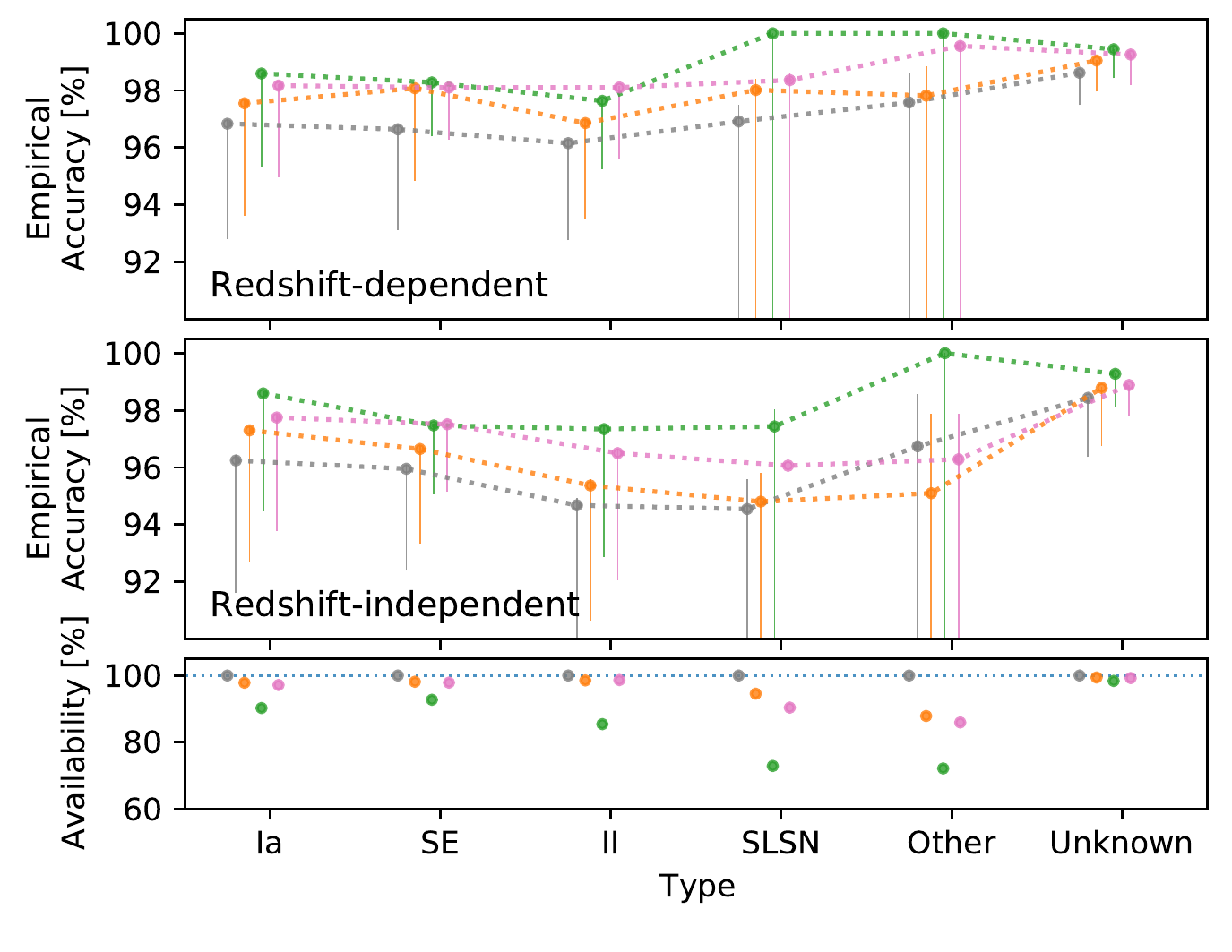} 
\includegraphics[width=0.45\linewidth]{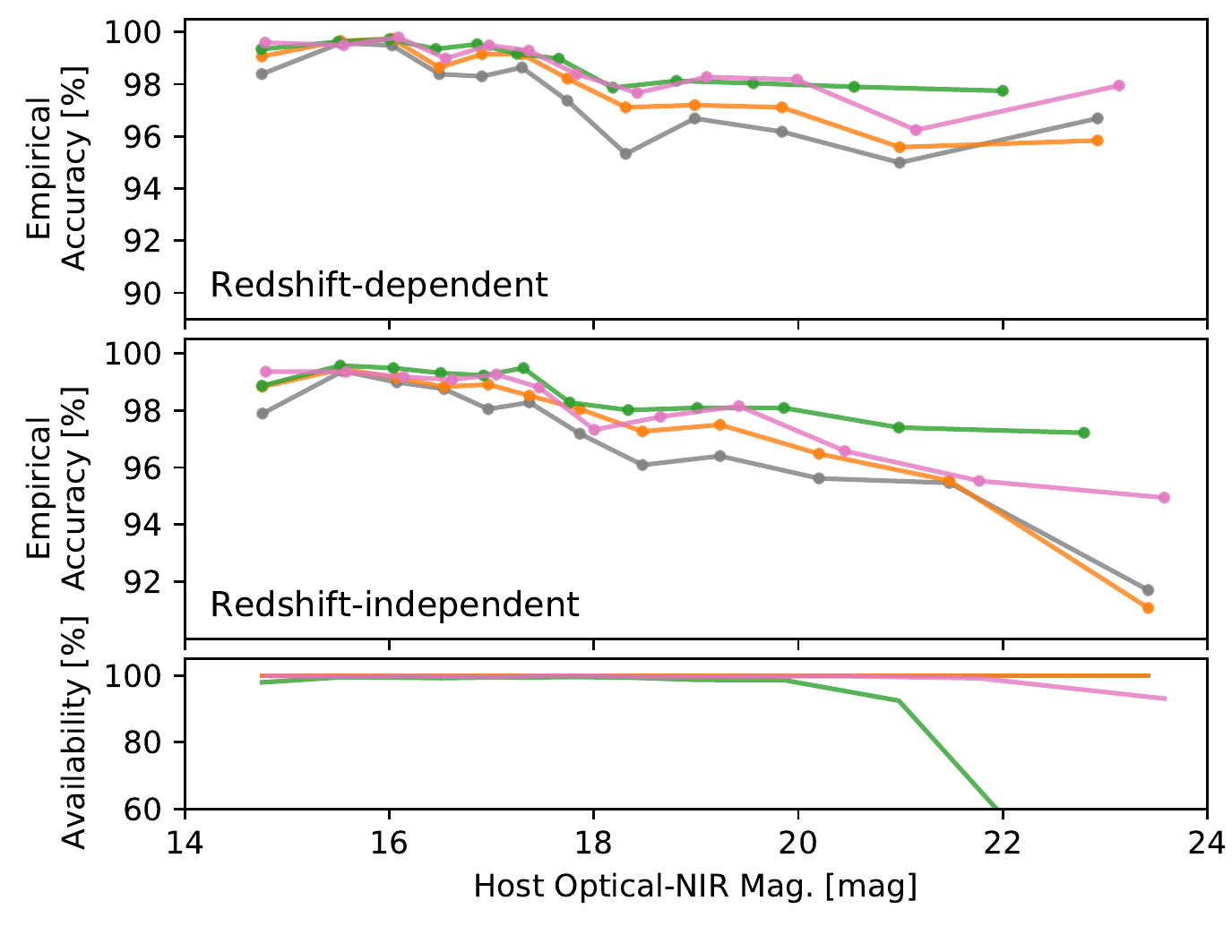} \\
\includegraphics[width=0.90\linewidth]{fig-vis-accuracy-legend.pdf}
\caption{
{The empirical accuracy of new hosts from visual inspection. Here the accuracy is the fraction of correctly identified hosts in cases where the host, either correctly identified or not, is clearly visible and unambiguous.
We show the dependence of accuracy on transient redshift, transient-host angular distance, transient type, and host optical-NIR magnitude in four separate panels. Except for the panel of transient type, we divide the sample into bins of equal sample size over the axis of interest and plot the per-bin accuracy and mean value there.
Each colored accuracy curve represents the performance of an input feature set using the default Logistic Regression classifier. Redshift-dependent and redshift-independent feature sets are grouped into two sub-panels for clarity.
We calculate the accuracy for feature sets using detailed source properties with the subset of events whose best hosts (either correctly identified or not) have the catalogs cross-matched.
Since these properties are not universally available, we further plot the fraction of best hosts with the required features available in the bottom sub-panels to indicate the possible loss of accuracy.
Furthermore, as transient type and redshift are independent of the best host, we also estimated the uncertainty of accuracy contributed by indecisive cases (``Unclear'' and ``Confusing'' cases). The optimistic and pessimistic limits of accuracy are indicated as ``error bars'' of these points.
Similar as Figure \ref{fig:ridgeaccuracy}, for events without redshift, we plot their accuracy in the rightmost gray shaded area of the redshift panel.}
\label{fig:visxidaccuracy}}
\end{figure*}

{We summarize the results of visual inspection in Table \ref{tab:empiricalaccuracy}, Figure \ref{fig:ridgeaccuracy} and Figure \ref{fig:visxidaccuracy}.}

{We use \textit{empirical} accuracy (Section \ref{sec:empiricalaccuracy} to evaluate the performance of ranking functions in new hosts. The empirical accuracy here refers to the fraction of clearly visible and unambiguous hosts that the ranking function has successfully recovered. Assuming that the best host is inside the search radius, our default ranking functions achieve $97.4\%$ when redshift-dependent parameters are used for ranking, and $96.8\%$ when only redshift-independent parameters are used (Table \ref{tab:empiricalaccuracy}).}

{There are a considerable fraction of indecisive cases (``Unclear,'' ``Confusing'') in which the best host is unidentifiable, ambiguous, or even absent. The empirical accuracy must take this potential source of uncertainty into account.
We quote the range of empirical accuracy by making assumptions about the performance of ranking functions in these indecisive cases. Assuming that these indecisive cases are all successfully identified, we may estimate an optimistic upper limit of accuracy; on the other hand, in the improbable worst situation where all their hosts are misidentified, we can estimate a pessimistic lower limit of accuracy.
The upper and lower limits of accuracy are also listed in Table \ref{tab:empiricalaccuracy}.
Considering the uncertainty due to these indecisive cases, the accuracy of the redshift-dependent default ranking function ranges from an optimistic case of $97.5\%$ to an unlikely pessimistic case of $92.5\%$. For the redshift-independent default ranking function, the limits are $97.0\%$ and $88.9\%$, respectively. These indecisive cases are more relevant to high-redshift events (Figure \ref{fig:ridgeaccuracy}, Figure \ref{fig:visxidaccuracy}) and are less of a concern for mid and low-redshift transients.}

{We checked the accuracy of the default ranking functions, while the accuracy of other ranking functions can be derived once the visually-identified most likely host is known in each field.
We noticed that the inclusion of detailed source properties in PS1 or DESI LS catalogs could clearly improve the results, as we pointed out earlier in Section \ref{appendix:rankingfunctionaccuracy}. The improved accuracy has similar dependence on key transient and host parameters as the default ranking functions due to the overlap in their input parameters or training sets.
Any percent-level improvement of accuracy in this range implies a significant reduction of misassociation rates compared to the default ranking functions. However, the improvement comes with the cost that the algorithm could ignore some true hosts if they do not have the required parameters measured.
For example, \texttt{PS1}/\texttt{PS1\_z} feature sets, which rely on source properties in PS1 catalog, have close to $99\%$ accuracy. However, the sensitivity of the PS1 catalog may not detect some fainter hosts, which leads to a decreased feature availability towards higher redshift. Higher redshift transients will likely not benefit from the improved accuracy here.}

\begin{deluxetable}{lcccc}
\tablecaption{Empirical Accuracy of Ranking functions \label{tab:empiricalaccuracy}}
\tablenum{10}
\tablehead{\colhead{Name} & \colhead{\texttt{Basic}(\texttt{\_z})} & \colhead{\texttt{AnyMag}(\texttt{\_z})} & \colhead{\texttt{PS1}(\texttt{\_z})} & \colhead{\texttt{LS}(\texttt{\_z})}}
\startdata
\rule{0pt}{4ex}    
         & \multicolumn{4}{c}{With redshift-dependent parameters} \\
\cline{2-5}
Accuracy    & $97.4$ & $98.0$ & $98.8$ & $98.6$ \\ 
Upper Limit & $97.5$ & $98.1$ & $98.9$ & $98.7$ \\ 
Lower Limit & $92.5$ & $93.5$ & $95.6$ & $95.0$ \\ 
\rule{0pt}{4ex}    
         & \multicolumn{4}{c}{Only redshift-independent parameters} \\
\cline{2-5}
Accuracy    & $96.8$ & $97.5$ & $98.7$ & $97.9$ \\ 
Upper Limit & $97.0$ & $97.6$ & $98.7$ & $98.0$ \\ 
Lower Limit & $88.9$ & $90.7$ & $93.9$ & $93.2$ \\ 
\enddata
\tablecomments{{The empirical accuracy is estimated for the default ranking functions (Appendix \ref{appendix:rankingfunction}). We use the fraction of clearly visible and unambiguous hosts that have been successfully recovered as the accuracy here.
The upper and lower limits of accuracy due to indecisive cases are also quoted.}}
\end{deluxetable}

\onecolumngrid \clearpage
\twocolumngrid

\section{Confidence Scores of Host Candidates} \label{appendix:confidencescore}

{To evaluate the reliability of individual newly identified hosts, besides visual inspection, we also derived a separate set of numeric metrics for each cross-matched group, which we refer to as the ``confidence score.''}

{The confidence score is a metric for outliers or anomalies, which distinguish likely host galaxies from the bulk of nearby non-host objects by comparing their ranking scores. We create a reference sample of nearby non-host objects by generating 16384 mock transients with randomized positions (i.e., without association with any galaxies).
Galaxies and stars near these mock transients are always nearby non-host objects rather than their ``true hosts.'' The highest-ranking ``host candidates'' of these mock transients then outline the possible range of ranking scores, represented as a baseline distribution, that a nearby non-host object may reach by chance under the best cases.
For a true transient, if the ranking score of a host candidate is at the higher value tail of the baseline distribution, then this candidate is likely an outlier among nearby non-host objects, and we can be more confident about this new candidate. Conversely, if the ranking score of this host candidate is inside or even below the range of the baseline distribution, then this candidate could be a nearby non-host object that ranked high in the field simply by chance.
When no host candidate in a field achieves an outstanding ranking score compared to the baseline distribution, the transient itself could be a hostless one.}

{We use the same source searching and cross-matching workflow for these mock transients under a constant search radius of $45''$, where the ranking scores are calculated using the same trained ranking functions as we used for those true transients. For each feature set and classifier combination, we construct the baseline distribution using the highest-ranking score of each mock transient. The confidence score is defined as the percentile rank of a new ranking score in the baseline distribution of the same feature set and classifier. A value closer to $100\%$ indicates a more outstanding and possibly more reliable candidate. Clearly, if the ranking scores depend on transient redshift and catalog coverage, then the baseline distributions should also be adjusted for these factors.
To compensate the redshift dependence of some feature sets (with ``\texttt{\_z}'' suffix), we generate a series of baseline distributions by setting these mock transients at some fixed redshifts on a pre-defined axis so that we can choose the nearest baseline distribution on the axis later.
To ensure similar catalog coverage of mock transients as our existing true transients, we generate mock transients following the density of true transients on the celestial sphere, represented in a \texttt{healpix} grid smoothed by a Gaussian kernel of FWHM$\sim$1 degree. Furthermore, we also calculate local confidence scores using only the nearest 1024 mock transients, rather than the entire sample of 16384 mock transients, for baseline distributions. Despite being noisy due to smaller sample sizes, this localized version of confidence scores may better offset the dependence of baseline distributions on the coverage of external catalogs.}

\begin{figure}
\centering
\includegraphics[width=\linewidth]{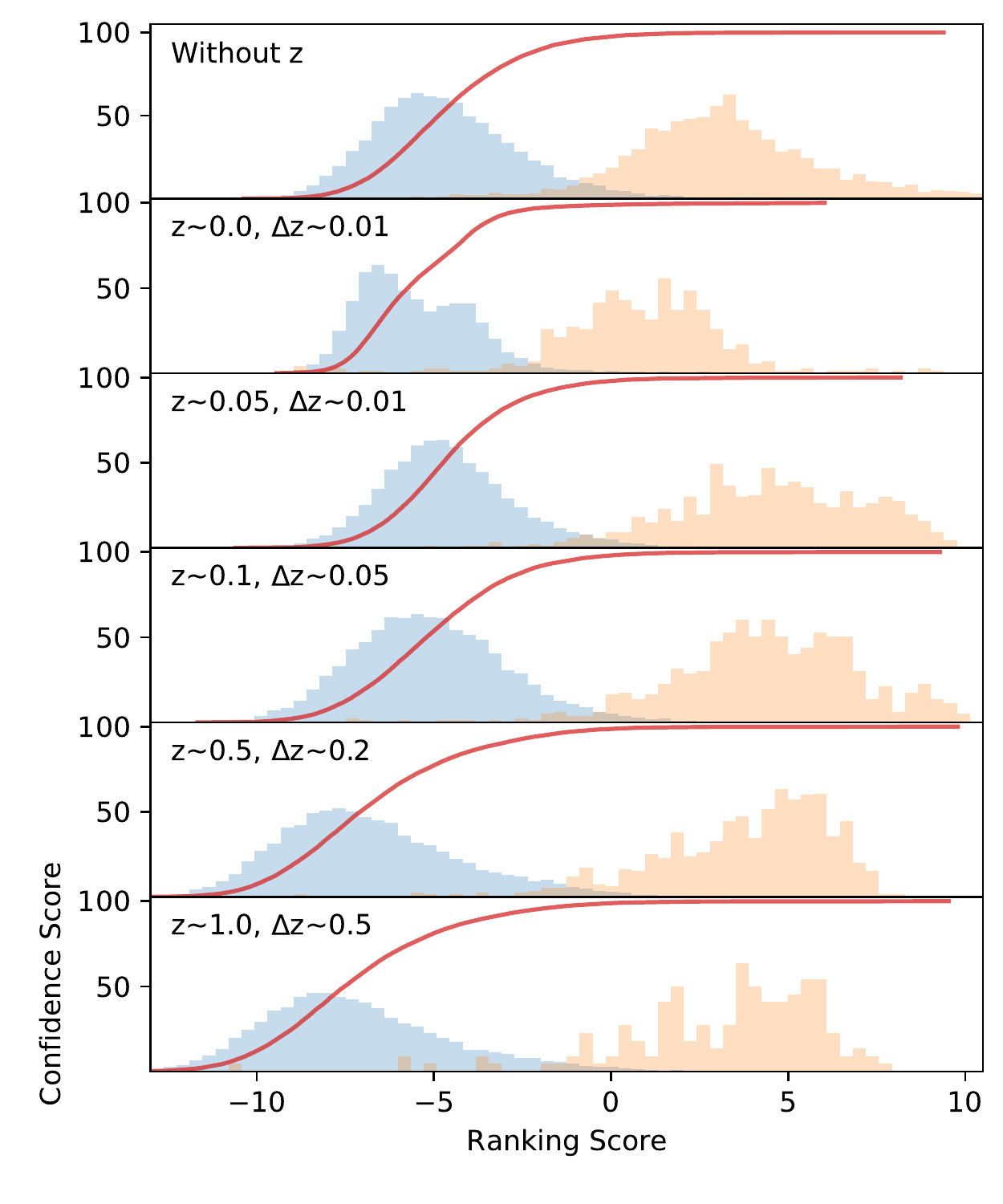} \\
\includegraphics[width=\linewidth]{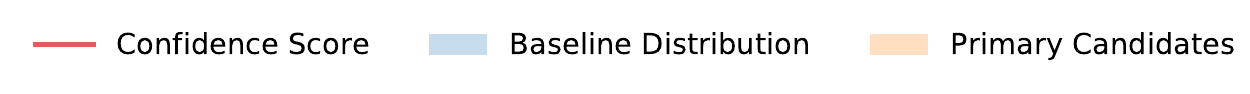}
\caption{{Examples of estimating confidence scores from baseline distributions.
We compare the ranking scores of new host candidates (orange) to a baseline distribution (blue), i.e., the highest-ranking scores that nearby \textit{non-host} objects can achieve by chance in a randomized mock transient sample.
We use the cumulative percentage curve of the baseline distribution (red) to quote the ranking scores of new host candidates.
True hosts, ideally, should have ranking scores at the positive tail of the baseline distribution. Their confidence scores are closer to $100\%$.
Here we show the case for the redshift-independent default ranking function (top panel) and the redshift-dependent default ranking function in several redshift intervals (centered at $z$, with bin width $\Delta z$).} \label{fig:confidencescore}}
\end{figure}

\begin{figure}
\centering
\includegraphics[width=\linewidth]{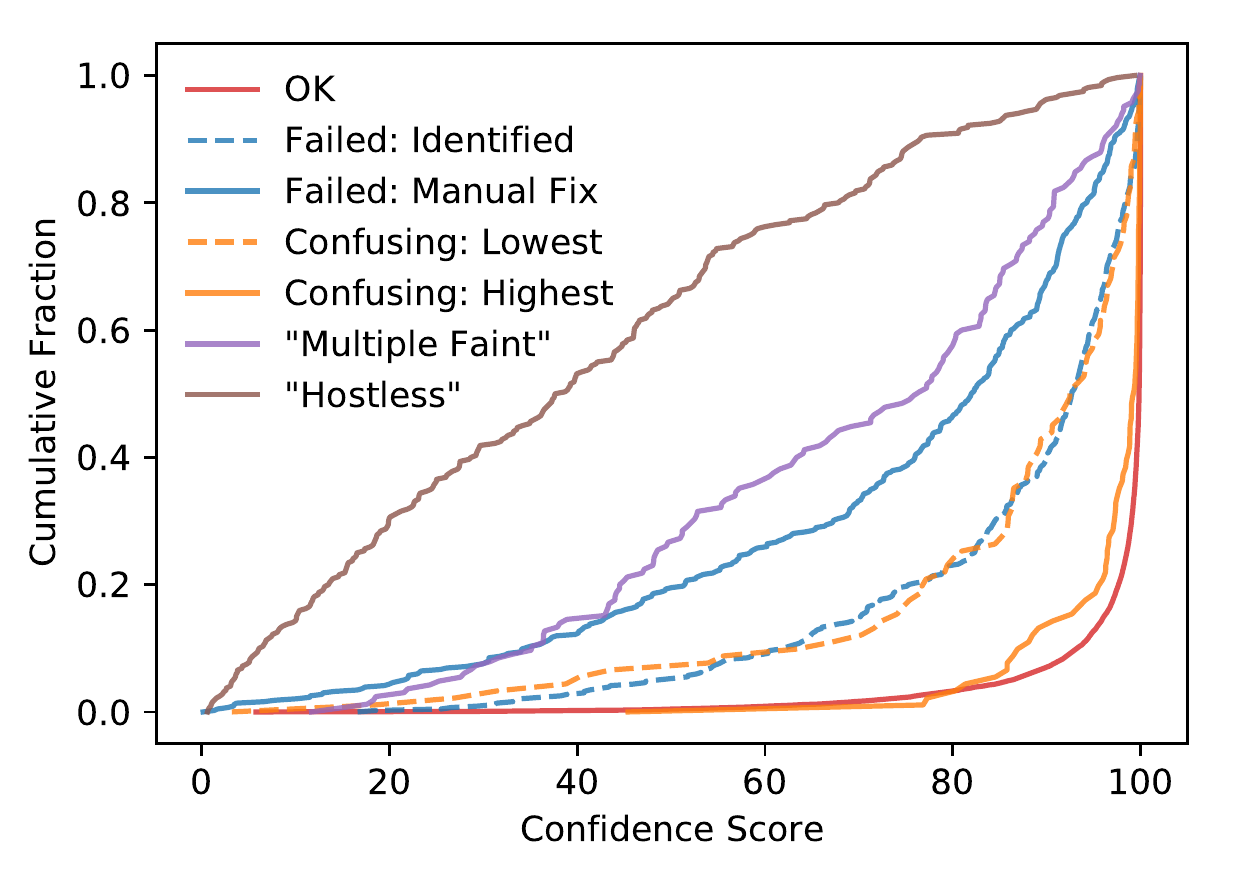}
\caption{
{The cumulative distribution of confidence scores for some host candidates identified during our visual inspection of new hosts (Appendix \ref{appendix:inspection}). Successfully identified hosts (``OK'') usually have very high confidence scores. However, higher confidence scores do not guarantee more reliable hosts; misidentified hosts (``Failed: Identified'') still have higher confidence scores than the manually reassigned hosts in the same field (``Failed: Manual Fix''). 
When there are multiple possible hosts, the other equally-possible host (``Confusing: Lowest'') also has high confidence scores, although not as high as the ``best'' host selected by the algorithm in the same field (``Confusing: Highest'').
We see much lower confidence scores for the selected ``host'' when the vicinity contains multiple faint sources, yet no one stands out to be a clear and unambiguous host (``Multiple Faint'').
Finally, the ``best candidate'' of possible hostless transients (``Hostless'') have much lower confidence scores than other candidates. They basically follow the baseline distribution.
Results here are global confidence scores using the default ranking function in each field.}
\label{fig:confidencescore2}}
\end{figure}

{In Figure \ref{fig:confidencescore}, we show examples of baseline distributions, ranking scores of newly identified hosts, and their derived confidence scores with our default ranking functions. For the redshift-dependent ranking function, we also show the results in several redshift bins.
The baseline distributions are clearly redshift-dependent. Towards higher redshift, although angular distance-related parameters may still distinguish true hosts from other objects, other properties of hosts become similar to those ubiquitous galaxies in the field.
However, the ranking scores of true hosts remain well above the ranges of baseline distributions, and the behavior of confidence scores remains stable.
In Figure \ref{fig:confidencescore2}, we show the cumulative distributions of confidence scores for a few kinds of host candidates identified during our visual inspection. Successfully identified new hosts (``OK''), as expected, have confidence scores close to $100\%$.
Misidentified hosts (``Failed: Identified''), as a comparison, have lower confidence scores than successful cases.
Higher confidence scores usually, but not always, mean better candidates. The manually reassigned hosts (``Failed: Manual Fix'') of ``Failed'' cases have lower confidence scores than galaxies misidentified in the same field (``Failed: Identified'').
Similarly, in ``Confusing'' cases, the equally-possible hosts (``Confusing: Lowest'') have lower confidence scores than ones chosen by the ranking functions in the same field (``Confusing: Highest'').
There are ``Unclear'' cases with multiple faint sources in the vicinity of transient coordinates, in which no one appears to be a clearly visible and unambiguous host, while any transient-host association cannot be fully excluded. Their highest-ranking ``hosts'' (``Multiple Faint'') have much lower confidence scores, although the distribution overlaps with manually reassigned hosts (``Failed: Manual Fix'').
Finally, the primary candidates of possible hostless transients (``Hostless'') have very low confidence scores, following the baseline distributions.}

{As a summary, the confidence score here serves as a quantitative metric for the reliability of host candidates, which clearly differentiates robust hosts from those less reliable ones. However, the confidence score is based on the same training framework as our ranking algorithm, which does \textit{not} directly indicate the correctness of individual host identifications. Cross-checking with visual inspection is encouraged.}

\section{Database Schema} \label{appendix:dataschema}

Here we show the schema of host collection using an example with comments. The output is abbreviated for clarity.

\onecolumngrid 
\pythonstyle
\begin{lstlisting}
# transient info
"event_name":   "SN2011fe",                     # preferred name
"event_alias":  ["SN2011fe", "PTF11kly", ...],  # other alternative names
"event_ra":     210.7737,                       # RA of event (degree)
"event_dec":    54.2736,                        # Dec of event (degree)
"event_z":      0.0008,                         # reported transient or host redshift
"event_dxy":    [-59.8553, -270.9878],          # event position in the field center (arcsec)
"revised_type": ["_SN", "Ia", ...],             # curated type labels

# as-reported host info
"reported_host": {
    "name":     ["M101", "NGC 5457"],           # reported host name(s)
    "ra":       210.8042,                       # reported host RA (degree)
    "dec":      54.3492,                        # reported host Dec (degree)
},

# summary of host name resolving,
"resolved_host_coord" : {
    "valid_names" : ["M101", "NGC 5457"],       # list of valid host names
    "resolved_names" : {                        # resolved host coord. of valid names
        # ** key: input host name; value: results of name resolving
        "M101" : {
            # selected coord. for this name
            "vac" : "NED",                      # selected reference for host coord
            "ra" : "14:03:12.5448",             # resolved RA
            "dec" : "+54:20:56.22",             # resolved Dec
            "ra_deg" : 210.8022,                # resolved RA (degree)
            "dec_deg" : 54.3489                 # resolved Dec (degree)
            "src" : "MESSIER 101/PREC:1.20e-01,1.20e-01", # resolved name and coord. precision

            "SIMBAD" : {                        # results from SIMBAD
                "src" : "M 101/QUAL:C", "ra" : "14:03:12.583", "dec" : "+54:20:55.50",
                "ra_deg" : 210.8024, "dec_deg" : 54.3487,  "raw_rec" : [...] # raw results in SIMBAD
            },
            "NED" : {...},                      # results from NED
        },
        "NGC 5457": {...},
        ...
    }

    # consistency check of name-resolved coord.
    "resolved_coord_inconsistent" : False,  # flag for any inconsistent pairs
    "inconsistent_pairs" : [...],           # list of inconsistent pairs, if any
    # ** list of tuple: (name, the other host name, distance in arcsecond)

    # best name-resolved coord., same structure as values under "resolved_names"
    "resolved_coord_best" : {
        "vac" : "...", "ra" : "14:03:12.5448", "dec" : "+54:20:56.22",
        "ra_deg" : 210.8022,   "dec_deg" : 54.3489, "src" : "MESSIER 101/PREC:1.20e-01,1.20e-01",
        "SIMBAD" : {...}, "NED" : {...},
    }
}

# flags for input host coord.
"host_coord_known" : True,              # host coord known or not
"host_coord_matched" : True,            # host coord cross-matched or not
"host_coord_type" : "name_resolved",    # type of host coord
# "name_resolved" or "reported_host_coord"

# queried coordinates and projection
"field_coord": {
    "ra":   210.8022,                           # field center RA (degree)
    "dec":  54.3489,                            # field center Dec (degree)
    # field centers at the queried coordinate (host or transient).

    "coord_src": "name_resolved",               # source of queried coordinate
    # "name_resolved", "reported_host_coord", "event_coord"

    "coord_comm" : ["NED", "MESSIER 101", ...], # flags for provided coordinates

    "radius": 15.0,                             # search radius (arcsec
    "radius_src" : "name_resolved_default",     # source of search radius
    # "name_resolved_default", "reported_host_coord_default",
    # "redshift", "grb_err_circle", "event_coord_default"

    # unit vectors in Cartesian frame
    "cvec":   [-0.5006, -0.2984, 0.8125],       # queried coordinate, or field center
    "vn_ra":  [ 0.5120, -0.8589, 0.0000],       # unit vector in RA direction
    "vn_dec": [ 0.6979,  0.4161, 0.5828],       # unit vector in Dec direction
},

# cross-matched groups.
"N_groups": 2,                                  # Number of cross-matched groups
"groups": [
    # ** list of dictionaries, one for each cross-matched group
    # confirmed host or primary candidate comes the first, followed by other groups
    {
        # ** key: catalog name or ancillary info (begin with underscore)
        "TwoMASSXSC": {                 # sources matched in 2MASS XSC catalog (example)
            "srcs": [
                # ** list of dictionaries, one for each source
                # the representative comes the first, followed by others, if any
                {
                    # ** key: name of source property; value: measured value
                    "2MASX": "14031258+5420555"
                    "J.ext": 6.517,
                    "H.ext": 5.805,
                    ...
                },
                ...
            ],
            "_confusion": False         # flag for confusion in this catalog
        },
        "MPAJHU": {                     # sources matched in MPA-JHU catalog (example)
            "srcs": [{...}, ...],       # detailed source properties
            "_confusion": False,
        },
        ... # other catalogs, and cross-matched sources

        "_xid_flag": "confirmed",               # type of cross-matched group
        # "confirmed", "primary", "secondary", "event", "other"

        # summary of cross-matched sources
        "_group_uid" : "2ebb53f01d24345f...",   # unique group id from cross-mateched sources
        "_includes_queried_coord" : True,       # queried coord cross-matched this group or not
        "_group_srcid" : {
            # ** key: catalog; value: list of source names
            "HyperLEDA" : ["NGC5457"],
            "TwoMASSXSC" : ["14031258+5420555"],
            ...
        }
        "_confusion": True,                     # flag for confusion in this group
        "_confusion_cats": ["LS8pz", "Gaia2"],  # list of catalogs with confusion

        # source positions
        "_avr_radec": [210.8023, 54.3489],      # mean RA/Dec of sources (degree)
        "_avr_dxy": [0.1064, -0.0938],          # mean offsets w.r.t. the field center (arcsec)
        "_std_dxy": [0.3033, 0.3943],           # std. dev. of offsets (arcsec)
        "_cov_dxy": [[0.0991, 0.0429], [0.0429, 0.1674]], # covariance of offsets (arcsec^2)

        # quality metrics for cross-matching
        "_shape_r": 0.3332,                 # Pearson's r
        "_shape_q": 0.6455,                 # q = sigma_b / sigma_a
        "_shape_p": 0.2154,                 # p = (1 - q) / (1 + q)
        "_shape_e": 0.7637,                 # "eccentricity"
        "_avr_dist": 0.4288,                # "average distance" or "mean offset" (arcsec)
        "_N_max_conn": 87.0,                # maximal possible number of connections
        "_N_conn": 82.0,                    # actual number of connections
        "_F_conn": 0.9318,                  # degree of connectivity

        # star-galaxy separation
        "_is_stellar": False,               # flag, any representative source is a star
        "_stellar_srcs": [                  # stellar sources
            # ** list of (catalog, index) pairs; index in the "srcs" list of each catalog.
            ["LS8n", 2], ...
        ],
        "_stellar_frac_all" : 0.0,          # fraction of sources marked as stars
        "_stellar_frac_repr" : 0.0,         # fraction of representative sources marked as stars
        # denominator does not include sources in catalogs without star-galaxy separation metrics

        # summary of host candidate ranking
        "_rank_score" : {                       # ranking scores
            # ** key: feature set or ancillar info.
            "Basic8Z" : {
                # ** key: classifier; value: results
                "Logistic_v3" : {
                    "score" : -0.6959,          # raw ranking score
                    "rank" : 0,                 # rank in the field
                    "cs_global" : 98.3947,      # global confidence score
                    "cs_local" : 98.3398,       # local confidence score
                },
                "LSVM_v3": {...}, "RF_v3": {...},   # other classifiers and results
                "_X" : [...],                       # raw input parameters
            }
            "LSZ" : {...}, "PS1Z": {...}, ...   # other feature sets and results

            "_default" : {                      # results from default ranking function
                "score" : -0.6959,              # raw ranking score
                "rank" : 0,                     # rank in the field
                "features" : "Basic8Z",         # feature set used
                "ranker" : "Logistic_v3",       # classifier used
                "cs_global" : 98.3947,          # global confidence score
                "cs_local" : 98.3398,           # local confidence score
            },
            "_cscore_local" : {                 # ancillary info. for local confidence score
                "avr_dist" : 11.5686,           # average distance to mock transients (degrees)
                "max_dist" : 18.8843,           # maximal distance to mock transients (degrees)
                "N" : 1024                      # number of mock transients used
            }
        },
        "_rank_features" : {                        # input parameters for ranking functions
            # ** key: group features; value: list of values
            "object_count" : [...], "angular_distance" : [...], "ls8_size_dRe" : [...], ...
        }
    }, ...
],

# coordinate of confirmed host or primary candidate, after cross-identification
"host_ra":          210.8024,                           # cross-matched host RA (degree)
"host_dec":         54.3487,                            # cross-matched host Dec (degree)
"host_dist":        84.6571,                            # distance to transient position (arcsec)
"host_dxy":         [0.1064, -0.0938],                  # position w.r.t. field center (arcsec)
"host_offset":      [59.8684, 270.7856],                # position w.r.t. transient (arcsec)
"host_srcid" : ["NED", "object_name", "MESSIER 101"],   # source id of cross-matched host
# ** (catalog, column name, source name)

# results from visual inspection, if applicable (copied from another event for example)
"vis_insp" : {
    # flags and comments
    "status" : "Other",             # status: "OK", "Failed", "Confusing", "Unclear", "Other"
    "flags" : ["H", ...],           #
    "comment" : "possible_host_beyond_radius prefer_alternative_host",
    # other flags and comments besides the "status" above

    # manual reassignment
    "reassigned" : {                # reassigned host, if within the radius
        # ** key: host name ("G0", "G1", "G2", ...)
        "G0": {
            # ** key: index of cross-matched group
            "1": [
                1,                  # index of selected group under "groups"
                [-5.4725, 9.2440],  # marked position in the image, RA Dec offset in arcsec
                [-6.0513, 8.9101]   # '_avr_dxy' of selected group
            ], ...
        },
        # each host may contain multiple cross-matched groups
    },
    "pin_pos" : [(-38.8769, -30.1731), ...],    # manually marked host(s), if outside the radius
    # ** list of (delta_RA, delta_Dec) pairs, w.r.t. field center, in arcseconds

    # summary of visual inspection
    "resolved_by" : "use_existing",             # source of compiled host data
    "case_kind" : ["known_host", "ok", ...],    # automatically generated descriptive flags
    "case_code" : "D1"                          # automatically generated case code
}

# cross-matching threshold
"xid_thres": {
    "N_cps": 3,                 # number of groups under optimal threshold,
    "thres_axis": [...],        # trial values of matching threshold,
    "conn_score": [...],        # number of valid pairs under each trial value
    "thres": 1.0                # the optimal matching threshold
},

# coverage of catalogs.
"coverage": {
    "field":   ["NED", "MPAJHU", ...],  # names of catalogs covering this field
    "primary": ["NED", "MPAJHU", ...],  # catalogs matched in the primary group
    "primary_unmatched": {              # catalogs not matched in the primary group
        # key: catalog name, value: list, [group_id, source_id, dist]
        "GALEXAIS67" : [
            1,      # zero-indexed id of the group including source in this catalog
            0,      # index of the source in that group
            2.9453  # distance of primary group to that source (arcsec)
        ],
        ... # ... other unmatched catalogs
    }
}

# excluded sources in the field
"excluded_sources": {
    # ** key: catalog or ancillary info (begin with underscore)
    "PS1v2": {                          # excluded sources in PS1 DR2
        "N_src": 3,                     # N of sources in this catalog
        "N_excl": 3,                    # N excluded
        "excl_srcs": [                  # info of excluded source,
            # ** list of dictionaries, one for each excluded source
            {
                "src": {...},           # source info
                "reason": "FILTER FUNC" # reason, "FILTER FUNC" or "BEYOND RADIUS"
            }, ...
        ]
    }, ...
    "_N_src": 21,                       # Number of sources in the field
    "_N_excl": 3,                       # N of excluded sources.
},

"last_update": "2021-03-06T02:04:36.303Z",  # time stamp for last update
"vcc": {                                    # version control code for each data source
    # ** key: data source ("tde", "grb", "rare"); value: version control code
    "sn": "d147bb8861a7e838..."
},
"rand_id": 0.1157,                          # random number within (0, 1)
"_id": "vtdrakibitozjnpg",                  # global unique id of this event
\end{lstlisting}
\twocolumngrid

\bibliography{ms.bib}

\end{document}